\documentclass[iop]{emulateapj}
\usepackage{amsmath}
\usepackage{url}
\usepackage{natbib}
\usepackage[finalnew]{trackchanges}

\begin{document}

\title{An evaluation of cosmological models from expansion and growth of structure measurements}

\author{Zhongxu~Zhai\altaffilmark{1},
 Michael~Blanton\altaffilmark{1},
An\v{z}e~Slosar\altaffilmark{2},
Jeremy~Tinker\altaffilmark{1}
 }
 \email{Email: zz681@nyu.edu}
\altaffiltext{1}{Center for Cosmology and Particle Physics, Department of Physics, New York University, 4 Washington Place, New York, NY 10003, USA.}
\altaffiltext{2}{Brookhaven National Laboratory, Upton, NY 11973, USA}

\begin{abstract}
We compare a large suite of theoretical cosmological models to
observational data from the cosmic microwave background, baryon
acoustic oscillation measurements of expansion, Type Ia SNe 
measurements of expansion, redshift space distortion measurements
of the growth of structure, and the local Hubble constant. 
Our theoretical models include parametrizations of 
dark energy as well as physical models of dark energy and 
modified gravity. We determine the constraints on the 
model parameters, incorporating the redshift space distortion
data directly in the analysis. To determine whether models can
be ruled out, we evaluate the $p$-value (the probability 
under the model of obtaining data as bad or worse than the 
observed data). In our comparison, we find the well known 
tension of H$_0$ with the other data; no model resolves this
tension successfully. Among the models we consider, the large 
scale growth of structure data does not affect the modified gravity 
models as  a category particularly differently than dark 
energy models; it matters for some modified gravity models but
not others, and the same is true for dark energy models.
We compute predicted observables for each model under
current observational constraints, and identify models 
for which future observational constraints will be particularly
informative.
\end{abstract}


\section{Introduction}

Since it was first definitively discovered \citep{Riess_1998, Perlmutter_1999}, 
the accelerated expansion of the universe has been one of the greatest 
mysteries in modern physics. The attempt to explain this phenomena has 
motivated physicists to propose numerous theoretical models. 
Among these candidates, the cosmological constant is the 
mathematically simplest solution. Although the cosmological constant 
is able to fit the observations well, it has a critical problem: 
that its value indicated from cosmology differs from that estimated 
by quantum field theory by about $\sim120$ orders of magnitude 
(\citealt{Weinberg_1989}). 

Alternative approaches to explain cosmic acceleration without the 
cosmological constant include a number that introduce an exotic 
energy component in the universe, the so-called dark energy. 
Proposed models include dynamical vacuum energy, cosmic fluids, 
scalar fields and others, as we explain in detail below. Other 
approaches alter the geometrical structure of spacetime, i.e. 
they modify gravity. These models, some of them motivated by 
string theory or generalizations of General Relativity, 
introduce modifications of late time evolution of the universe, 
such as $f(R)$ gravity, $f(T)$ gravity, extra dimensions, 
Galileon cosmology, and a number of others described below. In this work, we focus on the models which are numerically economic and do not explicitly assume an exotic matter energy component. Recent reviews outline the theoretical motivation for this zoo of 
cosmological models (\citealt{Peebles_2003, Copeland_2006, Linder_2008, Silverstri_2009, Caldwell_2009, Weinberg_2013, Joyce_2015}).
Confronting this large family of theoretical scenarios, it is important to 
compare their predictions with observations of the expansion rate, large 
scale structure, and the cosmic microwave background, to determine whether 
the theories are viable and to constrain their parameters. Such comparisons 
are standard practice for both theorists and experimentalists, and well-established
machinery exists to perform them.

Our work builds on these efforts in two major ways. First, the 
disparate investigations regarding the current models, conducted over a 
number of years and by many different groups, inevitably have used different 
data sets with different assumptions. In this work, we apply
a consistent set of constraints to a large set of theories, to better 
understand them when analyzed within the same framework of observational data. 

Second, while it is standard to constrain cosmological parameters with data, 
surprisingly not all investigations measure a ``goodness-of-fit'' of the models
they study --- that is, how statistically incompatible the data are with each 
model. Here we do so for all models in a consistent fashion. We use the $p$-value 
as a measure of goodness-of-fit 
(\citealt{Ronald_2016, Davis_2007}). The $p$-value is the probability within 
the model that one would observe data that, under some summary metric, 
is more different  from the model than the actual data is. In evaluating 
the $p$-value we will  account for the fact that we are  free to set 
the model parameters.   The particular metric we will use for 
the $p$-value is a global $\chi^2$  statistic.  This approach 
requires care in interpretation, as we describe  below, but does 
allow us to declare models whose $p$-values are 
intolerably low to be incompatible with the data.

We will use a combination of data to evaluate the models.
The first convincing evidence of the cosmic acceleration was from 
the observation of supernovae, soon followed by the measurement of 
degree-scale cosmic microwave background (CMB) anisotropies 
(\citealt{Hinshaw_2009, Planck_2013, Planck_2015}), and 
later by data on the baryon acoustic oscillation (BAO) at 
lower redshifts (\citealt{Sakharov_1966, Peebles_1970,Sunyaev_1970, Blake_2003, Seo_2003}). 
We combine the latest versions of these measurements with that of
the local Hubble constant H$_{0}$ (\citealt{Riess_2016}).
The combination of these measurements yields tight constraints 
on the parameters of the standard $\Lambda$CDM model, and some 
dark energy models. Under most such models, the  
local Hubble constant H$_{0}$ measurements are in tension
with the higher redshift constraints. 
The models that can resolve this tension are thus of importance 
to investigate, and this is one of the goals in our work. 

We also take advantage of the recent measurements of the linear 
growth of perturbations, which complements the above geometrical 
probes. This observation provides information on the dynamics of the cosmology which
might play an important role in dark energy model selection. 
Our work introduces a simple technique to incorporate these measurements 
of growth directly into  cosmological constraints in a manner 
compatible with the analysis of most modified gravity and dark 
energy models. We will use this technique to explore in detail the 
interplay of these different datasets, and evaluate the performance 
of different models given the data.

Our paper is organized as follows. In Section \ref{sec:intro_LCDM}, 
we introduce the dark energy and modified gravity models briefly, 
as well as the datasets we use. Section \ref{sec:pvalue} details 
our approach to evaluate the models by their $p$-values. Section 
\ref{sec:mcmc} presents the cosmological constraints for each 
model \add{with} the predictions of the geometrical 
and dynamical observable from current data, and compares with 
future experiments. We discuss and list our conclusions 
in Section \ref{sec:dis}.

\section{Cosmological models and data sets}\label{sec:intro_LCDM}

A homogeneous and isotropic universe is described by the Friedman-Robertson-Walker metric
\begin{equation}
ds^2 = -dt^2+a^2(t)\left[\frac{dr^2}{1-kr^2}+r^2d\Omega^2\right],
\end{equation}
where $a(t)$ is the scale factor related to redshift $z$ as $a(t)=(1+z)^{-1}$ and $k$ is the curvature parameter. In General Relativity (GR), the evolution of the universe is governed by the field equation which connects the Einstein tensor and energy momentum tensor
\begin{equation}
G_{\mu\nu}=8\pi G T_{\mu\nu},
\end{equation}
where $G$ is the Newton gravitational constant. This equation gives the expansion of the universe as
\begin{equation}\label{eq:Friedmann}
\frac{H^2}{H_{0}^{2}}=E^{2}(z)=\frac{\rho(a)}{\rho_{0}}+\Omega_{k}a^{-2},
\end{equation}
where $H\equiv \dot{a}/a$ is the Hubble parameter, $E(z)$ is a dimensionless function which is sometimes called expansion factor, $\rho(a)$ is the total energy density, the subscript `0' denotes the value at present ($z=0$). The density parameter of a given energy density is defined as 
\begin{equation}
\Omega_{x} = \frac{\rho_{x}}{\rho_{\rm{crit}}}=\frac{8\pi G}{3H^2}\rho_{x}.
\end{equation}
The normalization of the Hubble parameter gives the relation between the density parameter and curvature parameter
\begin{equation}
\sum_{x} {\Omega_{x}} + \Omega_{k} = 1,
\end{equation}
where the summation is over all the energy and matter components in the universe. Throughout this paper, the density parameters refer to the current values at $z=0$ unless otherwise denoted.

In the standard $\Lambda$CDM model, the dark energy is a constant in space and time, therefore we can write the Hubble parameter as 
\begin{equation}\label{eq:LCDM}
\frac{H^2}{H_{0}^{2}}=E^{2}(z) = \Omega_{m}(1+z)^3+\Omega_{r}(1+z)^4+\Omega_{\Lambda}+\Omega_{k}(1+z)^2,
\end{equation}
where $\Omega_{m}$ includes the contribution from cold dark matter (CDM) and baryons, $\Omega_{r}$ represents radiation and relativistic matter, and $\Omega_{\Lambda}$ is the cosmological constant. For simplicity and the efficiency of the calculation, we assume the neutrinos are massless and contribute the same as radiation.

Below, we briefly introduce the dark energy and modified gravity models 
that are explored in this paper. Table \ref{tab:models} summarizes the models 
we consider \remove{paper} and their compatibility with the linear growth data and 
local H$_{0}$ measurement.

\subsection{Cosmological models}\label{sec:model}

We divide the alternative models under consideration into two categories: dark energy models, 
motivated by new components or physical effects, and modified gravity models, which alter
the nature of general relativity.

\subsubsection{Dark energy models}
\paragraph{Parameterizations of the equation of state $w$ (XCDM, CPL, JBP $w_{\rm Linear}$)}\label{sec:intro_para_w}

The equation of state $w$ of a perfect fluid is defined as the ratio of its pressure and density
\begin{equation}
w = p/\rho.
\end{equation}
For cosmological constant, we have $w=-1$. Therefore the simplest generalization is $w={\mathrm constant}$ (\citealt{Turner_1997, Chiba_1997}) which changes the third term on the right hand side of Eq.(\ref{eq:LCDM}) to be
\begin{equation}\label{eq:XCDM}
\Omega_{\Lambda}(1+z)^{3(1+w)}.
\end{equation}
In this kind of cosmology (hereafter XCDM), the DE models with a conical kinetic term have $-1<w<-1/3$, while models with $w<-1$ are phantom dark energy (\citealt{Faraoni_2002, Caldwell_2002, Caldwell_2003}). Thus $w=-1$ is a critical value to differentiate the phantom property of the dark energy.

The generalization of constant $w$ is to assume it is a function of time. The first model considered in this paper is the CPL parameterization (\citealt{Chevallier_2001, Linder_2003}), which is the first order Taylor expansion of the scale factor at the present epoch
\begin{equation}\label{eq:CPL}
w=w_{0}+w_{1}(1-a)=w_{0}+w_{1}\frac{z}{1+z},
\end{equation}
where $w_{0}$ and $w_{1}$ are two parameters. The corresponding term for dark energy Eq. (\ref{eq:XCDM}) needs to be
\begin{equation}\label{eq:CPL}
\Omega_{\Lambda}\exp\Big(3\int_{0}^{z}\frac{1+w}{1+z'}dz'\Big).
\end{equation}

Another type of parameterization is proposed in \cite{Jassal_2005} (hereafter JBP): 
\begin{equation}\label{eq:JBP}
w=w_{0}+w_{1}\frac{z}{(1+z)^2}.
\end{equation}
Note that this model has quite different asymptotic behavior at high redshift than does the CPL parameterization.

The last model we consider here is a simple linear function 
of redshift $z$ (hereafter $w_{\rm Linear}$,  \citealt{Cooray_1999, Weller_2002, Astier_2001})
\begin{equation}\label{eq:Linear}
w=w_{0}+w_{1}z,
\end{equation}
These three parameterizations have 
similar behavior in the late time universe. Note that the integral in 
Eq. (\ref{eq:CPL}) can be calculated analytically, therefore no numerical 
routine is needed, which is important in the following computations.

\paragraph{Pseudo-Nambu Goldstone Boson (PNGB)}\label{sec:intro_PNGB}
This model is motivated by the potential $V(\phi\propto[1+\cos(\phi/F)])$ of the scalar field (\citealt{Frieman_1995, Dutta_2007}). The dark energy in this model is parameterized by its equation of state (\citealt{Basilakos_2010})
\begin{equation}
w = -1 + (1+w_0)(1+z)^{-F},
\end{equation}
where $w_{0}$ and $F$ are free parameters. When $w_{0}=-1$, this model becomes standard $\Lambda$CDM model. Because $z>0$, the second term also vanishes as $F$ becomes large, so we restrict the range of $F$ to be within [0, 8] as adopted in literature. 

\paragraph{Casimir effect (CE)}\label{sec:intro_CE}
This kind of cosmology involves a negative radiation-like term in the expression of Hubble parameter Eq. (\ref{eq:LCDM}):
\begin{equation}
\Omega_{c}(1+z)^4,
\end{equation}
where $\Omega_{c}<0$ is a constant (\citealt{Gaodlowski_2006}). 
In principle, there are different interpretations of this term. 
The first is the Casimir effect due to the vacuum energy and its
relationship to the curvature and topology of the universe
(\citealt{Bordag_2001}). The second is called ``dark radiation'' 
and is introduced by the extra dimension in the universe. For 
instance, in the  \citet{Randall_1999} scenario, the restriction 
of the field equation on the brane induces this term. The third 
interpretation that it arises from the global rotation of the 
universe (\citealt{Senovilla_1998, Gaodlowski_2006}).

\paragraph{Cardassian Ansatz (CA)}\label{sec:intro_CA}
The Cardassian expansion model was first proposed by \cite{Freese_2002}. 
It is a modification of the Friedmann equation from $H^2 = A\rho$ to 
\begin{equation}\label{eq:card}
H^2 = A\rho + B\rho^n,
\end{equation}
with the parameter $n<2/3$. 
Assuming the curvature and radiation are ignored,
this equation can be written in  terms of the density parameters as 
\begin{equation}
E^{2}(z) = \Omega_{m}(1+z)^3+(1-\Omega_{m})(1+z)^{-3n}.
\end{equation}
Therefore, it is equivalent to the XCDM model with $w=n-1$. 
This model is generalized in \cite{Wang_2003} by introducing 
one more free parameter $q>0$. The resulting Hubble parameter 
is expressed as:
\begin{equation}
E^{2}(z)=\Omega_{m}(1+z)^3\Big(1+\frac{\Omega_{m}^{-q}-1}{(1+z)^{3q(1-n)}}\Big)^{\frac{1}{q}}.
\end{equation}
This model is also called ``Modified Polytropic Cardassian'' when 
it is treated as a fluid. However, this model can also arise 
from the self-interaction of dark matter, or embedding the 
observable brane into a higher dimensional universe 
(\citealt{Davis_2007}). This is the model (CA hereafter) tested in this paper.

\paragraph{Early dark energy (EDE)}\label{sec:intro_EDE}
In many models, dark energy is important only in the late 
time universe. However, in some models 
the dark energy tracks the evolution of the dominant 
component in the universe. In those cases, dark energy can 
produce an imprint during the early time of the universe. We 
consider a generic parameterization of such early dark energy models 
as proposed by \cite{Doran_2006}. In this model, the density 
parameter of dark energy is given as a function of $a=(1+z)^{-1}$
\begin{equation}\label{eq:EDE}
\Omega_{\rm{de}} = \frac{\Omega_{\rm{de}}^{0}-\Omega_{\rm{de}}^{\rm{e}}(1-a^{-3w_{0}})}{\Omega_{\rm{de}}^{0}+\Omega_{m}a^{3w_{0}}}+\Omega_{\rm{de}}^{\rm{e}}(1-a^{-3w_{0}}),
\end{equation}
where $\Omega_{\rm{de}}^{\rm{e}}$ and $\Omega_{\rm{de}}^0$ are 
the dark energy density parameter at early times and at present 
respectively. Therefore this model is fully characterized by these 
two parameters and $w_{0}$, which is the effective value of 
the equation of state. The Hubble parameter is given by
\begin{equation}
E^{2}(a) = \frac{\Omega_{m}a^{-3}+\Omega_{r}a^{-4}}{1-\Omega_{\rm{de}}}.
\end{equation}
Note that this model approaches $\Lambda$CDM model when $w_{0}=-1$ and $\Omega_{\rm{de}}^{\rm{e}}$ goes to zero.

\paragraph{Slow roll dark energy (SR)}\label{sec:intro_SR}

The late time acceleration of the universe is similar to inflation. 
Therefore, it is possible that these two phenomena arise from the 
same physical mechanism. Based on this idea, \cite{Gott_2011, Slepian_2014} 
propose a simple dark energy model. This model class is a slowly rolling 
scalar field with potential $\frac{1}{2}m^2\phi^2$. The dark energy is 
parameterized by the equation of state:
\begin{equation}
w(z) = -1+\delta w(z) \approx \delta w_{0}\times H_{0}^2/H^2(z),
\end{equation}
and therefore 
\begin{equation}
E^{2}(z) = \Omega_{m}(1+z)^3+\Omega_{\rm{de}}\Big[\frac{(1+z)^{3}}{\Omega_{m}(1+z)+\Omega_{\rm{de}}}\Big]^{\delta w_{0}/\Omega_{\rm{de}}},
\end{equation}
where $\delta w_{0}$ is the first order term in the deviation at $z=0$ 
and represents the free parameter in this class of model. 

\paragraph{Parameterization of the Friedmann equation (PolyCDM, HLG)}\label{sec:intro_para_H}

The energy density of dark energy can be described by a quadratic 
polynomial form with non-zero space curvature. This model is 
highly flexible regarding the evolution of dark energy at low 
redshift (\citealt{Aubourg_2015}). The Hubble parameter is given as
\begin{eqnarray}\label{eq:poly}
E^{2}(z) = & &\Omega_{m}(1+z)^3+\Omega_{k}(1+z)^2+ \Omega_{1}(1+z)^2 \\ \nonumber
                                     & &+\Omega_{2}(1+z)^1+(1-\Omega_{m}-\Omega_{k}-\Omega_{1}-\Omega_{2}),
\end{eqnarray}
where $\Omega_{1}$ and $\Omega_{2}$ are two extra parameters.
This Polynomial CDM (PolyCDM) model can also be thought of as a parameterization of 
the Hubble parameter. 

For comparison, we consider the Logarithmic Hubble model (hereafter HLG)
inspired from $f(R)$ gravity (\citealt{Capozziello_2014}). We defer the 
discussion of $f(R)$ gravity in general to Sec. \ref{sec:intro_fR}, and 
treat the HLG model as a simple parameterization of the Hubble parameter
\begin{equation}\label{eq:log}
E^{2}(z) = \Omega_{m}(1+z)^3+\Omega_{k}(1+z)^2+\log(\alpha+\beta z),
\end{equation}
where $\alpha$ and $\beta$ are parameters with $\alpha=\exp(1-\Omega_{m}-\Omega_{k})$, 
so only $\beta$ is the free parameter.

\paragraph{Chaplygin gas (CG, GCG, MCG)} \label{sec:intro_CG}

The Chaplygin gas (CG) was first introduced in aerodynamics in 
1904. It was applied to cosmology by \cite{Kamenshchik_2001} 
as an unified fluid of dark matter and dark energy, which has an
equation of state
\begin{equation}
p = -\frac{A}{\rho},
\end{equation}
where $p$ and $\rho$ are pressure and energy density in 
a comoving reference frame, and $A$ is a positive constant. This model can also arise in string theory models (\citealt{Bordemann_1993}). 

This model can be extended to the Generalized 
Chaplygin gas (GCG) which has the equation of state 
(\citealt{Bento_2002, Bilic_2002})
\begin{equation}
p = -\frac{A}{\rho^{\alpha}},
\end{equation}
where $\alpha$ is a new parameter with $0<\alpha<1$. In this model, 
the pressure in high redshift is negligible, while at late times 
both the pressure and energy density are constant. Therefore 
this single fluid can describe the matter-dominated universe until 
late time acceleration. This model can also arise from a complex 
scalar field whose action can be written as a generalized Born-Infeld 
action corresponding to a ``perturbed" $d$-brane in a $(d + 1, 1)$ 
spacetime (\citealt{Bento_2002}).

GCG model can also be extended by adding a term, yielding
the Modified Chaplygin gas (MCG)
model (\citealt{Benaoum_2002}, hereafter MCG), with the equation of state
\begin{equation}
p = B\rho-\frac{A}{\rho^{\alpha}},
\end{equation}
where $B$ is the a new constant. As mentioned above, these classes of 
Chaplygin gas model were first proposed as a unified description 
of dark matter and dark energy. However, previous study of perturbation 
theory finds that growth of structure in this unified description
is inconsistent with cosmological data on perturbations 
(\citealt{Bean_2004, Gaiannantonio_2006}). Therefore, 
here we assume that the Chaplygin gas only contributes to the 
dark energy component. After redefinition of the parameter 
$A_{s} = \frac{A}{1+B}\frac{1}{\rho_{0}^{1+\alpha}}$, where 
$\rho_{0}$ is the current energy density of dark energy, the Hubble parameter for MCG model can be written as 
\begin{eqnarray}
E^{2}(z) = & &\Omega_{m}(1+z)^3+\Omega_{k}(1+z)^2+ \\ \nonumber
& & (1-\Omega_{m}-\Omega_{k})  \times \\ \nonumber
& & [A_{s}+(1-A_{s})(1+z)^{3(1+B)(1+\alpha)}]^{\frac{1}{1+\alpha}}. 
\end{eqnarray}
The GCG model corresponds to $B=0$, and the CG model 
corresponds to $B=0$ and $\alpha=1$.

\paragraph{Interacting DE and DM (IDE$_{\rm 1}$, IDE$_{\rm 2}$)}\label{sec:intro_IDE}

In typical cosmological models, including all of those 
discussed above, the dark matter and dark energy are assumed 
to evolve independently. However, some scalar field models can 
couple to the ordinary matter or dark matter 
(\citealt{Amendola_2000, Farrar_2004}). This kind of interaction 
of the dark sector can help resolve the coincidence problem. 
Such models have been widely studied (\citealt{Zimdahl_2001,Dalal_2001,Tocchini-Valentini_2002, Chimento_2003, Caldera-Cabral_2009, Bean_2008}, and references therein).
In general, the interaction between the dark sector components affects
observables. While some interaction models can mimic 
the expansion history of the $\Lambda$CDM model, 
the coupled dark energy model cannot in general mimic the expansion 
and growth of structure at the same time (\citealt{Huterer_2015}). 

In this paper, we consider the models proposed in \cite{Fay_2016}, which have. 
\begin{gather}
\bar{H}^2\propto \bar{\rho}_{m}+\bar{\rho}_{\rm{de}}  \\
\bar{\rho}_{m}'+3\bar{\rho}_{m}=\bar{Q}/\bar{H} \\
\bar{\rho}_{\rm{de}}'+3(1+\bar{w})\bar{\rho}_{\rm{de}} = -\bar{Q}/\bar{H},
\end{gather}
where the bar denotes the interacting quantity, $\bar{Q}$ characterizes 
the interaction term, and the prime means the derivative with respect 
to $\ln{a}$. The particular models are specified by the equation 
of state of interaction dark energy $\bar{w}$, which together with 
$\rho_{\rm{de}}$ and $\bar{H}$ determines $\bar{Q}$. We consider 
two models in this paper as detailed in \cite{Fay_2016}:
\begin{eqnarray}
\text{Model I ($IDE_{1}$)}: & &\quad \bar{w} = {\rm constant}  \\
\text{Model II ($IDE_{2}$)}: & &\quad \bar{w} = \bar{w}_{0}+\bar{w}_{1}\ln{a}, \\
\end{eqnarray}
where $\bar{w}_{0}$ and $\bar{w}_{1}$ in Model II are constants. 

\add{In order to investigate the possibility of the interacting model to solve the tension of H$_{0}$, we also consider a model without requiring it to mimic the expansion history as the $\Lambda$CDM model. This model has a interaction term proportional to the dark energy component $Q=\xi H \rho_{DE}$ with the coupling strength $\xi$. Note this model (hereafter IDE$_{3}$) has the specific use for the H$_{0}$ problem and not examined by the $p-$value test.}

\paragraph{Weakly-coupled canonical scalar field (WCSF$_{\rm 1D}$, WCSF$_{\rm 2D}$)}\label{sec:intro_WCSF}

Similar to inflation, the late time acceleration of the universe can 
be explained by a scalar field with some potential function. 
\cite{Huang_2011} introduced a three-parameter approximation of 
the equation of state $w(z; \epsilon_s, \epsilon_{\infty}, 
\zeta_{s})$ which allows trajectories from a wide class of scalar 
field potentials
\begin{equation}\label{eq:WCSF}
w = -1 + \frac{2}{3}\Big[\sqrt{\epsilon_{\infty}}+(\sqrt{\epsilon_s}-\sqrt{2\epsilon_{\infty}})\big[F\big(\frac{a}{a_{\rm{eq}}}\big)+\zeta_s F_2\big(\frac{a}{a_{\rm{eq}}}\big)\big]\Big]^2,
\end{equation}
where $\epsilon_s$ is the ``slope parameter" defined as $\epsilon_s\equiv\epsilon_{V}|_{a=a_{\rm{eq}}}$ which measures the slope of the potential. $\epsilon_{\infty}$ is the ``tracking parameter" which is defined from  $\epsilon_{\infty}\equiv\epsilon_{V}\Omega_{\rm{de}}|_{a\rightarrow0}$. The third parameter $\zeta_{s}$ captures the time-dependence of $\epsilon_{V}$ as a higher order correction, which is poorly constrained by the current and forecasted data. Therefore we set it to be zero in the following discussion. The function $F(x)$ is defined as 
\begin{equation}
F(x) = \frac{\sqrt{1+x^3}}{x^{3/2}}-\frac{\ln(x^{3/2}+\sqrt{1+x^3})}{x^3}.
\end{equation}
$a_{\rm{eq}}$ is the scale factor of ``matter-DE" equality which can be approximated by a fitting formula (\citealt{Huang_2011}). In this paper, we discuss the one-parameter ($\epsilon_{\infty}=0$) and two-parameter classes,  identified as $\rm{WCSF}_{\rm{1D}}$ and $\rm{WCSF}_{\rm{2D}}$ respectively,.

\paragraph{Holographic dark energy (HDE, ADE)}\label{sec:intro_HDE}

The holographic principle plays an important role in modern theoretical physics. In quantum gravity, the entropy of a system scales with its surface area instead of volume (see \citealt{tHooft_1993, Susskind_1995, tHooft_2001, Bekenstein_1973, Bekenstein_1981, Bekenstein_1994, Hawking_1976} and references therein), which contradicts the prediction from effective field theory where the entropy is an extensive quantity.  The reconciliation is suggested in \cite{Cohen_1999} by introducing a relationship between the UV and IR cut-off and thus an energy bound. However, this approach fails to produce the current cosmic acceleration as pointed out in \cite{Hsu_2004}. The solution of this problem is suggested in \cite{Li_2004} which is to use the future event horizon as the characteristic length scale. The resulting holographic dark energy model (HDE) has energy density
\begin{equation}
\rho_{\rm{de}} = 3c^2 M^2_{\rm{PL}} L^{-2},
\end{equation}
where $c$ is a numerical constant, $M_{\rm{PL}} = 1/\sqrt{8\pi G}$ is the reduced Planck mass and $L$ is related to the characteristic length scale. The Friedmann equation can be written as 
\begin{equation}
E^{2}(z) = \frac{\Omega_{m}(1+z)^3}{1-\Omega_{\rm{de}}},
\end{equation}
where $\Omega_{\rm{de}}$ is determined through (\citealt{Huang_2004, Zhang_2007})
\begin{equation}
\frac{d\Omega_{\rm{de}}}{dz} = -\Omega_{\rm{de}}\frac{1-\Omega_{\rm{de}}}{1+z}\Big(1+\frac{2}{c}\sqrt{\Omega_{\rm{de}}}\Big)
\end{equation}

Similar to HDE, \cite{Cai_2007} proposes the agegraphic dark energy model (ADE) motivated by the Karolyhazy relation (\citealt{Karolyhazy_1966, Maziashvili_2007, Maziashvili_2007b}). The dark energy density is determined by a time scale $T$
\begin{equation}
\rho_{\rm{de}} = 3n^2 M^2_{\rm{PL}} T^{-2},
\end{equation}
where $n$ is a constant to parameterize uncertainties. The original ADE 
model applies the age of the universe as the time scale. However, in this 
model the agegraphic dark energy in fact never dominates the universe's 
dynamics (\citealt{Wei_2008}). Therefore we consider the new agegraphic 
dark energy which uses the ``conformal time'' instead of the age of 
the universe. The solution is also expressed as a differential equation:
\begin{equation}
\frac{d\Omega_{\rm{de}}}{dz} = -\Omega_{\rm{de}}\frac{1-\Omega_{\rm{de}}}{1+z}\Big(3-\frac{2(1+z)}{n}\sqrt{\Omega_{\rm{de}}}\Big).
\end{equation}

Rather than using the length scale or time scale as the IR cut-off, 
\cite{Gao_2009} propose another possibility: the length scale is 
given by the average radius of Ricci scalar curvature, $R^{1/2}$ 
(hereafter RDE). Unlike the HDE and ADE models, the Hubble 
parameter of this model can be expressed analytically, which is 
convenient in numerical solutions:
\begin{equation}
E^{2}(z) = \frac{2\Omega_{m}}{2-\alpha}(1+z)^3+\Omega_{k}(1+z)^2+\big(1-\Omega_{k}-\frac{2\Omega_{m}}{2-\alpha}\big)(1+z)^{4-\frac{2}{\alpha}},
\end{equation}
where $\alpha$ is a constant to be determined. 

\paragraph{Quintessence scalar field model (QPL, QEX)}\label{sec:intro_Quint}

Scalar fields can naturally arise in particle physics. They are 
also a simple generalization of the cosmological constant, and can play 
the role of dark energy. A quintessence field is such a scalar field, with 
standard kinetic energy and minimally coupling to gravity. In a 
spatially flat FRW universe, the evolution of a scalar field 
$\phi$ is governed by the Friedmann equation Eq. (\ref{eq:Friedmann}) and 
the Klein-Gordon equation
\begin{equation}
\ddot{\phi}+3H\dot{a}+\frac{dV}{d\phi} = 0,
\end{equation}
where the dot denotes derivative with respect to cosmic time $t$. 
This model is completed by specifying the potential. 
In this paper, we consider two types of potential
\begin{eqnarray}
\text{Model I ($QPL$)}:  & & V \propto \phi^{-n} \nonumber \\
\text{Model II ($QEX$)}: & & V \propto \exp{(-\lambda\phi)},
\end{eqnarray}
where the proportionality can be determined by the initial conditions, 
and $n$ and $\lambda$ are the free parameters. Model I was originally 
proposed by several authors (\citealt{Ratra_1988, Peebles_1988, Caldwell_1998}). 
The solution  from this model can alleviate the fine-tuning problem 
(\citealt{Watson_2003}). Model II was first motivated by considering
the consequences of an anomaly in the dilatation symmetry within particle 
physics (\citealt{Wetterich_1988}). This potential can give rise to 
accelerated expansion, and yields a scaling solution 
(\citealt{Copeland_1998, Barreiro_2000}) in which the energy density 
of dark energy is proportional to the matter.

\paragraph{QCD ghost dark energy}\label{sec:intro_QCD}

The QCD ghost dark energy (hereafter QCD) was proposed in 
\cite{Urban_2010, Urban_2009, Urban_2009JCAP, Urban_2010NuPhB, Ohta_2011}. 
The key ingredient of this model is the Veneziano ghost field 
which is required to exist for the resolution of the U(1) 
problem in QCD (\citealt{WITTEN_1979,VENEZIANO_1979, Rosenzweig_1980, KAWARABAYASHI_1980, Kawarabayashi_1981, Ohta_1981}). 
This field \remove{makes} yields a vacuum energy density when it exists
in a curved space or time-dependent background. In a flat FRW 
universe, the energy density of DE in this model is given by 
$\rho_{\rm{DE}}=\alpha H+\beta H^2$, where $\alpha$ and $\beta$ 
are constants (\citealt{Cai_2012}). After a redefinition of the 
parameters, we can write the Friedmann equation of the QCD ghost 
dark energy model as:
\begin{equation}
E^{2}(z)= \kappa + \sqrt{\kappa^2+\frac{\Omega_{m}(1+z)^3+\Omega_{r}(1+z)^4}{\gamma}},
\end{equation}
where $\kappa = (1-(\Omega_{m}+\Omega_{r})/\gamma)/2$. Therefore 
this model has one more free parameter $\gamma$ compared with the 
$\Lambda$CDM model.

\subsubsection{Modified gravity}
\paragraph{DGP cosmology}\label{sec:intro_DGP}

As a modified gravity theory, DGP model is proposed by \cite{Davli_2000} as a braneworld model where our universe is a 4-dimensional brane embedded in a 5-dimensional bulk. It differs from the RS braneworld model (\citealt{Randall_1999, Randall_1999b}) by a curvature term on the brane. We follow the treatment 
of this model as detailed in \cite{Lombriser_2009}, wherein the Hubble parameter can be written as 
\begin{eqnarray}
E^{2}(z) = & & \Big( \sqrt{\Omega_{m}(1+z)^3+\Omega_{r}(1+z)^4+\Omega_{\Lambda}+\Omega_{r_{c}}}+\sigma\sqrt{\Omega_{r_{c}}}\Big)^2 \nonumber \\
& & +\Omega_{k}(1+z)^2,
\end{eqnarray}
where the density parameters have the same meaning as $\Lambda$CDM model. $\sigma=+1$ refers to the self-accelerating branch (sDGP) which has late time acceleration without a cosmological constant. $\sigma=-1$ is the normal branch (nDGP) where the DGP modifications slow the expansion rate. In the normal branch, 
a cosmological constant is then required to achieve late-time acceleration. Here 
\begin{equation}
\sqrt{\Omega_{r_c}}=\frac{1}{2H_{0}r_{c}}=\sigma\frac{\Omega_{\rm{DGP}}}{2\sqrt{1-\Omega_{k}}},
\end{equation}
where $\Omega_{\rm{DGP}}=1-\Omega_{m}-\Omega_{r}-\Omega_{k}-\Omega_{\Lambda}$, and $r_{c}$ is the crossover distance which governs the transition from 5D to 4D scalar-tensor gravity.

\paragraph{f(R) gravity} \label{sec:intro_fR}

$f(R)$ gravity is a non-trivial modification of GR which replaces the Ricci scalar in the Einstein-Hilbert action by a non-linear function. A detailed introduction of this theory is given in \cite{Sotiriou_2010} and references therein. This model has the action
\begin{equation}
S = \int d^{4}x\sqrt{-g}\left[\frac{R+f(R)}{2\mu^2}+\mathcal{L}_{\rm{m}}\right],
\end{equation}
where $R$ is the Ricci scalar, $\mu^2\equiv8\pi G$ and $\mathcal{L}_{\rm{m}}$ is the Lagrangian of matter. This model is completed once the functional form of $f(R)$ is specified. 
Much attention has been paid to this class of models, 
both theoretically and observationally (\citealt{Capozziello_2002, Nojiri_2003, Capozziello_2003, Cognola_2005, Amendola_2007, Amendola_2007b, Hu_2007, Starobinsky_2007, Li_2007, Zhao_2011}). 

In this paper, we consider the designer model introduced by \cite{Song_2007} 
and \cite{Lombriser_2012}. This model mimics by construction
the expansion history of the $\Lambda$CDM model. Imposing this
constraint yields a form for $f(R)$ but also some freedom. 
Measurements of the growth of perturbations constrain the 
form $f(R)$ can take further. The form of $f(R)$ can be parameterized 
in terms of the Compton wavelength parameter
\begin{equation}
B = \frac{f_{RR}}{1+f_{R}}R'\frac{H'}{H},
\end{equation}
evaluated at $B_{0}\equiv B(\ln{a}=0)$, where $f_{R}\equiv df/dR$, $f_{RR}\equiv d^{2}f/dR^{2}$, and the prime denotes the derivative with respect to $\ln{a}$.This parameter indicates the modification of gravity, and $B_{0}=0$ recovers the standard gravity. The condition of stability requires $B_{0}\geq0$ (\citealt{Song_2007, Sawicki_2007, Lombriser_2012}).

\remove{ NEED TO EXPLAIN SCALE CHOICE HERE}

\paragraph{f(T) gravity}\label{sec:intro_fT}

$f(T)$ gravity is based on the teleparallel equivalent of general relativity 
(\citealt{Einstein_1928, Unzicker_2005, Hayashi_1979, Arcos2004, Maluf_1994}). A 
comprehensive review of this theory can be found in \cite{Cai_2015}. The key ingredient in this formulation 
is the torsion tensor which plays the role as Ricci tensor in general relativity. The 
natural extension of this formulation is to generalize the Lagrangian to be a function of $T$, 
which is the equivalent quantity of $R$ in general relativity and $f(R)$ gravity 
(\citealt{Ferraro_2007, Ferraro_2008, Bengochea_2009, Linder_2010}). Unlike $f(R)$ gravity, 
the field equation in $f(T)$ theory is second-order rather than fourth-order, and 
this may cause pathological behaviors. In general, the corresponding action of $f(T)$ gravity is
\begin{equation}
S = \frac{1}{16\pi G}\int d^4x e[T+f(T)],
\end{equation}
where $e=\rm{det}(e_{\mu}^{A})=\sqrt{-g}$, and $e_{\mu}^A$ is the vierbein fields. Note that this theory gives $\Lambda$CDM model when $f(T)=$ constant.

In a flat FRW universe, the resulting Friedmann equation can be written as (\citealt{Nesseris_2013})
\begin{equation}
E^{2}(z) = \Omega_{m}(1+z)^3+\Omega_{r}(1+z)^4+(1-\Omega_{m}-\Omega_{r})y(z),
\end{equation}
where $y(z)$ is a function of redshift $z$ dependent on the particular $f(T)$ model. 
In this paper, we consider the following specific $f(T)$ models from the literature.

Model I ($f(T)_{PL}$): the power-law model (\citealt{Bengochea_2009})
\begin{equation}
f(T) = \alpha(-T)^b,
\end{equation}
where $\alpha$ and $b$ are parameters. The function in the Friedmann equation is 
\begin{equation} \label{eq:fT_PL}
y(z) = E^{2b}(z).
\end{equation}

Model II ($f(T)_{Exp1}$): the exponential model as $f(R)$ gravity (\citealt{Nesseris_2013, Linder_2009})
\begin{equation}
f(T) = \alpha T_{0}(1-e^{-T/(bT_{0})}),
\end{equation}
with $\alpha$ and $b$ are parameters and 
\begin{equation}
y(z) = \frac{1-(1+\frac{2E^2}{b})e^{-E^2/b}}{1-(1+\frac{2}{b})e^{-1/b}}.
\end{equation}

Model III ($f(T)_{Exp2}$): the exponential model proposed by \cite{Linder_2009}
\begin{equation}
f(T) = \alpha T_{0}(1-e^{-\sqrt{T/T_{0}}/b}),
\end{equation}
where $\alpha$ and $b$ are parameters, and 
\begin{equation}
y(z) = \frac{1-(1+\frac{E}{b})e^{-E/b}}{1-(1+\frac{1}{b})e^{-1/b}}.
\end{equation}

Model IV ($f(T)_{tanh}$): the hyperbolic-tangent model proposed by \cite{Wu_2011}
\begin{equation}
f(T) = \alpha(-T)^n\tanh{\frac{T_0}{T}},
\end{equation}
where $\alpha$ and $n$ are model parameters. We also obtain
\begin{equation}
y(z) = E^{2(n-1)}\frac{2\text{sech}^2(\frac{1}{E^2})+(1-2n)E^2\tanh(\frac{1}{E^2})}{2\text{sech}^2(1)+(1-2n)\tanh(1)}.
\end{equation}
This model is different from the previous 3 models in that it 
cannot return to $\Lambda$CDM for any value of the parameters. 

Another model we do not consider here is that of \cite{Bamba_2011}:
\begin{equation}
f(T) = \alpha T_{0}\sqrt(\frac{T}{qT_{0}})\ln{(\frac{qT_{0}}{T})},
\end{equation}
where $\alpha$ and $q$ are model parameters. The Hubble parameter in 
this model is independent from $\alpha$ or $q$, and it coincides 
with the flat sDGP model without cosmological constant. Therefore 
the analysis of sDGP model is equivalent to this model. 
This model has different dynamical behavior than sDGP which is represented 
by the perturbation equation (\citealt{Nesseris_2013}). However,
we do not consider this model here, as it is already ruled out
by the expansion and CMB data for the same reasons sDGP is.

\paragraph{Galileon cosmology: Tracker solution (GAL)}\label{sec:intro_Gal}

Galileon theory is a scalar field model introduced in \cite{Nicolis_2009, Deffayet_2009, Deffayet_2009b}. This model is inspired by the DGP model and its ability to produce the current acceleration without dark energy. This model is invariant under the Galileon symmetry in the Minkowski space-time, and keeps the field equation to be second order. The Galileon model considered in this paper is detailed in \cite{Nesseris_2010}, \cite{de_Felice_2010}, \cite{de_Felice_2011}, and \cite{de_Felice_2011b}. 
For numerical purposes, we consider the tracker solution, which has the Hubble parameter (hereafter $GAL$):
\begin{eqnarray}\label{eq:Gal}
& & E^{2}(z) = \frac{1}{2}\Omega_{k}(1+z)^2+\frac{1}{2}\Omega_{m}(1+z)^3+\frac{1}{2}\Omega_{r}(1+z)^4  \nonumber \\
& & +\sqrt{\Omega_{g}+\frac{(1+z)^4}{4}\big[\Omega_{m}(1+z)+\Omega_{k}+\Omega_{r}(1+z)^2\big]^2},
\end{eqnarray}
where $\Omega_{g} = 1-\Omega_{m}-\Omega_{k}-\Omega_{r}$. Note that this model has the same parameters as $\Lambda$CDM model.

\paragraph{Kinetic gravity braiding model (KGBM, KGBM$_{n=1}$)}\label{sec:intro_KGBM}

The kinetic braiding model is inspired from Galileon model (\citealt{Deffayet_2010c}), which introduces the extended self-interaction term $G(\phi, X)\Box\phi$ minimally coupled to gravity, where $G(\phi, X)$ is a function of $\phi$ and $X$ with $X=-g^{\mu\nu}\nabla_{\mu}\phi\nabla_{\nu}\phi/2$. This model is further generalized in \cite{Kimura_2011} as $G(\phi, X)\propto X^n$ and the original model corresponds to $n=1$. In a flat FRW universe, the Friedmann equation is
\begin{equation}
E^{2}(z)  = (1-\Omega_{m}-\Omega_{r})E^{-\frac{2}{2n-1}}+\Omega_{m}(1+z)^3+\Omega_{r}(1+z)^4.
\end{equation}
For $n=1$, the expansion in this model is the same as the tracking solution in Galileon model 
Eq. (\ref{eq:Gal}). If $n$ is taken as an arbitrary parameter, the expansion is equivalent to 
the power-law $f(T)$ theory Eq. (\ref{eq:fT_PL}), despite the fact that
these two models have different physical mechanisms. However, the KGBM and $f(T)_{\rm PL}$ 
growth rate predictions do differ, which can be used to distinguish them.

\begin{table*}\label{tab:models}
\centering
\begin{tabular}{llllllll}
\hline
Model   & Acronym & parameter\footnote{For models other than $\Lambda$CDM and $o\Lambda$CDM model, only the different parameters are shown}  &  Section  & Not in tension with: & $f\sigma_{8}$  & $H_{0}$ & LyaF BAO \\
\hline
Flat $\Lambda$CDM model     & $\Lambda$CDM  & $\Omega_{m}$, $h$, $\sigma_{8}$  &  \ref{sec:intro_LCDM} &  &  Yes  &  No  &  No\\
Non-flat $\Lambda$CDM model & $o\Lambda$CDM  & $\Omega_{m}$, $\Omega_{k}$, $h$, $\sigma_{8}$    &  \ref{sec:intro_LCDM} & & Yes & No    &  No  \\
Constant parameterization        & XCDM    &  $w$    & \ref{sec:intro_para_w} & & Yes  & No  &  No \\
CPL parameterization       & CPL     & $w_{0}, w_{1}$    & \ref{sec:intro_para_w}   & &  Yes  &  No  &  No \\
JBP parameterization       & JBP     & $w_{0}, w_{1}$   &  \ref{sec:intro_para_w}  & & Yes  &  No  &  No \\
Linear parameterization   &  $w_{\rm{Linear}}$  &  $w_{0}, w_{1}$   & \ref{sec:intro_para_w}   & &  Yes  &  No  &  No\\
Pseudo-Nambu Goldstone Boson  &  PNGB  &  $w_{0}, F$  &  \ref{sec:intro_PNGB}  & & Yes  & No  &  No \\
Cassimir effect                  &  CE        & $\Omega_{c}$ &  \ref{sec:intro_CE}   & &  Yes  &  No  &  No\\
Cardassian ansatz         & CA      & $q, n$     &  \ref{sec:intro_CA}  & &  Yes  & No   &  No\\
Early dark energy            & EDE     & $w_{0}, \Omega_{\rm{de}}^{\rm{e}}$   & \ref{sec:intro_EDE}   & &   Yes  & No  &  No\\
Slow roll dark energy      & SR       &  $\delta w_{0}$  &  \ref{sec:intro_SR}  & &  Yes  & No  &  No\\
Polynomial CDM             &  PolyCDM  & $\Omega_{1}, \Omega_{2}$    &  \ref{sec:intro_para_H}  & &  Yes  & No  &  Yes\\
Logarithmic Hubble parameter  &  HLG  & $\beta$  & \ref{sec:intro_para_H}  & &  Yes  & No   &  No \\
Chaplygin gas    & CG   &   $A_{s}$  &  \ref{sec:intro_CG}   & &  Yes  & No   &  No \\
Generalized Chaplygin gas &  GCG   & $A_{s}, \alpha$ & \ref{sec:intro_CG}  & &  Yes  & No   &  No\\
Modified Chaplygin gas      &   MCG   & $A_{s}, \alpha, B$  & \ref{sec:intro_CG}  & &  Yes  & No  &  No\\
Interacting DE (model I)    &  IDE$_{1}$  & $\bar{w}$ &  \ref{sec:intro_IDE}  & &   Yes  & No   &  No \\
Interacting DE (model II)   &  IDE$_{2}$  &  $\bar{w}_{0}, \bar{w}_{1}$ &  \ref{sec:intro_IDE} & &   Yes  & No  &  No\\
Weakly-coupled scalar field (1D)  & WCSF$_{1D}$  &  $\epsilon_{s}$ & \ref{sec:intro_WCSF}  & &  Yes  & No  &  No\\
Weakly-coupled scalar field (2D)  & WCSF$_{2D}$  &  $\epsilon_{s}, \epsilon_{\infty}$  & \ref{sec:intro_WCSF}  & &  Yes  & No  &  No\\
Holographic dark energy    &   HDE   & $c$  & \ref{sec:intro_HDE}  &   & Yes  & No   &  No\\
Agegraphic dark energy     &  ADE    & $n$  & \ref{sec:intro_HDE}   & &  Yes  & No   &  No\\
Ricci scalar dark energy               &  RDE    & $\alpha$  & \ref{sec:intro_HDE}  & &  No  & No  &  No\\
Quintessence: Power-law potential  &  QPL  &  $n$  & \ref{sec:intro_Quint}  & &  Yes  & No   &  No \\
Quintessence: Exponential potential &  QEX  &  $\lambda$  & \ref{sec:intro_Quint}  & &   Yes  & No  &  No \\
QCD Ghost dark energy  &  QCD   &  $\gamma$ &  \ref{sec:intro_QCD}  & &  No  & No   &  Yes\\
DGP cosmology                & DGP   &  --   & \ref{sec:intro_DGP}  & &  No  & No   &  No\\
$f(R)$ gravity (k=0.1)        & $f(R)_{1}$  &  $B_{0}$ & \ref{sec:intro_fR}  & &  Yes  & No  &  No\\
$f(R)$ gravity (k=0.02)      & $f(R)_{2}$  &  $B_{0}$ &  \ref{sec:intro_fR}   & &  Yes  & No  &  No\\
$f(T)$ gravity (model I)     & $f(T)_{PL}$  &  $b$  &  \ref{sec:intro_fT}   & &  Yes  & No   &  No \\
$f(T)$ gravity (model II)     & $f(T)_{Exp1}$  & $b$ & \ref{sec:intro_fT}     & &  Yes  & No  &  No \\
$f(T)$ gravity (model III)    & $f(T)_{Exp2}$  & $b$  &  \ref{sec:intro_fT}    & &  Yes  & No   &  No \\
$f(T)$ gravity (model IV)    & $f(T)_{tanh}$   & $n$ &  \ref{sec:intro_fT}    & &  No  & No   &  No \\
Galileon cosmology: tracker solution & GAL  &  --  &  \ref{sec:intro_Gal}    & &   No  & Yes   &  No\\
Kinetic gravity braiding model  &  KGBM  &  $n$ &  \ref{sec:intro_KGBM}    & &  Yes  & No    &  No \\
Kinetic gravity braiding model with $n=1$  &  KGBM$_{n=1}$  & --  &  \ref{sec:intro_KGBM}    & &  No  & No   &  No \\
\hline
\end{tabular}
\caption{Models considered in the paper, including their names, parameters and section in the paper where they are introduced. The last three columns also present their performance and the tension with the linear growth data, local H$_{0}$ measurement and BAO measurements from LyaF. For LyaF BAO, we classify the model as ``Yes" if the $p$-value difference $\Delta p$ is smaller than 0.1 and ``No" when $\Delta p$ is larger than 0.1 when the dataset is added. For H$_{0}$ and linear growth, the tension is assumed as existed if the model does not perform much better than $\Lambda$CDM model.}
\label{Table:Model}
\end{table*}

\subsection{Data sets}\label{sec:data}

We now discuss the data sets we use in this paper. We will consider
various combinations of the following five data sets.

\subsubsection{BAO data}

The BAO data arise from the measurements of the correlation function 
or power spectrum of large scale structure tracers. As an absolute distance 
measurement, the determination of the BAO scale is based on an 
fiducial cosmology, which translates the angular and redshift separations 
to comoving distances. In an anisotropic analysis, the measurement 
of the BAO scale constrains the comoving angular diameter distance 
$D_{M}(z)$ and the Hubble parameter $H(z)$ through
\begin{equation}
D_{M}(z)/r_{d} = \alpha_{\perp} D_{M, \rm{fid}}/r_{d,\rm{fid}},
\end{equation}
and
\begin{equation}
D_{H}(z)/r_{d} = \alpha_{\parallel} D_{H, \rm{fid}}/r_{d, \rm{fid}},
\end{equation}
where $D_{H}(z)=c/H(z)$ and $r_{d}$ is the sound horizon at the drag 
epoch $z_d$ when photons and baryons decouple:
\begin{equation}
r_d = \int_{z_{d}}^{\infty}\frac{c_{s}(z)}{H(z)}dz,
\end{equation}
with the sound speed in the photon-baryon fluid 
$c_{s}(z) = 3^{-1/2}c[1+\frac{3}{4}\rho_{b}(z)/\rho_{\gamma}(z)]^{-1/2}$. 
The subscript ``fid" refers to the quantity in the assumed 
fiducial model. $\alpha_{\perp}$ and $\alpha_{\parallel}$ are the 
ratios of the distances perpendicular and parallel to the line of sight.

An isotropic BAO analysis can be interpreted as constraining 
an effective distance that is a combination of $D_{M(z)}$ and 
$D_{H}(z)$ (\citealt{Eisenstein_2005})
\begin{equation}
D_{V} = [zD_{H}(z)D_M^2(z)]^{1/3}
\end{equation}
through
\begin{equation}
D_V(z)/r_{d} = \alpha D_{V, \rm{fid}}(z)/r_{d, \rm{fid}},
\end{equation}
where $\alpha$ is the ratio of the BAO scale to that predicted 
by the fiducial model.

The measurements of BAO come from galaxy surveys 
and Lyman $\alpha$ Forest (LyaF) surveys. In our calculation, 
the data adopted are taken from 6dFGS (\citealt{Beutler_2011}), 
the SDSS main galaxy sample (MGS, \citealt{Ross_2015}), 
BOSS galaxies (\citealt{Anderson_2014}), the BOSS LyaF auto-correlation 
(\citealt{Delubac_2015}), and the BOSS LyaF cross-correlation 
(\citealt{Font-Ribera_2014}). The likelihood calculations of these 
data are the same as \cite{Aubourg_2015} (as summarized in their 
Table II), and we refer the readers to that paper for more details. 

\subsubsection{Linear growth data}

The growth of structure is an important probe to 
test dark energy and modified gravity models, especially when  
geometrical measurements are not able to distinguish them 
from $\Lambda$CDM. For scales well within the Hubble radius, 
the growth of structure is governed by the equation
\begin{equation}\label{eq:growth}
\ddot{\delta}+2H\dot{\delta}-4\pi G\rho_{m}\delta=0,
\end{equation}
where the dots are derivatives with respect to time, and 
$\delta\equiv\delta\rho_{m}/\rho_{m}$ is the matter density contrast. 
Note that this equation should be modified accordingly in the modified 
gravity theories, or when the interaction between dark energy and 
dark matter is taken into account. Examples can be found in 
\cite{Fay_2016, Tsujikawa_2007}. 

\begin{figure}[htbp]
\begin{center}
\includegraphics[width=9cm, height=7cm]{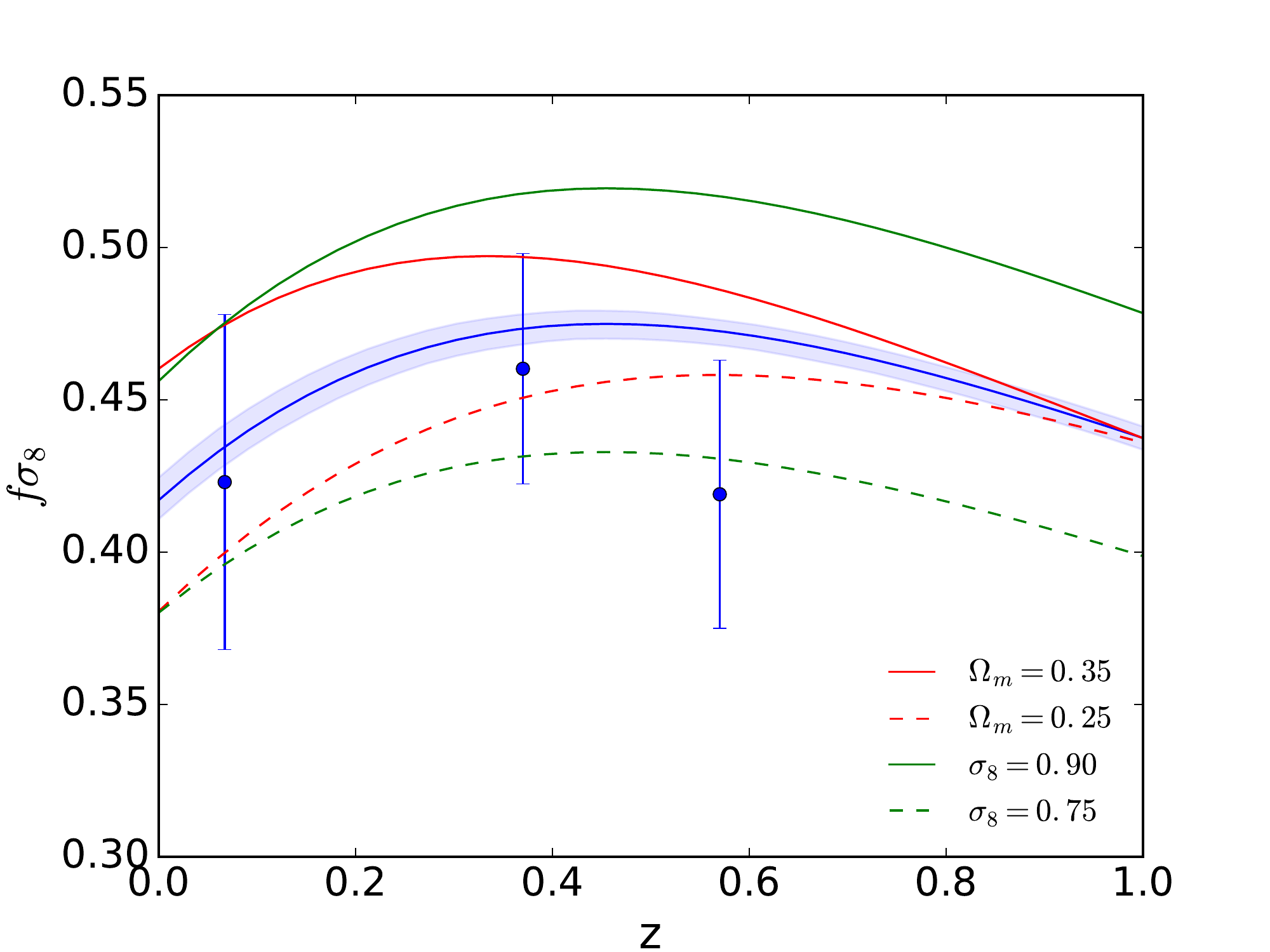}
\caption{The linear growth data $f\sigma_{8}$ as a function of $z$. The solid blue line corresponds to the best-fit flat $\Lambda$CDM model from all the datasets, while the shaded region is the $1\sigma$ error. Green and red lines represent the effect of the parameter $\Omega_{m}$ and $\sigma_{8}$ respectively (other parameters are kept fixed at the best-fit values).}
\label{fig:fsigma8_data}
\end{center}
\end{figure}

Here we consider measurements of this growth rate from redshift
space distortions (\citealt{Kaiser_1987, Scoccimarro_2004}). This measurement of 
the growth is often represented by $f\sigma_{8}(z)$, where 
$f\equiv d \ln{D}/d\ln{a}$, with $D(z)\equiv\delta(z)/\delta(0)$, 
and $\sigma_{8}(z) = \sigma_{8}D(z)$ is the power spectrum amplitude. 
The current value of $\sigma_{8}$ is a parameter to be fit, 
$f$ varies according to the cosmological model, and their 
combination is the observable that can be extracted from redshift
space distortions. We calculate $f$ starting from initial conditions 
at $z\approx30$, when the universe is dominated by matter and 
(for the models considered here) modified gravity effects have 
not become significant, and  thus we have $\delta\propto a$ 
(\citealt{Bertschinger_2008, Rapetti_2013}). The data we used in the calculation include measurements from 6dFGRS (\citealt{Beutler_2012}), SDSS galaxy (\citealt{Samushia_2012}) and BOSS CMASS galaxy (\citealt{Beutler_2014}) samples, as shown in Figure \ref{fig:fsigma8_data}. These measurements span a redshift range of $0.1\lesssim z\lesssim 0.7$ with limited overlap in the target samples, making them  independent data points. Specifically, the 6dFGRS survey is a Southern sky program, with a median redshift of $z\sim 0.07$. The \cite{Samushia_2012} data measure clustering for galaxies in the SDSS-I/II Legacy survey, with median redshift of $z\sim 0.37$. These galaxies were also included in the overall target sample for the BOSS data analyzed in \cite{Beutler_2014}, but only 1.5\% of CMASS targets were obtained from legacy SDSS data. This low fraction is due to the difference in redshift of the two samples, with CMASS probing structure at $z\sim 0.55$, and that $1/3$ of the BOSS footprint covers new area not mapped in SDSS. We note that the linear growth at this redshift was also measured by \cite{Reid_2012, Chuang_2013}, the results are in good agreement with the one used in our analysis and the latter has better accuracy due to the increased survey area (\citealt{Beutler_2014}) . These data at $z\gtrsim 0.1$ have been superceded by more recent analysis of the completed BOSS dataset (\citealt{Alam_2016}), published after this work was in an advanced stage. The error bars on  $f\sigma_{8}(z)$ are $\sim 10-20\%$ smaller in \cite{Alam_2016} than those used here, but the impact of replacing our current data with these new data would not significantly change our results.

\subsubsection{Cosmic Microwave Background data}

The Cosmic Microwave Background measurements contain important 
cosmological information. In this paper we focus on the expansion 
history and linear perturbation of the universe; therefore,
we can use an approximation that avoids computation of the full 
CMB power spectrum for each model evaluation. Therefore we adopt 
the same strategy as \cite{Aubourg_2015} and compress the CMB 
measurements to variables related to the expansion and linear 
growth. The geometrical aspect of this CMB compression is represented 
by the distance scale $D_{M}(1090)/r_{d}$ (a BAO measurement 
at $z=1090$), $\Omega_{m}h^{2}$ and $\Omega_{b}h^{2}$, the matter 
and baryon fractions of the universe, which determine the absolute 
length of the BAO ruler. 

For our work, we quantify the amplitude of 
perturbations using $\sigma_{8}(z=30)$, the amplitude of the matter 
power spectrum at the initial epoch of the perturbation 
Eq.(\ref{eq:growth}). Using this quantification of $\sigma_8$ allows
us to avoid integrating the growth equation from $z\sim 1090$ to 
low redshift for every set of model parameters. $z\sim 30$ is after the point
at which radiation matters to the expansion but before any of 
our dark energy or modified gravity models start to deviate from
$\Lambda$CDM, so it is a convenient starting point. 

The compressed data vector for the CMB is then: 
\begin{equation}
    \mathbf{v}= \begin{pmatrix}
        \Omega_{b}h^2 \\
        \Omega_{m}h^2  \\
        D_{M}(1090)/r_d   \\
        \sigma_{8}(30)  \\
     \end{pmatrix}.
\end{equation}
In this paper, we use the publicly available \texttt{base\_Alens} 
chains with the \texttt{planck\_lowl\_lowLike} dataset from the 
Planck dataset with low-$l$ WMAP polarization 
(\citealt{Planck_2013})\footnote{The Planck 2015 chains (\citealt{Planck_2015}) are not available at this time.}. 
Note that in the computation of $\sigma_{8}(30)$ for each sample in 
the Planck chains, we start from the late time variables in a Planck 
cosmology instead of the early time variables. This is only 
for the numerical considerations, and the accuracy compared 
from \texttt{camb} (\citealt{Lewis_2002ah, Lewis:1999bs}) is found to be better than 1\%. 

The resulting compression of the CMB data is represented by a simple Gaussian distribution with mean of the data vector
\begin{equation}
    \mu_{\mathbf{v}}= \begin{pmatrix}
        0.02245 \\
        0.1393   \\
        94.27  \\
        0.03448   \\
     \end{pmatrix}
\end{equation}
and its covariance

\begin{widetext}
\begin{equation*}
    C_{\mathbf{v}}= \begin{pmatrix}
        1.29\times10^{-7}  & -6.04\times10^{-7}  & 1.43\times10^{-5}  & 3.45\times10^{-8}  \\
       -6.04\times10^{-7}  &  7.55\times10^{-6}  &  -3.41\times10^{-5} &  -2.36\times10^{-7}  \\
        1.43\times10^{-5}  &  -3.41\times10^{-5}  &  4.24\times10^{-3} &  -6.60\times10^{-7}  \\
        3.45\times10^{-8}  &  -2.36\times10^{-7}  &  -6.60\times10^{-7} &  2.20\times10^{-7}
     \end{pmatrix}.
\end{equation*}
\end{widetext}


For more details regarding this type of compression of the CMB 
data, we refer the readers to \cite{Aubourg_2015}.

\subsubsection{Supernovae data}

Type-Ia supernovae (SNe) are standardizable candles that can be used 
to probe the  expansion history of the universe by measuring the luminosity distance 
as a function of redshift. In this paper, we use the Joint Light-curve 
Analysis (JLA) sample (\citealt{Betoule_2014}), which is constructed 
from the SDSS-II Supernova Survey (\citealt{Sako_2014}) and the Supernova 
Legacy Survey (SNLS) 3-year data set (\citealt{Conley_2011}) combined 
with several samples of low redshift SNe. For simplicity, we here 
use the compressed representation of the relative distance constraints 
rather than the full \texttt{cosmomc} module. The compressed data 
is described by a vector in 31 redshift bins, and a covariance 
matrix. Note that absolute luminosity of SNe is considered uncertain and 
therefore in our analysis we marginalize over the fiducial absolute 
magnitude. 

\subsubsection{Hubble constant $H_{0}$}

The Hubble constant (H$_{\rm{0}}$) measures the local expansion of the 
universe. The CMB measurements depend on early universe conditions and 
about the integral of light propagation across the Universe. Therefore, 
the CMB-inferred measurement of $H_{0}$ is model dependent, so the 
comparison with the local measurement is a test of the standard 
cosmological model. In this paper, we employ the analysis of \cite{Riess_2016},
which yields a 2.4\% determination of the quantity
\begin{equation}
H_{0} = 73.24 \pm1.74 \rm{km s}^{-1} \rm{Mpc}^{-1}.
\end{equation}
This value is 3.4$\sigma$ higher than the result from Planck data, and 
we will see its effect on the goodness of fit of the cosmological data
in detail below. 

\section{Methodology of goodness of fit}\label{sec:pvalue}

As mentioned earlier in the introduction, one of the aims in this paper 
is to evaluate the goodness of fit for each model introduced in 
Section \ref{sec:model}. Doing so will allow us to eliminate models that 
are incompatible with the data. 

Evaluating a $p$-value requires the choice of a summary metric for the 
data. For a summary metric, we choose a $\chi^2$-like global statistic. 
For linear models and Gaussian uncertainties, the distribution of 
this statistic is just the $\chi^2$ distribution with $N-M$ degrees
of freedom, where $N$ is the number of measurements and $M$ is the number
of linear parameters. In such a case, the $p$-value for a given set
of data, given that we have used the data to fit the parameters of 
the model, is  trivial to calculate. However, the models described above 
do not make predictions that vary linearly with the model parameters, 
and the uncertainties in the data are not always modeled 
as Gaussian.  The non-Gaussianity mainly comes from the LyaF, BOSS 
CMASS and SDSS MGS BAO measurements. Thus, we will estimate the $p$-value 
through a Monte Carlo experiment in which for a large number of simulated 
data sets drawn from  the model we explicitly fit the model to the data 
and then calculate the  $\chi^2$ statistic. In this procedure, fitting 
the model to each simulated data set effectively penalizes models which 
have extra degrees of  flexibility that are relevant to the observed data.

We do not attempt in this investigation to select between models whose
$p$-values are sufficiently compatible with the data (i.e. around $p \sim 0.05$
or better). A number of well-developed techniques exist for model
selection, including the evaluation of the Bayes Factor and other 
simpler techniques such as the Akaike Information Criterion 
(AIC, \citealt{Akaike_1974}), defined as AIC$=-2\mathcal{L}+2k$, where $\mathcal{L}$ is the maximum likelihood and $k$ is the number of parameters,  and the Bayesian  Information Criterion (BIC, \citealt{Schwarz_1978}) defined as BIC=$-2\mathcal{L}+k\ln{N},$ where $N$ is the number of data points. Our experiments with the latter two 
criteria indicate that they yield in practice for our problem similar 
results to comparing $p$-values. In the test with information criterion, the model with lower values of AIC or BIC is favored, and only the relative value between different models is of interest (\citealt{Liddle_2004}). For instance, a difference of 2 for BIC is considered as positive evidence, and of 6 or more as strong evidence against the model with larger value (\citealt{Liddle_2004}). A similar rule is also applicable to the outcome of AIC.

To calculate the $p$-value, in detail we perform the following steps:
\begin{enumerate}
\item
For a given dataset and model, we find the maximum likelihood 
$\mathcal{L}$ with respect to the model parameters $\theta$. 
We take the best-fit of the parameters $\hat{\theta}$ as our 
point estimate of the true but unknown parameters $\Theta$. 
For linear models and Gaussian uncertainties, the quantity
$-2 \ln \mathcal{L}(\hat{\theta})$ has a $\chi^2$ distribution, 
and we will refer to it below as ``$\chi_{\rm min}^2$.''
In general, either $\mathcal{L}(\hat{\theta})$ or $\chi_{\rm min}^2$ is a global measure 
of how different the model and the data are.
\item We generate an ensemble of simulated data sets (specifically,
the SN and BAO distances, compressed CMB parameters, $H_0$ measurements, 
and $f\sigma_8$) predicted under the model with the estimated $\hat{\theta}$,
and assuming the estimated uncertainties (which in some cases are non-Gaussian). 
\item For each simulated data set, we find the maximum likelihood 
$\mathcal{L}$ under the model while varying the parameters. 
\item The $p$-value is defined as the fraction of the simulated data sets 
with smaller values of $\mathcal{L}(\hat{\theta})$ than the data. 
\end{enumerate}
The resulting $p$-value then reflects how infrequently we expect 
data with lower $\mathcal{L}(\hat{\theta})$ or higher $\chi^2$, given
the model, if the model under consideration is correct. 

As a side note, a more robust test of the model acceptability might 
let $\hat{\theta}$ vary over an acceptable range in the above 
simulations. Since the parameters of our models are generally 
well-constrained, we do not explore the effect of doing so. 

\begin{figure}[htbp]
\begin{center}
\includegraphics[width=9cm, height=7cm]{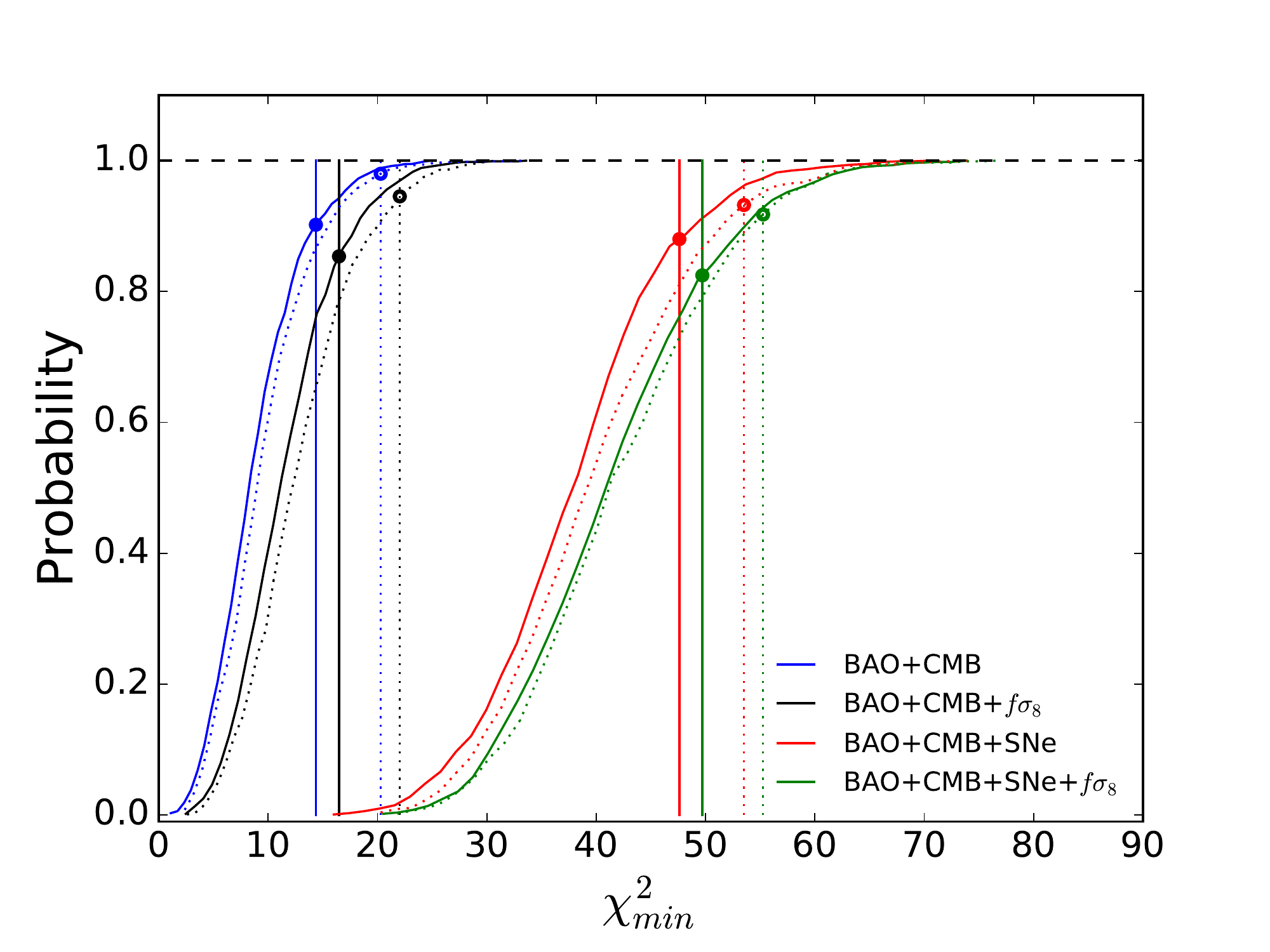}
\caption{The $p$-value test of the flat $\Lambda$CDM model with different data combinations. The vertical lines correspond to the $\chi^{2}_{\rm min}$ (corresponding to the maximum likelihood $\mathcal{L}_{max}$) from the real observations. The curved lines correspond to the cumulative probability from the distribution of $\mathcal{L}_{max}$ from the random ensemble, which has one thousand realizations in this figure. The intersection points represented by the dots correspond to $1-p$, where $p$ is the probability under the model of observing data with a lower value of the maximum likelihood than the actual data. The dotted lines and empty dots correspond to the datasets plus the local Hubble constant $H_{0}$. }
\label{fig:Flat_LCDM}
\end{center}
\end{figure}

Fig. \ref{fig:Flat_LCDM} shows the application of this method 
to the spatially flat $\Lambda$CDM model, which is the simplest 
model. The vertical lines show $\chi_{\rm min}^2$ for eight different 
combinations of the data. The curves show the cumulative probability
of finding a lower $\chi_{\rm min}^2$ estimated using the simulations
described above. The points indicate the position of the 
real data in this distribution. Adding more data raises $\chi_{\rm min}^2$,
but also raises the expected distribution. In this plot the vertical
axis is $1$ minus the $p$-value.

The $p$-values range from 0.15 to about 0.01 depending on the data
combination. That is, all data combinations experience a slight 
tension, and some of them are fairly severe, but none of them 
invalidate the underlying model at a significance comparable to
the ordinarily desired 5-$\sigma$ criterion. In general, adding 
$H_0$ always lowers the $p$-value, due to the well-known tension 
between $H_0$ and other data sets. The lowest $p$-value among our 
data combinations is achieved by combining BAO, CMB, and $H_0$. 
Adding $f\sigma_8$ or SNe always raises the $p$-value; this 
rise results because adding these data effectively dilutes the 
impact of the $H_0$ tension on the global $\chi^2_{\rm min}$
statistic. 

In detail, the addition of $H_0$ increases the $\chi^2_{\rm min}$ 
by at least 5.5. This is consistent with previous results reported
by the Planck team (\citealt{Planck_2013,Planck_2015}) and 
\citet{Riess_2016}.

\begin{figure}[htbp]
\begin{center}
\includegraphics[width=9cm, height=7cm]{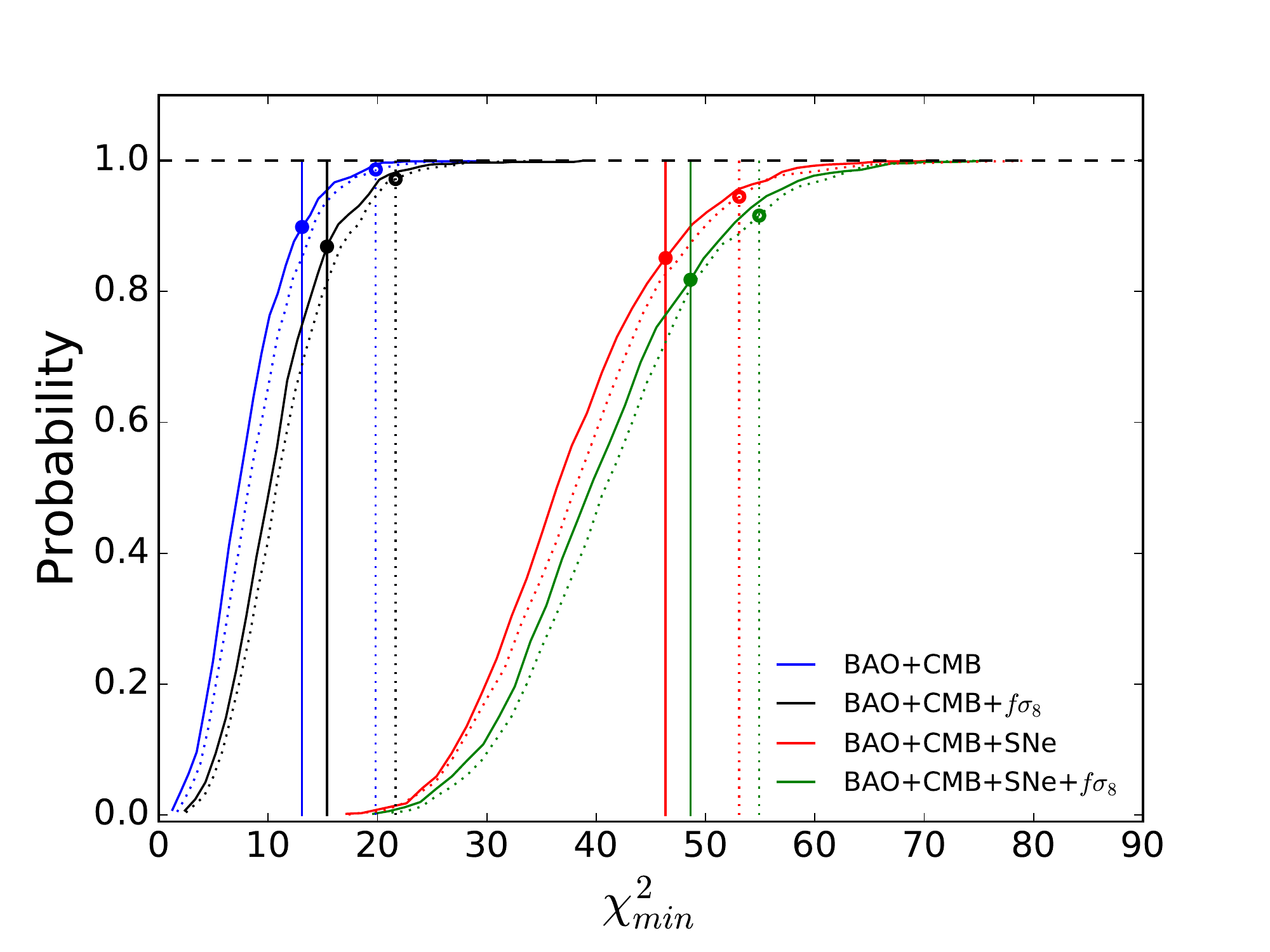}
\caption{Similar to Fig.\ref{fig:Flat_LCDM}, but for $\Lambda$CDM with spatial curvature $\Omega_{k}$.}
\label{fig:LCDM}
\end{center}
\end{figure}

Fig. \ref{fig:LCDM} shows the result with the addition of spatial 
curvature $\Omega_{k}$. The values of $\chi^{2}_{\rm min}$ from each 
real data set is smaller than in Fig. \ref{fig:Flat_LCDM}, which is 
consistent with expectation due to a small but non-zero curvature. 
The cumulative probability from the random ensemble is also shifted 
to the left because of the increased flexibility of the model. The 
amounts of the shift are not far from what one would expect for a 
$\chi^2$ distribution with one fewer degree of freedom. The resulting
$p$-values are similar to those found for the spatially flat 
$\Lambda$CDM model; in some data combinations the data is a slightly 
greater outlier than for the flat model.

\begin{figure}[htbp]
\begin{center}
\includegraphics[width=9.8cm, height=7cm]{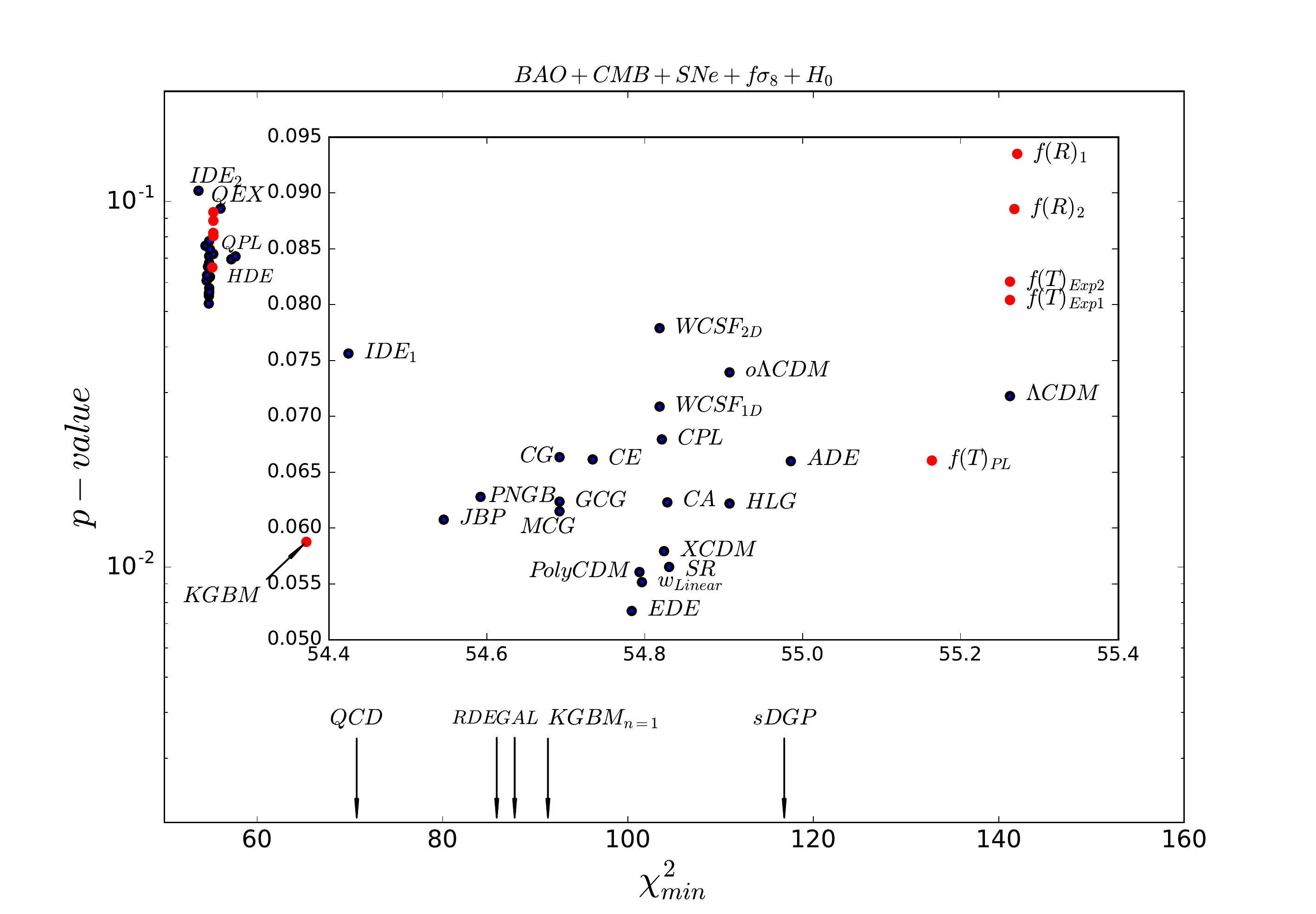}
\includegraphics[width=9.8cm, height=7cm]{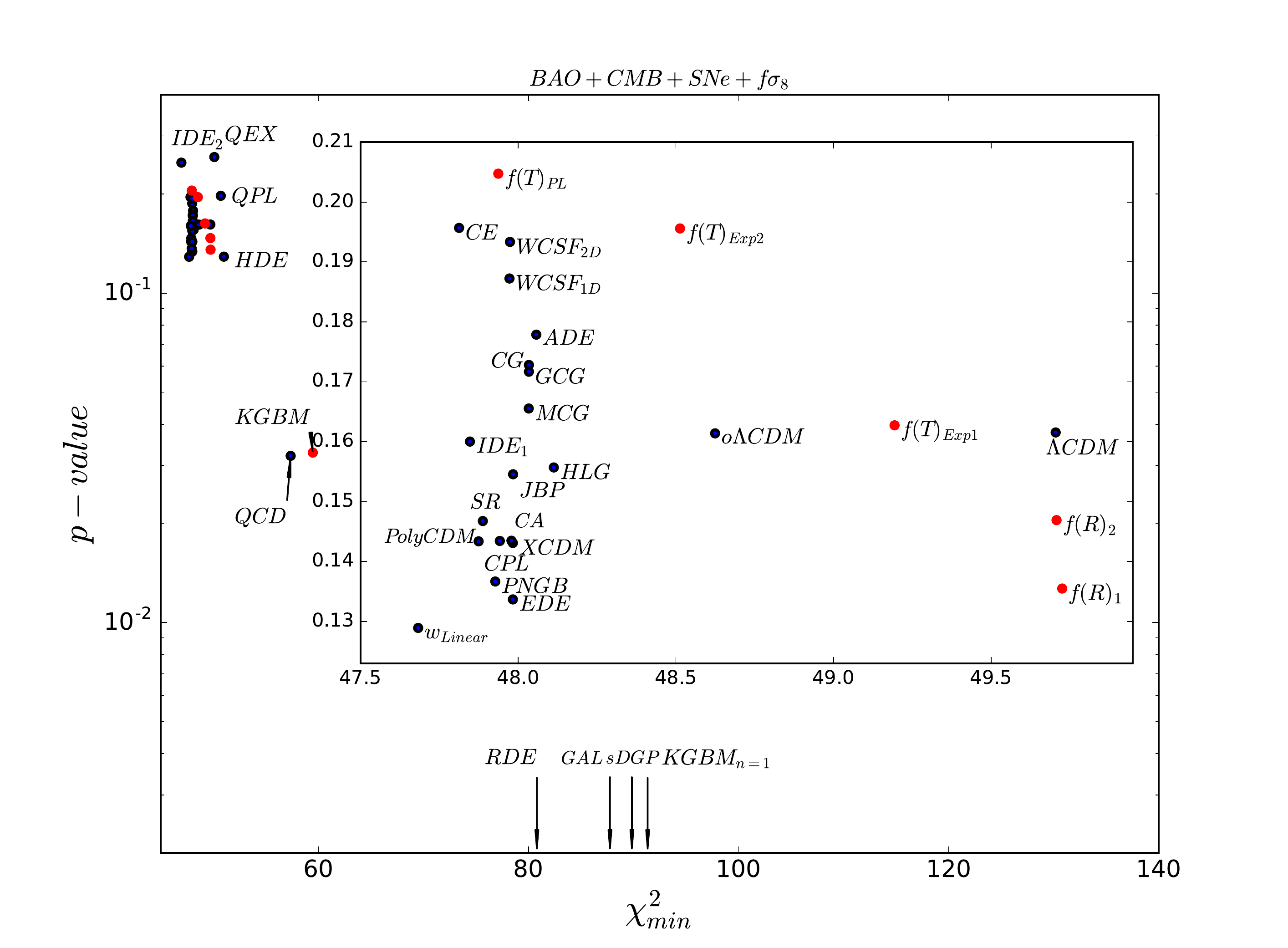}
\caption{The $p$-value test for the dark energy (blue dots) and modified gravity (red dots) models by the use of the datasets: $Top:$ BAO+CMB+SNe+$f\sigma_{8}$+H$_{0}$; $Bottom:$ BAO+CMB+SNe+$f\sigma_{8}$. The box inside the figure is the zoom-in region around the unidentified points. Note that the $f(T)_{\rm tanh}$ model has a very high $\chi^2$ and a $p$-value that is effectively zero and does not appear here. The vertical arrows show the upper limit of the $p$-value for models which has none in the ensemble of the simulated data sets have a $\chi^2$ larger than the real data (\citealt{Gehrels_1986}).}
\label{fig:comp1}
\end{center}
\end{figure}

\begin{figure}[htbp]
\begin{center}
\includegraphics[width=9.8cm, height=7cm]{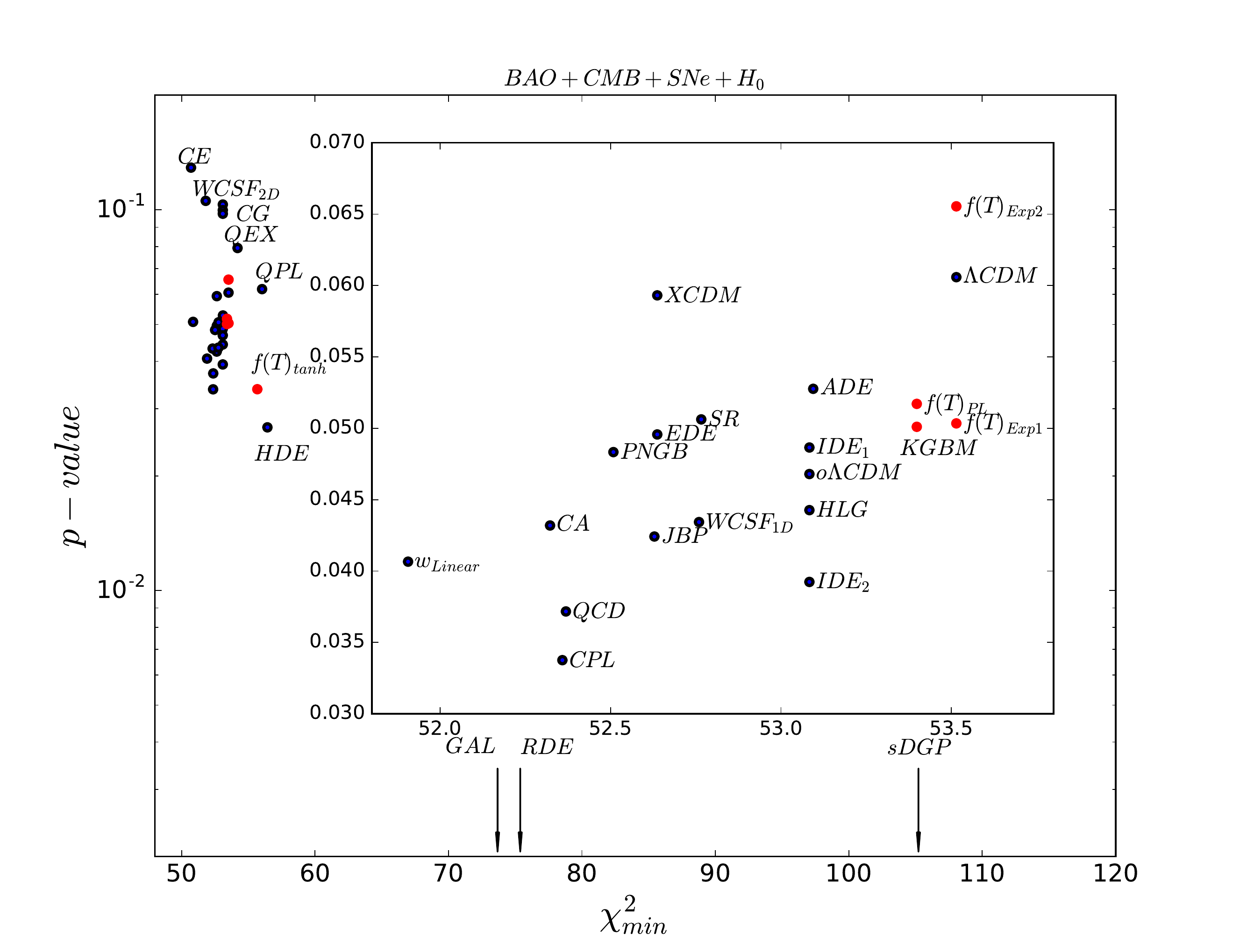}
\includegraphics[width=9.8cm, height=7cm]{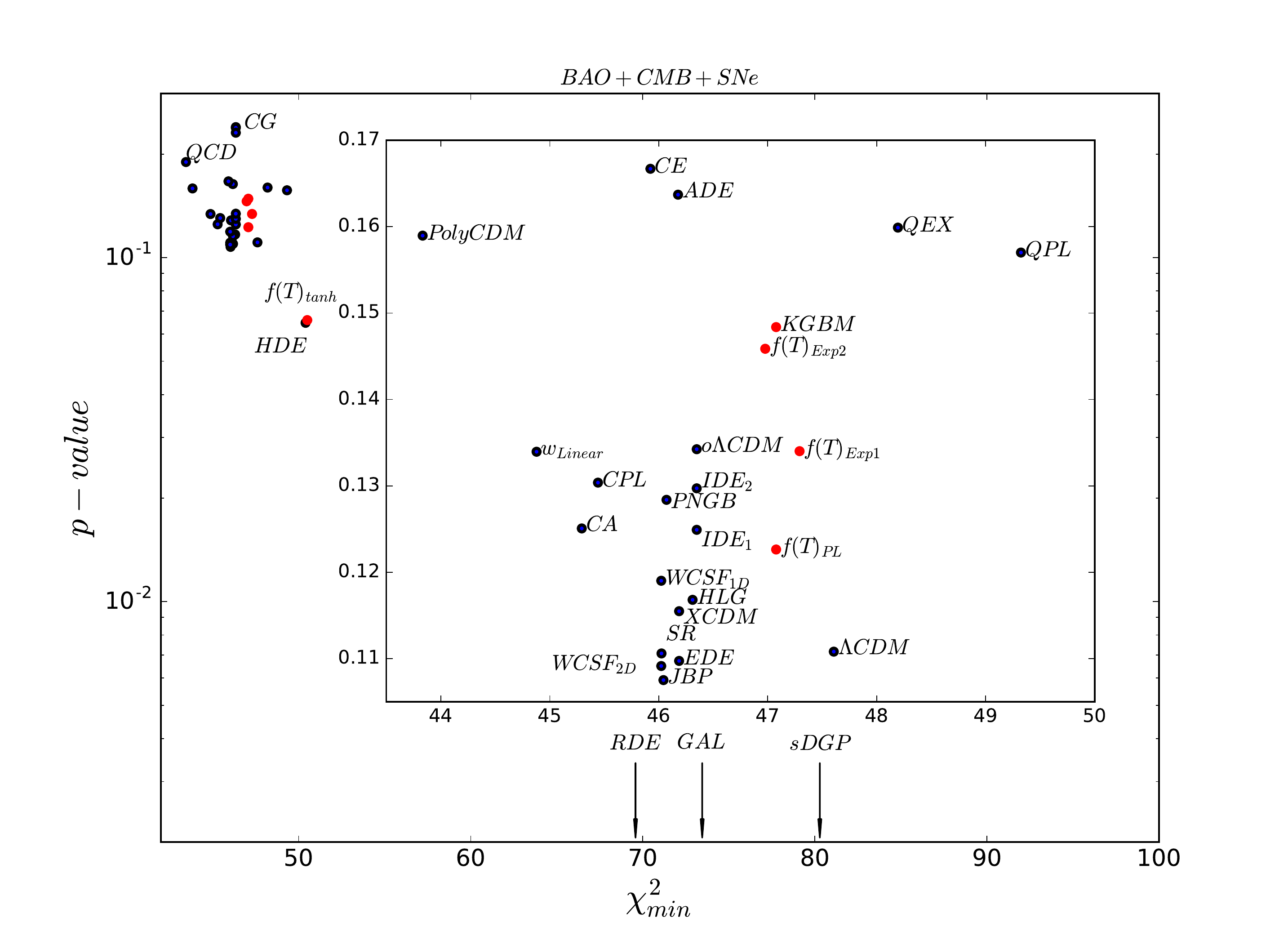}
\caption{The $p$-value test for the dark energy (blue dots) and modified gravity (red dots) models by the use of the datasets: $Top:$ BAO+CMB+SNe+H$_{0}$; $Bottom:$ BAO+CMB+SNe.}
\label{fig:comp2}
\end{center}
\end{figure}

We apply this method to the models introduced in Section 
\ref{sec:model} and compare the inferred $p$-value from 
each data set. Figures \ref{fig:comp1} and \ref{fig:comp2}
summarize the results of the $p$-value from different data
combinations, in particular focusing on the effects of 
$H_0$ and linear growth. 

In the top panel of Figure \ref{fig:comp1} all of the data 
is included.  In the bottom panel of Figure \ref{fig:comp1} we exclude
$H_0$ measurements.  In the top panel of Figure \ref{fig:comp2} we exclude 
$f\sigma_8$ data.  In the bottom panel of Figure \ref{fig:comp2} we exclude
both H$_{0}$ and $f\sigma_8$. In these figures, the red points
indicate modified gravity models.

The bottom of Figure \ref{fig:comp2} shows that even
just with BAO, CMB, and SNe there are a handful of 
models fully incompatible with the data: RDE, GAL, 
and sDGP. Although GAL and sDGP are occasionally used
as examples for which $f\sigma_8$ can yield extra
information (because they are modified gravity models),
in fact these particular examples are already strongly
ruled out without it. The rest of the models exist 
between about $p\sim 0.1$ and $0.25$ --- they mostly have
some mild tension with the data, but nothing alarming. The $f(R)$
models we consider here do not appear in this figure, because they yield
identical expansion rate predictions to $\Lambda$CDM by 
construction.

On the top of Figure \ref{fig:comp2} we see the effect 
of adding back in H$_0$. This makes all of the fits
less compatible with the data, quite substantially, to
$p \sim 0.03$--$0.13$. Given that this is a global statistic,
it is indicative of how difficult that single data point
is to fit. The Casimir Effect (CE) model and a handful of
other models suffer the tension less than others, though
not by actually fitting the H$_0$ much better (see Section
\ref{sec:results:CE}). Again, the $f(R)$ models do not appear.

On the bottom of Figure \ref{fig:comp1} we see the effect
of adding back in $f\sigma_8$ (but not H$_0$). $f\sigma_8$ 
fully rules out $f(T)_{\rm tanh}$ and KGBM$_{n=1}$ and 
furthermore makes KGBM and QCD decidedly
less well favored. Notably, for a number of modified
gravity models ($f(R)$ and the exponential and power law $f(T)$ 
models) it has virtually no effect on their compatibility 
with the data. 

On the top of Figure \ref{fig:comp1} we see the full data combination. 
The QCD, RDE, GAL, KGBM{$_{n=1}$}, $f(T)_{\rm tanh}$, and sDGP models 
are all now fully ruled out. Note that in the way we 
have quantified goodness of fit, many models have higher
$p$-values than without $f\sigma_8$ --- this result means that
$f\sigma_8$'s large error bars are diluting the goodness
of fit, possibly because they are overestimated but equally
plausibly because they lie closer to most of these 
models by accident. 

The KGBM model survives with $p\sim 0.01$. The remaining models
all imply that the data is a marginal outlier with 
$p\sim 0.05$--$0.11$. The data is most compatible with 
the IDE$_2$ model (though we remind the reader that this 
fact does not imply that this model is most probable,
which is not something we can determine, at least using
our methodology). 

It is also well known that some tension exists between the LyaF
BAO and Planck $\Lambda$CDM model (\citealt{Planck_2013, Aubourg_2015}). We investigate this problem for the models introduced here and the resulting $p-$value test is summarized in the last column of Table \ref{Table:Model}. We compare the $p-$value for the dataset BAO+CMB with and without LyaF BAO. The addition of the LyaF BAO typically decreases the $p-$value by $0.1\sim0.3$, and indicates mild tension between the datasets. The difference of the $p$-values $\Delta p$ when a particular dataset is added can be used to characterize this tension. In the test of LyaF BAO, PolyCDM and QCD ghost models have a $p$-value difference smaller than 0.1, which we believe indicates a mitigation of the tension, due to achieving a better $\chi^2$ than other models when the LyaF BAO is added (this is how we define ``Yes" for these two models and ``No" for all the other models in the last column of Table \ref{Table:Model}). In the test of H$_{0}$ tension, we note that adding H$_{0}$ typically decreases the $p$-value by $0.05\sim0.1$. Although some models have $\Delta p$ smaller than 0.1 as a criterion for LyaF BAO test, none of these models performs much better than $\Lambda$CDM, which implies that the tension of H$_{0}$ still exists in these models and this is represented as ``No" in Table \ref{Table:Model} \footnote{The Galileon model with tracker solution is ruled out by $p$-value, but the best-fit parameter indicates the H$_{0}$ discrepancy does not exist, therefore we fill it with ``Yes".}. The result for linear growth data is obtained in a similar way.

We should emphasize here the point made in \cite{Ronald_2016}, 
that whereas $p$-values can indicate how incompatible the 
data are with a specified statistical model, they do not measure 
the probability that the model is true, or the probability that 
the disagreement was produced by random chance alone. 
An extremely low $p$-value for a specific model does indicate an 
important incompatibility with the data, which can be reasonably
used to rule out it as a successful theory (assuming that
there is not some systematic issue affecting the data or its 
uncertainty estimates). 

\section{Analysis of individual models}\label{sec:mcmc}

\begin{figure}[t!]
\begin{center}
\includegraphics[width=9cm, height=8cm]{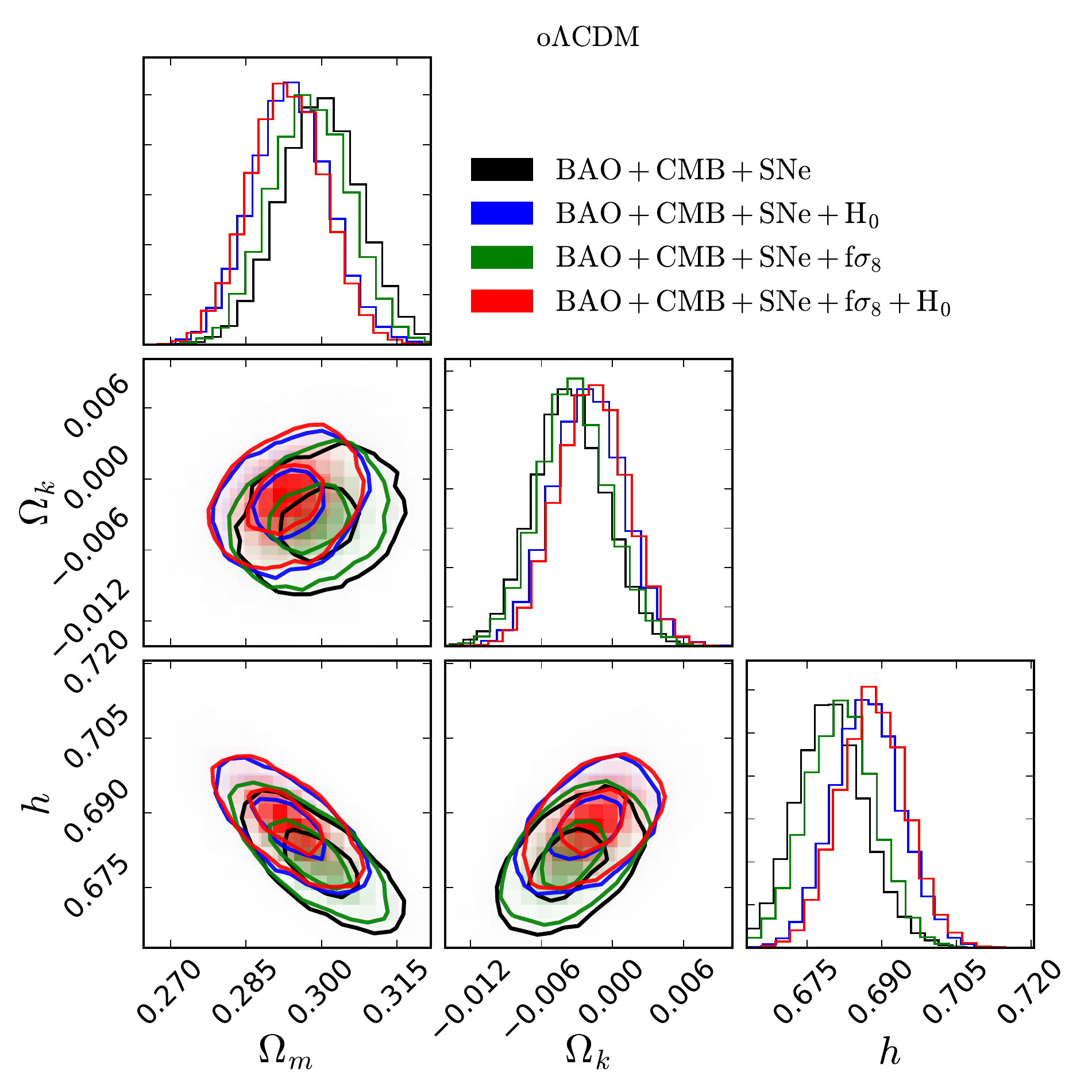}
\caption{The 68.7\% and 95.0\% confidence regions of the 
parameters for the non-flat $\Lambda$CDM. The diagonal 
panels show the one-dimensional probability distribution 
functions. The data combinations used are: 
BAO+CMB+SNe (black), BAO+CMB+SNe+H$_{0}$ (blue), 
BAO+CMB+SNe+$f\sigma_{8}$ (green), and
BAO+CMB+SNe+$f\sigma_{8}$+H$_{0}$ (red). The color coding for 
the constraint results in the following sections is the same.}
\label{fig:LCDM_mcmc}
\end{center}
\end{figure}

\begin{figure}[t!]
\begin{center}
\includegraphics[width=0.49\textwidth]{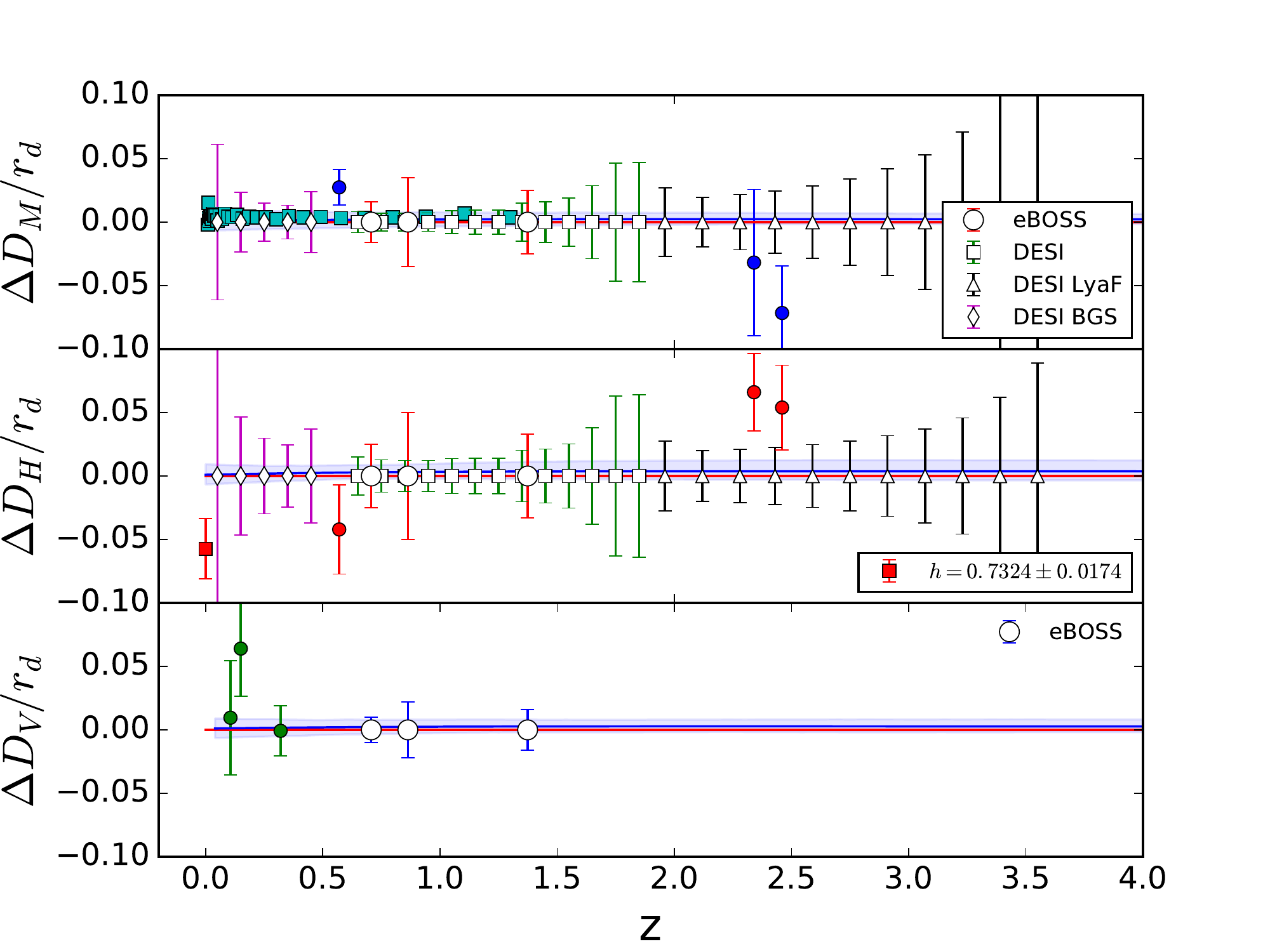}
\includegraphics[width=0.49\textwidth]{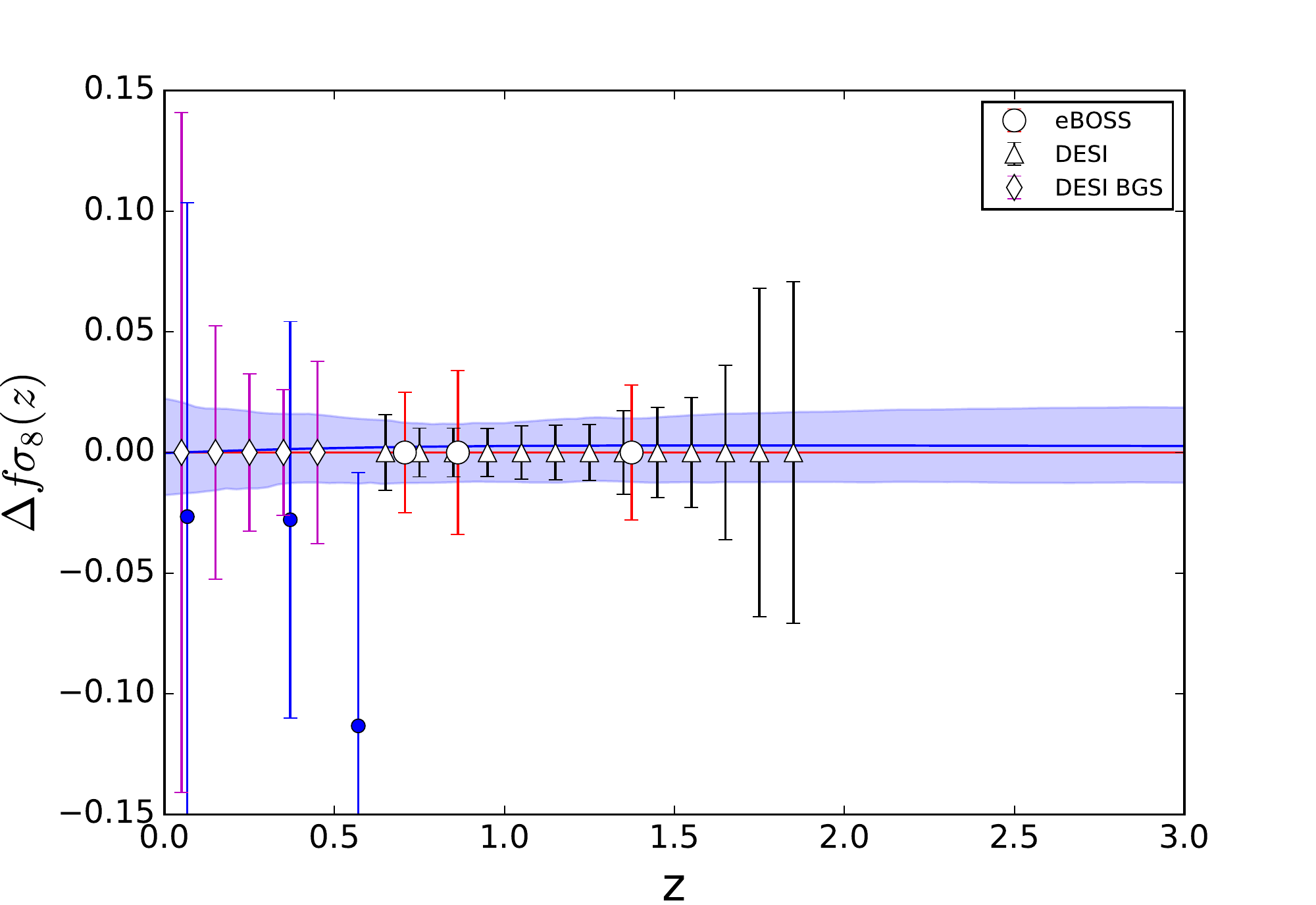}
\caption{\label{fig:LCDM_obs} Predicted observables for non-flat $\Lambda$CDM model based
  on current data. {\it Top~panel:} the BAO distance scales; {\it
    Bottom~panel:} $f\sigma_{8}$ from the linear perturbation theory.
  Quantities shown are the fraction differences from the flat
  $\Lambda$CDM model.  The horizontal red line at
  zero is the prediction from flat $\Lambda$CDM model.  Filled symbols
  with error bars represent current observations: red circles are BAO
  measurements, cyan squares are SNe, and the red square is
  H$_0$. Open symbols with error bars are the forecasts for future
  experiments: circles are eBOSS (LRGs, ELGs, and QSOs), diamonds are
  the DESI Bright Galaxy Survey (DESI BGS), squares are the DESI LRG
  and ELG surveys, and triangles are the DESI Lyman $\alpha$ Forest
  survey (DESI LyaF).  The blue line is the best-fit prediction from
  the non-flat $\Lambda$CDM model and the shaded region shows the
  $1\sigma$ distribution given the MCMC constraints. }
\end{center}
\end{figure}

\begin{table*}
\centering
\begin{tabular}{llllll}
\hline
\multicolumn{6}{c}{$\Lambda$CDM} \\
\cline{1-6}
Data    & $\Omega_{m}$   &   $h$   &    $\sigma_{8}$   & BIC  & AIC\\
\hline
BAO+CMB+SNe                                        & 0.301$\pm$0.008    & 0.684$\pm$0.006  &  -    & 58.9  & 53.6 \\
BAO+CMB+SNe+H$_0$                           & 0.296$\pm$0.007    & 0.689$\pm$0.006  &  -    &  64.9  & 59.5 \\
BAO+CMB+SNe+$f\sigma_{8}$                & 0.300$\pm$0.008    & 0.685$\pm$0.006  & 0.825$\pm$0.011      & 65.0  & 57.7\\
BAO+CMB+SNe+$f\sigma_{8}$+H$_0$   & 0.294$\pm$0.007    &  0.690$\pm$0.006  & 0.823$\pm$0.011   & 70.7  & 63.3  \\
\hline
\end{tabular}
\caption{Cosmological parameter constraints for the flat $\Lambda$CDM model, the values of the BIC and AIC are also shown in the last two columns.}
\label{tab:Flat}
\end{table*}

\begin{table*}
\centering
\begin{tabular}{lllllll}
\hline
\multicolumn{7}{c}{$o\Lambda$CDM} \\
\cline{1-7}
Data    & $\Omega_{m}$   &  $\Omega_{k}$  &   $h$   &    $\sigma_{8}$     & $\Delta$BIC  &  $\Delta$AIC  \\
\hline
BAO+CMB+SNe                                        & 0.300$\pm$0.008    &   -0.003$\pm$0.003    & 0.680$\pm$0.007  &  -    & 2.5  &  0.7 \\
BAO+CMB+SNe+H$_0$                           & 0.294$\pm$0.008    &   -0.002$\pm$0.003    &  0.687$\pm$0.007  &  -    & 3.4   & 1.6   \\
BAO+CMB+SNe+$f\sigma_{8}$                & 0.298$\pm$0.008    &   -0.003$\pm$0.003    & 0.682$\pm$0.007  & 0.828$\pm$0.012     &  2.7  &  0.9  \\
BAO+CMB+SNe+$f\sigma_{8}$+H$_0$   & 0.293$\pm$0.007    &   -0.002$\pm$0.003    & 0.688$\pm$0.007  & 0.825$\pm$0.011    & 3.5  & 1.6 \\
\hline
\end{tabular}
\caption{Cosmological parameter constraints for the non-flat $\Lambda$CDM model. The last two columns show the difference of BIC and AIC compared with spatially flat $\Lambda$CDM model: $\Delta \text{BIC} = \text{BIC}_{o\Lambda \text{CDM}}-\text{BIC}_{\Lambda \text{CDM}}$, $\Delta \text{AIC} = \text{AIC}_{o\Lambda \text{CDM}}-\text{AIC}_{\Lambda \text{CDM}}$.}
\label{tab:LCDM}
\end{table*}

We now turn to consider each of the cosmological 
models described in Section \ref{sec:model}, both its 
goodness of fit and its cosmological constraints. 

As we showed above, the simplest $\Lambda$CDM model explains the data
roughly as well as any other model, at least vis-a-vis our chosen
summary metric.  Nevertheless it is interesting to explore what the
constraints are on both the conventional cosmological parameters, and
the additional parameters introduced in the more exotic models.

To examine these constraints, we use a Monte Carlo Markov Chain
technique.  Specifically, we use the python package \texttt{emcee}
(\citealt{Foreman-Mackey_2013}), which is based on an affine-invariant
ensemble sampling algorithm \citep{GW_2010}.

Figure \ref{fig:LCDM_mcmc} shows the confidence regions and 
one-dimensional probability functions of the parameters for 
the non-flat $\Lambda$CDM model, under our four chosen 
parameter combinations. The impact of the $H_{0}$  data is 
evident from this figure; the best-fit value of $h$ increases 
by about $1\sigma$ when the H$_{0}$ measurement is added. 
The constraint on the spatial curvature $\Omega_{k}$ doesn't 
show strong deviation from flatness; $\Omega_{k}=0$ is well 
within the $1\sigma$ region for all the data sets combined. 
Table \ref{tab:Flat} and \ref{tab:LCDM} summarize the mean 
values and uncertainties of the parameters for the flat 
$\Lambda$CDM and non-flat $\Lambda$CDM model respectively. The BIC and AIC calculations are also shown. For models different than flat $\Lambda$CDM model, only the difference is shown in the following calculation.

Figure \ref{fig:LCDM_obs} shows the predicted observables under the
non-flat $\Lambda$CDM models. These observables shown are the distance
measures and $f\sigma_8$ across redshift.  The predictions for these
observables are shown as fractional deviations from the best-fit flat
$\Lambda$CDM model.  The band shown is the 1$\sigma$ variation of the
observable under the MCMC sampling of the model parameters.  We
include current data and uncertainties as the filled symbols.  Clearly
in this model space the range of variation in observed quantities is
quite small.  In this figure, we also include projections for eBOSS
(\citealt{Dawson_2016, Zhao_2016}) and DESI (\citealt{Levi_2013,
  DESI_2016}) as open symbols.

\subsection{Parameterization of equation of state $w$ (XCDM, CPL, JBP, $w_{\rm Linear}$ and PNGB)}

\begin{figure}[htbp]
\begin{center}
\includegraphics[width=9cm, height=8cm]{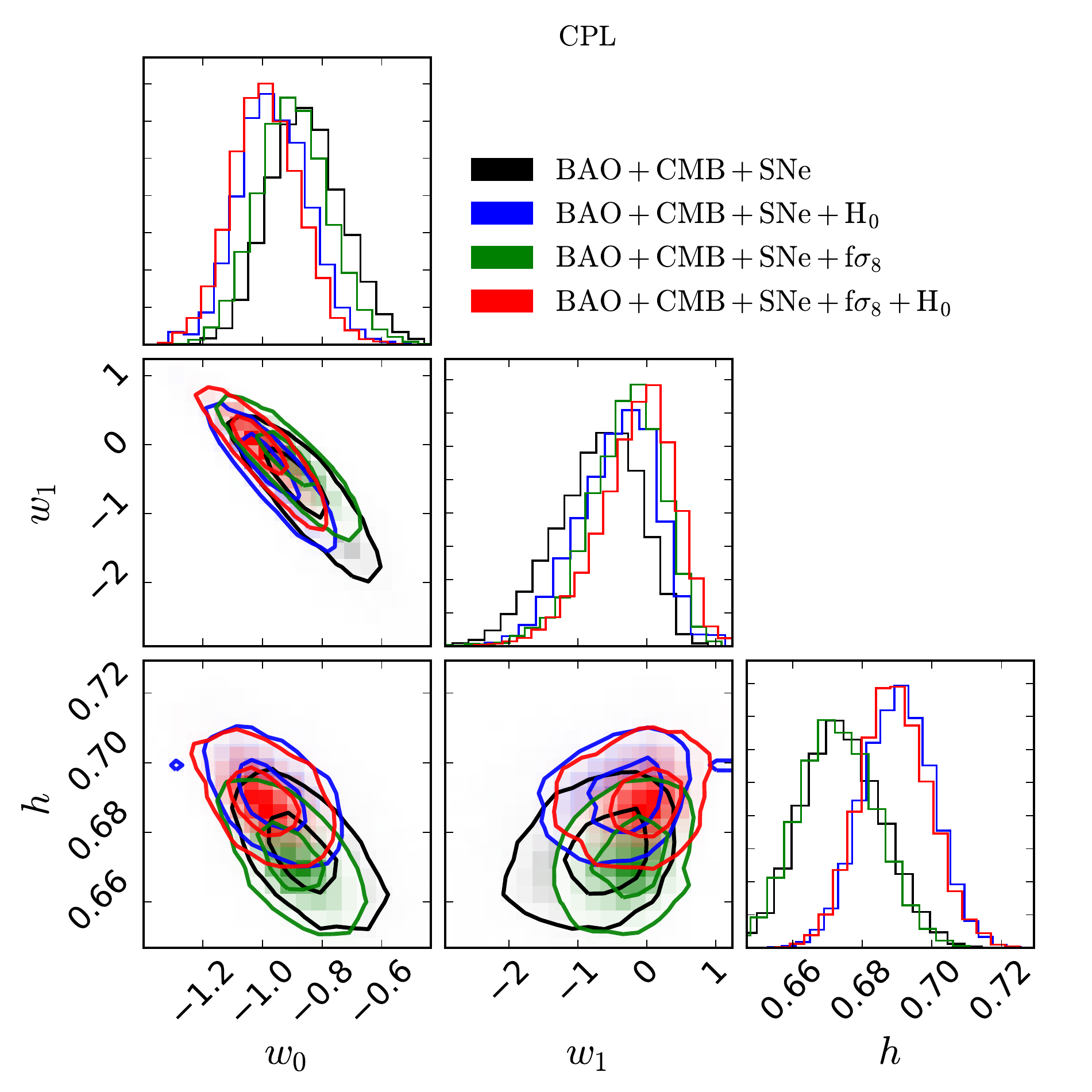}
\caption{The 68.7\% and 95.0\% confidence regions of the parameters for non-flat CPL parameterization of $w$. The diagonal panels show the one-dimensional probability distribution functions.}
\label{fig:CPL_mcmc}
\end{center}
\end{figure}

Simple parameterizations of the equation of state of dark energy $w$ allow
us to study aspects of the observed acceleration without explicit assumptions
about its physical cause. As an example, Figure \ref{fig:CPL_mcmc} presents 
the constraints on the CPL parameterization. A constant equation of state ($w_1=0$)
lies within the 95\% distribution for all data combinations. The 
distribution of $h$ is strongly affected by the $H_0$ data. 
The other parametrizations (XCDM, JBP, and $w_{\rm Linear}$)  behave very similarly.

\begin{figure}[htbp]
\begin{center}
\includegraphics[width=9cm, height=8cm]{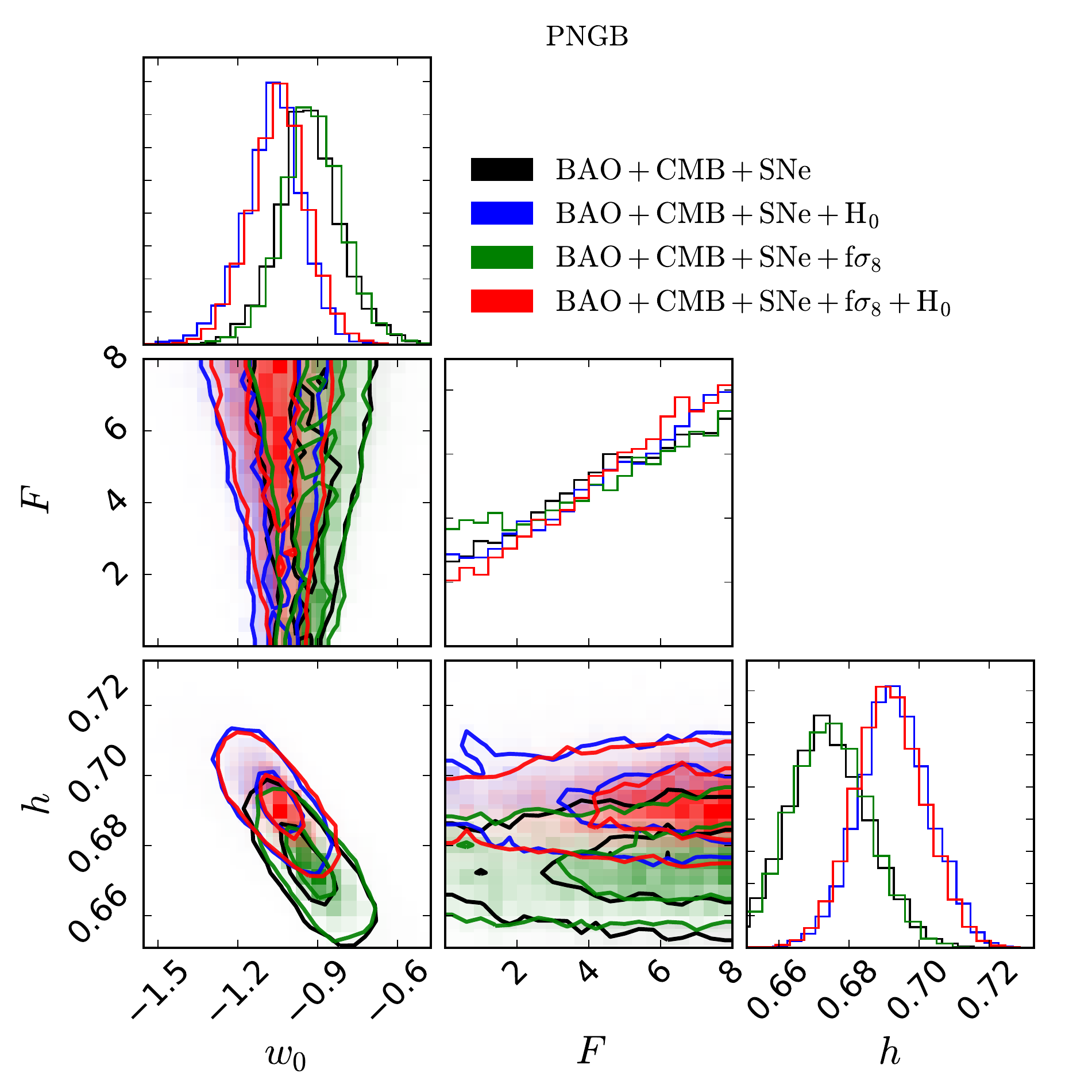}
\caption{The 68.7\% and 95.0\% confidence regions of the parameters $w_{0}$, $F$ and $h$ for non-flat PNGB model. The diagonal panels show the one-dimensional probability distribution functions.}
\label{fig:PNGB_mcmc}
\end{center}
\end{figure}

The Pseudo-Nambu Goldstone Boson (PNGB) model has a well motivated physical 
background, but the cosmological solution is represented by a parameterization 
of $w$, which is a perturbation around $-1$. In this model, we find similar 
results as the CPL model. The constraints on the parameters are displayed in 
Figure \ref{fig:PNGB_mcmc}. The parameter $F$ is restricted to a limited 
range due to the numerical considerations, but it is clear that the upper 
bound is much larger than illustrated. 
The constraint on $w_{0}$ is consistent with $-1$.

Tables \ref{tab:XCDM} to \ref{tab:PNGB} summarize the constraints on 
the XCDM, CPL, JBP, Linear parameterization, and PNGB models. The 
values of $w_{0}$ from different datasets prefer a quintessence-like 
result to a phantom-like result in the very late universe. The 
result doesn't show clear evidence of the dynamical evolution 
of dark energy. The well-known tension of $H_{0}$ is a little 
larger than $1\sigma$. 

The $p-$value test of these models all show similar behavior. 
The addition of the local measurement of H$_{0}$ degrades the 
$p-$value to around $p\sim 0.05$. Compared with the standard 
$\Lambda$CDM model, their flexibility allows them to lower 
$\chi^2$ by about 1 to 2, but not a much better $p-$value.

\begin{figure}[t!]
\begin{center}
\includegraphics[width=0.49\textwidth]{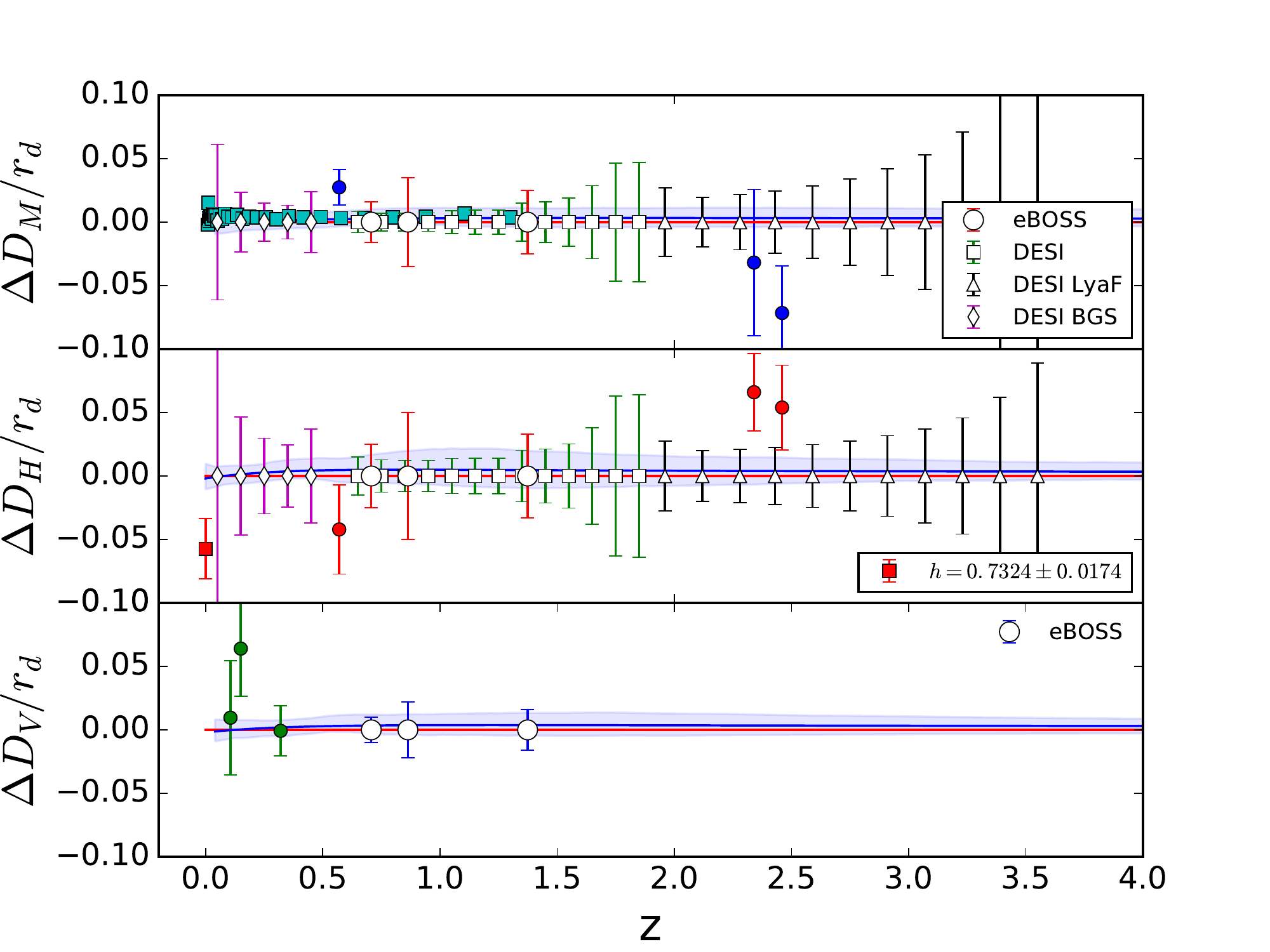}
\includegraphics[width=0.49\textwidth]{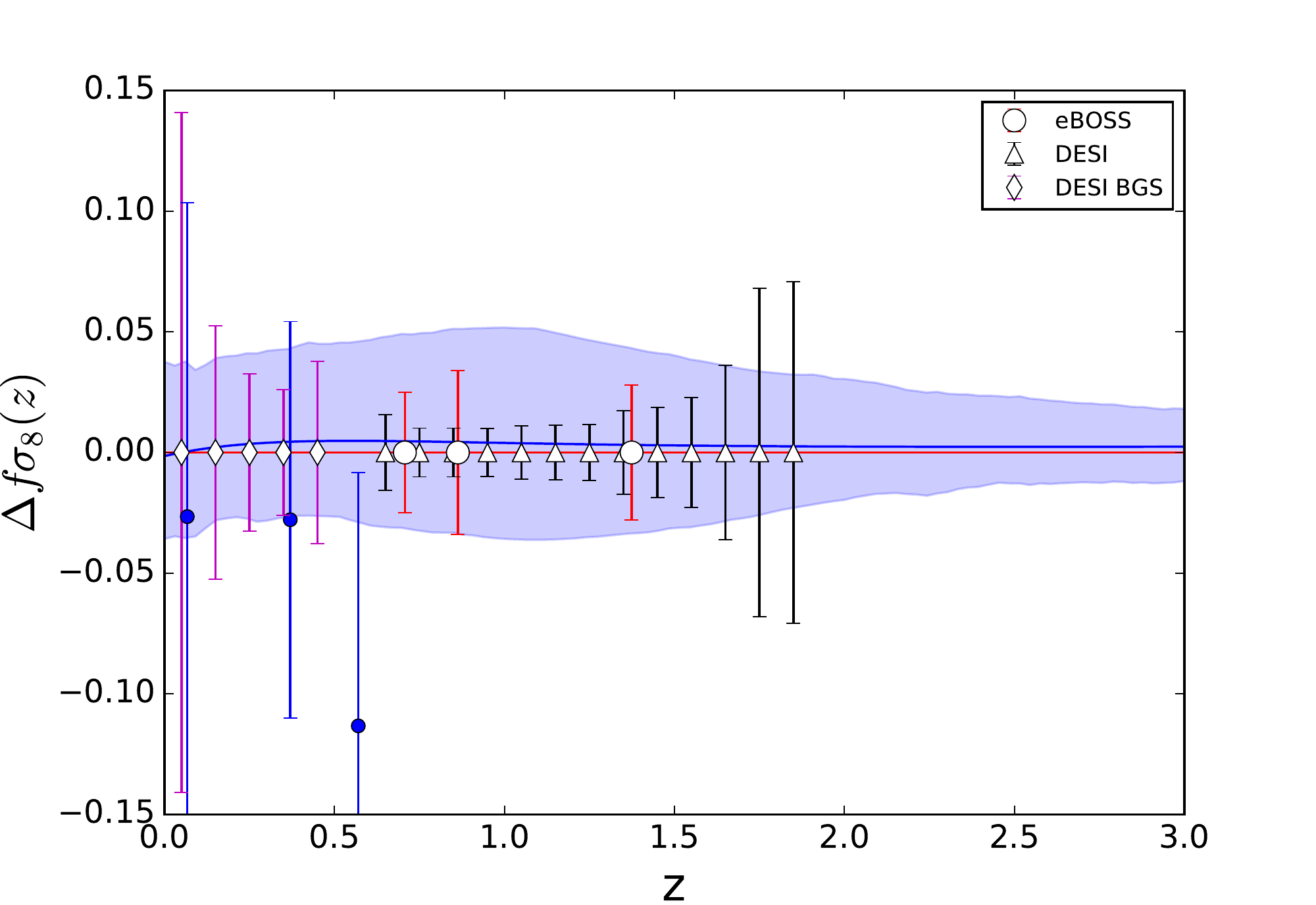}
\caption{\label{fig:CPL_obs} Similar to Figure \ref{fig:LCDM_obs}, for
the CPL parametrization.}
\end{center}
\end{figure}

Figure \ref{fig:CPL_obs} shows the predicted observables under
the CPL model, which are illustrative of most of these parametrizations 
aside from PNGB.
Whereas PNGB behaves similarly to $\Lambda$CDM, for the 
others the predicted observations have a broader range,
especially in $f\sigma_8$. The fractional variations do 
not appear preferentially high at any range of redshifts. 

\begin{table*}
\centering
\begin{tabular}{llllllll}
\hline
\multicolumn{8}{c}{XCDM} \\
\cline{1-8}
Data    & $\Omega_{m}$   &  $\Omega_{k}$  & $w$  &  $h$   &    $\sigma_{8}$   & $\Delta$BIC  &  $\Delta$AIC \\
\hline
BAO+CMB+SNe                                        & 0.302$\pm$0.010    &   -0.003$\pm$0.003    & -0.98$\pm$0.05  & 0.677$\pm$0.011  &  -    & 6.1  &  2.6 \\
BAO+CMB+SNe+H$_0$                           & 0.292$\pm$0.009    &   -0.003$\pm$0.003    & -1.03$\pm$0.05  & 0.690$\pm$0.010  &  -     & 6.7  &  3.1 \\
BAO+CMB+SNe+$f\sigma_{8}$                & 0.303$\pm$0.010    &   -0.002$\pm$0.003    & -0.96$\pm$0.05  & 0.675$\pm$0.011  & 0.822$\pm$0.013      & 5.9  &  2.3 \\
BAO+CMB+SNe+$f\sigma_{8}$+H$_0$   & 0.292$\pm$0.009    &   -0.002$\pm$0.003    & -1.01$\pm$0.05  & 0.690$\pm$0.010  & 0.826$\pm$0.014    & 7.3  & 3.6 \\
\hline
\end{tabular}
\caption{Cosmological parameter constraints for the XCDM model.}
\label{tab:XCDM}
\end{table*}

\begin{table*}
\centering
\begin{tabular}{lllllllll}
\hline
\multicolumn{9}{c}{CPL parameterization} \\
\cline{1-9}
Data    & $\Omega_{m}$   &  $\Omega_{k}$    &    $w_{0}$    &     $w_{1}$    &    $h$   &    $\sigma_{8}$   & $\Delta$BIC  &  $\Delta$AIC  \\
\hline
BAO+CMB+SNe                                        & 0.305$\pm$0.011    &   -0.006$\pm$0.004    & -0.86$\pm$0.12  &  -0.7$\pm$0.7        & 0.674$\pm$0.012  &  -      & 9.1  & 3.8\\
BAO+CMB+SNe+H$_0$                           & 0.293$\pm$0.009    &   -0.004$\pm$0.004    & -0.97$\pm$0.12  &  -0.4$\pm$0.5        & 0.690$\pm$0.010  &  -     & 10.2  & 4.8 \\
BAO+CMB+SNe+$f\sigma_{8}$                & 0.305$\pm$0.011    &   -0.004$\pm$0.004    & -0.90$\pm$0.12  &  -0.3$\pm$0.5        & 0.673$\pm$0.011  & 0.829$\pm$0.019    & 9.7  & 4.2  \\
BAO+CMB+SNe+$f\sigma_{8}$+H$_0$   & 0.293$\pm$0.009    &   -0.002$\pm$0.004    & -0.99$\pm$0.11  & -0.1$\pm$0.5         &0.690$\pm$0.010  & 0.829$\pm$0.020    & 11.1  & 5.6 \\
\hline
\end{tabular}
\caption{Cosmological parameter constraints for the CPL parameterization of the equation of state.}
\label{tab:CPL}
\end{table*}

\begin{table*}
\centering
\begin{tabular}{llllll}
\hline
\multicolumn{6}{c}{JBP parameterization} \\
\cline{1-6}
Data     &    $w_{0}$    &     $w_{1}$    &    $h$    & $\Delta$BIC  &  $\Delta$AIC  \\
\hline
BAO+CMB+SNe                                        & -0.90$\pm$0.18  &  -0.6$\pm$1.2        & 0.674$\pm$0.011   & 9.7  & 4.4 \\
BAO+CMB+SNe+H$_0$                           & -1.01$\pm$0.16  &  -0.2$\pm$1.2        & 0.690$\pm$0.010    & 10.5  & 5.1  \\
BAO+CMB+SNe+$f\sigma_{8}$                & -0.94$\pm$0.16  &  -0.1$\pm$1.1        & 0.675$\pm$0.011   &  9.8  & 4.3 \\
BAO+CMB+SNe+$f\sigma_{8}$+H$_0$    & -1.08$\pm$0.16  &  0.5$\pm$1.1         &0.692$\pm$0.010   & 10.8 & 5.3 \\
\hline
\end{tabular}
\caption{Cosmological constraints for a selection of parameters for the JBP parameterization of the equation of state.}
\label{tab:JBP}
\end{table*}

\begin{table*}
\centering
\begin{tabular}{llllll}
\hline
\multicolumn{6}{c}{Linear parameterization} \\
\cline{1-6}
Data     &    $w_{0}$    &     $w_{1}$    &    $h$     & $\Delta$BIC  &  $\Delta$AIC  \\
\hline
BAO+CMB+SNe                                        & -0.86$\pm$0.09  &  -0.4$\pm$0.3        & 0.673$\pm$0.010   & 8.6  &  3.3 \\
BAO+CMB+SNe+H$_0$                           & -0.93$\pm$0.09  &  -0.4$\pm$0.4        & 0.689$\pm$0.009   &  9.7  &  4.4  \\
BAO+CMB+SNe+$f\sigma_{8}$                & -0.87$\pm$0.08  &  -0.3$\pm$0.3        & 0.672$\pm$0.011   & 9.5  & 4.0 \\
BAO+CMB+SNe+$f\sigma_{8}$+H$_0$    & -0.95$\pm$0.07  &  0.2$\pm$0.2         &0.689$\pm$0.010   & 11.1  & 5.5 \\
\hline
\end{tabular}
\caption{Cosmological constraints for a selection of parameters for the Linear parameterization of the equation of state.}
\label{tab:Linear}
\end{table*}

\begin{table*}
\centering
\begin{tabular}{lllll}
\hline
\multicolumn{5}{c}{PNGB} \\
\cline{1-5}
Data     &    $w_{0}$    &     $F$   & $\Delta$BIC  &  $\Delta$AIC  \\
\hline
BAO+CMB+SNe                                        & -0.94$\pm$0.11  &  4.85$_{-2.79}$    & 9.7 & 4.5  \\
BAO+CMB+SNe+H$_0$                           & -1.06$\pm$0.10  &  5.06$_{-2.97}$   & 10.3 & 5.0   \\
BAO+CMB+SNe+$f\sigma_{8}$                & -0.92$\pm$0.10  &  4.79$_{-3.15}$   & 9.7  & 4.2   \\
BAO+CMB+SNe+$f\sigma_{8}$+H$_0$    & -1.04$\pm$0.11  &  5.24$_{-2.83}$   & 10.9 & 5.3    \\
\hline
\end{tabular}
\caption{Cosmological constraints for a selection of parameters for the PNGB model.}
\label{tab:PNGB}
\end{table*}

\subsection{Casimir effect (CE)}
\label{sec:results:CE}

\begin{figure}[htbp]
\begin{center}
\includegraphics[width=9cm, height=8cm]{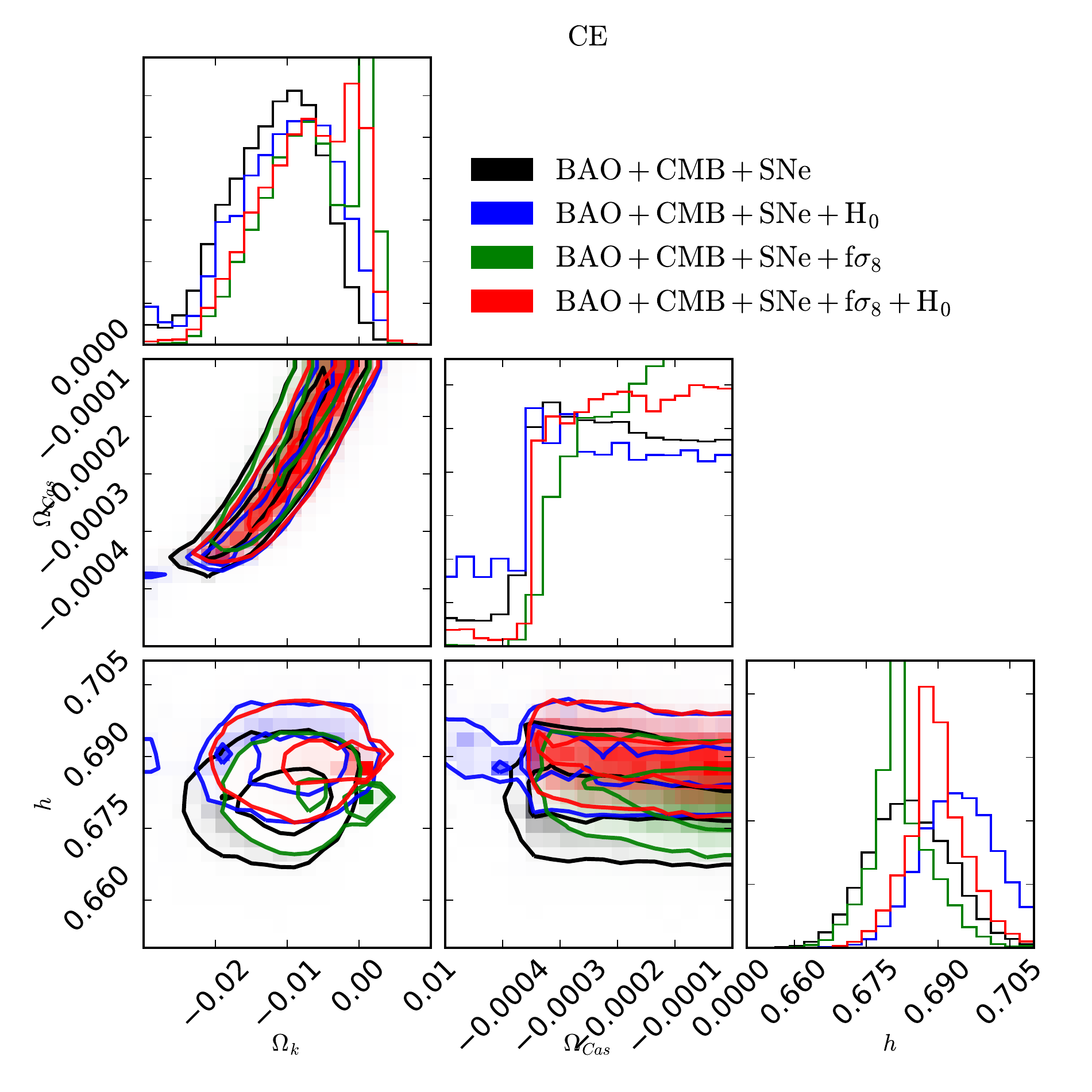}
\caption{The 68.7\% and 95.0\% confidence regions of the parameters for the Casimir-effect cosmology.}
\label{fig:Casi_mcmc}
\end{center}
\end{figure}

For the CE model, the constraint on the parameter $\Omega_{\rm{c}}$ together with 
the spatial curvature $\Omega_{k}$ (marginalizing the other parameters 
in the Casimir-effect model) is presented in Figure \ref{fig:Casi_mcmc} 
and Table \ref{tab:Casi}. A flat FRW cosmology is within the  
than 2$\sigma$ contours for all four data combinations. The geometrical 
probes favor a larger deviation of $\Omega_{k}$ from 0; this deviation is 
suppressed when including the linear growth data. 

The upper bound of the contribution from the negative radiation-like term is much smaller 
than 1.0\% at a 95\% confidence level. The results for $h$ show that 
despite the improved $p$-value, the disagreement between 
the large scale structure data and the local measurement of H$_{0}$ remains. 

In the $p-$value test, for the  geometrical probes CE gives better 
$p-$value than the $\Lambda$CDM  model, and in fact is the model 
under which the data is least discrepant when H$_0$ is included. 
However, adding linear growth  data suppresses the ability of CE 
to introduce curvature and reduces its advantage over other models.

When we consider the predicted observables under this model (the 
analog of Figure \ref{fig:LCDM_obs}) we find it is very similar to that
for non-flat $\Lambda$CDM --- there is not much more freedom allowed by
this model regarding the observables considered here.

\begin{table*}
\centering
\begin{tabular}{lllll}
\hline
\multicolumn{5}{c}{Casimir effect} \\
\cline{1-5}
Data     &    $\Omega_{k}$    &     $\Omega_{Cas}$   & $\Delta$BIC  &  $\Delta$AIC  \\
\hline
BAO+CMB+SNe                                        & $-0.016_{-0.036}^{+0.009}$   &  $0.0_{-0.001}$    & 5.8  & 2.3  \\
BAO+CMB+SNe+H$_0$                           & $-0.050_{-0.012}^{+0.040}$  &  $-0.001_{-0.002}$    & 4.8  &  1.2  \\
BAO+CMB+SNe+$f\sigma_{8}$                & $-0.006\pm0.007$                  &  $0.0_{-0.0002}$     & 5.8  &  2.1 \\
BAO+CMB+SNe+$f\sigma_{8}$+H$_0$    & $-0.007_{-0.008}^{+0.007}$   & $0.0_{-0.0001}$      & 7.2  & 3.5   \\
\hline
\end{tabular}
\caption{Cosmological constraints for a selection of parameters for the Casimir effect.}
\label{tab:Casi}
\end{table*}

\subsection{Cardassian expansion (CA)}

The Cardassian expansion tested here (specifically the version introduced
by \citealt{Wang_2003}) is a direct modification of the Friedmann 
equation Eq.(\ref{eq:card}). However, the model also can be thought of as 
a generalization of the XCDM model. A number of previous studies have
considered constraints on this  model (\citealt{Zhu_2002, Zhu_2003, Sen_2003, Zhu_2004,Frith_2004, Alcaniz_2005, Xu_2012}).

\begin{figure}[htbp]
\begin{center}
\includegraphics[width=9cm, height=8cm]{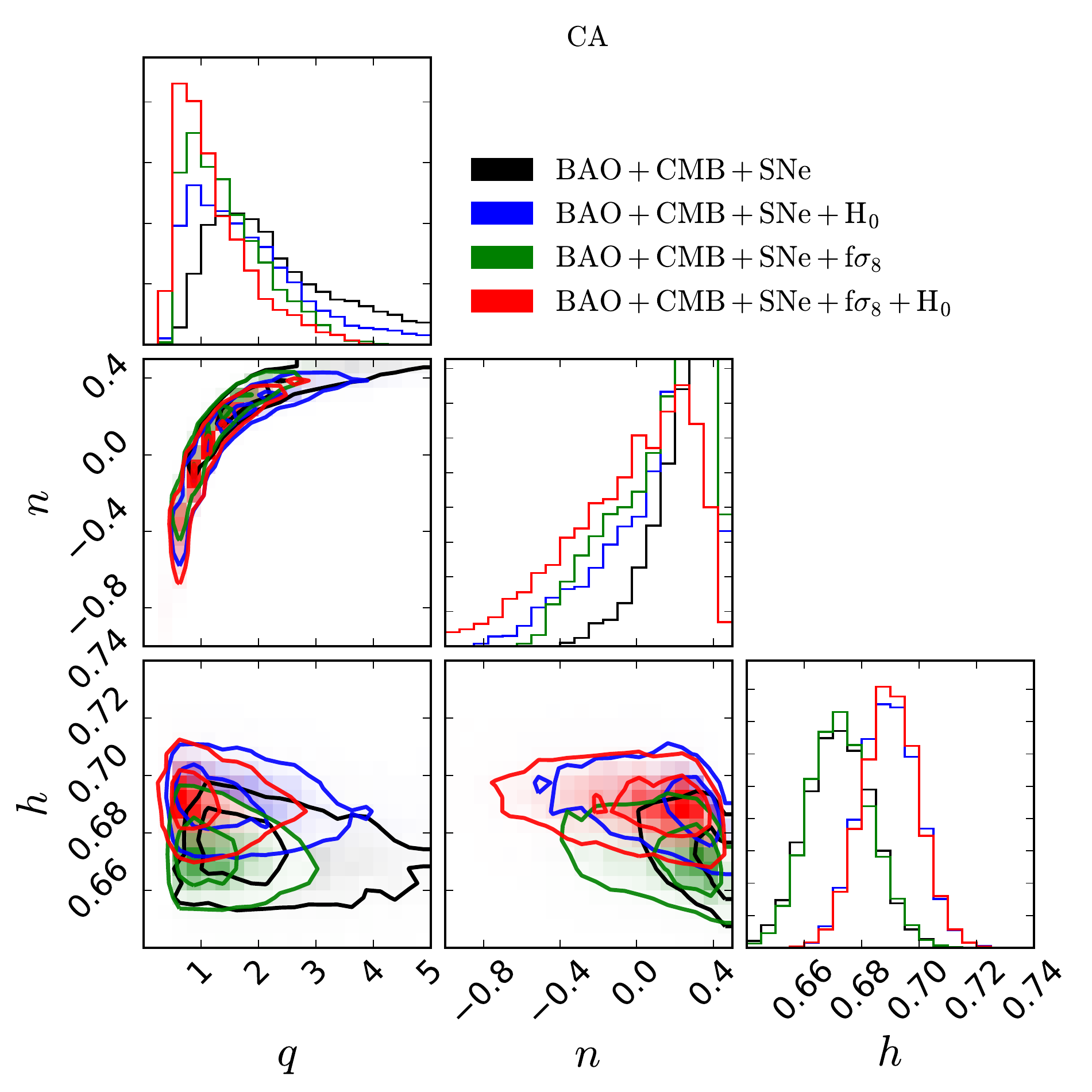}
\caption{The 68.7\% and 95.0\% confidence regions of the parameters for the Cardassian expansion. The diagonal panels show the one-dimensional probability distribution functions.}
\label{fig:Card_mcmc}
\end{center}
\end{figure}

The constraints on the parameters in this paper are shown in 
Figure \ref{fig:Card_mcmc} and Table \ref{tab:Card}. The deviation from 
$\Lambda$CDM model with $q=1$ and $n=0$ is not significant. 
The degeneracy between $q$ and $n$ is clear in various data combinations;
generally a broad range of each is allowed as long as the value 
for the other is such that the expansion and growth rate remain 
similar to $\Lambda$CDM. 

This property also appears in the $p-$value test. The tiny allowed deviation 
from the $\Lambda$CDM  model means that it behaves similarly in terms
of compatibility with the data.

This model behaves similarly to parametrizations such as CPL in terms
of its predicted observables given current data (the results look similar
to Figure \ref{fig:CPL_obs}). 


\begin{table*}
\centering
\begin{tabular}{lllll}
\hline
\multicolumn{5}{c}{Cardassian expansion} \\
\cline{1-5}
Data     &    $q$    &     $n$   & $\Delta$BIC  &  $\Delta$AIC  \\
\hline
BAO+CMB+SNe                                        & $2.3_{-1.0}^{+2.7}$   &  $0.36_{-0.20}^{+0.10}$     &  9.0  & 3.7 \\
BAO+CMB+SNe+H$_0$                           & $1.6_{-0.8}^{+1.3}$   &  $0.21_{-0.35}^{+0.16}$     &  10.2  & 4.8\\
BAO+CMB+SNe+$f\sigma_{8}$                & $1.3_{-0.5}^{+0.9}$   &  $0.19_{-0.34}^{+0.18}$    & 9.8  &  4.3 \\
BAO+CMB+SNe+$f\sigma_{8}$+H$_0$    & $1.1_{-0.4}^{+0.8}$   & $0.02_{-0.41}^{+0.26}$      &  11.1  & 5.6 \\
\hline
\end{tabular}
\caption{Cosmological constraints for a selection of parameters for the Cardassian expansion model.}
\label{tab:Card}
\end{table*}

\subsection{Early dark energy (EDE)}

\begin{figure}[htbp]
\begin{center}
\includegraphics[width=9cm, height=7cm]{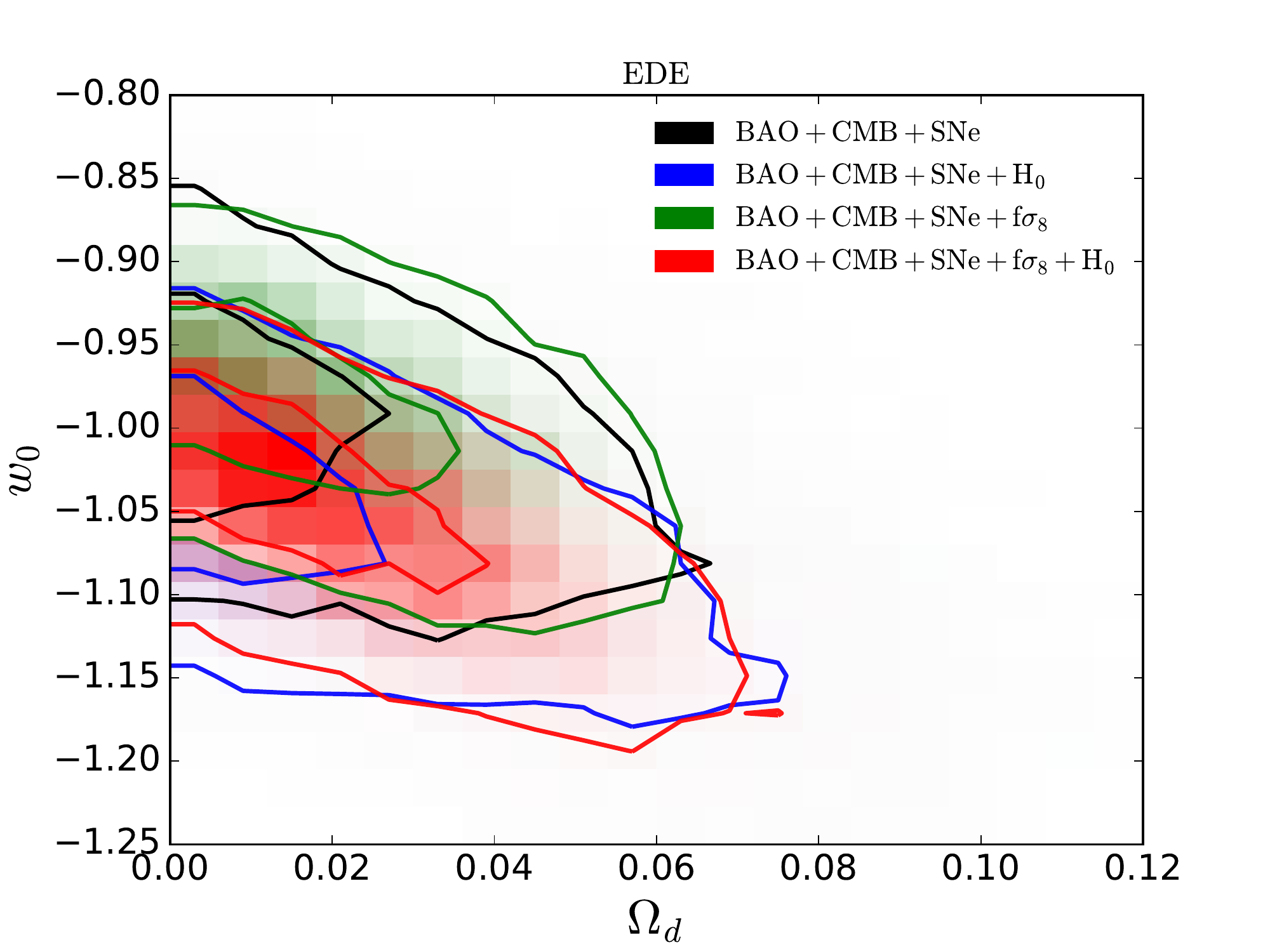}
\caption{The 68.7\% and 95.0\% confidence regions of the parameters $\Omega_{de}^{e}$ and $w_{0}$ for non-flat early dark energy model. }
\label{fig:EDE_mcmc}
\end{center}
\end{figure}

Figure \ref{fig:EDE_mcmc} and Table \ref{tab:EDE} present constraints 
on the early dark energy model from the data. The dark energy component 
has effects on early times and leaves its footprint on the matter 
perturbation. However, the data constraint suggests a small energy 
fraction, with $\Omega_{de}^{e}$ less than 0.07 at 95\% confidence
(in agreement with \citealt{Doran_2006}). Explaining 
the linear growth data does not require this early energy, though
all the measurements are compatible with some modest amount of it.
The different data combinations give consistent 
results regarding $\Omega_{de}^{e}$.  As emphasized in \cite{Aubourg_2015, 
Shi_2016}, a detailed analysis of the CMB power spectrum (not limited
to the ``compressed'' CMB data used here) can  impose much tighter 
constraints on early dark energy, at least for this specific model.

The nearly-negligible contribution of early dark energy doesn't improve 
the fit to the data significantly, therefore the limited improvement of 
$\chi^2$ is not able to offset the added model complexity, and the 
$p-$value for this model is comparable to that for the $\Lambda$CDM model. 

Early dark energy models constrained by current data have predicted 
observables with a range similar to that shown in Figure \ref{fig:CPL_obs}
for the CPL parametrization.

\begin{table*}
\centering
\begin{tabular}{lllll}
\hline
\multicolumn{5}{c}{Early dark energy} \\
\cline{1-5}
Data     &    $\Omega_{d}$    &     $w_{0}$  & $\Delta$BIC  &  $\Delta$AIC \\
\hline
BAO+CMB+SNe                                        & $0.0^{+0.025}$   &  $-1.01_{-0.07}^{+0.06}$     & 9.9  &  4.6\\
BAO+CMB+SNe+H$_0$                           & $0.0^{+0.029}$   &  $-1.06_{-0.07}^{+0.06}$      & 10.5  &  5.1\\
BAO+CMB+SNe+$f\sigma_{8}$                & $0.0^{+0.023}$   &  $-1.00\pm0.06$     & 9.8  &  4.3\\
BAO+CMB+SNe+$f\sigma_{8}$+H$_0$    & $0.0^{+0.025}$   & $-1.06_{-0.07}^{+0.06}$       &  11.1  & 5.5\\
\hline
\end{tabular}
\caption{Cosmological constraints for a selection of parameters for the Early dark energy model.}
\label{tab:EDE}
\end{table*}

\subsection{Slow roll dark energy (SR)}

\begin{figure}[htbp]
\begin{center}
\includegraphics[width=9cm, height=8cm]{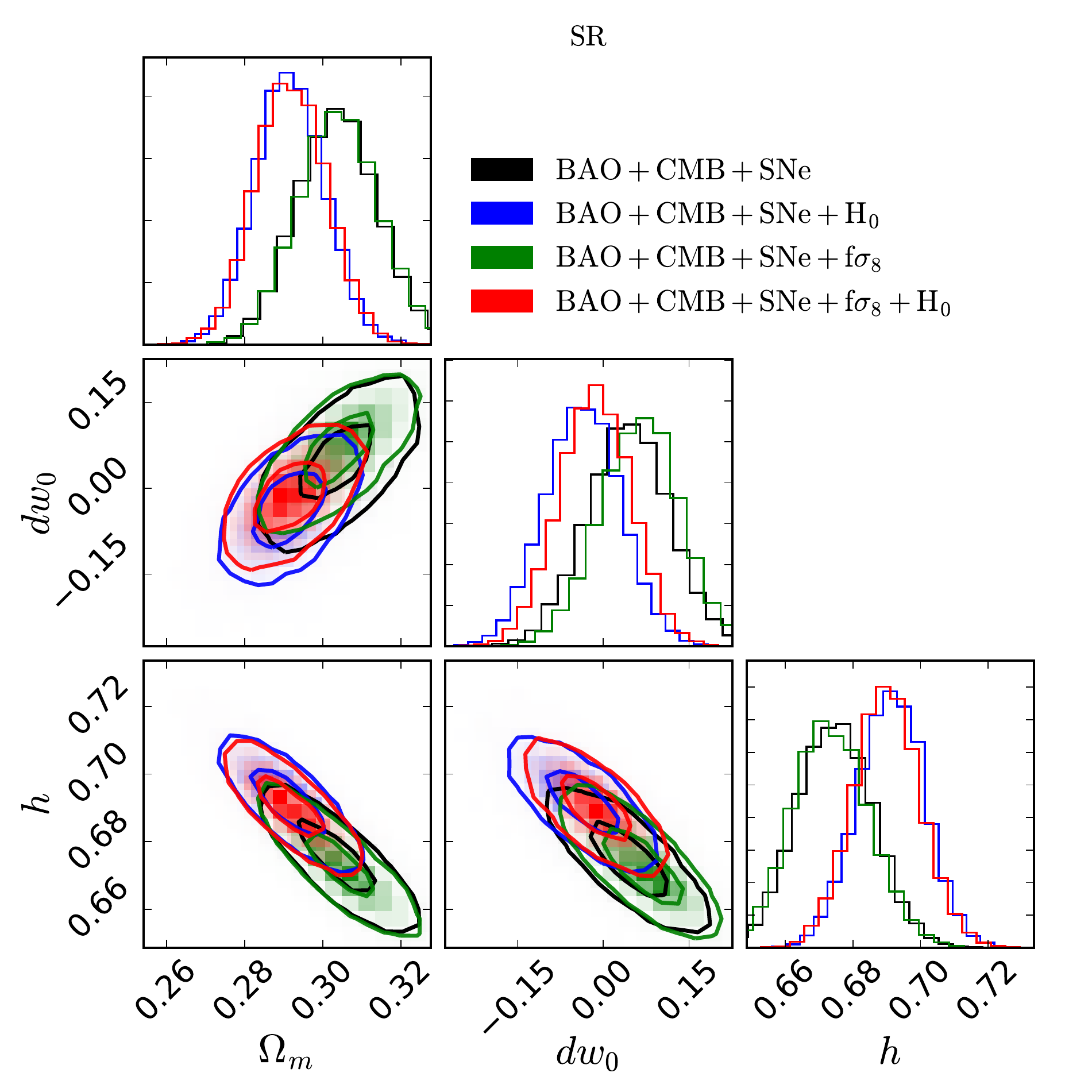}
\caption{The 68.7\% and 95.0\% confidence regions of the parameters $\Omega_{m}$ and $dw_{0}$ for non-flat slow roll dark energy model. The diagonal panels show the one-dimensional probability distribution functions.}
\label{fig:SL_DE_mcmc}
\end{center}
\end{figure}

The slow roll dark energy scenario has the same number of parameters as 
the XCDM model and is very similar in form, especially for small deviations
from $w=-1$. The constraints are shown in Figure \ref{fig:SL_DE_mcmc}, 
which yields $\delta w_{0}=0.01\pm0.06$ for all datasets combined, and
no significant departure from zero for any combination of data.
Table \ref{tab:SL_DE} presents the mean and uncertainties of the 
parameters for the slow roll dark energy model.

The $p-$value for this model is quite similar to the early dark energy 
model, as their deviation from the $\Lambda$CDM model is constrained to be small.
The predicted observables given current data are even tighter, very similar
to that found for non-flat $\Lambda$CDM in Figure \ref{fig:LCDM_obs}.

\begin{table*}
\centering
\begin{tabular}{lllll}
\hline
\multicolumn{5}{c}{Slow roll dark energy} \\
\cline{1-5}
Data     &    $\Omega_{m}$    &     $\delta w_{0}$  & $\Delta$BIC  &  $\Delta$AIC \\
\hline
BAO+CMB+SNe                                        & $0.304\pm0.011$   &  $0.04\pm0.07$    &  5.9  &  2.4 \\
BAO+CMB+SNe+H$_0$                           & $0.291\pm0.009$   &  $-0.04\pm0.07$   & 6.8  &  3.3   \\
BAO+CMB+SNe+$f\sigma_{8}$                & $0.304\pm0.010$   &  $0.07\pm0.07$    & 5.8  & 2.2 \\
BAO+CMB+SNe+$f\sigma_{8}$+H$_0$    & $0.292\pm0.009$   & $-0.01\pm0.06$     & 7.3  & 3.6  \\
\hline
\end{tabular}
\caption{Cosmological constraints for a selection of parameters for the Slow roll dark energy model.}
\label{tab:SL_DE}
\end{table*}

\subsection{Parameterization of the Hubble parameter (PolyCDM, HLG)}

\begin{figure}[htbp]
\begin{center}
\includegraphics[width=9cm, height=8cm]{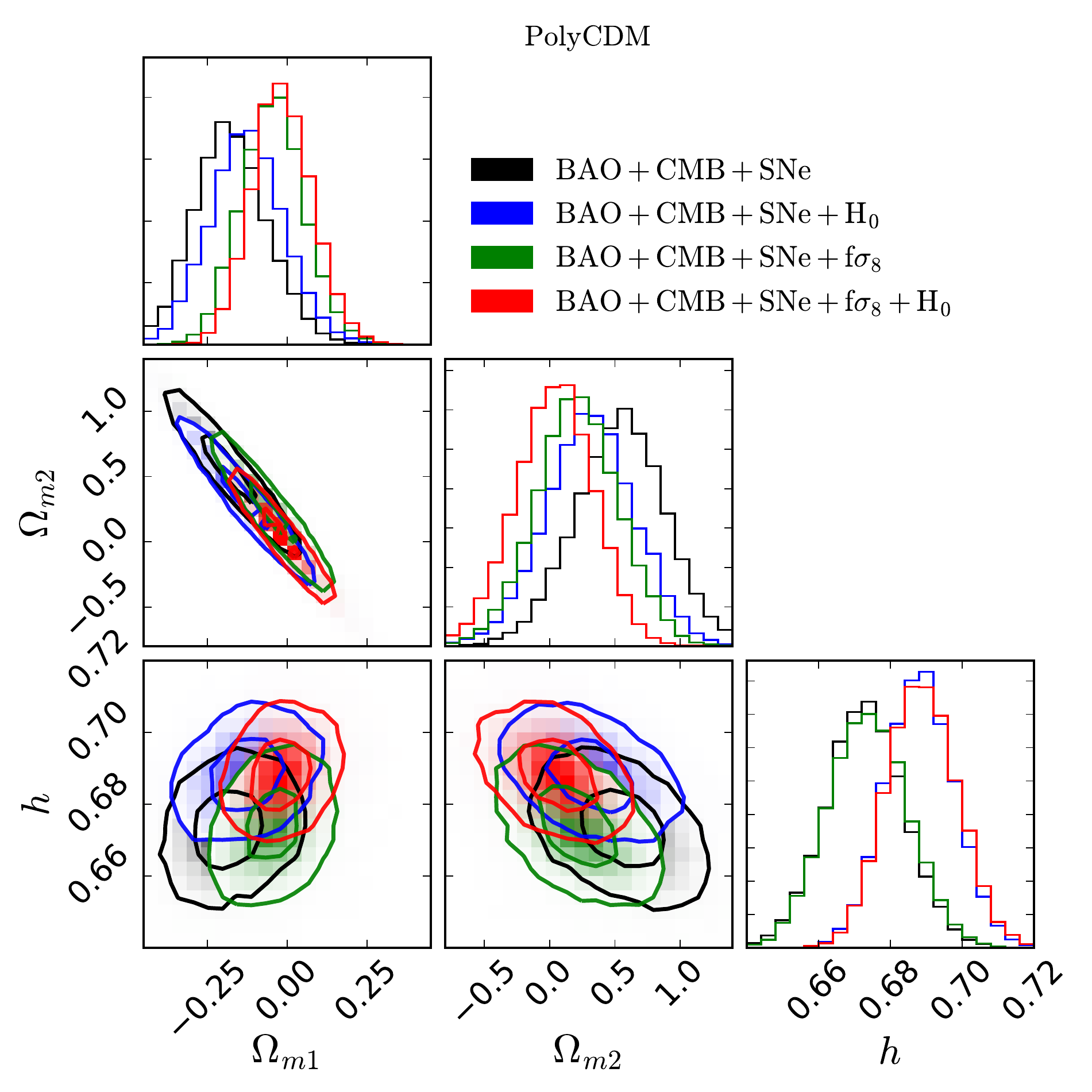}
\includegraphics[width=9cm, height=8cm]{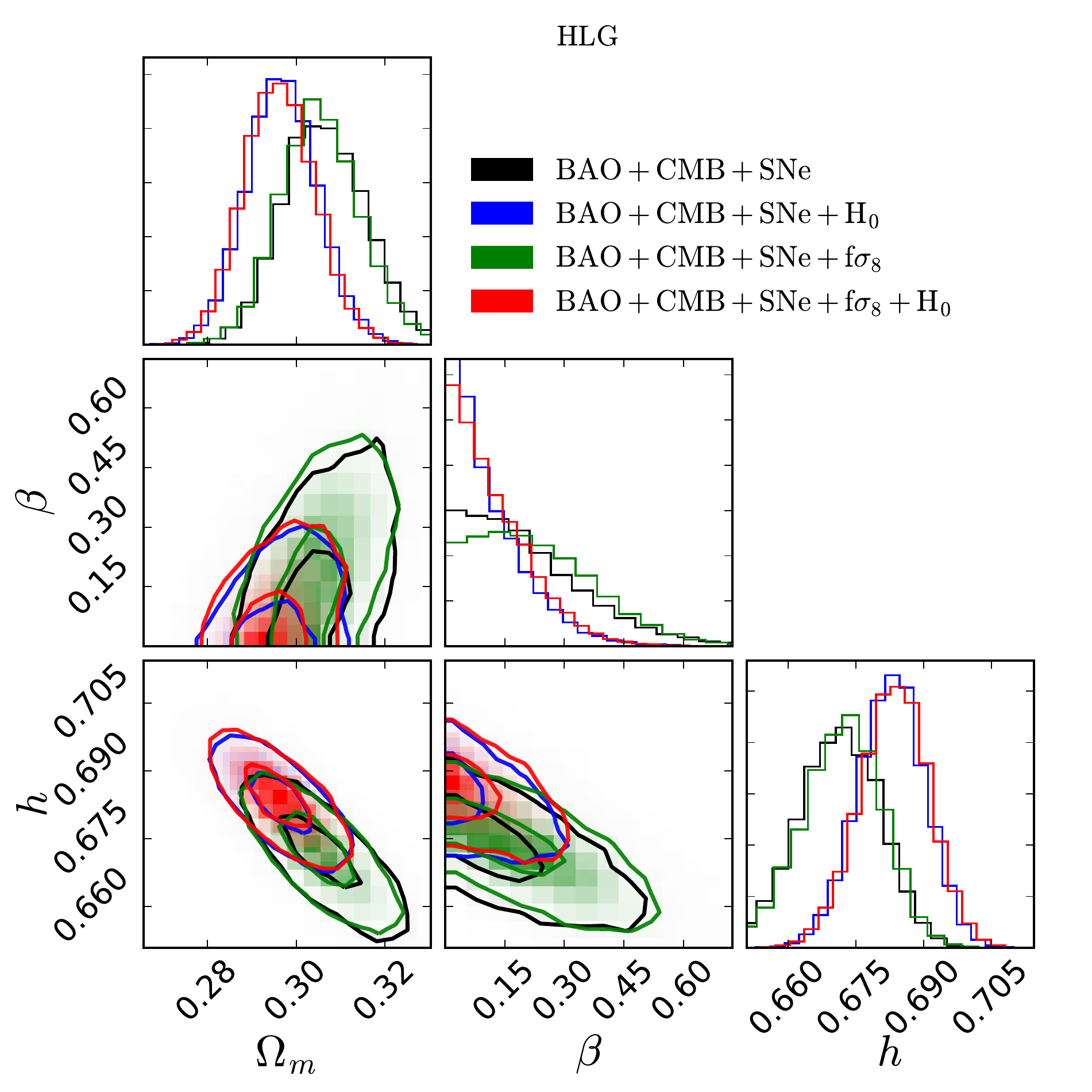}
\caption{The 68.7\% and 95.0\% confidence regions of the parameters for two parameterizations 
of the Hubble parameter: the PolyCDM model (Top panel) and the Logarithmic model 
(Bottom panel). The diagonal panels show the one-dimensional probability distribution 
functions.}
\label{fig:poly_mcmc}
\end{center}
\end{figure}

Figure \ref{fig:poly_mcmc} presents the constraints on the 
parameters of two phenomenological parameterizations of the 
Hubble parameter: the PolyCDM model (Eq. \ref{eq:poly}) and 
the HLG model (Eq. \ref{eq:log}). The figure shows 
clear differences in best fit parameters depending on the datasets 
used for the PolyCDM model; because the model is quite flexible, it results
in quite different parameters if some data is excluded (though none
that are more than $2\sigma$ away from $\Lambda$CDM). 
As the measurements of Hubble constant and linear growth are 
added, the parameters $\Omega_{m1}$ and $\Omega_{m2}$ approach 
0.0, which recovers the standard $\Lambda$CDM model. In other 
words, a non-zero value of these two parameters can 
fit certain datasets better than $\Lambda$CDM, but there is no
such choice when all the datasets are included. The same effect is found 
in the logarithmic model. For different data samples, the 
$\Lambda$CDM model is well within $1\sigma$ confidence region. 
Inclusion of H$_0$ pulls $h$ to higher values but does not 
resolve the tension of H$_0$ with the other data. Table 
\ref{tab:Poly} and \ref{tab:Log}  summarize the constraint results 
for these two models respectively.

The $p-$value test of these two models reveal similar patterns. For 
instance, when the local measurement of H$_{0}$ and linear growth 
data are excluded, the PolyCDM model gives a better $p-$value 
than $\Lambda$CDM model. As more data are added, PolyCDM is 
constrained to be closer to $\Lambda$CDM and this difference
is reduced. 

\begin{figure}[t!]
\begin{center}
\includegraphics[width=0.49\textwidth]{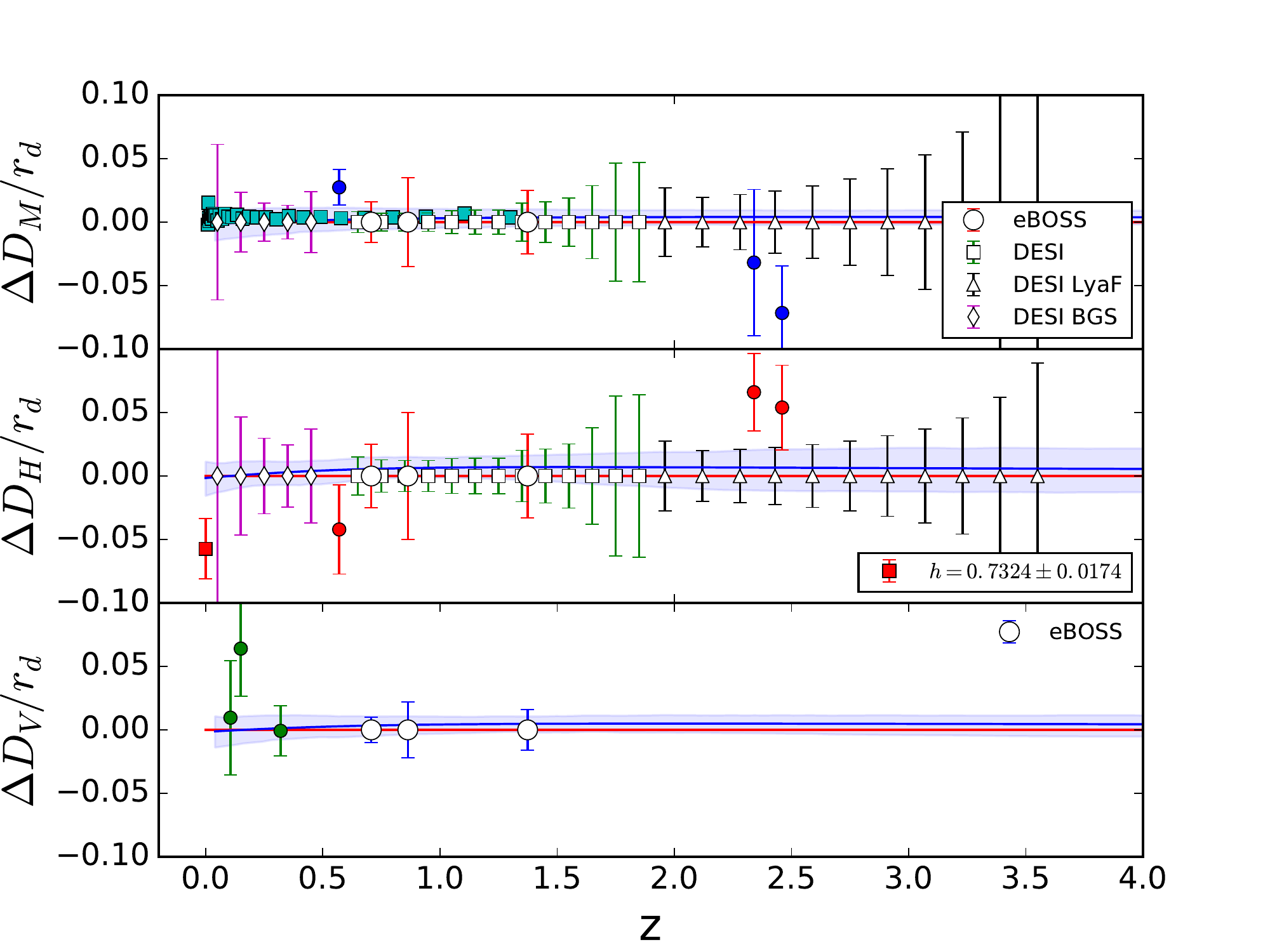}
\includegraphics[width=0.49\textwidth]{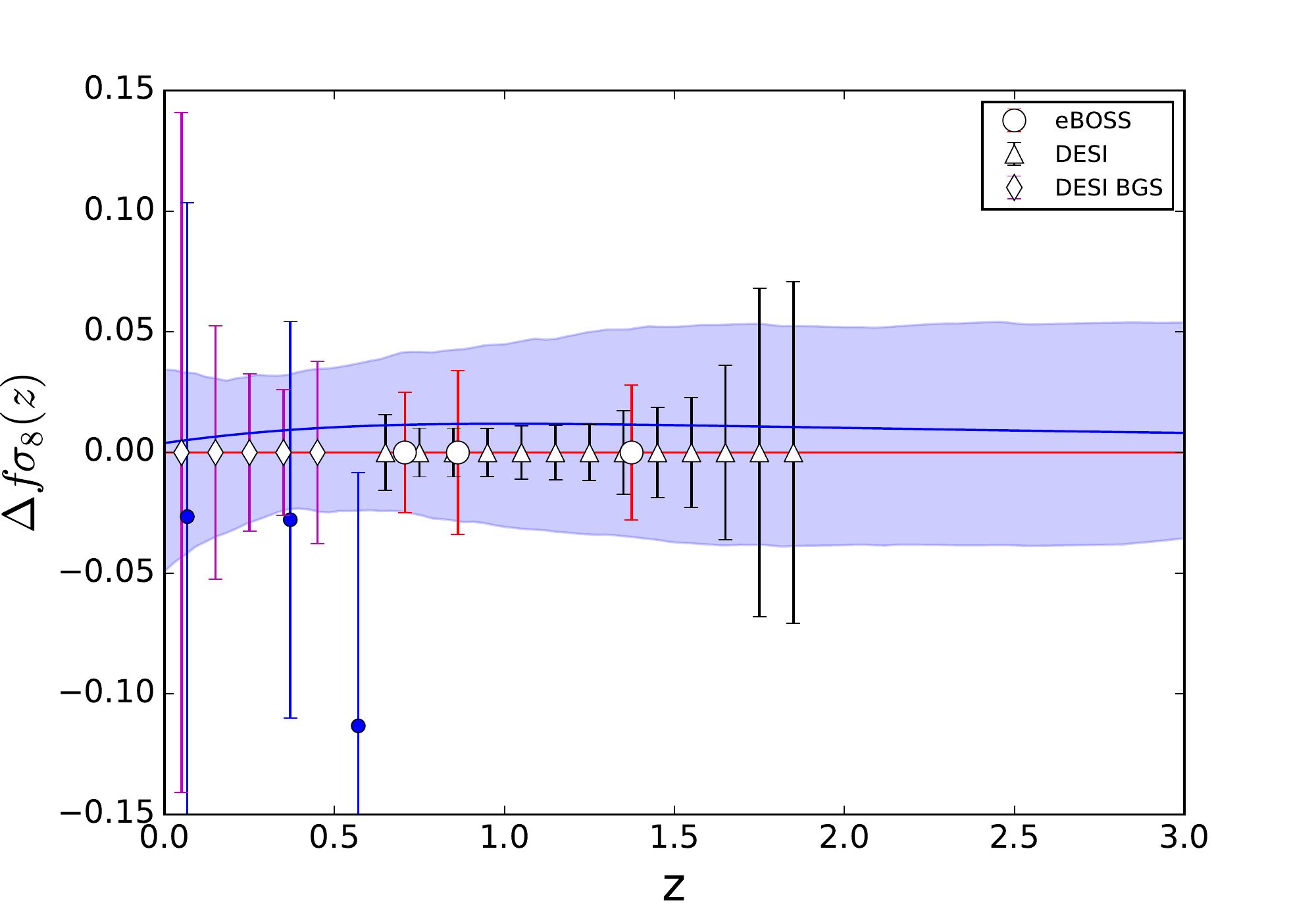}
\caption{\label{fig:PolyCDM_obs} Similar to Figure \ref{fig:LCDM_obs}, for
the PolyCDM model.}
\end{center}
\end{figure}

Because of the increased freedom in PolyCDM, the predicted 
observables are somewhat broader than other generalizations like
CPL. Figure \ref{fig:PolyCDM_obs} shows these predicted observables. 
On the other hand, HLG has predicted observables similar to those
for $\Lambda$CDM (Figure \ref{fig:LCDM_obs}).

\begin{table*}
\centering
\begin{tabular}{lllll}
\hline
\multicolumn{5}{c}{PolyCDM model} \\
\cline{1-5}
Data     &    $\Omega_{m1}$    &     $\Omega_{m2}$  & $\Delta$BIC  &  $\Delta$AIC \\
\hline
BAO+CMB+SNe                                        & $-0.19_{-0.11}^{+0.12}$   &  $0.55_{-0.35}^{+0.33}$    &  7.5  & 2.2 \\
BAO+CMB+SNe+H$_0$                           & $-0.13_{-0.11}^{+0.12}$   &  $0.31\pm0.34$      & 8.7   &  3.3\\
BAO+CMB+SNe+$f\sigma_{8}$                & $-0.05\pm0.10$              &  $0.21_{-0.31}^{+0.32}$    &  9.7  &  4.2 \\
BAO+CMB+SNe+$f\sigma_{8}$+H$_0$    & $-0.02_{-0.09}^{+0.10}$  & $0.04\pm0.30$       & 11.1  &  5.5\\
\hline
\end{tabular}
\caption{Cosmological constraints for a selection of parameters for the PolyCDM dark energy model.}
\label{tab:Poly}
\end{table*}

\begin{table*}
\centering
\begin{tabular}{lllll}
\hline
\multicolumn{5}{c}{Logarithmic parameterization} \\
\cline{1-5}
Data     &    $\Omega_{m}$    &     $\beta$  & $\Delta$BIC  &  $\Delta$AIC \\
\hline
BAO+CMB+SNe                                        & $0.306\pm0.009$   &  $0.04^{+0.18}$     & 6.2  & 2.7 \\
BAO+CMB+SNe+H$_0$                           & $0.296\pm0.008$   &  $0.00^{+0.12}$     &  7.1  &  3.6 \\
BAO+CMB+SNe+$f\sigma_{8}$                & $0.305\pm0.009$   &  $0.13^{+0.17}$    &  6.1  & 2.4 \\
BAO+CMB+SNe+$f\sigma_{8}$+H$_0$    & $0.296\pm0.008$  &  $0.00^{+0.12}$     & 7.3   & 3.6  \\
\hline
\end{tabular}
\caption{Cosmological constraints for a selection of parameters for the logarithmic parameterization of the Friedmann equation.}
\label{tab:Log}
\end{table*}

\subsection{Chaplygin gas (CG, GCG, MCG)}

\begin{figure}[htbp]
\begin{center}
\includegraphics[width=9cm, height=8cm]{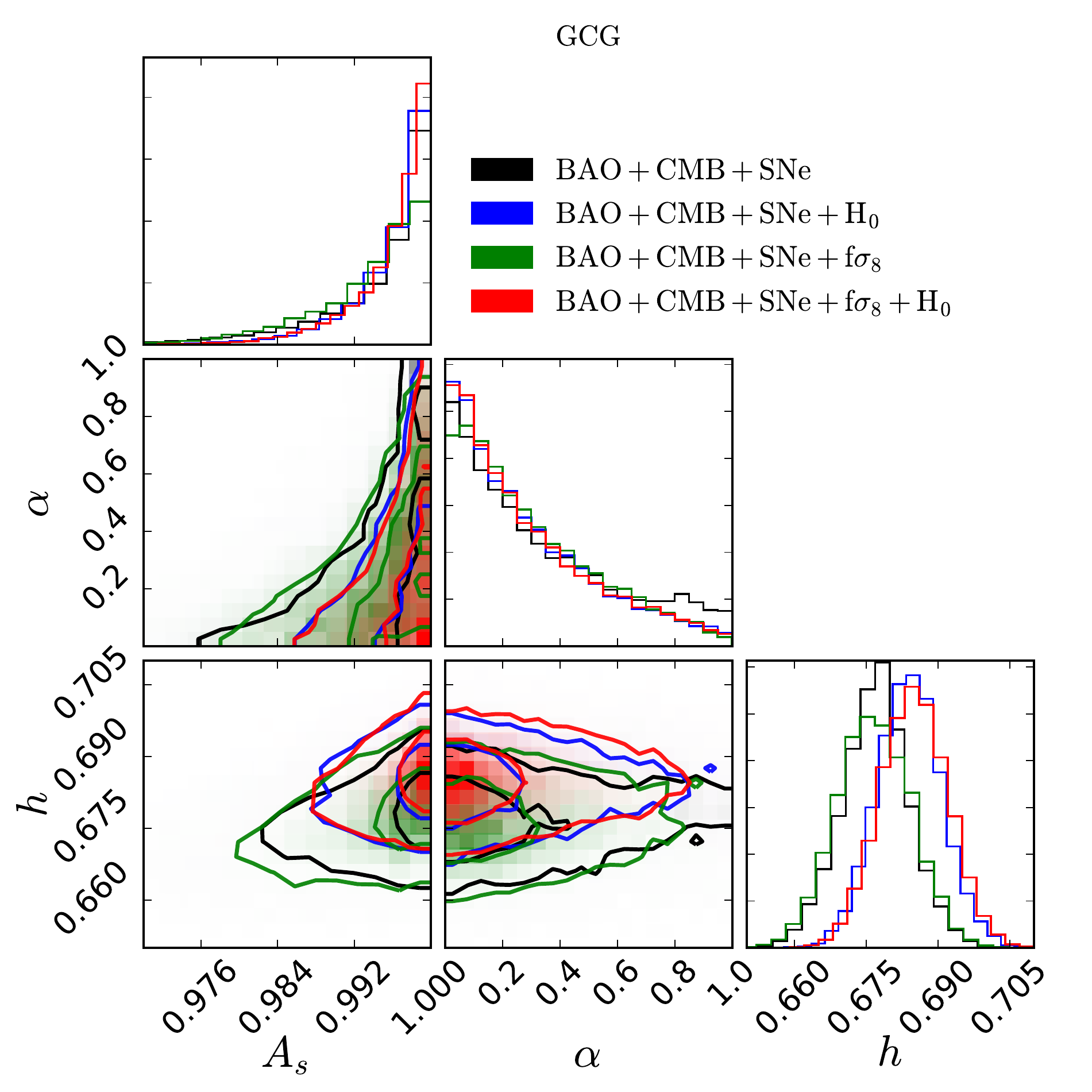}
\caption{The 68.7\% and 95.0\% confidence regions of the parameters for the Generalized 
Chaplygin gas model. The diagonal panels show the one-dimensional probability distribution functions.}
\label{fig:gcg_mcmc}
\end{center}
\end{figure}

The Chaplygin gas model has several generalizations that can produce 
cosmic acceleration.
We present constraints on the Generalized Chaplygin gas model (GCG) parameters
in Figure \ref{fig:gcg_mcmc}, but the results for the other Chaplygin 
gas models are similar. This model does not show significant deviations 
from the $\Lambda$CDM model. 
The allowed deviation from a cosmological constant 
is small and the $\Lambda$CDM model is well within the $1\sigma$ 
confidence regions. Table \ref{tab:CG}, \ref{tab:GCG} and \ref{tab:MCG} 
summarize the constraints on the parameters for these Chaplygin gas models. 
For recent discussions about the cosmological constraints of this model
see \cite{Park_2010}, \cite{Paul_2013}, \cite{Wang_2013}, \cite{Lu_2015}, 
\cite{Sharov_2016}, and references  therein.

The Chaplygin gas models give the most favorable
$p$-values relative to other models when only the geometrical data are 
included. Adding  either H$_{0}$ data or linear growth data makes the
$p$-value less favorable and more comparable to other models. 

The predicted observables for this model are similar to that for CPL (Figure
\ref{fig:CPL_obs}).  As expected, the GCG and MCG models show somewhat
more variation in their predictions.

\begin{table*}
\centering
\begin{tabular}{lllll}
\hline
\multicolumn{5}{c}{CG} \\
\cline{1-5}
Data     &    $\Omega_{m}$    &     $h$  & $\Delta$BIC  &  $\Delta$AIC \\
\hline
BAO+CMB+SNe                                        & $0.301\pm0.008$   &  $0.677\pm0.007$    &  6.3  &  2.7 \\
BAO+CMB+SNe+H$_0$                           & $0.295\pm0.008$   &  $0.684\pm0.007$     & 7.1  &  3.6 \\
BAO+CMB+SNe+$f\sigma_{8}$                & $0.302\pm0.008$   &  $0.677\pm0.007$    &  6.0  & 2.3 \\
BAO+CMB+SNe+$f\sigma_{8}$+H$_0$    & $0.295\pm0.008$  &  $0.684\pm0.007$      &  7.1  & 3.4 \\
\hline
\end{tabular}
\caption{Cosmological constraints for a selection of parameters for the Chaplygin gas model. The model parameter $A_s$ is found to be constrained tightly in the range (0.999, 1).}
\label{tab:CG}
\end{table*}

\begin{table*}
\centering
\begin{tabular}{lllll}
\hline
\multicolumn{5}{c}{GCG} \\
\cline{1-5}
Data     &    $\Omega_{m}$    &     $\alpha (68\% \rm{CL})$   & $\Delta$BIC  &  $\Delta$AIC\\
\hline
BAO+CMB+SNe                                        & $0.302\pm0.008$                  &  $0.0<\alpha<0.43$    &  10.0  &  4.7 \\
BAO+CMB+SNe+H$_0$                           & $0.296_{-0.007}^{+0.008}$   &  $0.0<\alpha<0.35$    & 10.9  &  5.6  \\
BAO+CMB+SNe+$f\sigma_{8}$                & $0.302_{-0.008}^{+0.009}$   &  $0.0<\alpha<0.38$    & 9.8  & 4.3  \\
BAO+CMB+SNe+$f\sigma_{8}$+H$_0$    & $0.295\pm0.008$                 &  $0.0<\alpha<0.35$    &  11.0  & 5.4   \\
\hline
\end{tabular}
\caption{Cosmological constraints for a selection of parameters for the Generalized Chaplygin gas model.}
\label{tab:GCG}
\end{table*}

\begin{table*}
\centering
\begin{tabular}{lllll}
\hline
\multicolumn{5}{c}{MCG} \\
\cline{1-5}
Data     &    $\alpha (68\% \rm{CL})$    &     $B (68\% \rm{CL})$   \\
\hline
BAO+CMB+SNe                                        & $0.0<\alpha<0.28$                  &  $0.0<B<0.16$   & 13.8  &  6.7  \\
BAO+CMB+SNe+H$_0$                           & $0.0<\alpha<0.33$                  &  $0.0<B<0.14$    &  14.7  & 7.6  \\
BAO+CMB+SNe+$f\sigma_{8}$                & $0.0<\alpha<0.36$                  &  $0.0<B<0.15$    &  13.6  &  6.3 \\
BAO+CMB+SNe+$f\sigma_{8}$+H$_0$    & $0.0<\alpha<0.36$                 &  $0.0<B<0.16$      & 14.8  & 7.4 \\
\hline
\end{tabular}
\caption{Cosmological constraints for a selection of parameters for the Modified Chaplygin gas model.}
\label{tab:MCG}
\end{table*}

\subsection{Interacting DE and DM (IDE$_{\rm 1}$, IDE$_{\rm 2}$, IDE$_{\rm 3}$)}

\begin{figure}[htbp]
\begin{center}
\includegraphics[width=8.5cm, height=7.5cm]{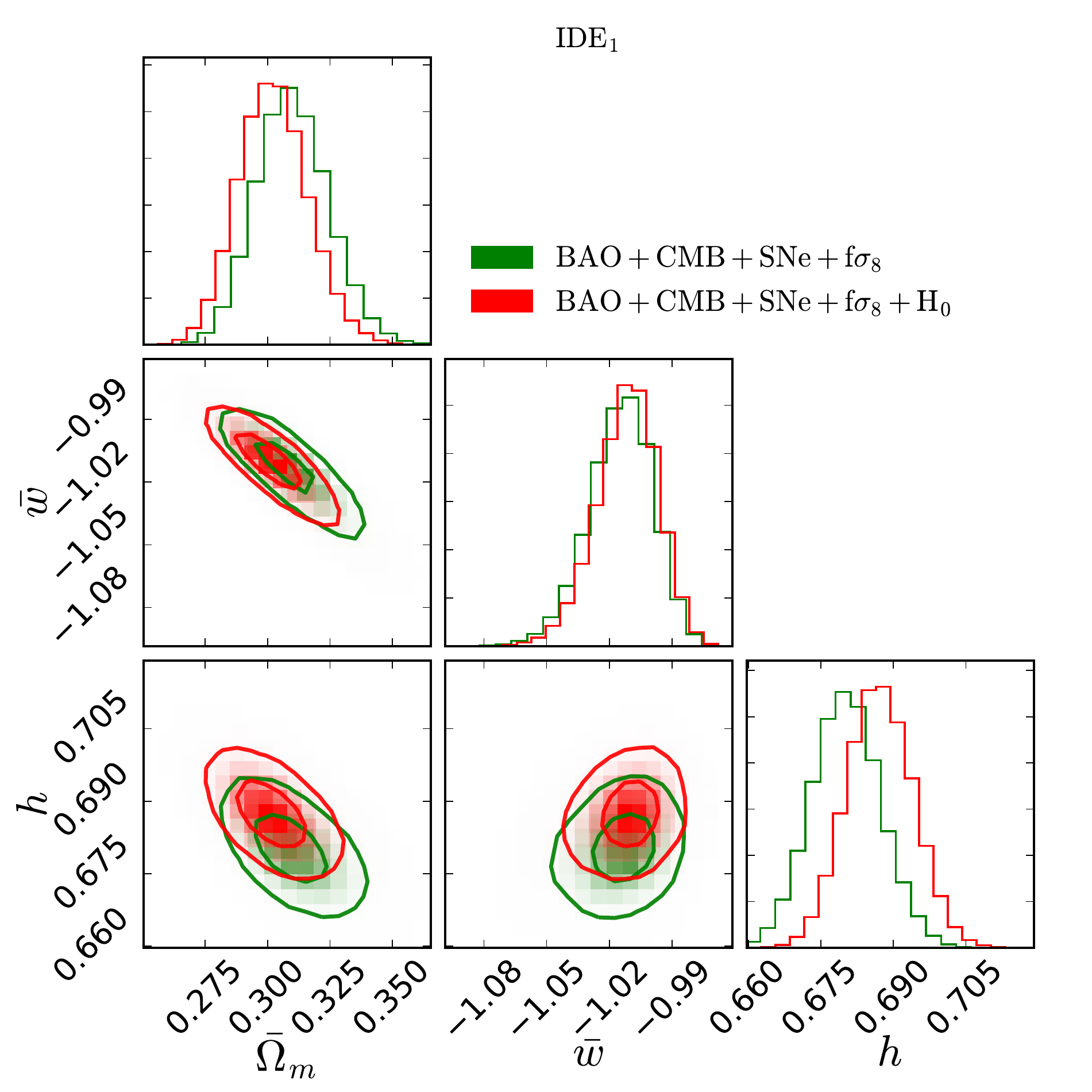}
\includegraphics[width=8.5cm, height=7.5cm]{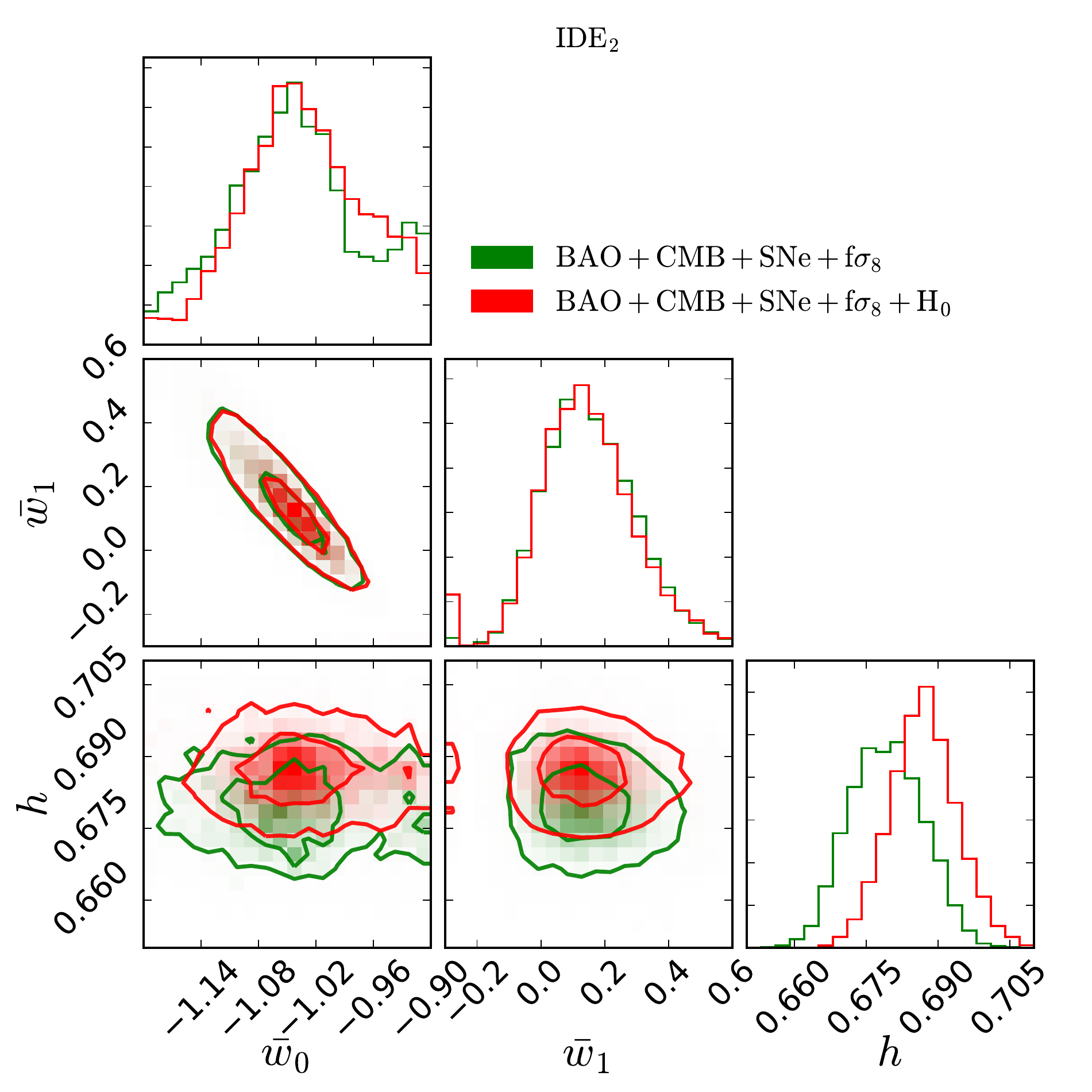}
\includegraphics[width=8.5cm, height=7.5cm]{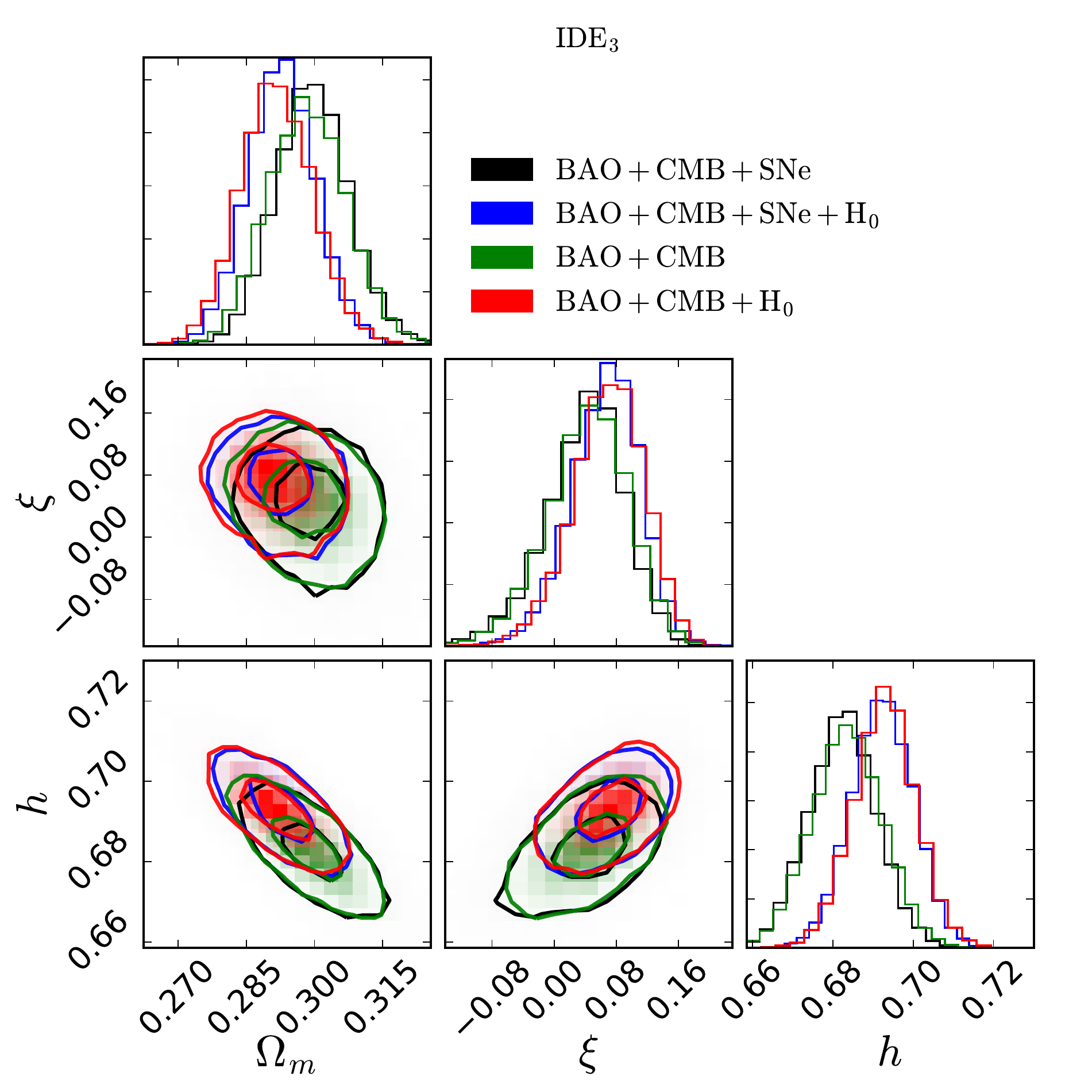}
\caption{The 68.7\% and 95.0\% confidence regions of the parameters for two coupled 
dark energy models: Model I (Top panel) and Model II (Middle panel). The bottom panel shows a coupled dark energy and dark matter model which is not required to mimic the expansion of $\Lambda$ model. The diagonal 
panels show the one-dimensional probability distribution functions.}
\label{fig:coup_mcmc}
\end{center}
\end{figure}

Figure \ref{fig:coup_mcmc} shows the constraints on the coupled dark energy 
models. This model mimics the expansion of the $\Lambda$CDM model.
The constraint on the coupling between DE and DM is through the 
linear growth data. We here consider two parameterizations of the equation 
of state of the coupled dark energy. Compared with \cite{Fay_2016}, which uses 
linear growth data only, the combination with the geometrical probes provide 
tighter constraints, as expected due to the better constraint on the dark 
matter component. Similarly, the degeneracy between the parameters is also 
consistent with the result using growth data only. The constraints on this 
model are shown in Table \ref{tab:coup1} and \ref{tab:coup2}. These models
do not relieve the tension of H$_0$ with the other data. 

\begin{figure}[t!]
\begin{center}
\includegraphics[width=0.49\textwidth]{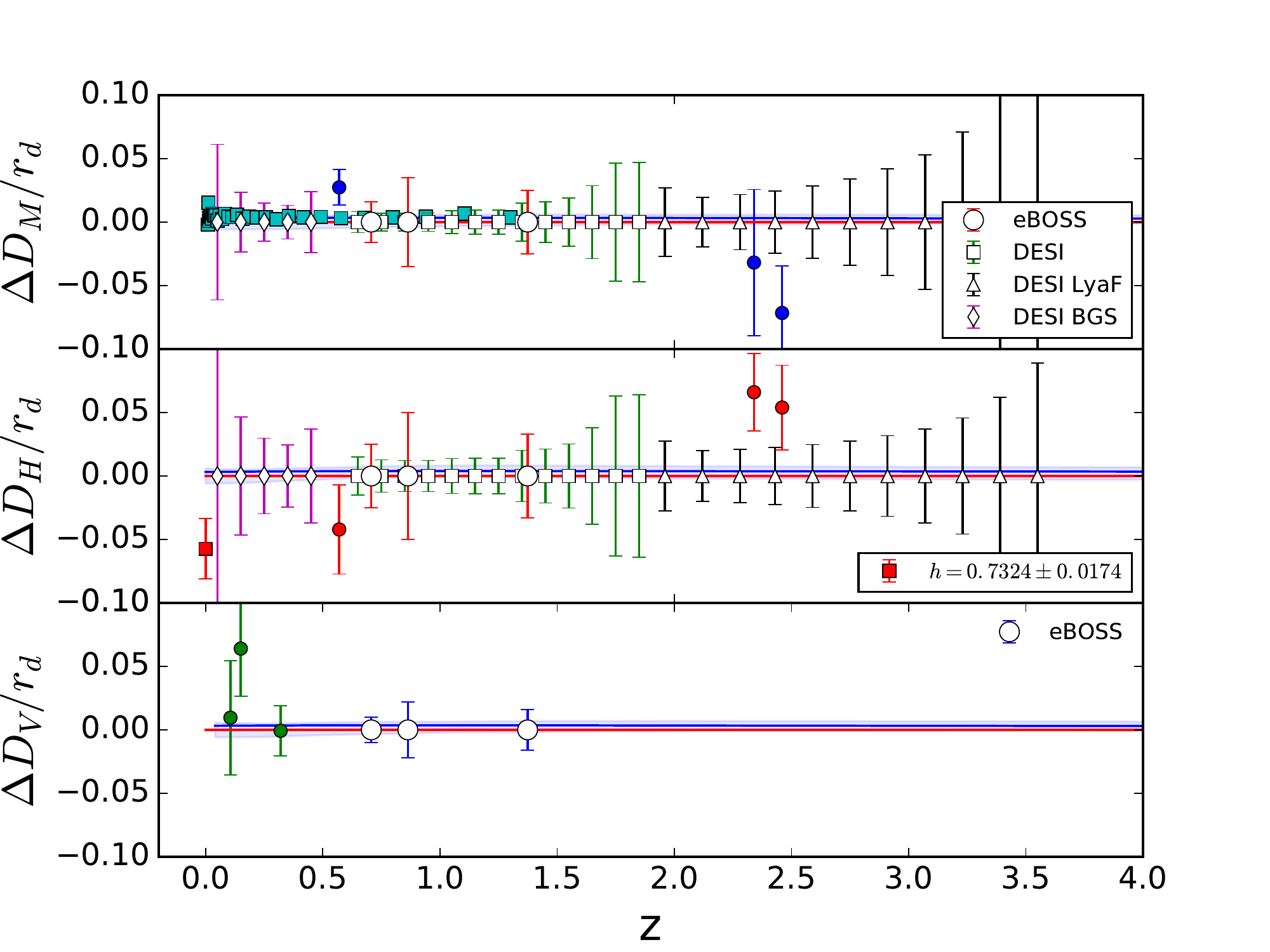}
\includegraphics[width=0.49\textwidth]{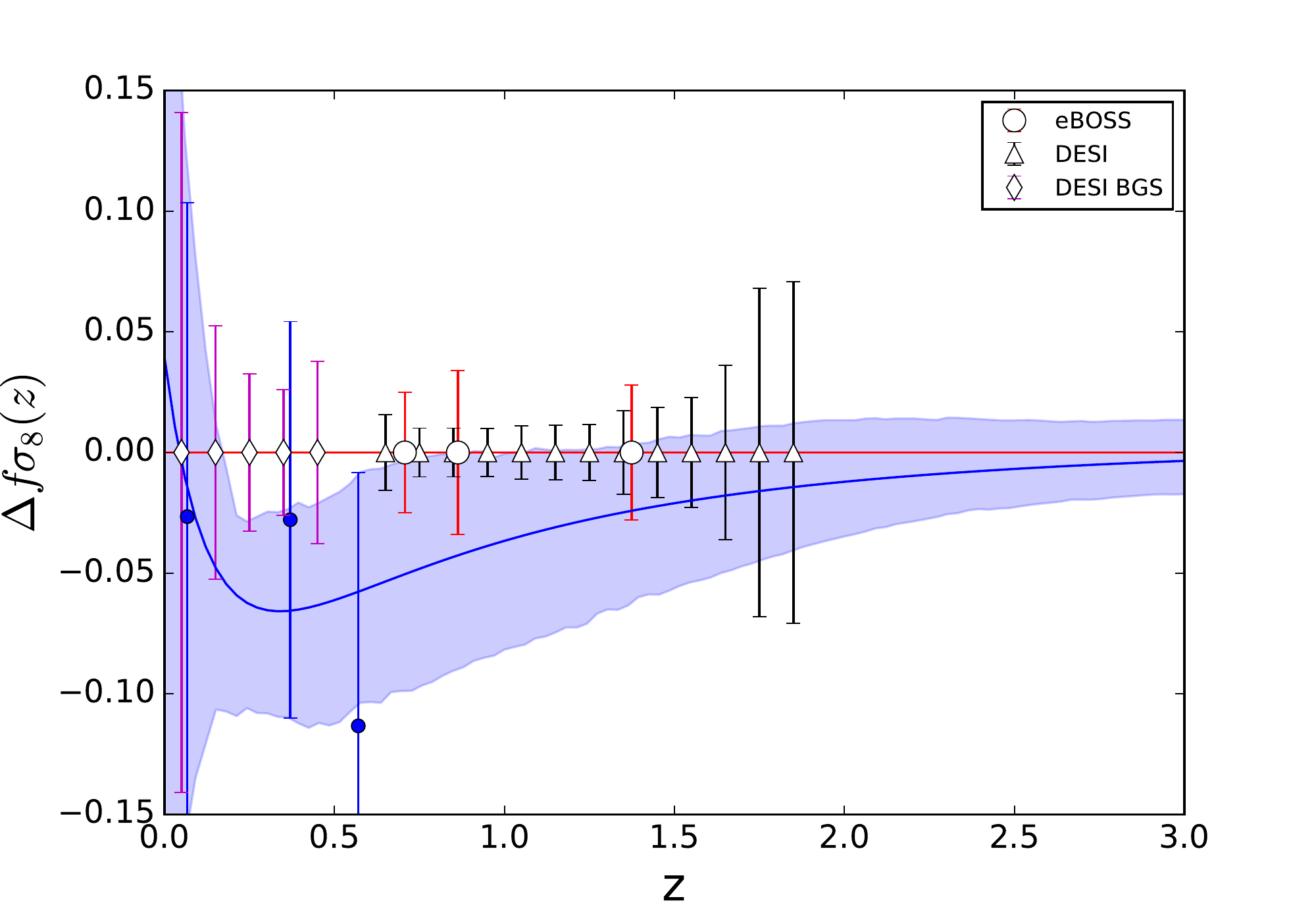}
\caption{\label{fig:IDE2_obs} Similar to Figure \ref{fig:LCDM_obs}, for
the IDE$_{\rm 2}$ model.}
\end{center}
\end{figure}

From the $p$-value analysis, without $f\sigma_8$ IDE$_2$ performs 
similarly to $\Lambda$CDM. However,  it is able
to fit the observed  $f\sigma_8$  a little better than other models,
which makes it one of the models with highest $p$-value when all 
data is included.
The predicted observables, shown in Figure \ref{fig:IDE2_obs}, demonstrate how
its fit improves over other models for $f\sigma_8$. Clearly additional
data from eBOSS and DESI on the growth of structure will tighten 
this model space further.

It has been suggested that a different dark sector interaction may
be able to resolve the tension
(\citealt{Salvatelli_2013, Salvatelli_2014, Costa_2014, Murgia_2016}). 
As a simple test, 
we apply the geometrical probes to a particular 
coupled dark energy and dark matter model which has the interaction 
term proportional to the dark energy component $Q=\xi H \rho_{DE}$. 
In this model, a larger H$_{0}$ can be obtained by increasing
the coupling $\xi$,
which implies a energy transfer from dark energy 
to dark matter (\citealt{Murgia_2016}).
Note that this model is not required to mimic the expansion as the 
$\Lambda$CDM model. Our result is presented in the bottom panel of Figure 
\ref{fig:coup_mcmc}. The result shows that the improvement of the 
tension of H$_{0}$ is marginal. The use of the non-compressed CMB 
data might resolve the tension by some amount; a more detailed analysis 
would be required to answer this question. The MCMC result of this model 
also presents a positive degeneracy between $h$ and the coupling 
constant $\xi$. \remove{ AWKWARD THAT THIS IS NOT INTRODUCED IN SECTION 2}

\begin{table*}
\centering
\begin{tabular}{lllll}
\hline
\multicolumn{5}{c}{Coupled dark energy: Model I} \\
\cline{1-5}
Data     &    $\bar{\Omega}_{m}$    &     $\bar{w}$   & $\Delta$BIC  &  $\Delta$AIC\\
\hline
BAO+CMB+SNe+$f\sigma_{8}$                & $0.310\pm0.015$   &  $-1.01\pm0.02$     & 5.8  & 2.1\\
BAO+CMB+SNe+$f\sigma_{8}$+H$_0$    & $0.302\pm0.014$  &  $-1.01\pm0.02$     &  6.9  & 3.2  \\
\hline
\end{tabular}
\caption{Cosmological constraints for a selection of parameters for the coupled dark energy model (Model I).}
\label{tab:coup1}
\end{table*}

\begin{table*}
\centering
\begin{tabular}{llllll}
\hline
\multicolumn{6}{c}{Coupled dark energy: Model II} \\
\cline{1-6}
Data     &    $\bar{\Omega}_{m}$    &     $\bar{w}_{0}$ & $\bar{w}_{1}$ & $\Delta$BIC  &  $\Delta$AIC \\
\hline
BAO+CMB+SNe+$f\sigma_{8}$                & $0.335_{-0.028}^{+0.030}$  &  $-1.05\pm0.04$                & $0.14_{-0.13}^{+0.16}$   & 8.7  &  3.2  \\
BAO+CMB+SNe+$f\sigma_{8}$+H$_0$    & $0.327_{-0.026}^{+0.030}$  &  $-1.05\pm0.04$               & $0.14_{-0.12}^{+0.15}$    & 10.0  &  4.4 \\
\hline
\end{tabular}
\caption{Cosmological constraints for a selection of parameters for the coupled dark energy model (Model II).}
\label{tab:coup2}
\end{table*}

\subsection{Weakly-coupled canonical scalar field (WCSF)}

\begin{figure}[htbp]
\begin{center}
\includegraphics[width=9cm, height=8cm]{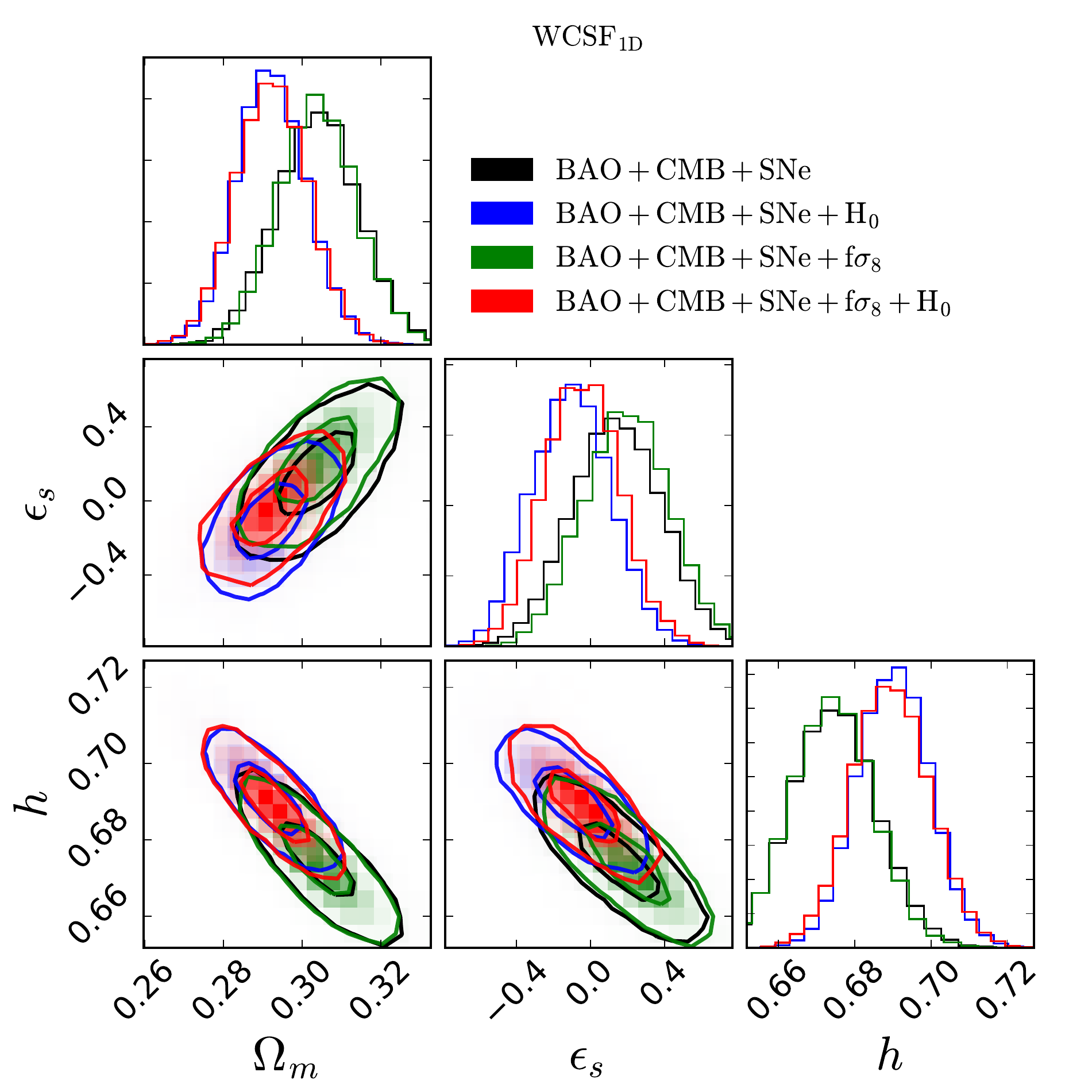}
\includegraphics[width=9cm, height=8cm]{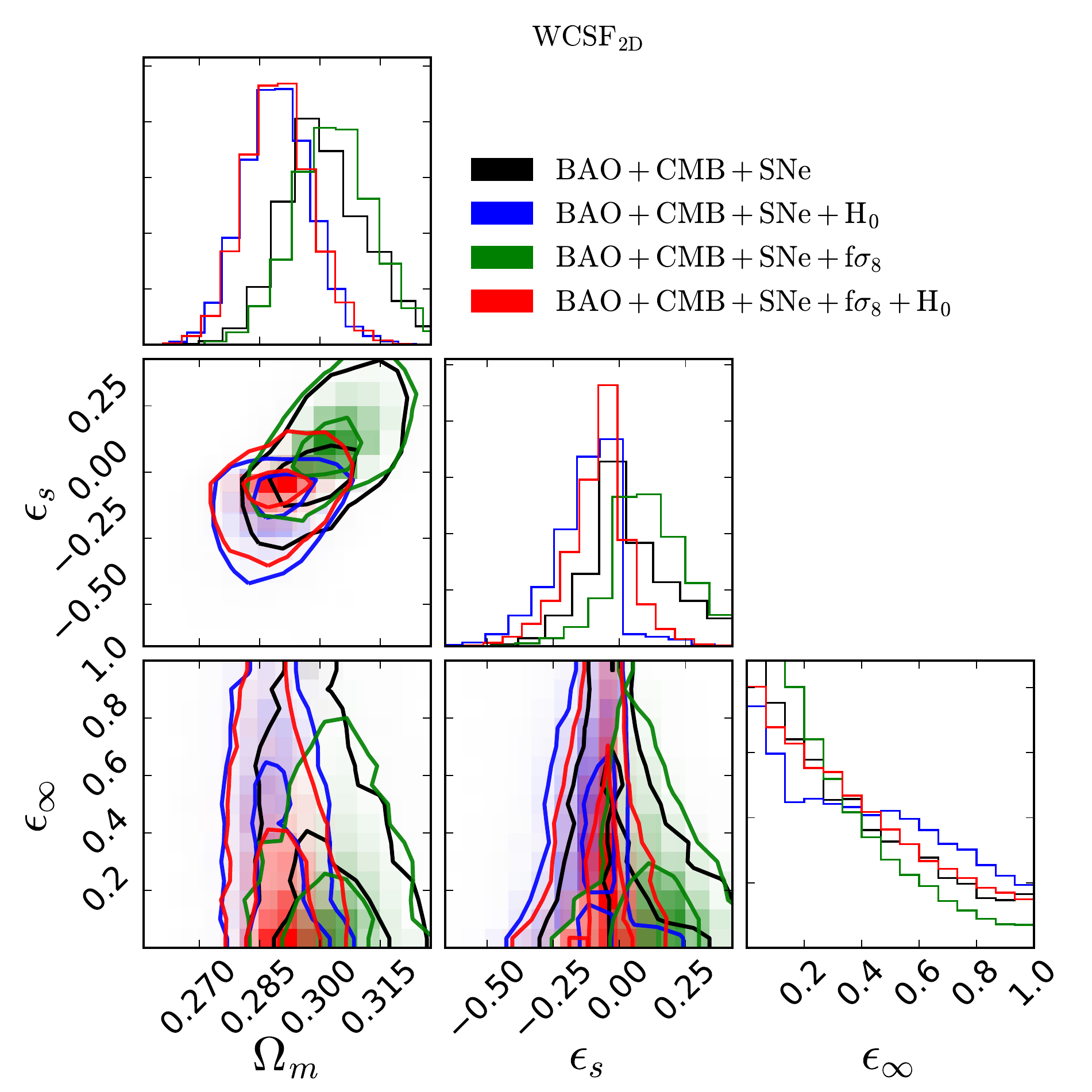}
\caption{The 68.7\% and 95.0\% confidence regions of the parameters for the parameterization of the weakly-coupled canonical scalar field: one-parameter trajectory (Top panel) and two-parameter trajectory (Bottom panel) respectively. The diagonal panels show the one-dimensional probability distribution functions.}
\label{fig:WCSF_mcmc}
\end{center}
\end{figure}

Figure \ref{fig:WCSF_mcmc} presents the constraints of the 
parameterizations for a weakly-coupled canonical scalar field (WCSF). 
This model parameterizes the equation of state $w(a)$ at late times 
instead of writing down explicitly the potential energy of the field.
Therefore various quintessence potentials can be mapped into the 
same parameter space. The figure shows constraints for one and 
two-parameter trajectories respectively.  The constraints include 
$\Lambda$CDM, which corresponds to setting all parameters 
($\epsilon_{\infty}, \epsilon_{s}, \zeta_{s}$) to be 0, though
allow some departure from it. Like \citealt{Planck_2015b}, 
we also find that the constraint on $\epsilon_{s}$ for the two-parameter case is 
tighter than the one-parameter case because the parameters are correlated. 
Table \ref{tab:WCSF1} and \ref{tab:WCSF2} present the results 
for this model. 

The $p-$value test of this model doesn't show significant difference 
from the $\Lambda$CDM model. The predicted observables are tightly constrained
under this model, comparable to those found in Figure \ref{fig:LCDM_obs} for 
the non-flat $\Lambda$CDM model.

\begin{table*}
\centering
\begin{tabular}{lllll}
\hline
\multicolumn{5}{c}{WCSF$_{1D}$: one-parameter trajectory} \\
\cline{1-5}
Data     &    $\Omega_{m}$    &     $\epsilon_{s}$   & $\Delta$BIC  &  $\Delta$AIC\\
\hline
BAO+CMB+SNe                                        & $0.304\pm0.010$   &  $0.14\pm0.24$   &  5.9  &  2.4  \\
BAO+CMB+SNe+H$_0$                           & $0.292\pm0.009$   &  $-0.10\pm_{-0.22}^{+0.21}$     &  6.8  &  3.2 \\
BAO+CMB+SNe+$f\sigma_{8}$                & $0.304\pm0.010$   &  $0.20_{-0.24}^{+0.23}$     &  5.9  &  2.3 \\
BAO+CMB+SNe+$f\sigma_{8}$+H$_0$    & $0.292\pm0.009$  &  $-0.04\pm0.20$     &  7.3  & 3.6  \\
\hline
\end{tabular}
\caption{Cosmological constraints for a selection of parameters for the weakly-coupled scalar field: one-parameter trajectory.}
\label{tab:WCSF1}
\end{table*}

\begin{table*}
\centering
\begin{tabular}{lllll}
\hline
\multicolumn{5}{c}{WCSF$_{2D}$: two-parameter trajectory} \\
\cline{1-5}
Data     &    $\epsilon_{s}$    &     $\epsilon_{\infty} (68\% \rm{CL})$   & $\Delta$BIC  &  $\Delta$AIC\\
\hline
BAO+CMB+SNe                                        & $0.01_{-0.11}^{+0.22}$   &  $<0.64$    &  9.7  &  4.4 \\
BAO+CMB+SNe+H$_0$                           & $-0.11_{-0.14}^{+0.08}$   &  $<0.73$   & 9.6 & 4.3   \\
BAO+CMB+SNe+$f\sigma_{8}$                & $0.11_{-0.11}^{+0.17}$   &  $<0.51$    & 9.8  & 4.3 \\
BAO+CMB+SNe+$f\sigma_{8}$+H$_0$    & $-0.06_{-0.12}^{+0.08}$  &  $<0.65$     & 11.1  & 5.6   \\
\hline
\end{tabular}
\caption{Cosmological constraints for a selection of parameters for the weakly-coupled scalar field: two-parameter trajectory.}
\label{tab:WCSF2}
\end{table*}

\subsection{Holographic dark energy (HDE, ADE, RDE)}

\begin{figure}[htbp]
\begin{center}
\includegraphics[width=9cm, height=8cm]{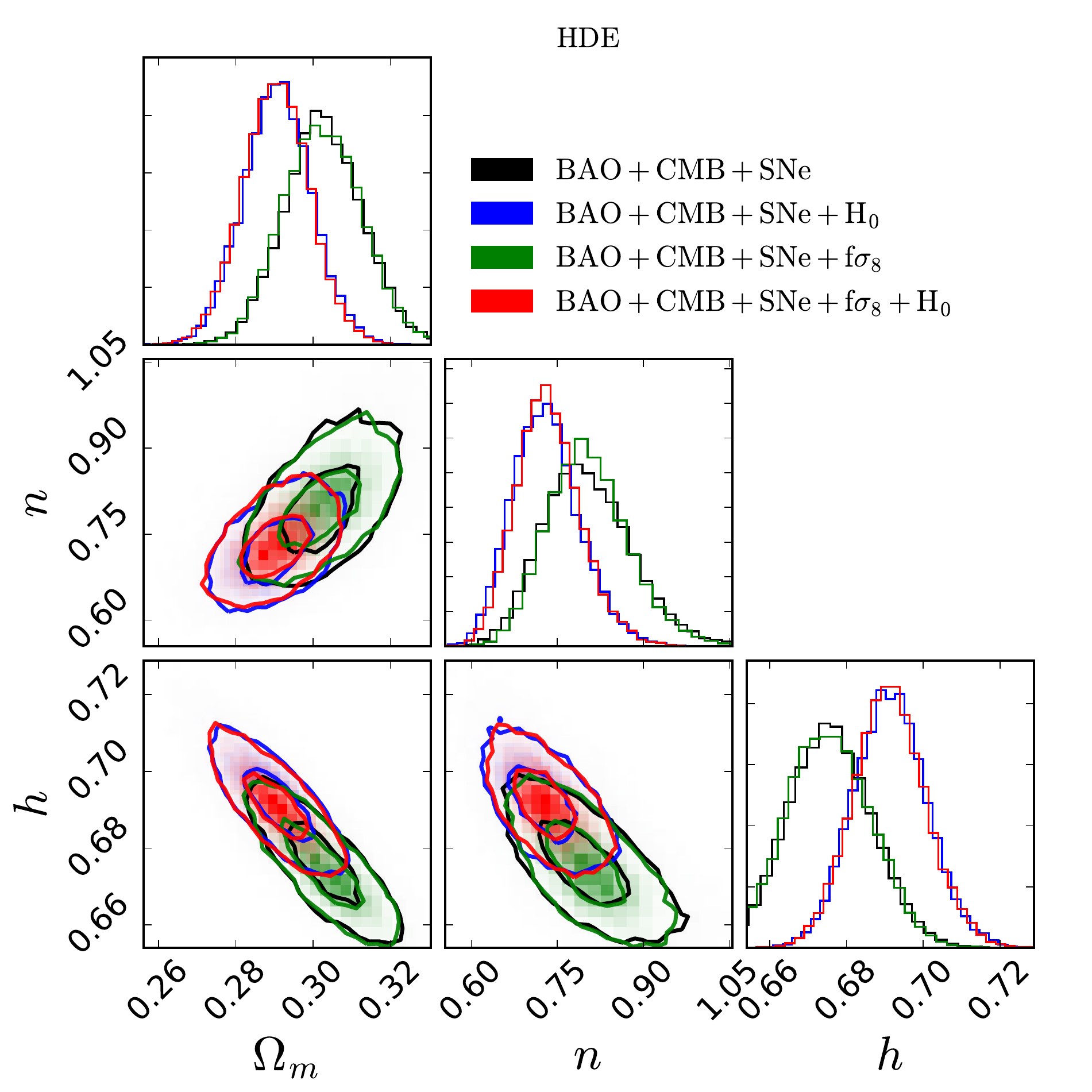}
\includegraphics[width=9cm, height=8cm]{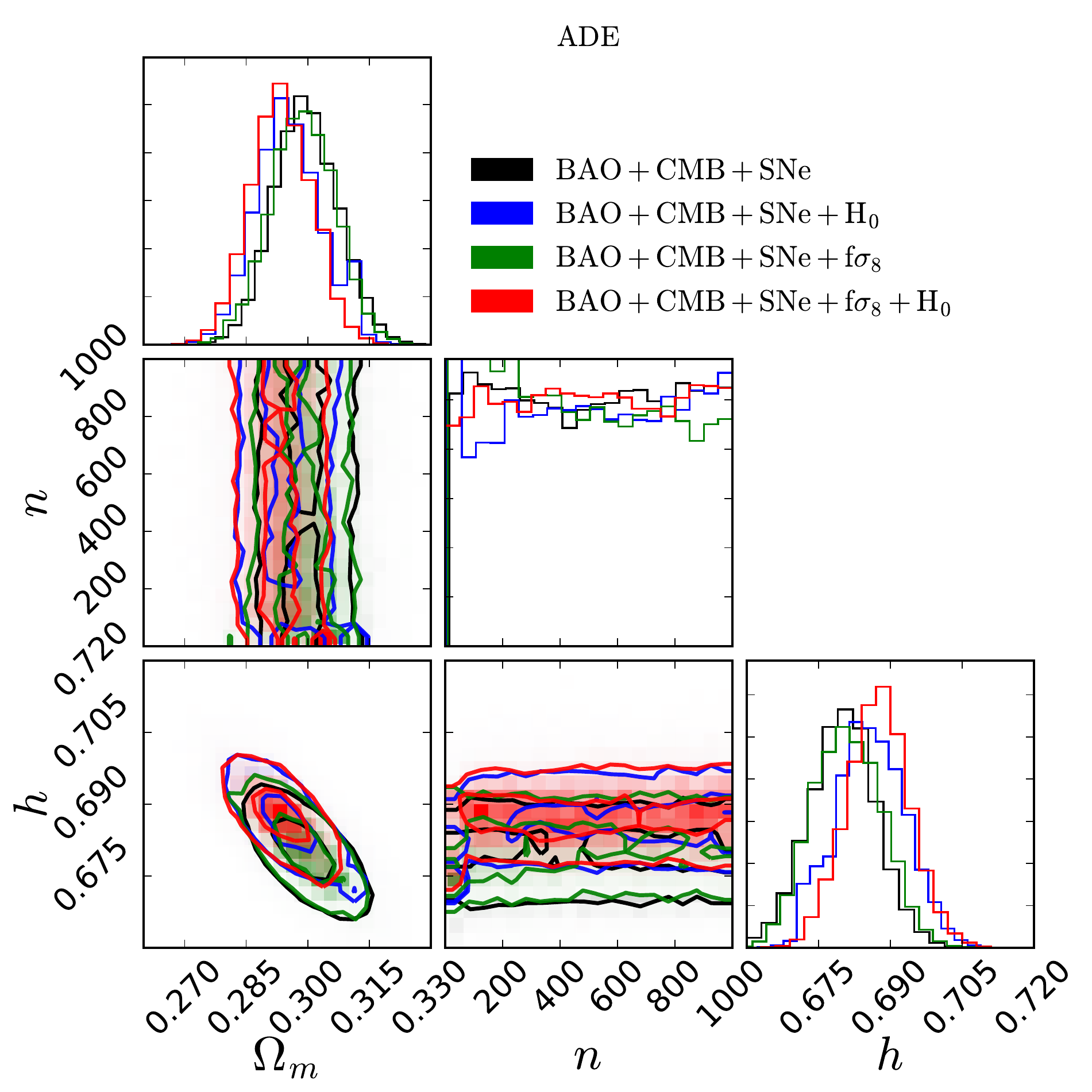}
\includegraphics[width=9cm, height=8cm]{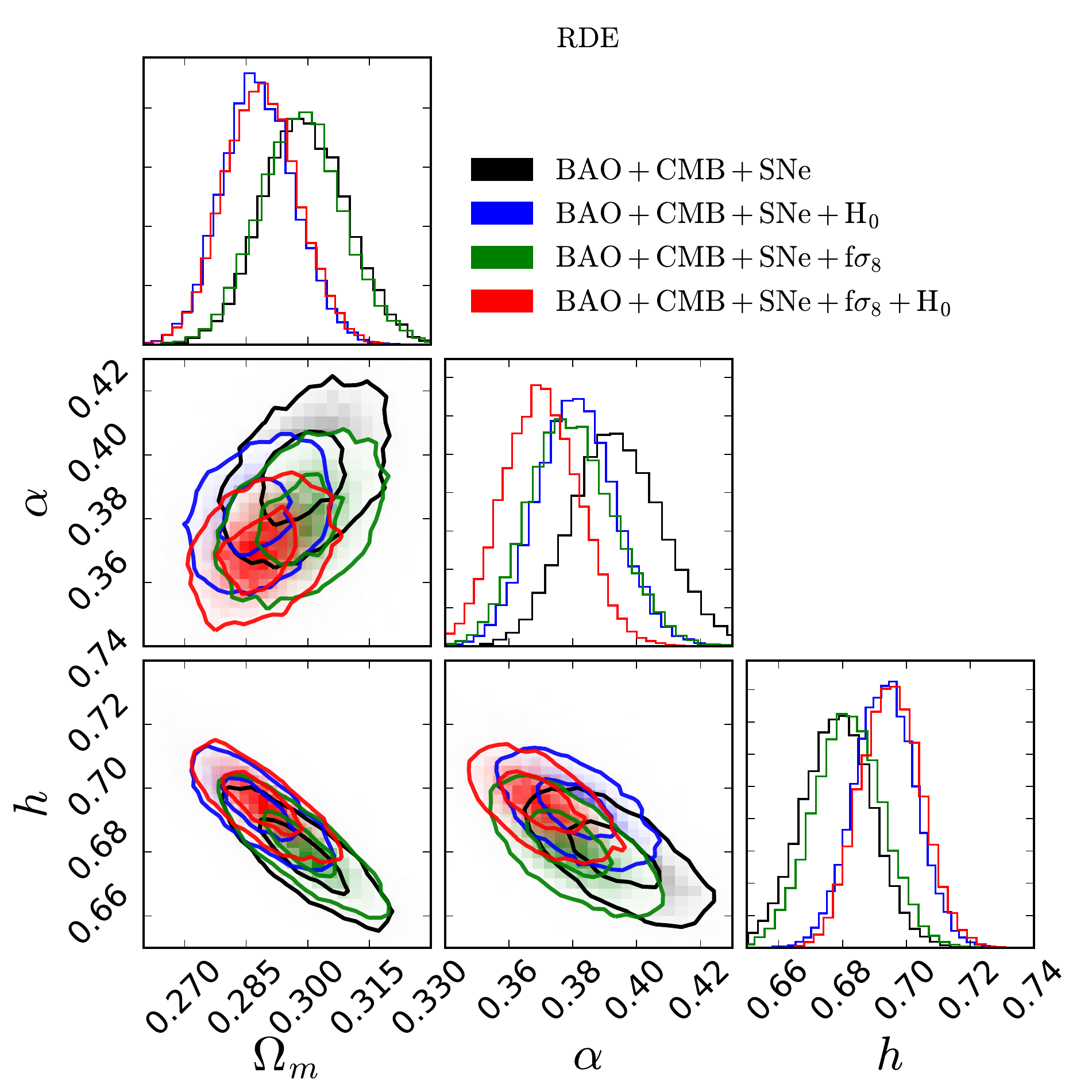}
\caption{The 68.7\% and 95.0\% confidence regions of the parameters for the holographic dark energy model: HDE(Top panel), ADE(Middle panel), RDE(Bottom panel). The diagonal panels show the one-dimensional probability distribution functions. }
\label{fig:HDE_mcmc}
\end{center}
\end{figure}

The constraints of the holographic dark energy models are 
displayed in Figure \ref{fig:HDE_mcmc}, and listed in 
Tables \ref{tab:HDE}, \ref{tab:ADE} and \ref{tab:RDE}. 

\begin{figure}[t!]
\begin{center}
\includegraphics[width=0.49\textwidth]{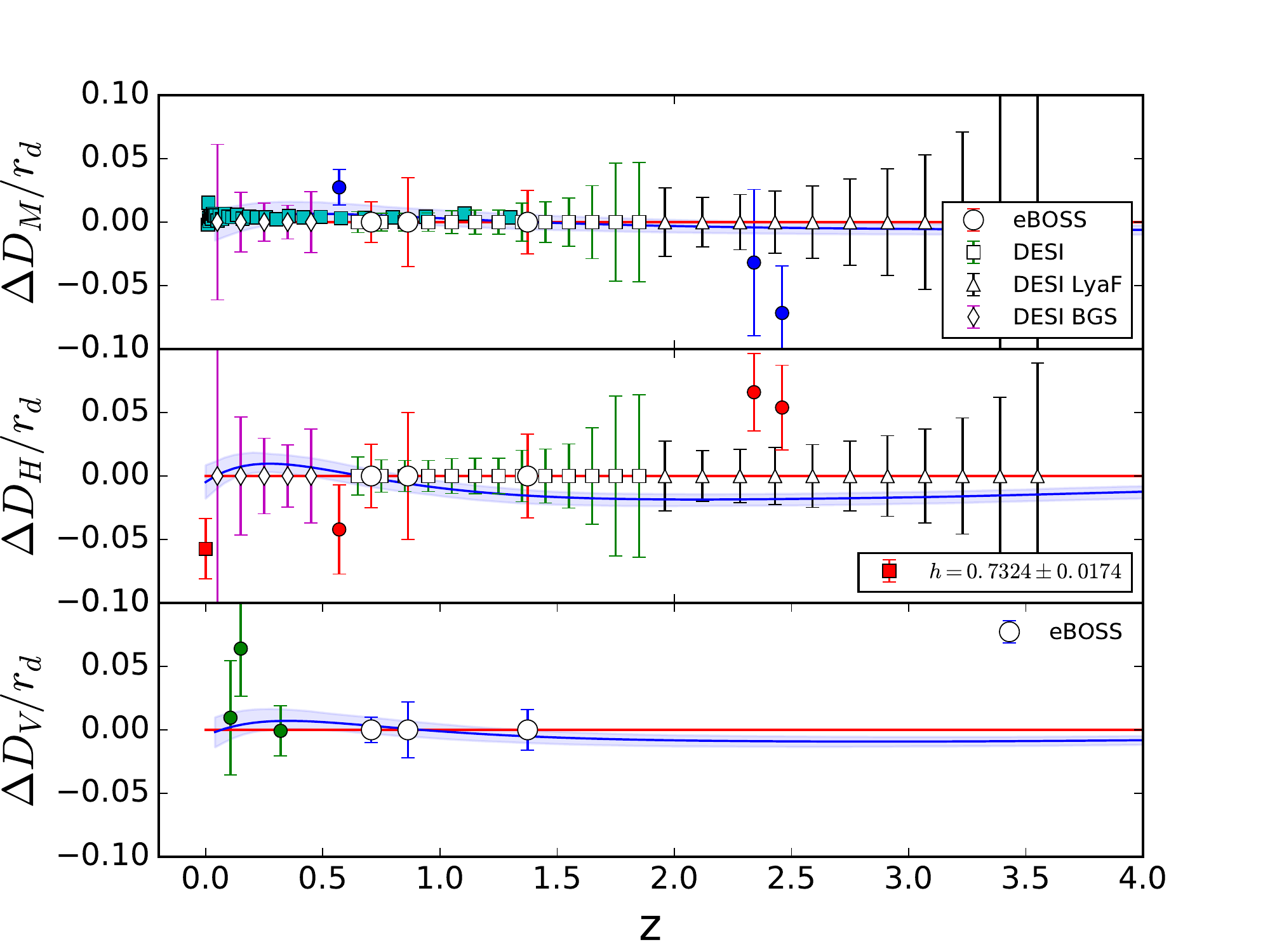}
\includegraphics[width=0.49\textwidth]{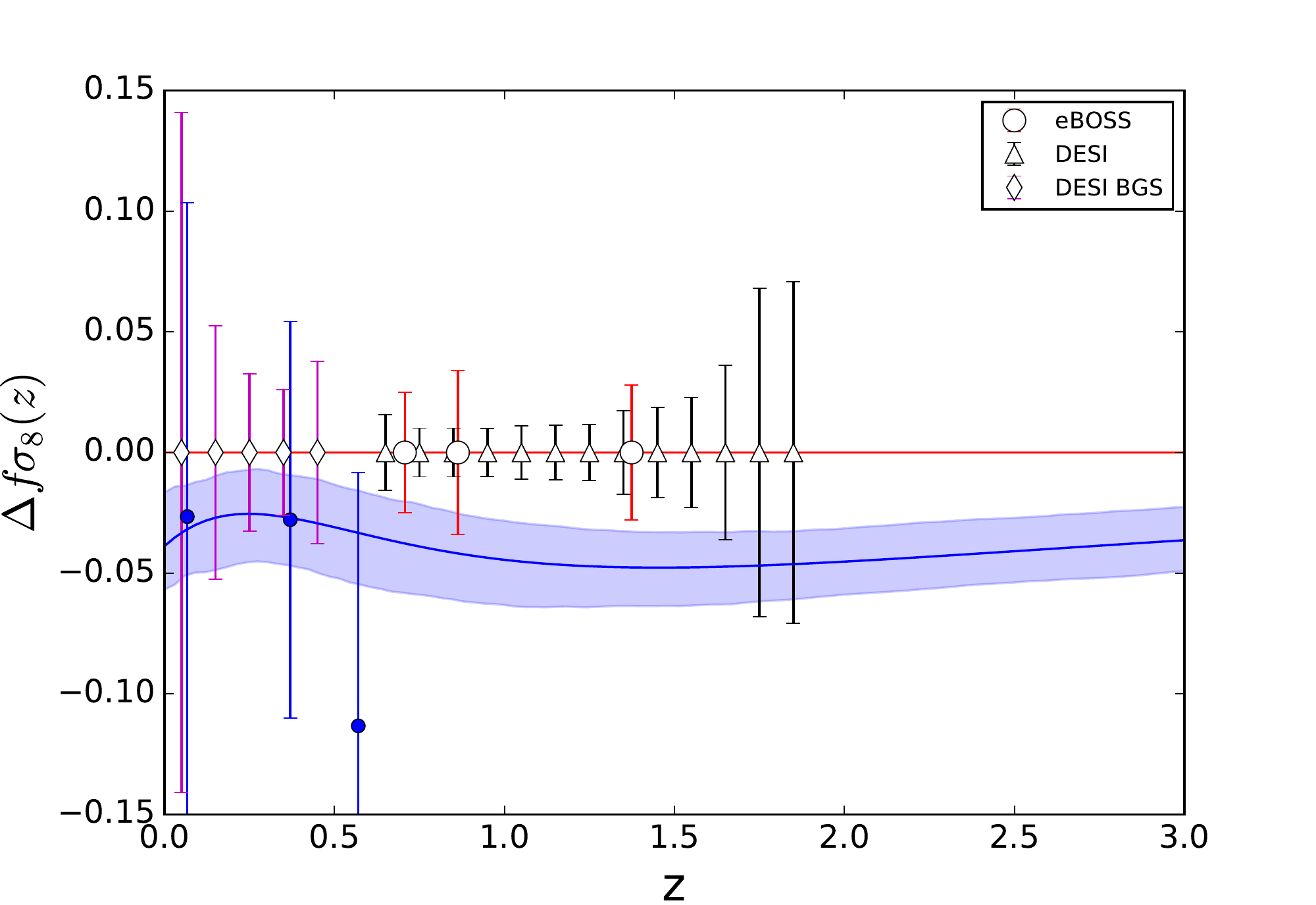}
\caption{\label{fig:HDE_obs} Similar to Figure \ref{fig:LCDM_obs}, for
the HDE model.}
\end{center}
\end{figure}

For the HDE model, the dependence of $h$ on the inclusion 
of H$_0$ data is stronger than in the standard $\Lambda$CDM 
model. The constraint on $n$ is tight, a result previously pointed 
out (\citealt{Li_2013, Li_2013_HDE, Zhang_2014}). The addition 
of the linear growth data does not change the constraints
significantly, revealing the consistency of the geometrical 
probes and the dynamical probes for this model (\citealt{Xu_2013,Zhang_2015}). 
Figure \ref{fig:HDE_obs} shows the predicted observables for this model, which
show that although they are not currently ruled out, this class of models can 
be ruled out by future experiments.

The importance of linear growth data increases somewhat for the 
agegraphic dark energy (ADE) model but is still a marginal effect. 
The ADE predictions for $f\sigma_8$ are more similar to $\Lambda$CDM
than HDE's predictions are; unlike HDE, the predicted observables
from ADE given current constraints fully contain $\Lambda$CDM (and
are very similar to those found for $\Lambda$CDM in Figure \ref{fig:LCDM_obs}).
An important difference between these two models is that the 
parameter $n$ in ADE is not constrained. Unlike other studies
in the literature, we are not applying the initial condition 
of the ADE model to reduce the number of parameters, so 
both the matter fraction $\Omega_{m}$ and $n$ are constrained 
in the MCMC test (\citealt{Zhang_2007ADE, Wei_2008, Li_2009}). 
In this case, the large value of $n$ preferred by the 
data gives $\Lambda$CDM-like  evolution in this model.
Further theoretical considerations would be required to 
limit $n$ to low values, and doing so would likely rule out 
this model (\citealt{Neupane_2009}).

\begin{figure}[t!]
\begin{center}
\includegraphics[width=0.49\textwidth]{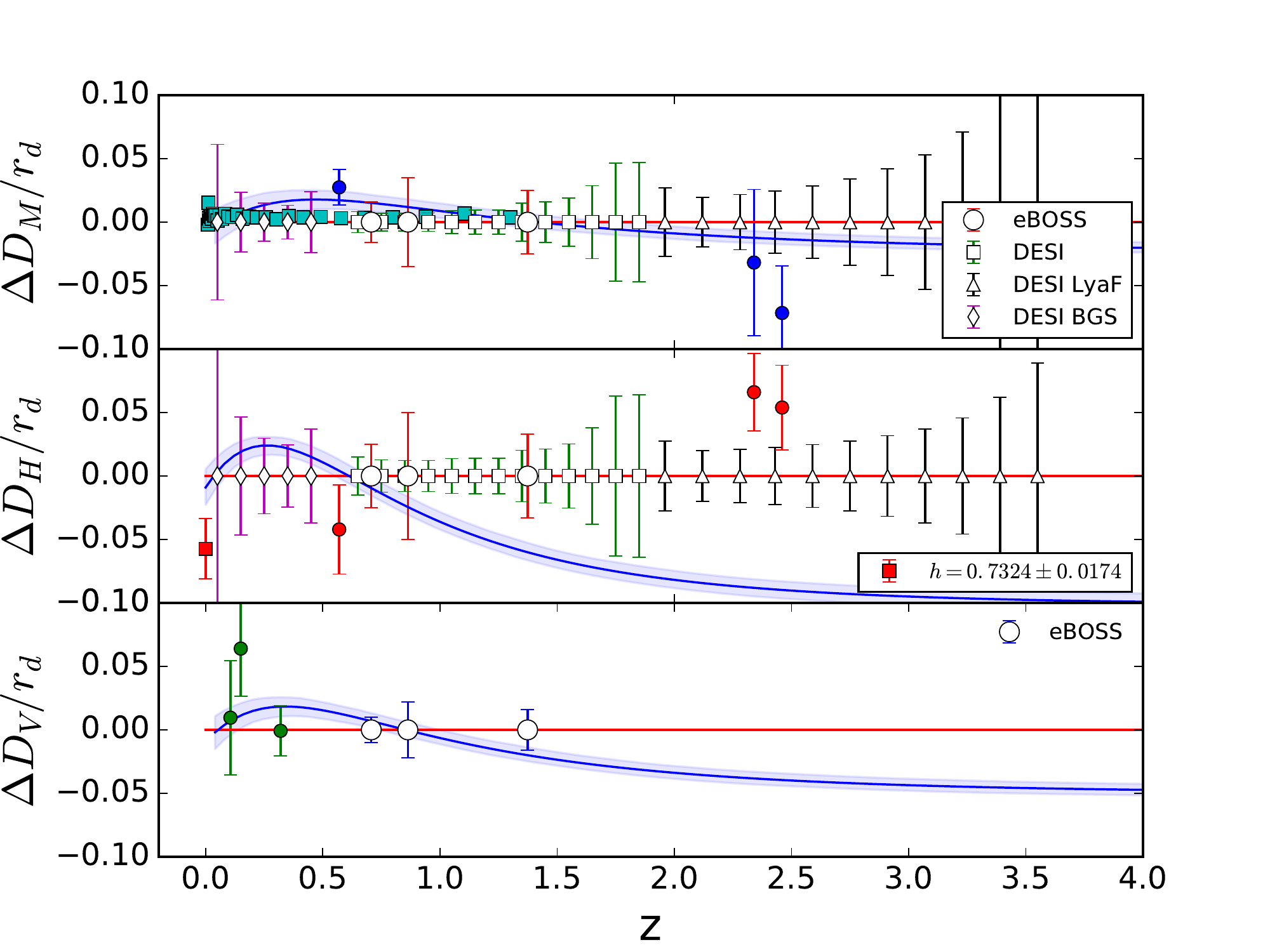}
\includegraphics[width=0.49\textwidth]{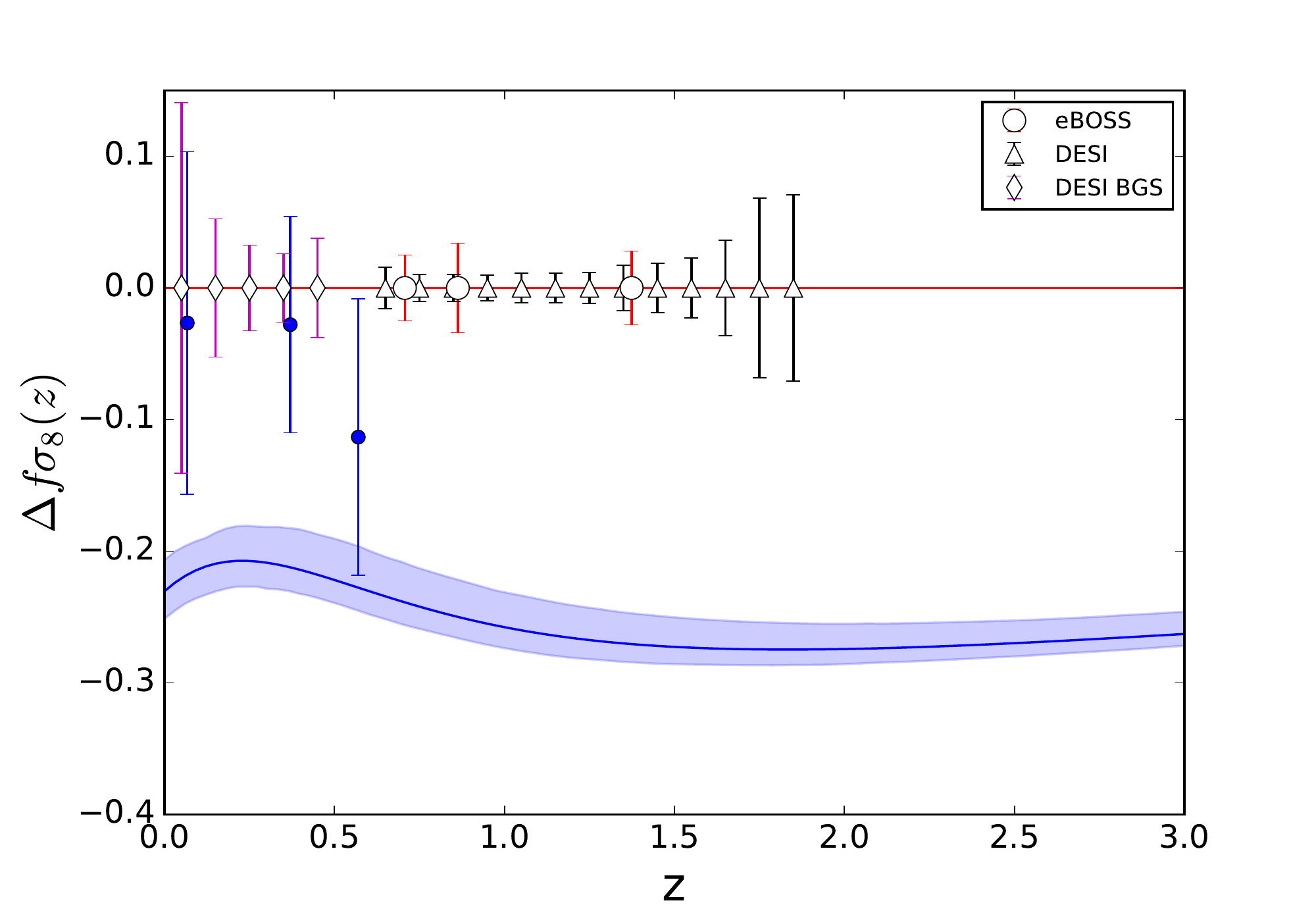}
\caption{\label{fig:RDE_obs} Similar to Figure \ref{fig:LCDM_obs}, for
the RDE model.}
\end{center}
\end{figure}

The bottom panel of Figure \ref{fig:HDE_mcmc} displays the 
constraints of the Ricci dark energy model (RDE). The $p$-value 
test from the last section shows that there is significant 
incompatibility between this model and the data, which 
implies that this model is not favored. The constraints 
of the parameters also reveal tensions in $\alpha$ among
different data combinations. This behavior is also found 
in the sDGP model, which is also not favored in the $p$-value test.
Figure \ref{fig:RDE_obs} shows the predicted observables
under the best fit RDE model; both the expansion rate and
the growth rate measurements clearly exclude it.

In HDE and RDE models, the degeneracy between $\Omega_{m}$ 
and $h$ is found to be similar to that of $\Lambda$CDM. 
Therefore the unsolved tension of H$_{0}$ also indicates 
a $1\sigma$ disagreement in $\Omega_{m}$ and $>1.5\sigma$ 
disagreement for $h$ among the datasets. However for ADE 
model, this tension is relaxed to be $0.5\sigma$ for 
$\Omega_{m}$ and $1.0\sigma$ for $h$. Nevertheless, our $p$-value 
analysis shows that this improvement comes at the expense
of fitting the other data sets so that it is not a much
different $\chi^2$ than many other models, and that the extra
parameters introduced mean that 
the $p$-value is  not much better for ADE or HDE.  

\begin{table*}
\centering
\begin{tabular}{lllll}
\hline
\multicolumn{5}{c}{HDE} \\
\cline{1-5}
Data     &    $\Omega_{m}$    &     $n$   & $\Delta$BIC  &  $\Delta$AIC\\
\hline
BAO+CMB+SNe                                        & $0.302\pm0.010$   &  $0.80_{-0.07}^{+0.08}$     &  10.3  &  6.8\\
BAO+CMB+SNe+H$_0$                           & $0.290\pm0.009$   &  $0.73\pm0.06$   &  10.5  & 6.9   \\
BAO+CMB+SNe+$f\sigma_{8}$                & $0.302\pm0.010$   &  $0.80\pm0.07$    & 8.9  &  5.3 \\
BAO+CMB+SNe+$f\sigma_{8}$+H$_0$    & $0.290\pm0.008$  &  $0.73_{-0.05}^{+0.06}$       & 9.6  & 5.9\\
\hline
\end{tabular}
\caption{Cosmological constraints for a selection of parameters for the holographic dark energy model.}
\label{tab:HDE}
\end{table*}

\begin{table*}
\centering
\begin{tabular}{lllll}
\hline
\multicolumn{5}{c}{ADE} \\
\cline{1-5}
Data     &    $\Omega_{m}$    &     $n (68\% \rm{CL})$  & $\Delta$BIC  &  $\Delta$AIC \\
\hline
BAO+CMB+SNe                                        & $0.299\pm0.008$                 &  $>154$   & 6.1 & 2.6  \\
BAO+CMB+SNe+H$_0$                           & $0.296_{-0.008}^{+0.009}$   &  $>126$   & 7.1 &  3.6   \\
BAO+CMB+SNe+$f\sigma_{8}$                & $0.298\pm0.008$                 &  $>134$  & 6.0  &  2.4 \\
BAO+CMB+SNe+$f\sigma_{8}$+H$_0$    & $0.293\pm0.008$                 & $>168$    & 7.4 &  3.7   \\
\hline
\end{tabular}
\caption{Cosmological constraints for a selection of parameters for the agegraphic dark energy model. The constraint on $n$ is an artifact due to the cutoff at around 1000 in our MCMC test. A less conservative choice of the this cutoff may change the result a little, but the estimation of other parameters here is still robust.}
\label{tab:ADE}
\end{table*}

\begin{table*}
\centering
\begin{tabular}{lllll}
\hline
\multicolumn{5}{c}{RDE} \\
\cline{1-5}
Data     &    $\Omega_{m}$    &     $\alpha$  & $\Delta$BIC  &  $\Delta$AIC \\
\hline
BAO+CMB+SNe                                        & $0.299_{-0.010}^{+0.011}$   &  $0.393_{-0.014}^{+0.015}$   &  29.5   &  26.0 \\
BAO+CMB+SNe+H$_0$                           & $0.288\pm0.009$                  &  $0.381\pm0.012$      & 29.4  &  25.9\\
BAO+CMB+SNe+$f\sigma_{8}$                & $0.298\pm0.010$                 &  $0.379_{-0.013}^{+0.014}$  &  38.7  &  35.1  \\
BAO+CMB+SNe+$f\sigma_{8}$+H$_0$    & $0.289\pm0.009$                 &  $0.370\pm0.012$      &  38.3  &  34.6 \\
\hline
\end{tabular}
\caption{Cosmological constraints for a selection of parameters for the Ricci scalar dark energy model.}
\label{tab:RDE}
\end{table*}

\subsection{Quintessence scalar field model}

\begin{figure}[htbp]
\begin{center}
\includegraphics[width=9cm, height=8cm]{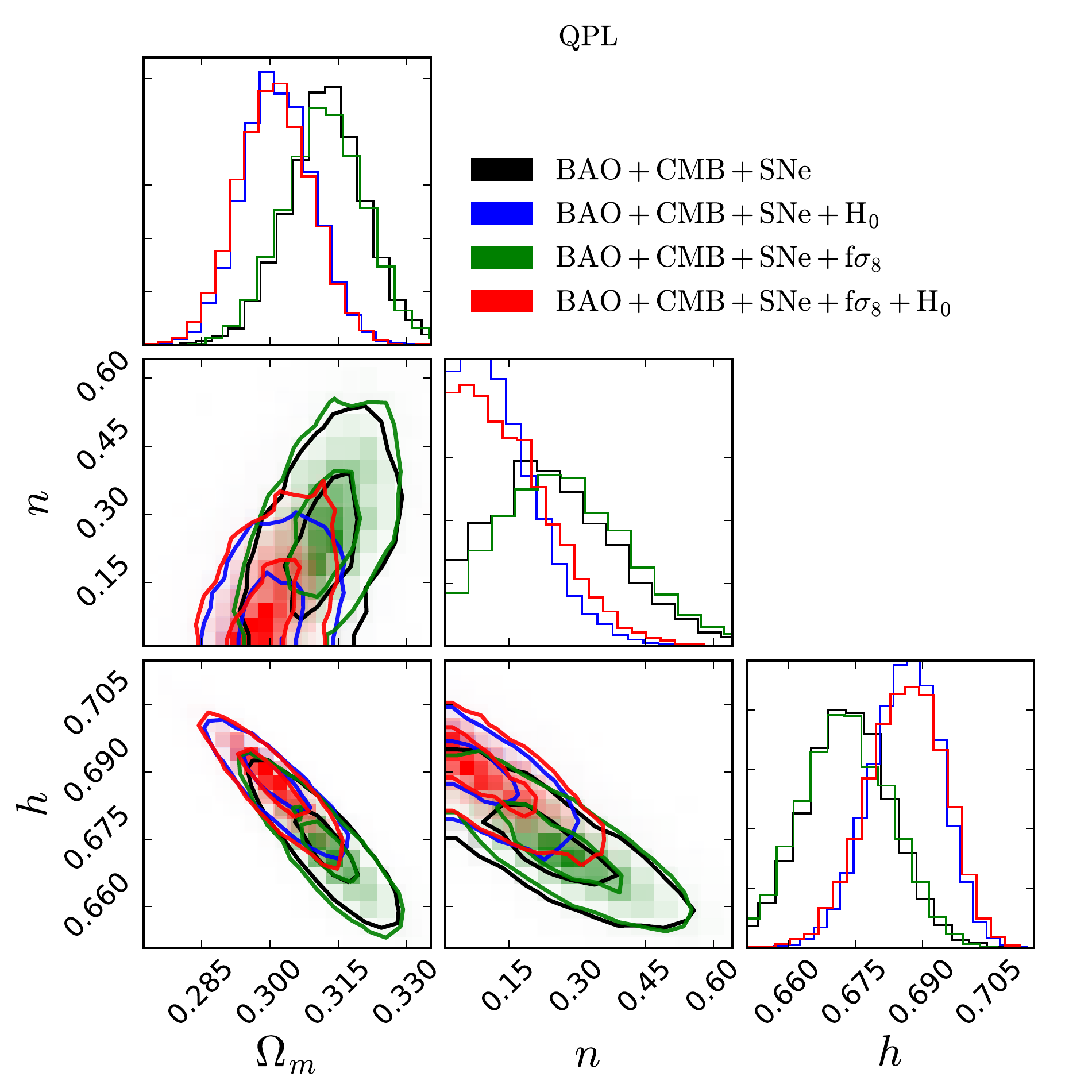}
\includegraphics[width=9cm, height=8cm]{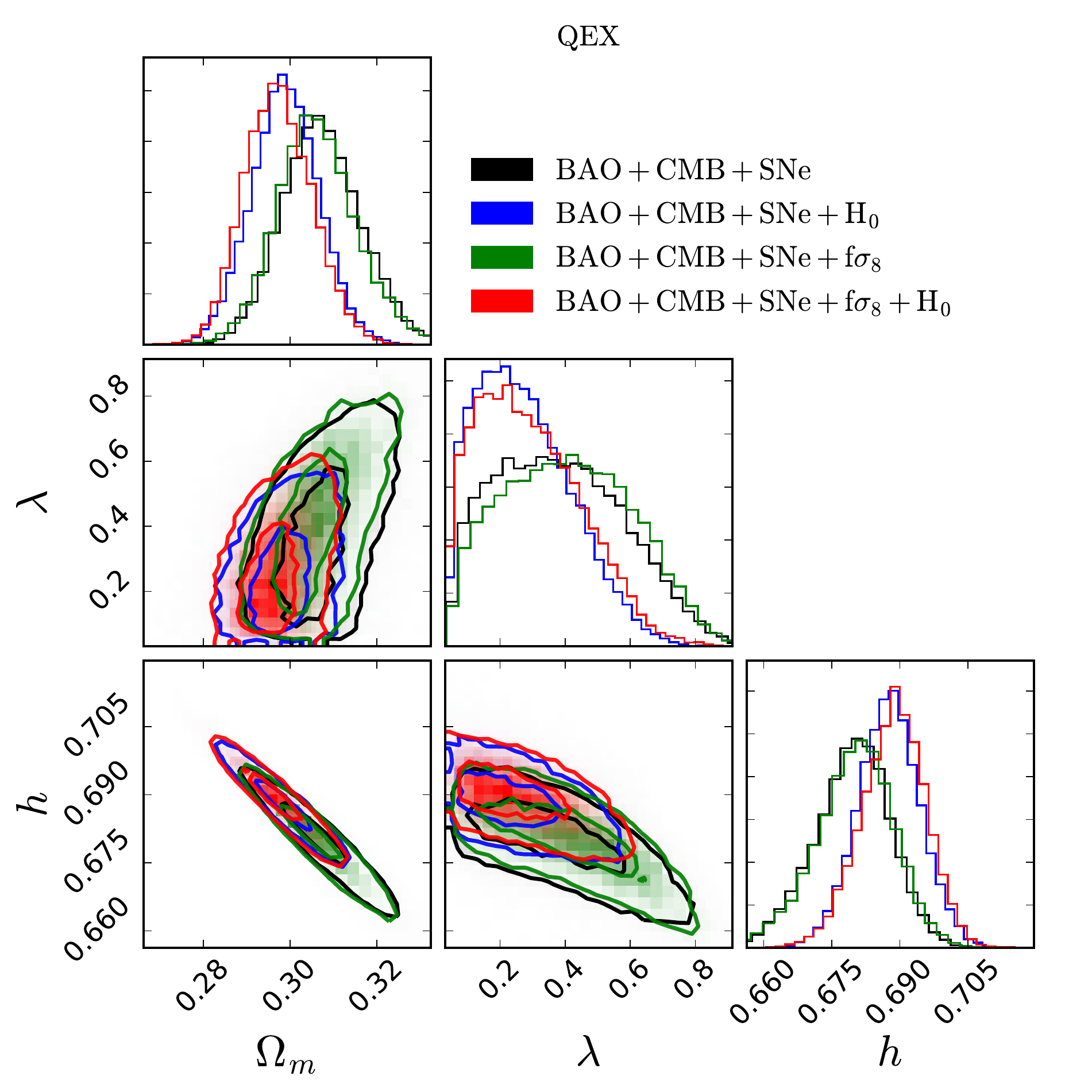}
\caption{The 68.7\% and 95.0\% confidence regions of the parameters for the Quintessence scalar field model: Model I (Top panel), Model II (Bottom panel). The diagonal panels show the one-dimensional probability distribution functions. }
\label{fig:Quint_mcmc}
\end{center}
\end{figure}

Cosmological constraints on the quintessence scalar field have 
been considered by a number of authors
(\citealt{Samushia_2008, Chen_2011, Farooq_2013, Chiba_2013,Pavlov_2014, Paliathanasis_2014}; 
and references therein). Figure \ref{fig:Quint_mcmc} shows the 
constraints on the two Quintessence scalar field models considered 
in this work: the power law potential and exponential potential. 
The results of these models show an anti-correlation between the 
potential parameter and the Hubble constant. When H$_0$ data is
excluded, the non-zero value  of $n$ for Model I and $\lambda$ 
for Model II implies a deviation from the $\Lambda$CDM model, 
which can be seen by  the green curves in the figure. 
The addition 
of H$_0$ data reduces this deviation and the standard model 
is well within $1\sigma$ confidence region. This result holds 
for both scalar field models; a difference is that 
the tension of H$_{0}$ in the power-law model is tighter 
than the exponential model. Table \ref{tab:QPL} and \ref{tab:QEX} 
summarize the constraint on the parameters for the quintessence 
scalar field models.

The $p$-value is relatively high for these models,
and remains so when H$_0$ is included, but this is a 
relatively small difference from $\Lambda$CDM and 
does not signify a very interesting reduction of the 
tension with H$_0$. The predicted observables 
for these models are very similar to those for $\Lambda$CDM
(Figure \ref{fig:LCDM_obs}).

\begin{table*}
\centering
\begin{tabular}{lllll}
\hline
\multicolumn{5}{c}{Quintessence scalar field: power law potential} \\
\cline{1-5}
Data     &    $\Omega_{m}$    &     $n (68\% \rm{CL})$  & $\Delta$BIC  &  $\Delta$AIC \\
\hline
BAO+CMB+SNe                                        & $0.312\pm0.008$                 &  $0.00<n<0.40$     &  5.5  &  3.7 \\
BAO+CMB+SNe+H$_0$                           & $0.302_{-0.007}^{+0.008}$   &  $0.00<n<0.22$     &  6.3  & 4.5 \\
BAO+CMB+SNe+$f\sigma_{8}$                & $0.312\pm0.009$                 &  $0.00<n<0.42$  &  4.8  &  3.0\\
BAO+CMB+SNe+$f\sigma_{8}$+H$_0$   & $0.301\pm0.008$                 &  $0.00<n<0.26$       &  6.3  &  4.4\\
\hline
\end{tabular}
\caption{Cosmological constraints for a selection of parameters for the quintessence scalar field with power law potential.}
\label{tab:QPL}
\end{table*}

\begin{table*}
\centering
\begin{tabular}{lllll}
\hline
\multicolumn{5}{c}{Quintessence scalar field: exponential potential} \\
\cline{1-5}
Data     &    $\Omega_{m}$    &     $\lambda  (68\% \rm{CL})$  & $\Delta$BIC  &  $\Delta$AIC \\
\hline
BAO+CMB+SNe                                        & $0.307\pm0.009$                 &  $0.00<\lambda<0.59$     &  4.3  &  2.6\\
BAO+CMB+SNe+H$_0$                           & $0.299\pm0.008$                 &  $0.00<\lambda<0.43$    & 4.5  & 2.7  \\
BAO+CMB+SNe+$f\sigma_{8}$                & $0.306\pm0.009$                 &  $0.00<\lambda<0.63$    & 4.2  & 2.4\\
BAO+CMB+SNe+$f\sigma_{8}$+H$_0$   & $0.297_{-0.007}^{+0.008}$   &  $0.00<\lambda<0.45$      & 4.6  & 2.8 \\
\hline
\end{tabular}
\caption{Cosmological constraints for a selection of parameters for the quintessence scalar field with exponential potential.}
\label{tab:QEX}
\end{table*}

\subsection{QCD ghost dark energy and DGP cosmology}

The QCD ghost dark energy model and the DGP model 
have similar expressions of the cosmic expansion. The 
$p$-value test shows that these two models are highly disfavored 
by the data, but they have different reasons for their 
incompatibility with the observations. 

\begin{figure}[htbp]
\begin{center}
\includegraphics[width=9cm, height=8cm]{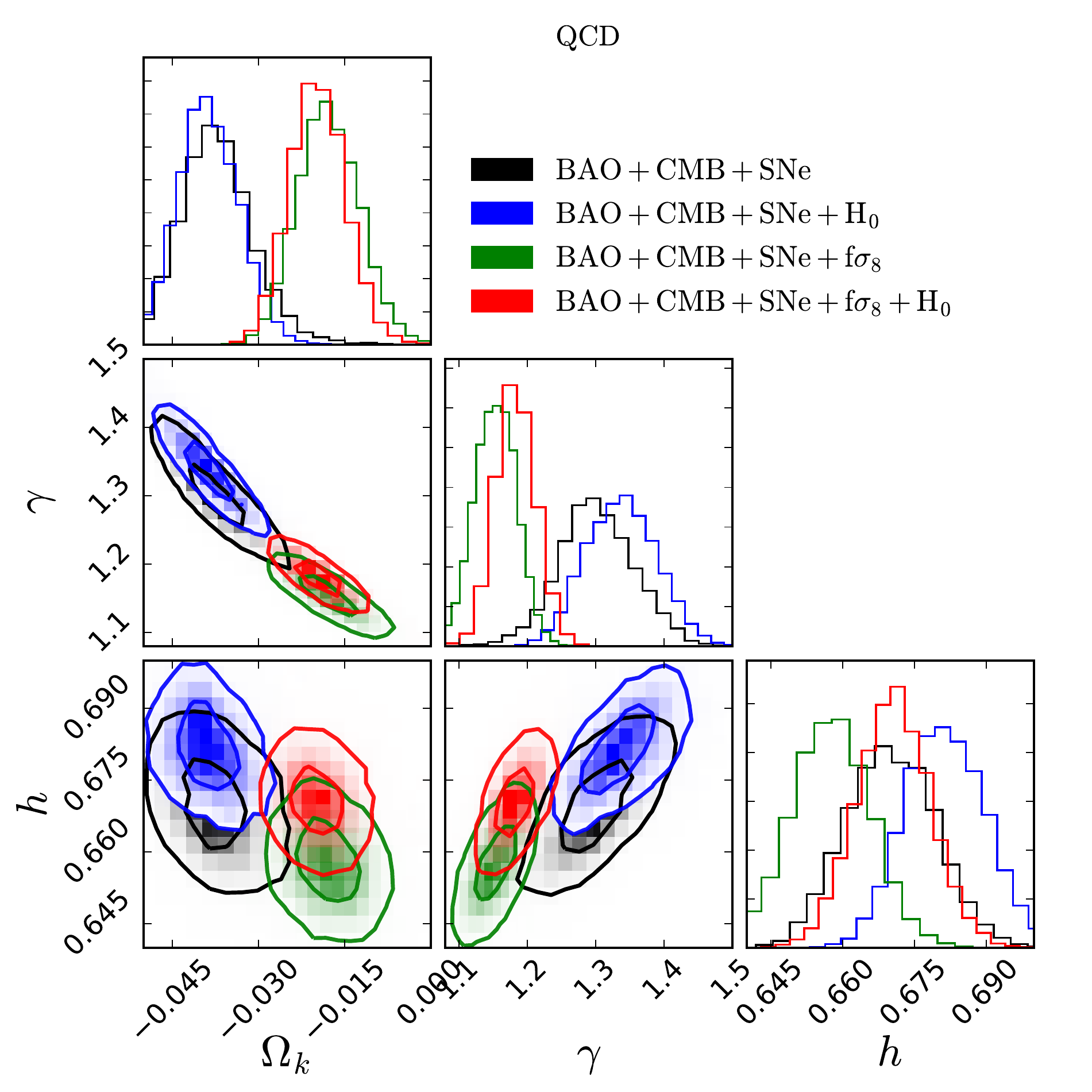}
\includegraphics[width=9cm, height=8cm]{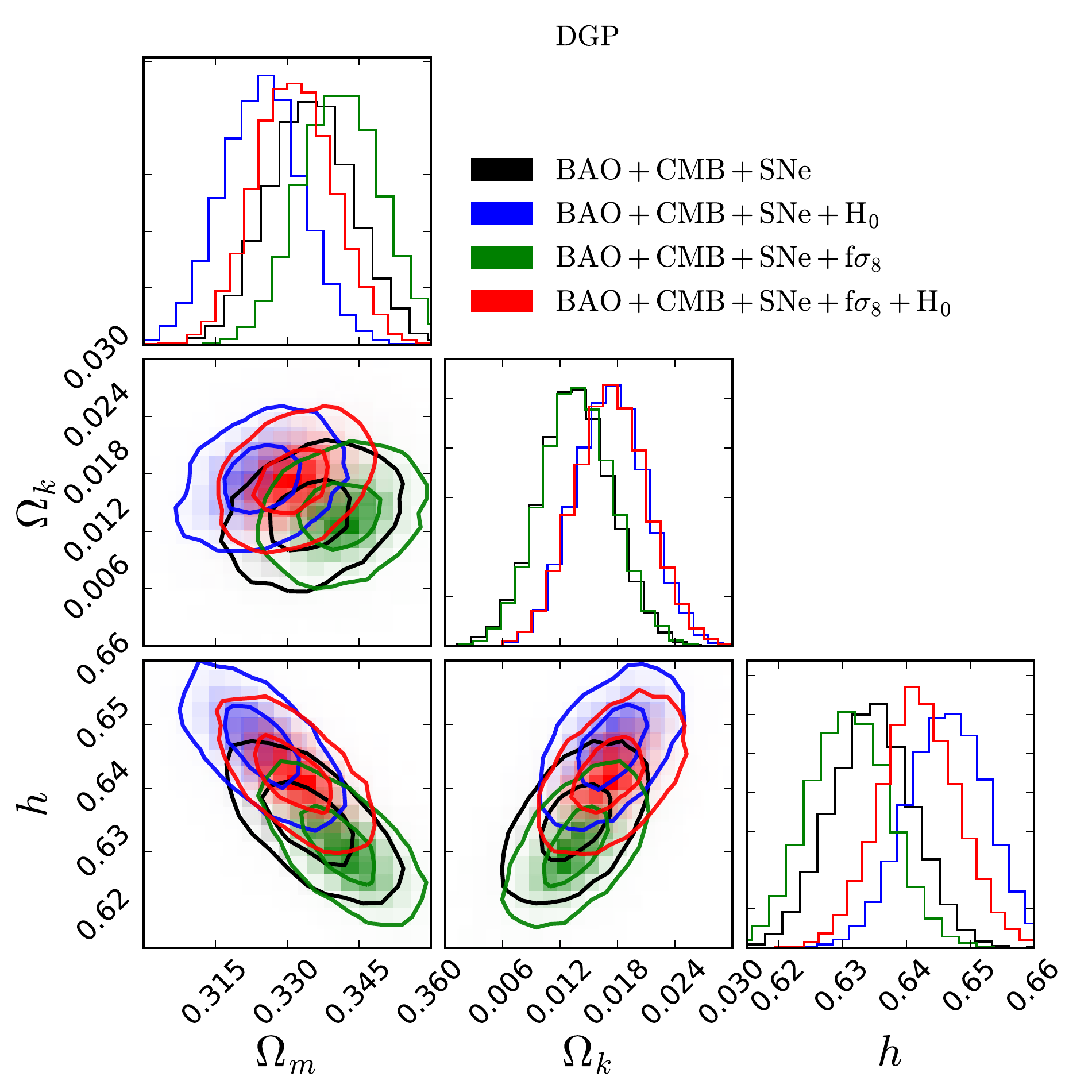}
\caption{The 68.7\% and 95.0\% confidence regions of the parameters for the QCD Ghost dark energy model (Top panel) and self-accelerating DGP model (Bottom panel). The diagonal panels show the one-dimensional probability distribution functions. }
\label{fig:QCD_mcmc}
\end{center}
\end{figure}

According to the $p$-value, the QCD ghost dark energy model
agrees with the data quite  well when only the BAO, 
CMB and SNe data are considered. The addition of the linear 
growth data or H$_0$ reduces the $p$-value significantly. 
The change of $\Delta\chi^2$ from H$_{0}$ is about twice 
the amount found for the $\Lambda$CDM model, which results in a 
significant decrease of the $p$-value. 
The top panel of  Figure \ref{fig:QCD_mcmc} displays the constraints on the 
QCD ghost dark energy model. The inclusion or exclusion of H$_0$
clearly has a strong effect on the best fit parameters.
The figure also shows a large difference made by $f\sigma_8$.
The addition of this data set changes the component contributed 
from spatial curvature by about 2\%, which is a source of
the tension measured in the $p$-value.

\begin{figure}[t!]
\begin{center}
\includegraphics[width=0.49\textwidth]{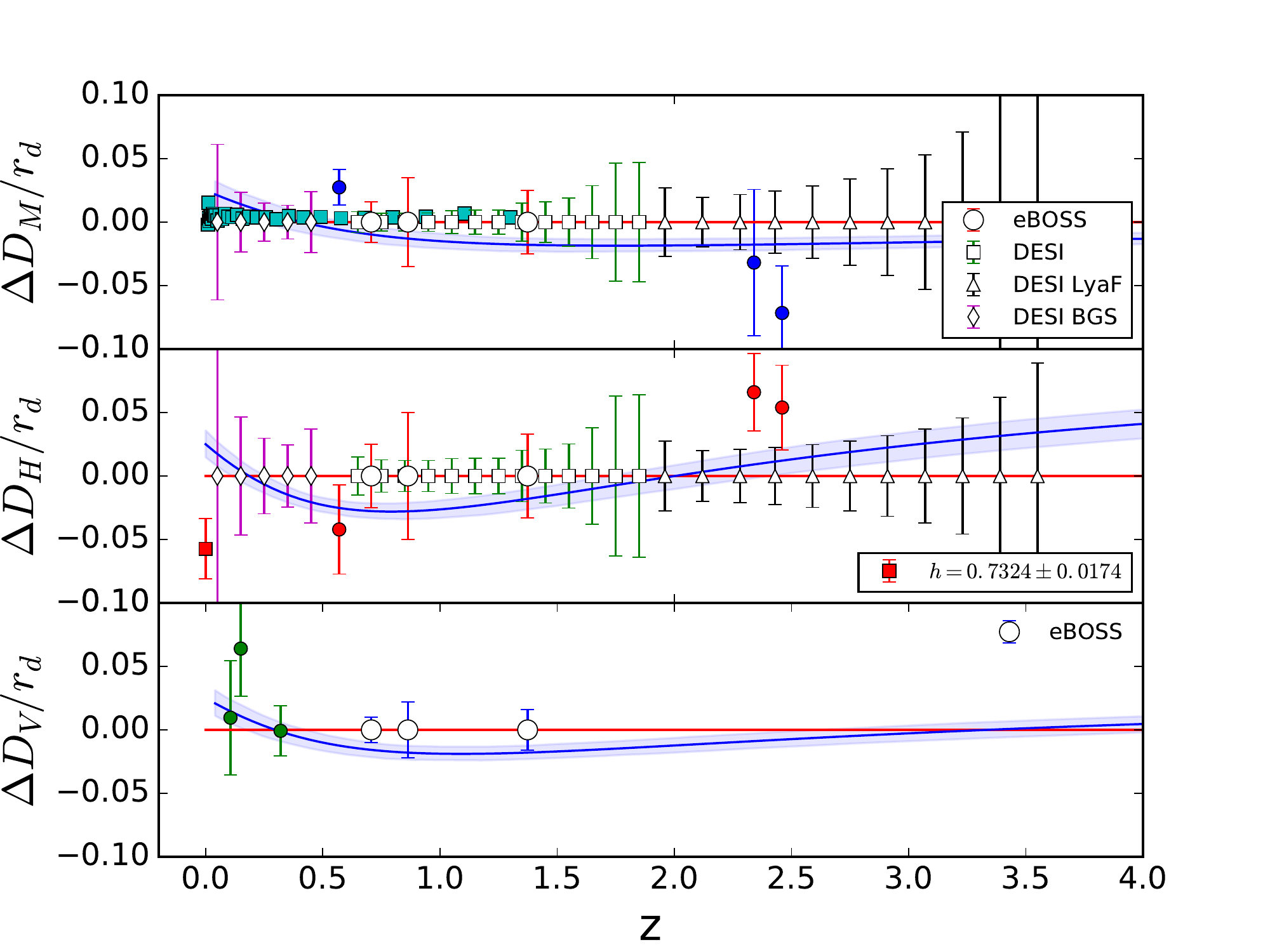}
\includegraphics[width=0.49\textwidth]{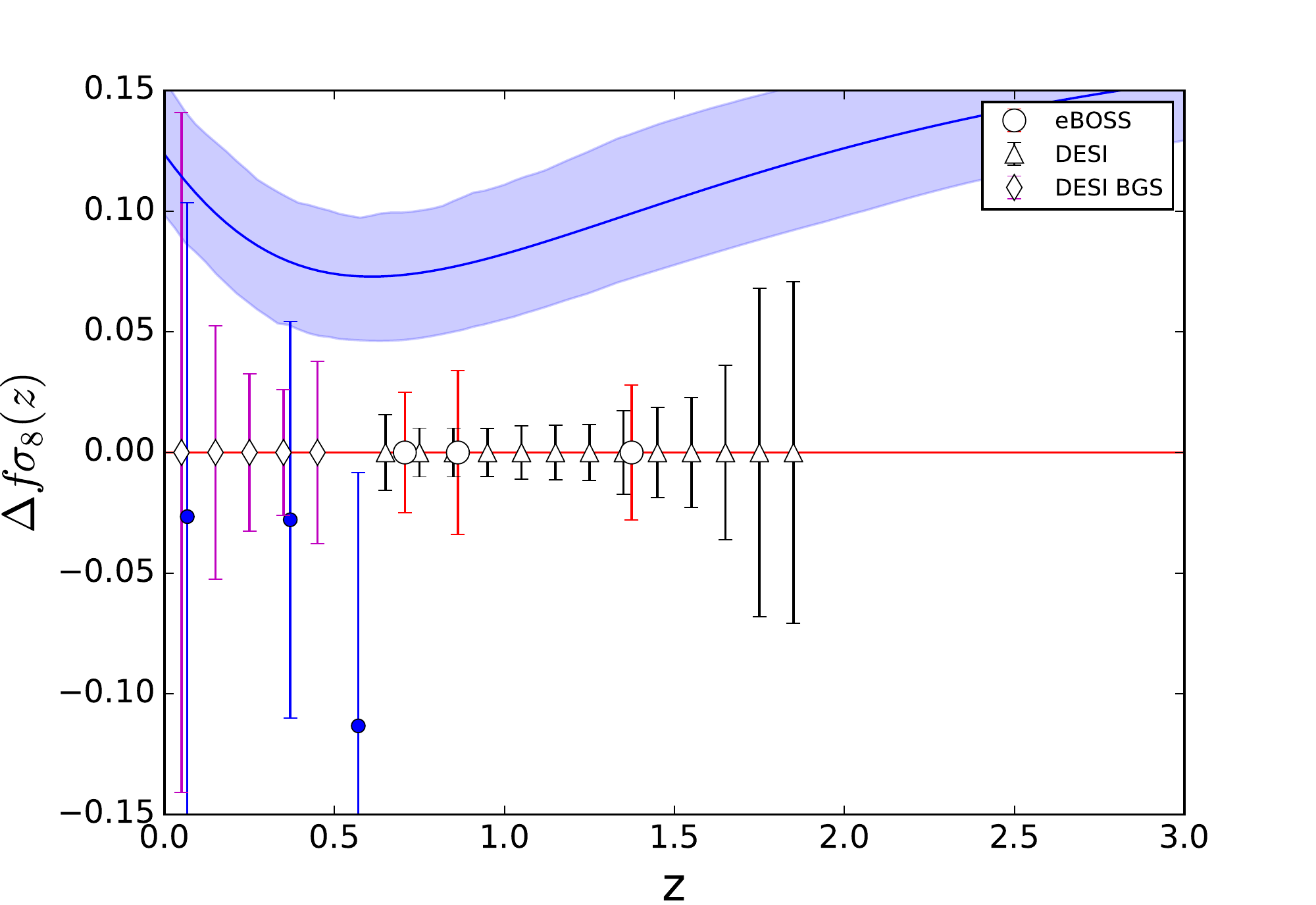}
\caption{\label{fig:QCD_obs} Similar to Figure \ref{fig:LCDM_obs}, for
the QCD model.}
\end{center}
\end{figure}

The reasons for this become clear when we consider the 
predicted observables from this model, in Figure \ref{fig:QCD_obs}. 
Relative to $\Lambda$CDM, the QCD ghost dark energy model 
eases several tensions within the BAO data while requiring
only slight compromises with the SNe and CMB. It slightly 
exacerbates the disagreement with H$_0$, and also with $f\sigma_8$.
Taken together, these two combine to make the model untenable.
With eBOSS and DESI measurements of expansion and growth of 
structure this model could be ruled out completely without
reference to H$_0$.

As a modified gravity theory with an extra dimension, the 
DGP model and its application to cosmology has been explored 
widely
(\citealt{Lombriser_2009, Azizi_2012, Shi_2012, Xu_2014, Santos_2016, Xu_2016} 
and references therein). 
The $p$-value test indicates that this model is not favored 
by the data and the disagreement is stronger than the QCD ghost 
model and Ricci dark energy model. 
The bottom panel of Figure \ref{fig:QCD_mcmc} shows the constraints 
of the sDGP model.  These constraints show clear tensions 
among different data sets. However, one important contribution 
to the $\chi^2$ is from the disagreement of DGP model with the 
CMB data alone, which implies that the expansion of the universe 
at early times is quite different than the prediction from the 
model. The combination with other data sets enhances the 
tension and results in a huge incompatibility between the 
model and observations. 

The DGP model allows the existence of the vacuum energy, 
therefore the normal branch is also able to predict the 
current acceleration. However, the addition of the vacuum 
energy in both the normal branch and self-accelerating branch 
just gives a $\Lambda$CDM-like model, and the modification 
of gravity from the extra-dimension is a weak effect 
(\citealt{Xu_2014}). Thus, we do not explore this possibility
here.

The constraints on the parameters for the QCD ghost and DGP 
model are summarized in Table \ref{tab:QCD} and \ref{tab:DGP}.

\begin{table*}
\centering
\begin{tabular}{lllll}
\hline
\multicolumn{5}{c}{QCD ghost dark energy model} \\
\cline{1-5}
Data     &    $\Omega_{k}$    &     $\gamma$   & $\Delta$BIC  &  $\Delta$AIC\\
\hline
BAO+CMB+SNe                                        & $0.037\pm0.006$                  & $1.30_{-0.05}^{+0.06}$    & 3.4  &  -0.2 \\
BAO+CMB+SNe+H$_0$                           & $0.039_{-0.005}^{+0.006}$   &  $1.34\pm0.05$       &  6.4  &  2.9\\
BAO+CMB+SNe+$f\sigma_{8}$                & $0.018_{-0.005}^{+0.006}$   &  $1.15\pm0.03$   & 15.3  &  11.6\\
BAO+CMB+SNe+$f\sigma_{8}$+H$_0$   & $0.020\pm0.005$                  &   $1.18\pm0.03$    & 23.2  &  19.5  \\
\hline
\end{tabular}
\caption{Cosmological constraints for a selection of parameters for the QCD ghost dark energy model.}
\label{tab:QCD}
\end{table*}

\begin{table*}
\centering
\begin{tabular}{lllll}
\hline
\multicolumn{5}{c}{DGP} \\
\cline{1-5}
Data     &    $\Omega_{m}$    &     $h$  & $\Delta$BIC  &  $\Delta$AIC \\
\hline
BAO+CMB+SNe                                        & $0.335\pm0.009$                  & $0.635\pm0.006$   &  36.4  &  34.7  \\
BAO+CMB+SNe+H$_0$                           & $0.325_{-0.009}^{+0.008}$   &  $0.646\pm0.006$    &  55.5  &  53.7  \\
BAO+CMB+SNe+$f\sigma_{8}$                & $0.341\pm0.009$                  &  $0.631\pm0.006$    &  44.0  &  42.1\\
BAO+CMB+SNe+$f\sigma_{8}$+H$_0$   & $0.331\pm0.008$                  &   $0.642\pm0.006$     & 65.4  &  63.6 \\
\hline
\end{tabular}
\caption{Cosmological constraints for a selection of parameters for the DGP model.}
\label{tab:DGP}
\end{table*}

\subsection{$f(R)$ gravity}

The designer model of $f(R)$ gravity has the same expansion history 
as the standard $\Lambda$CDM model; therefore, constraints on it 
come in this work from the linear growth data. Because the growth in 
$f(R)$ gravity is scale-dependent (\citealt{Pogosian_2008}), \add{the modifications to structure occur for some wavenumbers given the scale $B$} (\citealt{Linder_2009}), we 
test this model using two scaling wavenumbers as representatives, $k=0.1h$Mpc$^{-1}$ and 
$k=0.02h$Mpc$^{-1}$,  as suggested in \cite{Huterer_2015} (hereafter 
called the $f(R)_{1}$ and  $f(R)_{2}$ model respectively). 

\begin{figure}[htbp]
\begin{center}
\includegraphics[width=9cm, height=7cm]{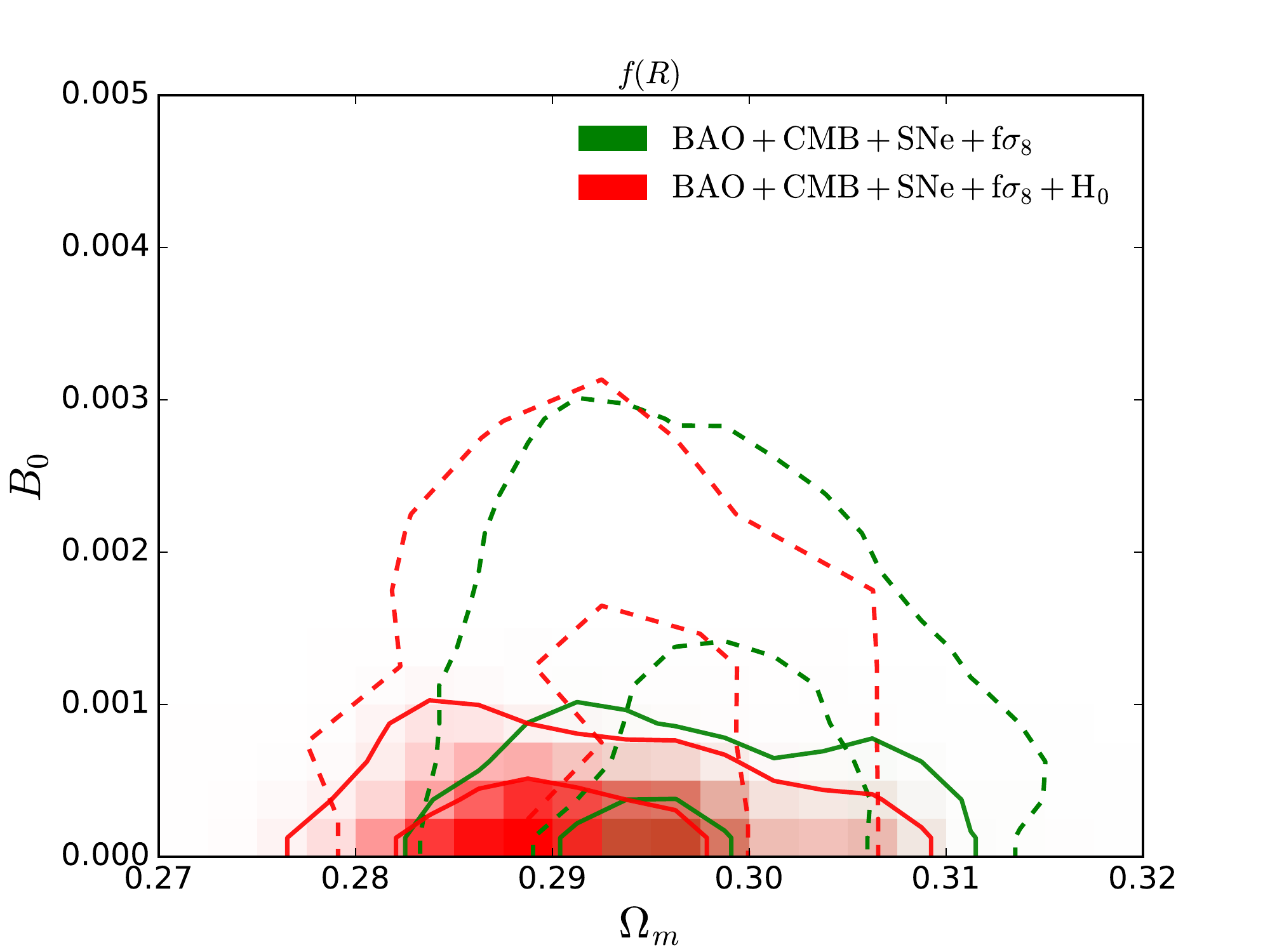}
\caption{The 68.7\% and 95.0\% confidence regions of the parameters for the designer model of $f(R)$ gravity, solid line for $k=0.1$ and dashed line for $k=0.02$.}
\label{fig:fR_mcmc}
\end{center}
\end{figure}

Figure \ref{fig:fR_mcmc} shows the constraints on these two models, with 
and without the inclusion of H$_{0}$. The two models give similar constraints 
on the matter fraction $\Omega_{m}$; the allowed region for the factor 
$B_{0}$ is different, and includes $B_{0}=0$ in both cases. 
For $k=0.1h$Mpc$^{-1}$, the constraint on the Compton wavelength parameter 
is $B_{0}<1.0\times10^{-3}$ at 95\% confidence level, while 
for $k=0.02h$Mpc$^{-1}$, the constraint is twice as bad. This property 
does not depend much on the H$_{0}$ data, since the parameter $B_{0}$ 
only affects the evolution of the perturbation in the designer model. 
Note that for the constraints on $B_{0}$, only the redshift space
distortion data  as a dynamical probe are used. Previous work using
more dynamical probes, such as the galaxy correlations and cluster data, shows consistent results 
(\citealt{Lombriser_2012, Cataneo_2015}). The constraint on the matter 
fraction $\Omega_{m}$ and the Compton wavelength parameter $B_{0}$ at 
the two different scales are summarized in Table \ref{tab:fR1} and \ref{tab:fR2}.

Since the constraints limit this model to be close to $\Lambda$CDM,
we expect that the $p$-value test will yield comparable results to $\Lambda$CDM. 
Indeed, the $p$-value test of $f(R)$ gravity model shows is marginally more 
consistent with the data than is the $\Lambda$CDM model when all the 
datasets are used, and somewhat less consistent when H$_0$ is 
included. The predicted observables under this model are tightly 
constrained to be near $\Lambda$CDM (similar to Figure \ref{fig:LCDM_obs}).

\begin{table*}
\centering
\begin{tabular}{lllll}
\hline
\multicolumn{5}{c}{$f(R)$ gravity: $k=0.1h$Mpc$^{-1}$} \\
\cline{1-5}
Data     &    $\Omega_{m}$    &     $B_{0} (95\% \rm{CL})$  & $\Delta$BIC  &  $\Delta$AIC \\
\hline
BAO+CMB+SNe+$f\sigma_{8}$                & $0.300\pm0.008$   &  $0<B_{0}<0.001$    & 3.8  & 2.0 \\
BAO+CMB+SNe+$f\sigma_{8}$+H$_0$    & $0.294\pm0.007$  &  $0<B_{0}<0.001$     &  3.9  & 2.0   \\
\hline
\end{tabular}
\caption{Cosmological constraints for a selection of parameters for the $f(R)$ gravity with $k=0.1h$Mpc$^{-1}$.}
\label{tab:fR1}
\end{table*}

\begin{table*}
\centering
\begin{tabular}{lllll}
\hline
\multicolumn{5}{c}{$f(R)$ gravity: $k=0.02h$Mpc$^{-1}$} \\
\cline{1-5}
Data     &    $\Omega_{m}$    &     $B_{0}  (95\% \rm{CL})$  & $\Delta$BIC  &  $\Delta$AIC \\
\hline
BAO+CMB+SNe+$f\sigma_{8}$                & $0.300\pm0.008$   &  $0<B_{0}<0.003$   & 3.8  & 2.0  \\
BAO+CMB+SNe+$f\sigma_{8}$+H$_0$    & $0.294\pm0.007$  &  $0<B_{0}<0.003$    & 3.9  & 2.0     \\
\hline
\end{tabular}
\caption{Cosmological constraints for a selection of parameters for the $f(R)$ gravity with $k=0.02h$Mpc$^{-1}$.}
\label{tab:fR2}
\end{table*}

\subsection{$f(T)$ gravity ($f(T)_{\rm PL}$, $f(T)_{\rm Exp1}$,
$f(T)_{\rm Exp2}$, $f(T)_{\rm tanh}$)}

\begin{figure}[htbp]
\begin{center}
\includegraphics[width=4.1cm, height=3.8cm]{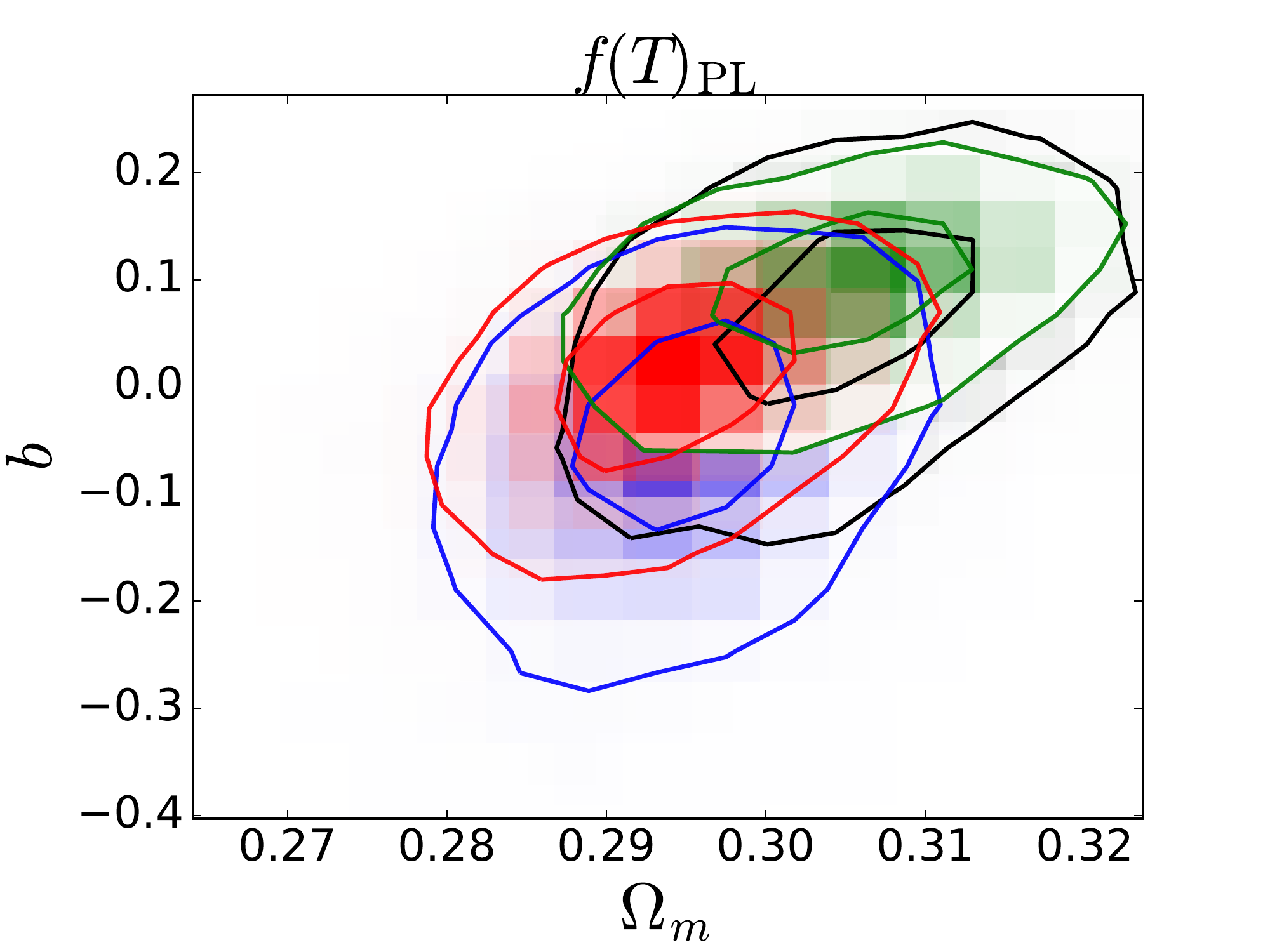}
\includegraphics[width=4.1cm, height=3.8cm]{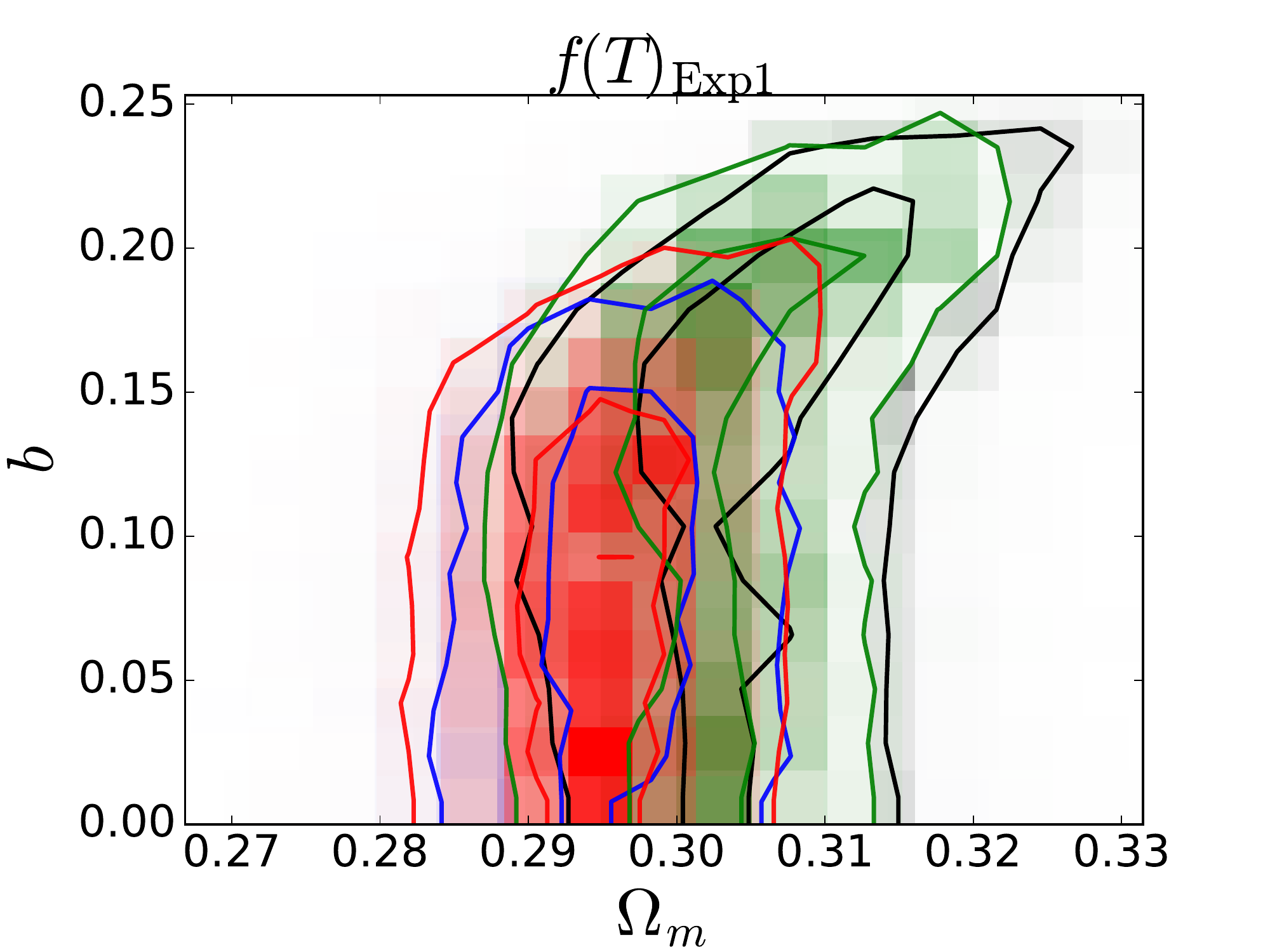}
\includegraphics[width=4.1cm, height=3.8cm]{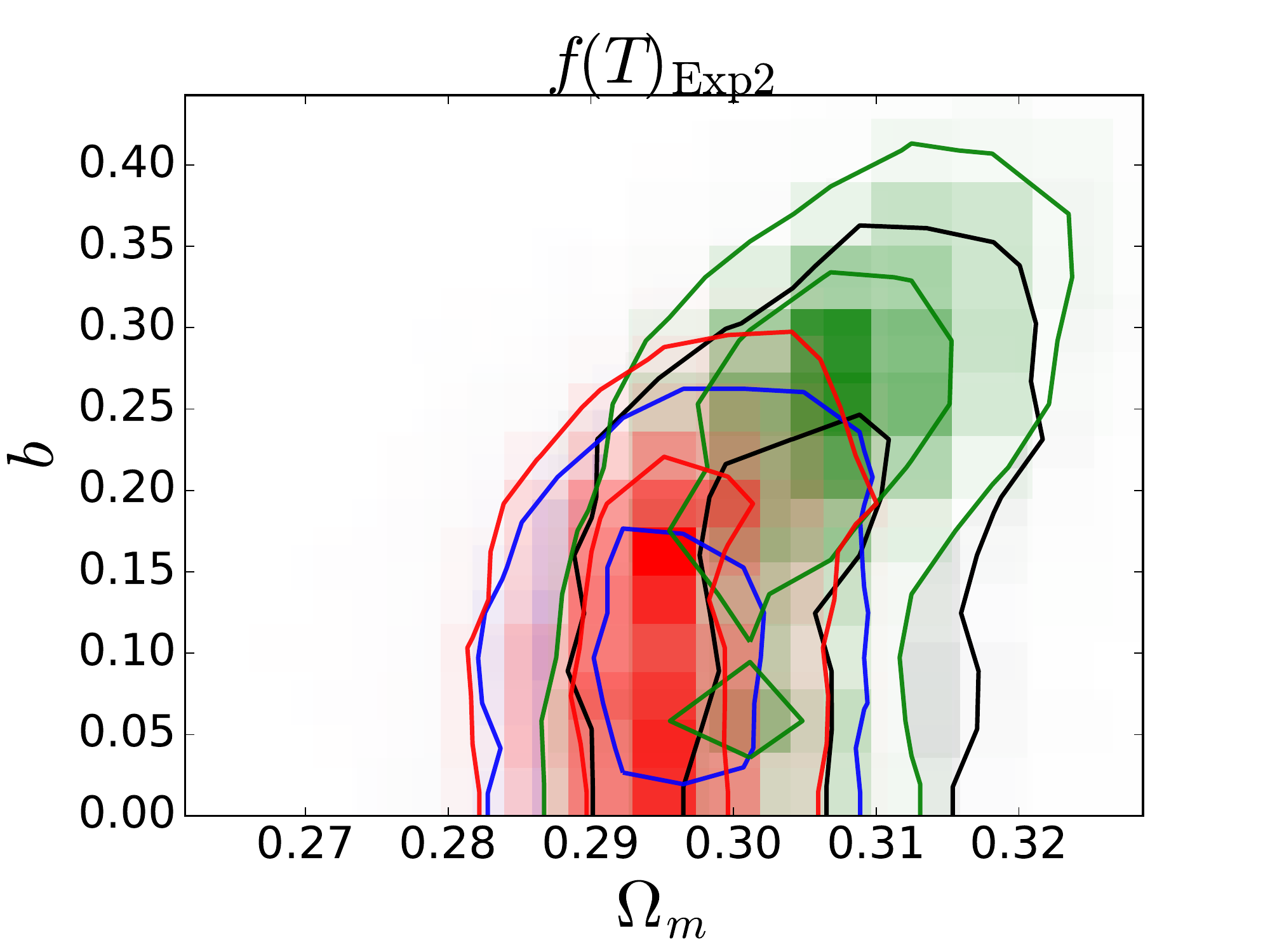}
\includegraphics[width=4.1cm, height=3.8cm]{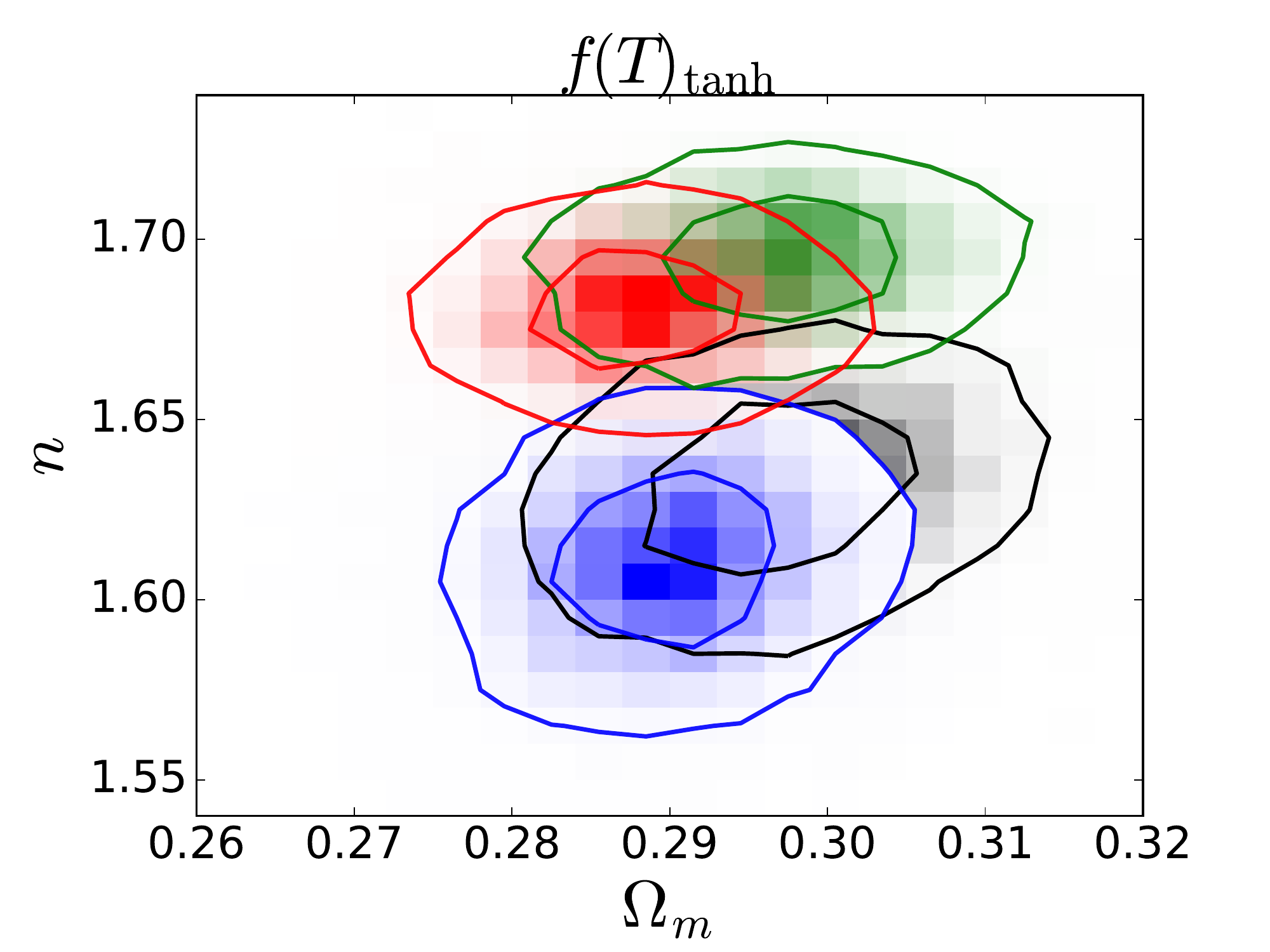}
\caption{The 68.7\% and 95.0\% confidence regions of the parameters for the $f(T)$ gravity: Model I (power-law model, upper left), Model II (exponential model, upper right), Model III (another exponential model, lower left),  Model IV (hyperbolic-tangent model, lower right). The diagonal panels show the one-dimensional probability distribution functions. }
\label{fig:fT_mcmc}
\end{center}
\end{figure}

Figure \ref{fig:fT_mcmc} displays the constraints on the four $f(T)$
gravity models considered in this paper. The first three models 
tend to the $\Lambda$CDM cosmology when their parameter $b$ approaches 0. 
When $f\sigma_8$ is included in the constraints, there is a slight 
deviation from $b=0$ for $f(T)_{\rm PL}$ (Model I), but still less than $2\sigma$.
This deviation is apparently reduced when the H$_{0}$ measurement is added.
All of these models perform similarly to $\Lambda$CDM in the $p$-value
tests, in some cases slightly worse or slightly better (the latter especially
when H$_0$ is excluded). In these models, the predicted observables
are fairly well constrained, similar to non-flat $\Lambda$CDM (Figure
\ref{fig:LCDM_obs}), with a 
little more freedom for the power-law $f(T)$ (Model III).

\begin{figure}[t!]
\begin{center}
\includegraphics[width=0.49\textwidth]{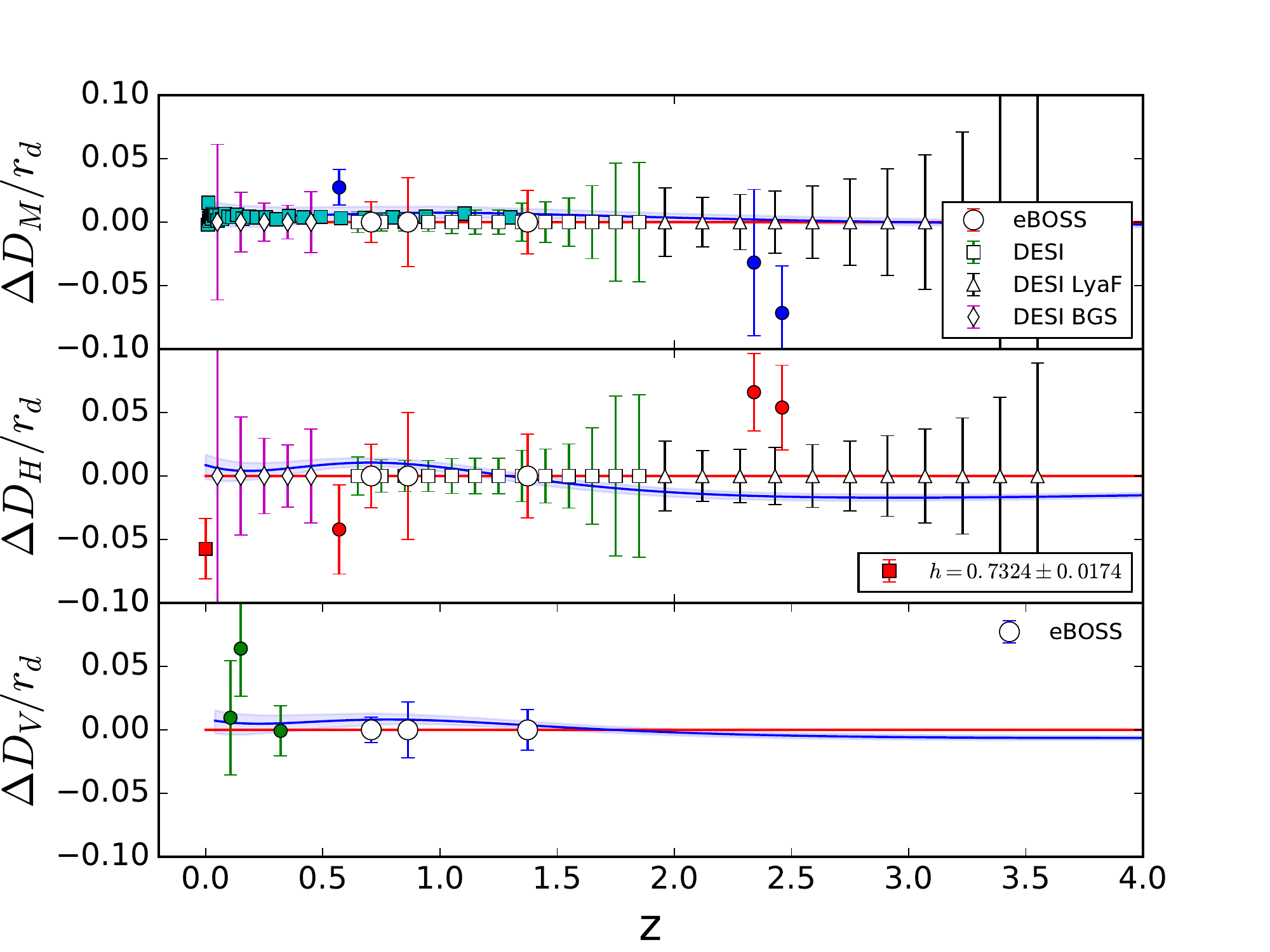}
\includegraphics[width=0.49\textwidth]{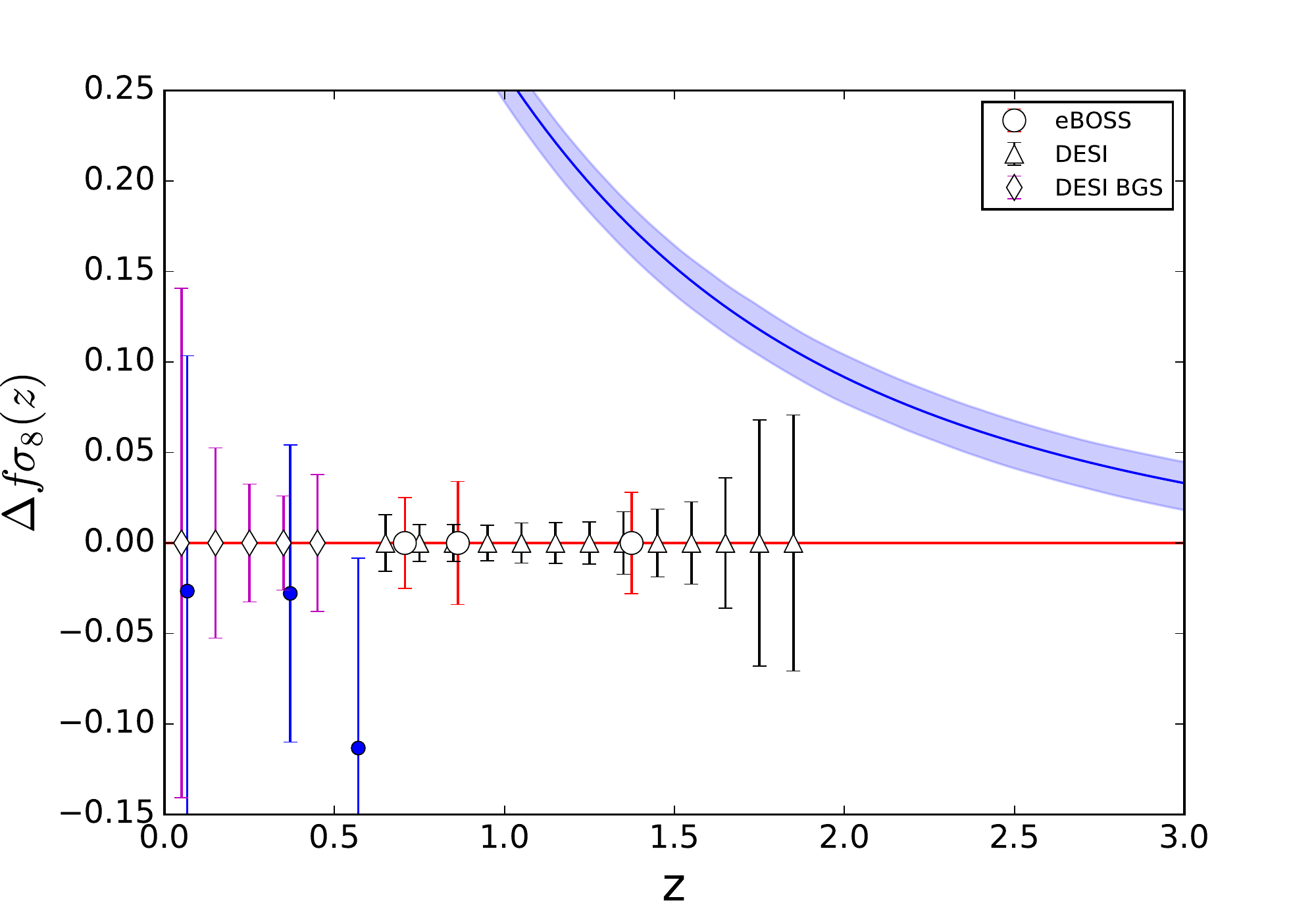}
\caption{\label{fig:fTtanh_obs} Similar to Figure \ref{fig:LCDM_obs}, for
the $f(T)_{\rm tanh}$ model.}
\end{center}
\end{figure}

$f(T)_{\rm tanh}$ (Model IV) is the hyperbolic-tangent model, which 
has the same number 
of parameters as the other $f(T)$ models. However, this model 
does not reduce to the $\Lambda$CDM cosmology for any value of its 
parameters. The constraints clearly depend strongly on whether the 
$f\sigma_8$ data are included. Correspondingly, the $p$-values for 
Model IV are only somewhat worse than $\Lambda$CDM when $f\sigma_8$ is 
included, but far worse when it is included. Due to its non-trivial 
modification of the linear perturbation function, this model shows 
a value of the perturbation amplitude $\sigma_{8}$ of around 1.1 
with about a 2\% uncertainty, much higher than indicated by the CMB.
This $f(T)$ model is therefore firmly ruled out by the growth of 
structure data, as also pointed out by \cite{Nesseris_2013}. Figure 
\ref{fig:fTtanh_obs} shows the predicted observables, which 
demonstrates clearly how the predictions for the growth of structure
deviate from that observed.

Note that none of these four different $f(T)$ models relax the 
tension with H$_0$ better than the $\Lambda$CDM model.

The parameter constraints for the models are presented in 
Tables \ref{tab:fT1}--\ref{tab:fT4}. 

\begin{table*}
\centering
\begin{tabular}{lllll}
\hline
\multicolumn{5}{c}{$f(T)$ gravity: Model I} \\
\cline{1-5}
Data     &    $\Omega_{m}$    &     $b$  & $\Delta$BIC  &  $\Delta$AIC \\
\hline
BAO+CMB+SNe                                        & $0.304\pm0.009$                  & $0.06_{-0.10}^{+0.09}$    &  3.2  &  1.5 \\
BAO+CMB+SNe+H$_0$                           & $0.295\pm0.008$                  &  $-0.05_{-0.12}^{+0.10}$    & 3.7  &  1.9  \\
BAO+CMB+SNe+$f\sigma_{8}$                & $0.304\pm0.009$                  &  $0.09_{-0.08}^{+0.06}$   & 2.1  &  0.2 \\
BAO+CMB+SNe+$f\sigma_{8}$+H$_0$   & $0.294_{-0.007}^{+0.008}$    &  $0.01_{-0.09}^{+0.08}$    &  3.8  & 1.9  \\
\hline
\end{tabular}
\caption{Cosmological constraints for a selection of parameters for the $f(T)$ gravity: power-law model.}
\label{tab:fT1}
\end{table*}

\begin{table*}
\centering
\begin{tabular}{lllll}
\hline
\multicolumn{5}{c}{$f(T)$ gravity: Model II} \\
\cline{1-5}
Data     &    $\Omega_{m}$    &     $b (68\% \rm{CL})$  & $\Delta$BIC  &  $\Delta$AIC \\
\hline
BAO+CMB+SNe                                        & $0.305\pm0.009$                  & $0.0<b<0.20$     &  3.4  & 1.7\\
BAO+CMB+SNe+H$_0$                           & $0.296\pm0.007$                  &  $0.0<b<0.15$      &  3.8  &  2.0\\
BAO+CMB+SNe+$f\sigma_{8}$                & $0.303_{-0.009}^{+0.010}$   &  $0.0<b<0.19$  &  3.3  & 1.5\\
BAO+CMB+SNe+$f\sigma_{8}$+H$_0$   & $0.295_{-0.007}^{+0.008}$    &  $0.0<b<0.15$      &  3.9  & 2.0\\
\hline
\end{tabular}
\caption{Cosmological constraints for a selection of parameters for the $f(T)$ gravity: exponential model.}
\label{tab:fT2}
\end{table*}

\begin{table*}
\centering
\begin{tabular}{lllll}
\hline
\multicolumn{5}{c}{$f(T)$ gravity: Model III} \\
\cline{1-5}
Data     &    $\Omega_{m}$    &     $b (68\% \rm{CL})$   & $\Delta$BIC  &  $\Delta$AIC\\
\hline
BAO+CMB+SNe                                        & $0.305_{-0.008}^{+0.009}$   &  $0.0<b<0.27$     & 3.1  &  1.4\\
BAO+CMB+SNe+H$_0$                           & $0.297\pm0.007$                  &  $0.0<b<0.20$      &  3.8  &  2.0\\
BAO+CMB+SNe+$f\sigma_{8}$                & $0.305_{-0.009}^{+0.010}$   &  $0.0<b<0.32$  &  2.6  & 0.8\\
BAO+CMB+SNe+$f\sigma_{8}$+H$_0$   & $0.295_{-0.007}^{+0.008}$    &  $0.0<b<0.22$      &  3.9  & 2.0\\
\hline
\end{tabular}
\caption{Cosmological constraints for a selection of parameters for the $f(T)$ gravity: the second exponential model.}
\label{tab:fT3}
\end{table*}

\begin{table*}
\centering
\begin{tabular}{lllll}
\hline
\multicolumn{5}{c}{$f(T)$ gravity: Model IV} \\
\cline{1-5}
Data     &    $\Omega_{m}$    &     $n$   & $\Delta$BIC  &  $\Delta$AIC\\
\hline
BAO+CMB+SNe                                        & $0.297\pm0.008$                 &  $1.63\pm0.02$     &  6.7  &  4.9\\
BAO+CMB+SNe+H$_0$                           & $0.290\pm0.007$                  &  $1.61\pm0.02$     &  5.9  & 4.2  \\
BAO+CMB+SNe+$f\sigma_{8}$                & $0.297\pm0.008$                  &  $1.69\pm0.02$  & 154.0  & 152.2 \\
BAO+CMB+SNe+$f\sigma_{8}$+H$_0$   & $0.288\pm0.007$                  &  $1.68\pm0.02$   &  159.2 &  157.3   \\
\hline
\end{tabular}
\caption{Cosmological constraints for a selection of parameters for the $f(T)$ gravity: hyperbolic-tangent model.}
\label{tab:fT4}
\end{table*}

\subsection{Galileon cosmology: Tracker solution (GAL)}

The Galileon theory has solutions with different branches, 
including the model conformally (disformally) coupled to matter, 
general solution, tracker solution and so on. 
For numerical simplicity, we only consider the tracker solution 
in this paper, and refer the reader to the literature for a 
comprehensive discussion of the Galileon cosmology 
(\citealt{Barreira_2014a, Barreira_2014b, Brax_2015, Neveu_2016}). 

\begin{figure}[htbp]
\begin{center}
\includegraphics[width=9cm, height=8cm]{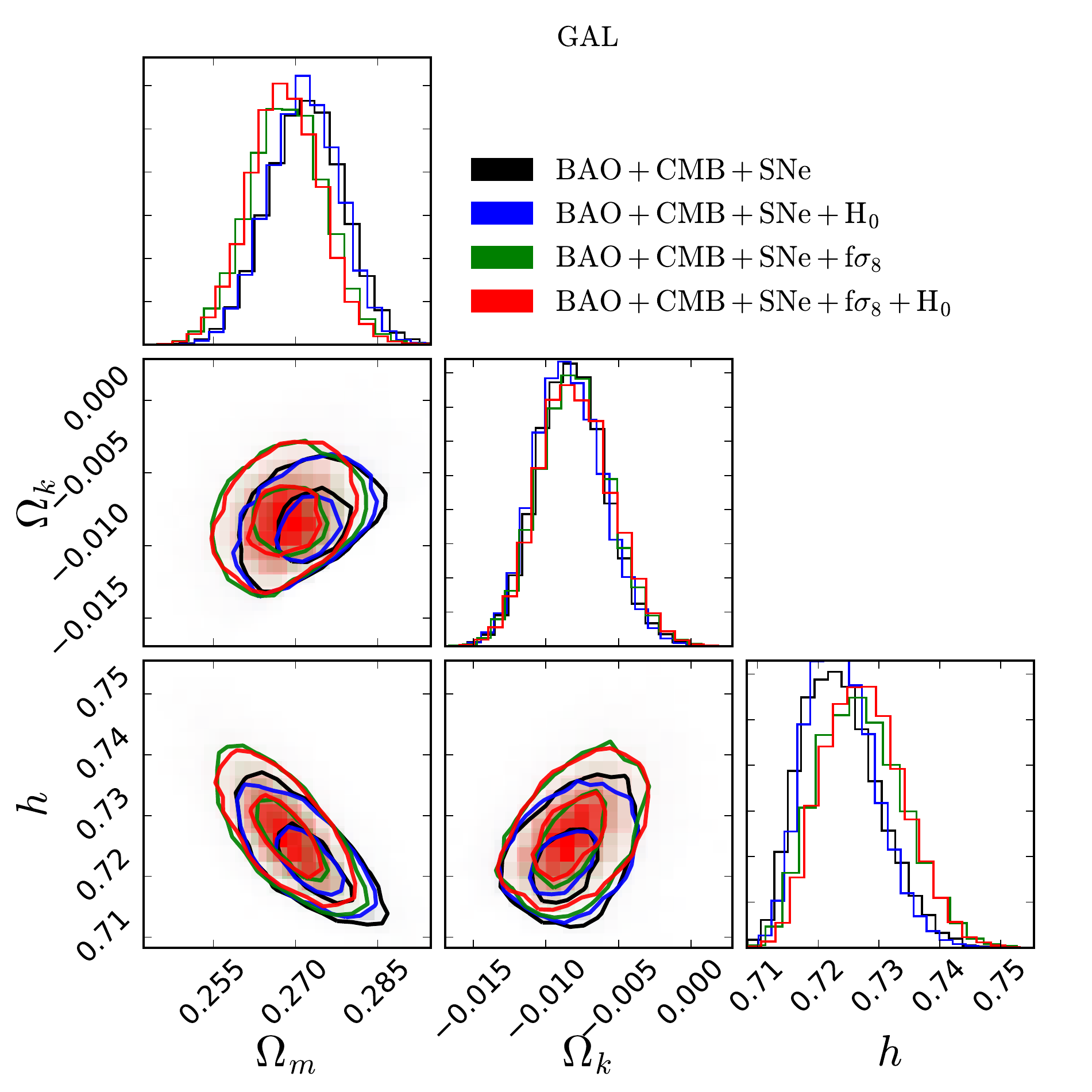}
\caption{The 68.7\% and 95.0\% confidence regions of the parameters for the tracker solution of the Galileon cosmology. The diagonal panels show the one-dimensional probability distribution functions. }
\label{fig:Gal_mcmc}
\end{center}
\end{figure}

Figure \ref{fig:Gal_mcmc} and Table \ref{tab:Gal} present the constraints on 
the tracker solution of the Galileon cosmology. The $p$-value test from the 
last section shows that this model is clearly ruled out by the data. 

\begin{figure}[t!]
\begin{center}
\includegraphics[width=0.49\textwidth]{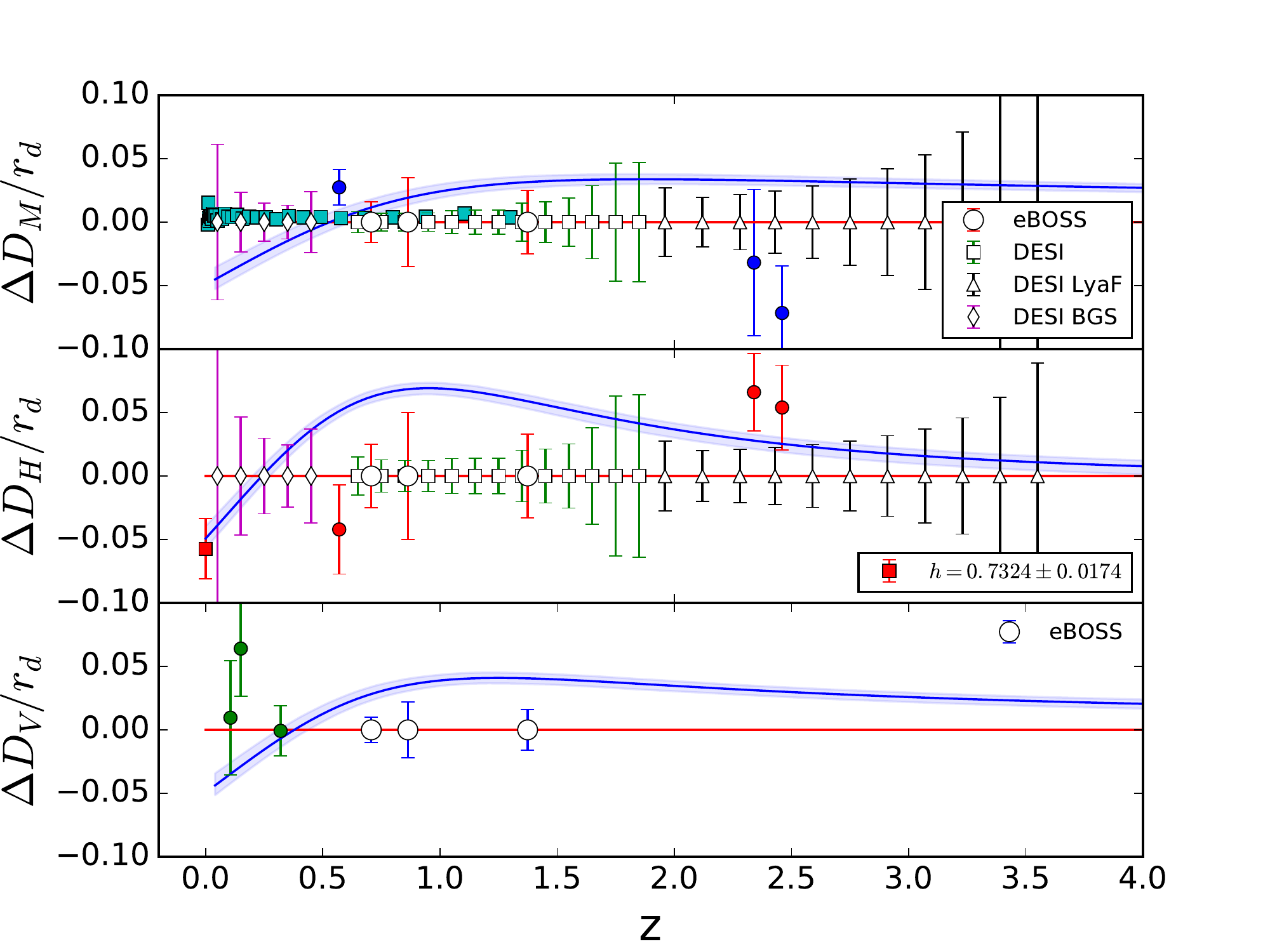}
\includegraphics[width=0.49\textwidth]{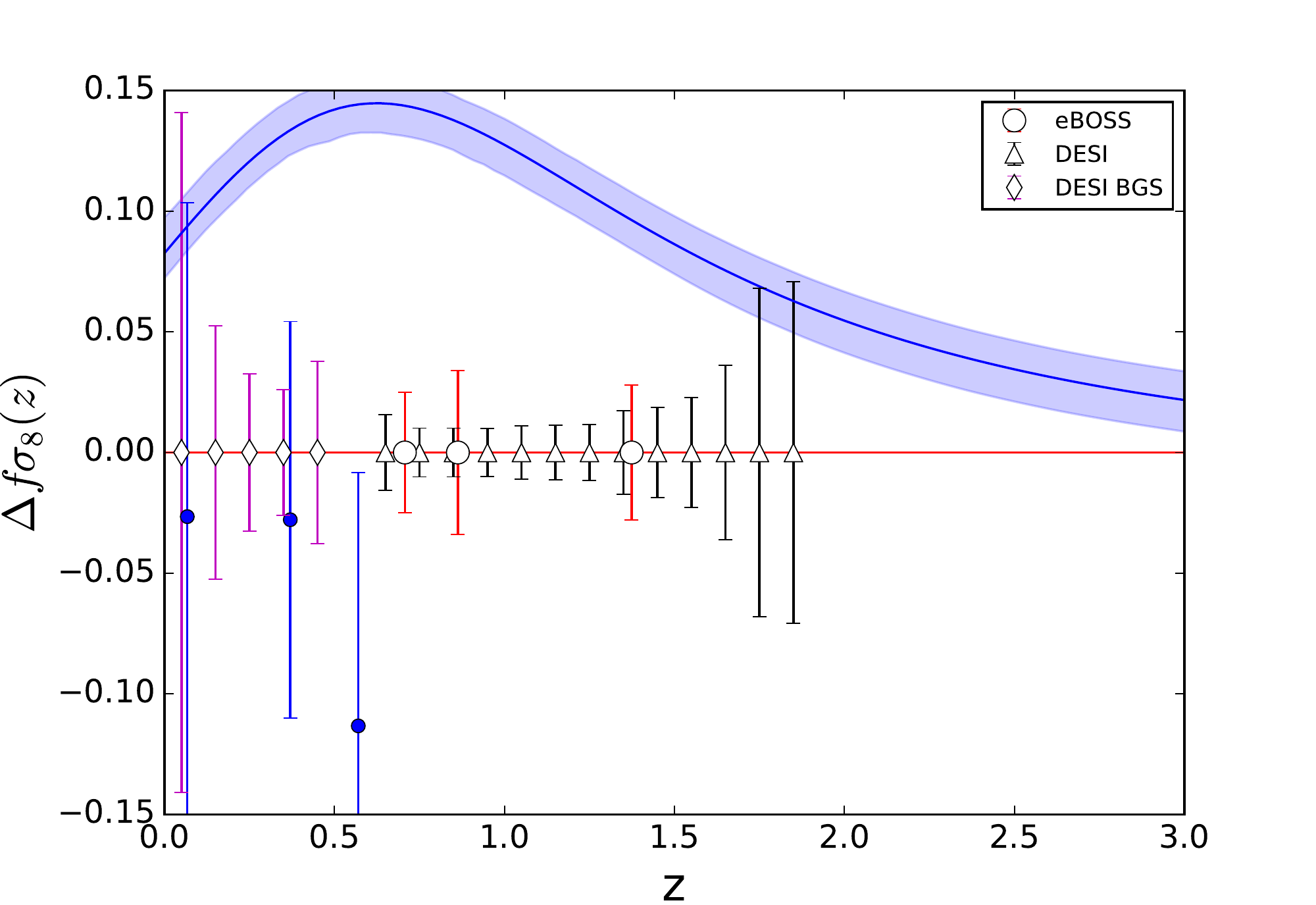}
\caption{\label{fig:Gal_obs} Similar to Figure \ref{fig:LCDM_obs}, for
the Galileon tracker model.}
\end{center}
\end{figure}

Figure \ref{fig:Gal_obs} shows the predicted observables for this 
model.  Interestingly, we see that in this model the tension with the Hubble 
constant H$_0$ is resolved, so that the best fit $h$ does not depend much
on whether H$_0$ is included (as found, for example, by 
\cite{Barreira_2014b}). However, this resolution comes at the 
expense of being able to fit well the BAO and 
especially the SNe data. As has been noted before, the growth of structure
measurements are very different than $\Lambda$CDM, but the expansion
measurements are already enough to rule out the model completely.

The $p$-value test shows that the incompatibility between the data 
and the Galileon theory is only for the tracker solution. It is still 
possible to have more flexible descriptions of the evolution of the universe 
in this theoretical frame (\citealt{Neveu_2016}).

\begin{table*}
\centering
\begin{tabular}{lllll}
\hline
\multicolumn{5}{c}{Galileon cosmology: tracker solution} \\
\cline{1-5}
Data     &    $\Omega_{m}$    &     $h$      & $\Delta$BIC  &  $\Delta$AIC \\
\hline
BAO+CMB+SNe                                        & $0.272\pm0.007$                 &  $0.723_{-0.006}^{+0.007}$   & 29.6  &  27.8  \\
BAO+CMB+SNe+H$_0$                           & $0.272\pm0.007$                  &  $0.723_{-0.005}^{+0.006}$    & 23.9  &  22.2  \\
BAO+CMB+SNe+$f\sigma_{8}$                & $0.268\pm0.007$                  &  $0.727\pm0.007$  &  41.9  & 40.0  \\
BAO+CMB+SNe+$f\sigma_{8}$+H$_0$   & $0.268\pm0.006$                  &  $0.727\pm0.007$   &  36.4  & 34.5   \\
\hline
\end{tabular}
\caption{Cosmological constraints for a selection of parameters for the tracker solution of the Galileon cosmology.}
\label{tab:Gal}
\end{table*}

\subsection{Kinetic gravity braiding model (KGBM, KGBM$_{n=1}$)}

\begin{figure}[htbp]
\begin{center}
\includegraphics[width=9cm, height=7cm]{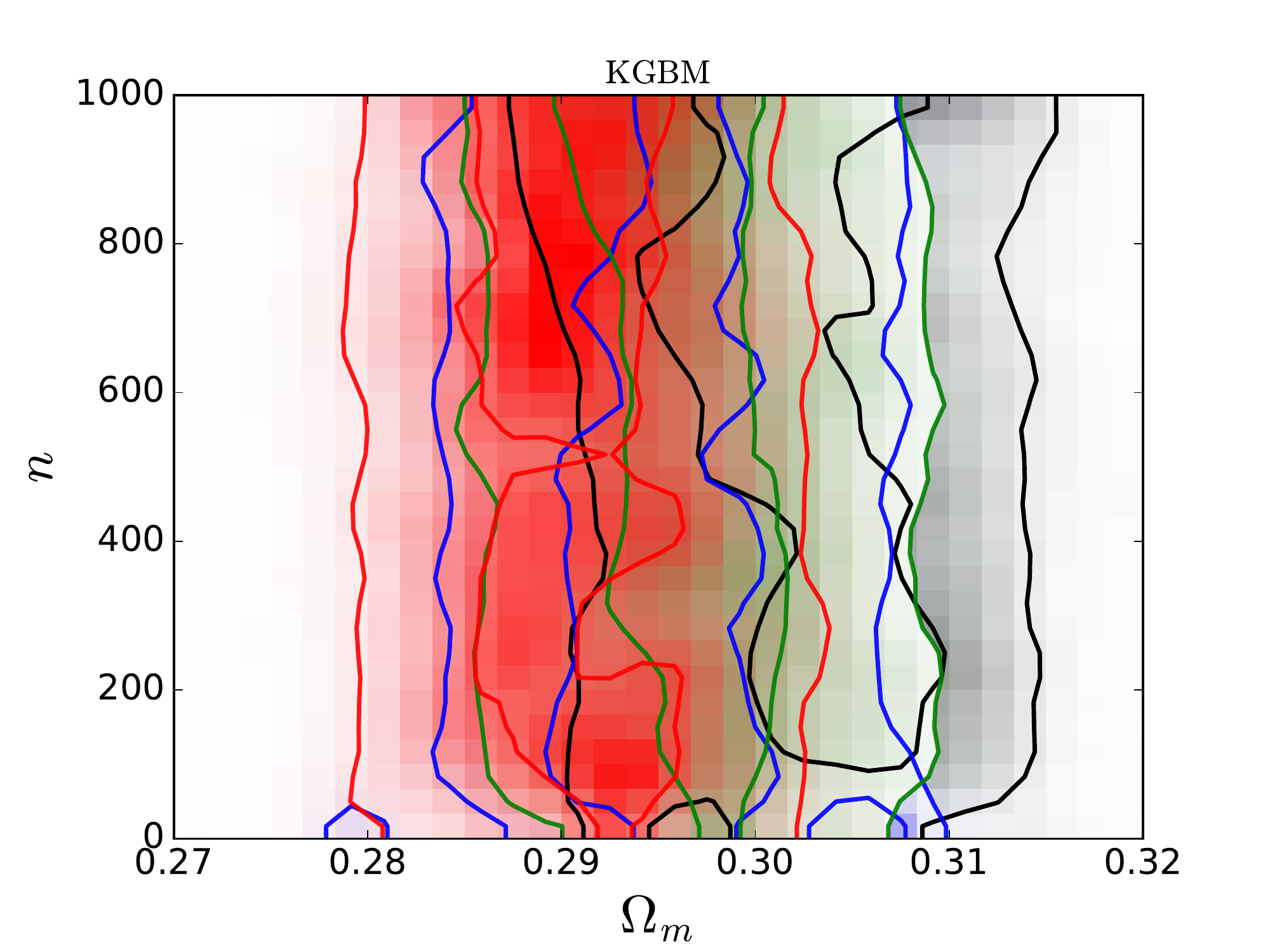}
\caption{The 68.7\% and 95.0\% confidence regions of the parameters for the tracker solution of the Kinetic gravity braiding model. The diagonal panels show the one-dimensional probability distribution functions. }
\label{fig:KGBM_mcmc}
\end{center}
\end{figure}

The kinetic gravity braiding model can be thought of as a 
generalization of the Galileon theory. The constraints on 
this model are presented in Figure \ref{fig:KGBM_mcmc}. As 
pointed out in Section \ref{sec:intro_KGBM}, the geometrical 
evolution of this model is equivalent to the power-law $f(T)$ 
theory (with a transformation relating $n$ and $b$). When $n$ 
approaches $\infty$, this model returns to the $\Lambda$CDM 
model. The constraint on $n$ has no useful upper bound for 
the data combinations considered here; in this way the KGBM model
and the  power-law $f(T)$ model give consistent results with each other. 

\begin{figure}[t!]
\begin{center}
\includegraphics[width=0.49\textwidth]{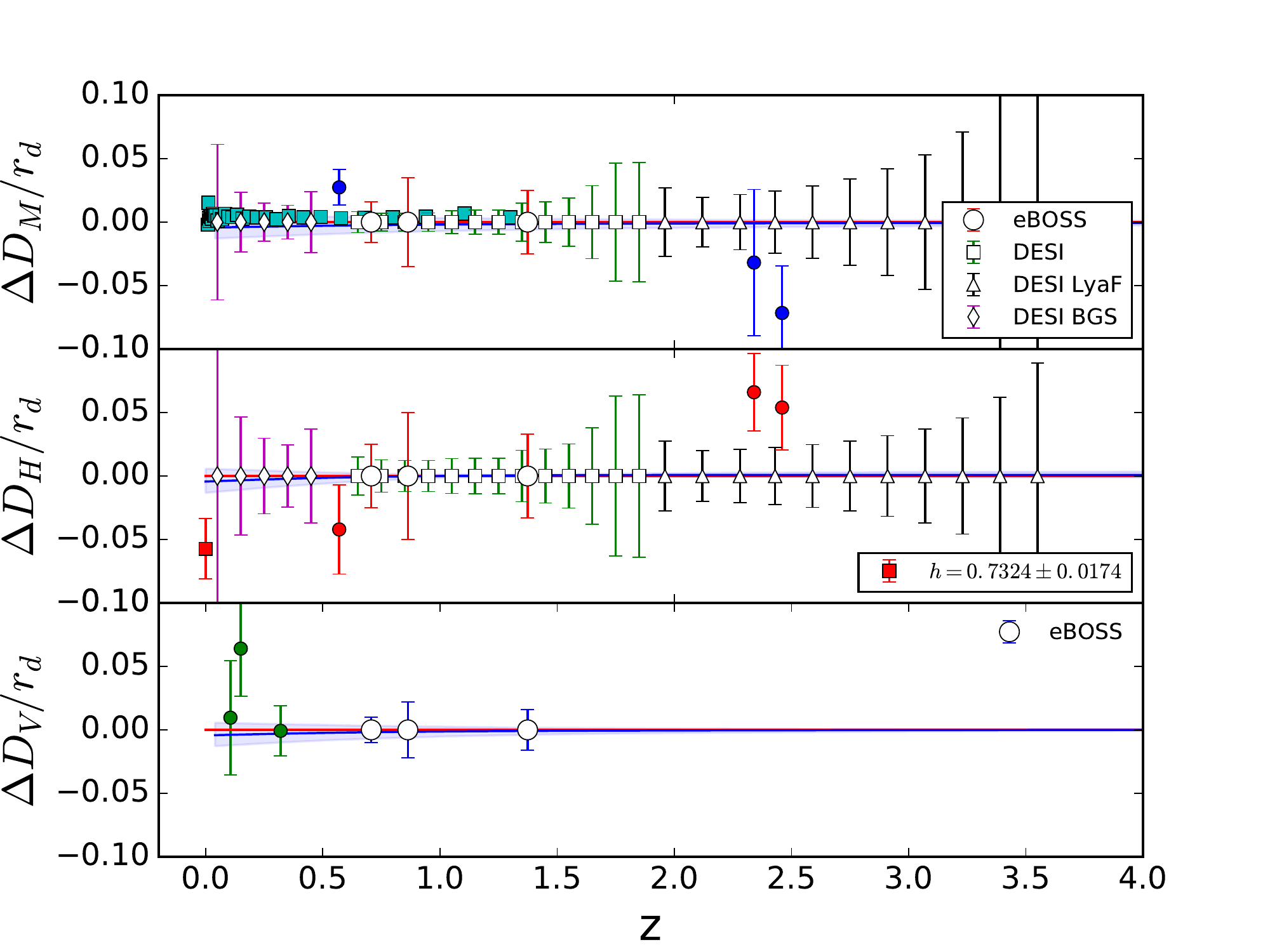}
\includegraphics[width=0.49\textwidth]{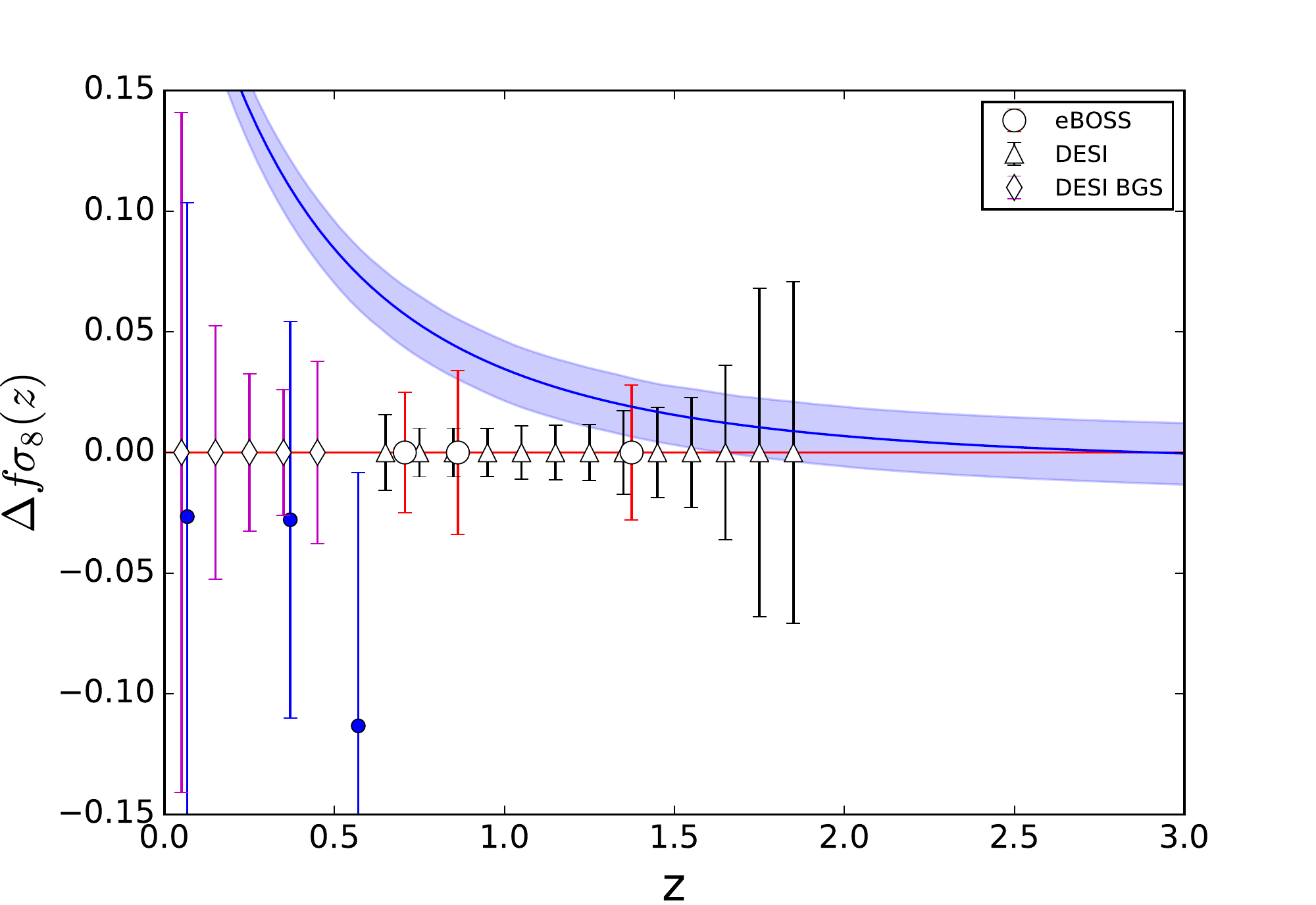}
\caption{\label{fig:KGBM_obs} Similar to Figure \ref{fig:LCDM_obs}, for
the KGBM model.}
\end{center}
\end{figure}

The difference of these two theories is reflected by the 
linear growth data, because KGBM
modifies the perturbation equation differently due to the 
time-dependent effective gravitational constant 
(\citealt{Kimura_2012}). This results in a fit worse even than the 
power-law $f(T)$ model and therefore not favored by the linear growth data. 
The reason for the bad fit is that at early times, the effective gravitational 
constant $G_{\rm{eff}}$ from the modification of the perturbation 
equation is nearly equal to $G$ in general relativity, and 
becomes larger than $G$ at late times (\citealt{Kimura_2011}). 
Figure \ref{fig:KGBM_obs} shows the predicted observables,
revealing how KGBM can recover expansion very similar to $\Lambda$CDM
but with a very different, and ruled out, rate of growth of structure.

The kinetic gravity braiding model has a special case when $n=1$, 
because its geometrical evolution is equivalent to the tracker solution 
of Galileon theory. However, these two models predict different growth
of structure; even for $n=1$, the modified perturbation equation for 
the kinetic gravity braiding model is not identical to the tracker 
solution of the Galileon theory. The comparison of the KGBM$_{n=1}$
predictions with linear growth data is however not better than the 
Galileon model; thus  both of these two models are ruled out in the
$p$-value test. Table 
\ref{tab:KGBM} and \ref{tab:KGBMn1} show the constraint results for 
these two cases of the kinetic gravity braiding model respectively.

\begin{table*}
\centering
\begin{tabular}{lllll}
\hline
\multicolumn{5}{c}{KGBM} \\
\cline{1-5}
Data     &    $\Omega_{m}$    &     $n  (68\% \rm{CL})$    & $\Delta$BIC  &  $\Delta$AIC \\
\hline
BAO+CMB+SNe                                        & $0.304\pm0.009$                 &  $>182$    &  3.2  &  1.5 \\
BAO+CMB+SNe+H$_0$                           & $0.295\pm0.008$                 &  $>138$     &  3.7  & 1.9 \\
BAO+CMB+SNe+$f\sigma_{8}$                & $0.296_{-0.006}^{+0.008}$  &  $>180$   &  13.6  &  11.8 \\
BAO+CMB+SNe+$f\sigma_{8}$+H$_0$   & $0.291_{-0.007}^{+0.008}$   &  $>169$   &  13.9  &  12.0    \\
\hline
\end{tabular}
\caption{Cosmological constraints for a selection of parameters for the kinetic gravity braiding model.}
\label{tab:KGBM}
\end{table*}

\begin{table*}
\centering
\begin{tabular}{lllll}
\hline
\multicolumn{5}{c}{KGBM with $n=1$} \\
\cline{1-5}
Data     &    $\Omega_{m}$    &     $h$    & $\Delta$BIC  &  $\Delta$AIC  \\
\hline
BAO+CMB+SNe                                        & $0.272\pm0.007$                 &  $0.723_{-0.006}^{+0.007}$     & 29.6  &  27.8  \\
BAO+CMB+SNe+H$_0$                           & $0.272\pm0.007$                  &  $0.723_{-0.005}^{+0.006}$    &  23.9  &  22.2  \\
BAO+CMB+SNe+$f\sigma_{8}$                & $0.268\pm0.007$                  &  $0.727\pm0.007$   &  45.5  &  43.6   \\
BAO+CMB+SNe+$f\sigma_{8}$+H$_0$   & $0.268\pm0.006$                  &  $0.728\pm0.007$    &  40.0  &  38.1   \\
\hline
\end{tabular}
\caption{Cosmological constraints for a selection of parameters for the tracker solution of the Galileon cosmology with $n=1$.}
\label{tab:KGBMn1}
\end{table*}

\section{Discussion and Conclusion}\label{sec:dis}

The underlying cause of cosmic acceleration is as yet unknown. The phenomenon has inspired
a large variety of theoretical models and considerable expenditure of observational 
resources to create samples designed to test these models. The cosmological experiments 
have provided measurements of the history of expansion and growth of structure to a 
precision of percent-level or better, which strongly constrain the cosmological parameters.
These developments motivate us to test the compatibility between the data and the various 
models for cosmic acceleration, in a consistent manner for all of them. We use measurements
of the CMB, BAO distances, $f\sigma_8$ using redshift space distortions, SNe Type Ia, and
the local Hubble constant.

We have evaluated the models in three ways. First, we have calculated the $p$-value 
associated with a global $\chi^2$ statistic for each model, yielding the probability 
under the model of obtaining data in at least as much disagreement as the observations 
are. This approach allows us to determine which models can be ruled out by the data. 
However, it does not provide a framework for model selection --- i.e. determining 
which of the models that are not ruled out should be favored. We have decided to 
avoid model selection because there is little  basis for assigning any prior 
probability for each theoretical model (\citealt{Efstathiou_2008}). 
Second, we have constrained the parameters of each model using MCMC. 
Third, we have used the distribution of parameters from the MCMC to 
compute the predicted observables within each model space, given current
constraints. These predicted observables yield insight into where tensions
are arising with respect to the model and how future observations could
constrain them.

The non-flat $\Lambda$CDM and its parametrized generalizations (XCDM, CPL,
JBP, $w_{\rm Linear}$, PolyCDM) exhibit some mild tension with the CMB, 
BAO, and SNe, but not much with $f\sigma_8$. They all have difficulty 
explaining H$_0$, which substantially reduces the $p$-value. In general,
the less constrained models (e.g. PolyCDM) exhibit a larger variation
in their predicted observables, but the range of behavior allowed with
these model spaces, under current observational constraints, is relatively
small. 

Most other physical or parametric models we consider behave very similarly 
in general to the $\Lambda$CDM-like ones. They 
do not exhibit specific redshift ranges that would be especially informative
and for the most part when the expansion history measurements agree with
$\Lambda$CDM the growth of structure is predicted to as well. 

However, some models behave in interesting ways.
Several models are firmly ruled out by just the CMB, BAO, and SNe (Ricci 
Dark Energy, the Galileon tracker solution, and sDGP). 

Adding $f\sigma_8$ 
rules out the $f(T)_{\rm tanh}$ modified gravity model, 
rules out a particular class of the kinetic
gravity braiding model (the $n=1$ solution), 
and puts great pressure even on the more general kinetic gravity braiding 
model and on the QCD ghost model. Holographic Dark Energy agrees very
well with current $f\sigma_8$ measurements; it is a model that can 
reproduce $\Lambda$CDM but could  be ruled out strongly by future 
expansion rate or $f\sigma_8$ measurements from eBOSS and DESI. 

Further adding H$_0$ firmly rules out the 
QCD ghost model. In general, H$_0$ adds substantial tension to all models.
The Galileon tracker solution explains it pretty well, but fails on 
multiple other data sets, particularly the Type Ia SNe.

The H$_{0}$ problem has been visited by many groups recently. By exploring an extended parameter space, \cite{Valentino_2016, Valentino_2017} find that the dynamical dark energy component can mitigate the tension on H$_{0}$ at some extent. A similar conclusion is also reached in \cite{Zhao_2017} which employ an non-parametric approach in the modeling of dark energy, while the Bayesian evidence is not sufficient to favor over $\Lambda$CDM model. A more physically well-motivated dark energy model is examined in \cite{Sola_2017a, Sola_2017b, Sola_2017c} which assume the dark energy arises as running vacuum energy instead of a rigid cosmological constant. On the other hand, there are some studies on the H$_{0}$ measurement. For instance, \cite{Zhang_2017} present a blinded SNe Ia-based H$_{0}$ determination and the result is found to be consistent with \cite{Riess_2016}. \cite{Wu_2017} calculate the sample variance of the local H$_{0}$ measurement with a large-volume cosmological N-body simulation. The result indicates that the sample variance is not able to explain the current tension. \cite{Feeney_2017} perform an analysis with a Bayesian hierarchical model for the local distance ladder and show that the odds against $\Lambda$CDM is less dramatic than the common $3-\sigma$ discrepancy. Their work assesses the discrepancy by considering the full likelihood instead of a Gaussian or least square approximation, and pays particular attention to the Bayesian odds in order to perform model selection. In this work, we are not aiming at judging different models to select the ones that are favored. The frequentist $p$-value test, rather than evaluating models against each other directly, only tells us whether how unusual our data would be under each model.

As a general rule, among the models considered here, the redshift 
space distortion data on the linear $f\sigma_8$ does not appear to 
yield constraints specifically tuned to modified gravity models. sDGP
and the Galileon tracker cosmology are ruled out already by 
expansion data, and the $f(R)$ designer model and several $f(T)$ models 
can reproduce the  growth rate measurements. The modified gravity
models where currently $f\sigma_8$ plays an important role are 
$f(T)_{\rm tanh}$ and the kinetic gravity braiding model. Among 
the dark energy models, the QCD ghost model and the Holographic Dark 
Energy are most affected by $f\sigma_8$ (for HDE this power will 
likely only become evident in the next generation of experiments
like eBOSS and DESI).

The models we consider in this paper are based on phenomenological parameterizations of different quantities or physically motivated theoretical framework. In general, this parametric approach has clear behavior and prediction of observables for properly chosen parameters and initial conditions, at different redshifts and scales. The low number of parameters simplifies our analysis and interpretation. On the other hand, there are also numerous non-parametric or model-independent approaches in the modeling of cosmology, such as methodologies based on Principal Component Approach (PCA, \citealt{Huterer_2003, Zhao_2017}),  Gaussian Process (GP, \citealt{Holsclaw_2010, Seikel_2012, Shafieloo_2012}) and so on. The non-parametric approach is much more flexible, and can capture possible features in the data that may be lost in particular parametric models, at the price of higher complexity and larger uncertainties and degeneracies in model parameters  (\citealt{Huterer_2003}).
In addition, the exact methodology may matter, since the best-fit of the reconstructed result may be inconsistent with different methods given the uncertainties. Therefore a detailed comparison between different non-parametric approaches with high dimensionality would be useful.

There are several limitations to our analysis that we note here.
First, we have included only a compressed version of the CMB data;
there are further important constraints from the full Planck and
other data sets. These constraints have informed the models we 
have chosen here (e.g. we have not considered the effects of 
extra relativistic degrees of freedom in the early universe) but 
not in a formal statistical way. Second, for the growth of structure
data we have not included measurements that might probe scale 
dependence of gravity on large scales or of nonlinear scales. 
The predictions for these observables are much less straightforward
to calculate for many of the models considered here. 

\acknowledgements{}

We thank Joel Bronwnstein, Matthew Kleban, Boris Leistedt, Lucas Lombriser and Roman Scoccimarro for their kind helps and valuable comments. This work is supported by National Science Foundation NSF-AST-1615997 and NSF-AST-1109432. 

\bibliographystyle{apj}
\bibliography{DE_Gof}

\begin{thebibliography}{}
\expandafter\ifx\csname natexlab\endcsname\relax\def\natexlab#1{#1}\fi

\bibitem[{Akaike(1974)}]{Akaike_1974}
Akaike, H. 1974, IEEE transactions on automatic control, 19, 716

\bibitem[{{Alam} {et~al.}(2016){Alam}, {Ata}, {Bailey}, {Beutler}, {Bizyaev},
  {Blazek}, {Bolton}, {Brownstein}, {Burden}, {Chuang}, {Comparat}, {Cuesta},
  {Dawson}, {Eisenstein}, {Escoffier}, {Gil-Mar{\'{\i}}n}, {Grieb}, {Hand},
  {Ho}, {Kinemuchi}, {Kirkby}, {Kitaura}, {Malanushenko}, {Malanushenko},
  {Maraston}, {McBride}, {Nichol}, {Olmstead}, {Oravetz}, {Padmanabhan},
  {Palanque-Delabrouille}, {Pan}, {Pellejero-Ibanez}, {Percival}, {Petitjean},
  {Prada}, {Price-Whelan}, {Reid}, {Rodr{\'{\i}}guez-Torres}, {Roe}, {Ross},
  {Ross}, {Rossi}, {Rubi{\~n}o-Mart{\'{\i}}n}, {S{\'a}nchez}, {Saito},
  {Salazar-Albornoz}, {Samushia}, {Satpathy}, {Sc{\'o}ccola}, {Schlegel},
  {Schneider}, {Seo}, {Simmons}, {Slosar}, {Strauss}, {Swanson}, {Thomas},
  {Tinker}, {Tojeiro}, {Vargas Maga{\~n}a}, {Vazquez}, {Verde}, {Wake}, {Wang},
  {Weinberg}, {White}, {Wood-Vasey}, {Y{\`e}che}, {Zehavi}, {Zhai}, \&
  {Zhao}}]{Alam_2016}
{Alam}, S., {Ata}, M., {Bailey}, S., {et~al.} 2016, ArXiv e-prints,
  arXiv:1607.03155

\bibitem[{{Alcaniz} {et~al.}(2005){Alcaniz}, {Dev}, \& {Jain}}]{Alcaniz_2005}
{Alcaniz}, J.~S., {Dev}, A., \& {Jain}, D. 2005, \apj, 627, 26

\bibitem[{{Amendola}(2000)}]{Amendola_2000}
{Amendola}, L. 2000, \prd, 62, 043511

\bibitem[{{Amendola} {et~al.}(2007{\natexlab{a}}){Amendola}, {Gannouji},
  {Polarski}, \& {Tsujikawa}}]{Amendola_2007b}
{Amendola}, L., {Gannouji}, R., {Polarski}, D., \& {Tsujikawa}, S.
  2007{\natexlab{a}}, \prd, 75, 083504

\bibitem[{{Amendola} {et~al.}(2007{\natexlab{b}}){Amendola}, {Polarski}, \&
  {Tsujikawa}}]{Amendola_2007}
{Amendola}, L., {Polarski}, D., \& {Tsujikawa}, S. 2007{\natexlab{b}}, Physical
  Review Letters, 98, 131302

\bibitem[{{Anderson} {et~al.}(2014){Anderson}, {Aubourg}, {Bailey}, {Beutler},
  {Bhardwaj}, {Blanton}, {Bolton}, {Brinkmann}, {Brownstein}, {Burden},
  {Chuang}, {Cuesta}, {Dawson}, {Eisenstein}, {Escoffier}, {Gunn}, {Guo}, {Ho},
  {Honscheid}, {Howlett}, {Kirkby}, {Lupton}, {Manera}, {Maraston}, {McBride},
  {Mena}, {Montesano}, {Nichol}, {Nuza}, {Olmstead}, {Padmanabhan},
  {Palanque-Delabrouille}, {Parejko}, {Percival}, {Petitjean}, {Prada},
  {Price-Whelan}, {Reid}, {Roe}, {Ross}, {Ross}, {Sabiu}, {Saito}, {Samushia},
  {S{\'a}nchez}, {Schlegel}, {Schneider}, {Scoccola}, {Seo}, {Skibba},
  {Strauss}, {Swanson}, {Thomas}, {Tinker}, {Tojeiro}, {Maga{\~n}a}, {Verde},
  {Wake}, {Weaver}, {Weinberg}, {White}, {Xu}, {Y{\`e}che}, {Zehavi}, \&
  {Zhao}}]{Anderson_2014}
{Anderson}, L., {Aubourg}, {\'E}., {Bailey}, S., {et~al.} 2014, \mnras, 441, 24

\bibitem[{{Arcos} \& {Pereira}(2004)}]{Arcos2004}
{Arcos}, H.~I., \& {Pereira}, J.~G. 2004, International Journal of Modern
  Physics D, 13, 2193

\bibitem[{{Astier}(2001)}]{Astier_2001}
{Astier}, P. 2001, Physics Letters B, 500, 8

\bibitem[{{Aubourg} {et~al.}(2015){Aubourg}, {Bailey}, {Bautista}, {Beutler},
  {Bhardwaj}, {Bizyaev}, {Blanton}, {Blomqvist}, {Bolton}, {Bovy},
  {Brewington}, {Brinkmann}, {Brownstein}, {Burden}, {Busca}, {Carithers},
  {Chuang}, {Comparat}, {Croft}, {Cuesta}, {Dawson}, {Delubac}, {Eisenstein},
  {Font-Ribera}, {Ge}, {Le Goff}, {Gontcho}, {Gott}, {Gunn}, {Guo}, {Guy},
  {Hamilton}, {Ho}, {Honscheid}, {Howlett}, {Kirkby}, {Kitaura}, {Kneib},
  {Lee}, {Long}, {Lupton}, {Maga{\~n}a}, {Malanushenko}, {Malanushenko},
  {Manera}, {Maraston}, {Margala}, {McBride}, {Miralda-Escud{\'e}}, {Myers},
  {Nichol}, {Noterdaeme}, {Nuza}, {Olmstead}, {Oravetz}, {P{\^a}ris},
  {Padmanabhan}, {Palanque-Delabrouille}, {Pan}, {Pellejero-Ibanez},
  {Percival}, {Petitjean}, {Pieri}, {Prada}, {Reid}, {Rich}, {Roe}, {Ross},
  {Ross}, {Rossi}, {Rubi{\~n}o-Mart{\'{\i}}n}, {S{\'a}nchez}, {Samushia},
  {Santos}, {Sc{\'o}ccola}, {Schlegel}, {Schneider}, {Seo}, {Sheldon},
  {Simmons}, {Skibba}, {Slosar}, {Strauss}, {Thomas}, {Tinker}, {Tojeiro},
  {Vazquez}, {Viel}, {Wake}, {Weaver}, {Weinberg}, {Wood-Vasey}, {Y{\`e}che},
  {Zehavi}, {Zhao}, \& {BOSS Collaboration}}]{Aubourg_2015}
{Aubourg}, {\'E}., {Bailey}, S., {Bautista}, J.~E., {et~al.} 2015, \prd, 92,
  123516

\bibitem[{{Azizi} {et~al.}(2012){Azizi}, {Movahed}, \& {Nozari}}]{Azizi_2012}
{Azizi}, T., {Movahed}, M.~S., \& {Nozari}, K. 2012, New Astronomy, 17, 424

\bibitem[{{Bamba} {et~al.}(2011){Bamba}, {Geng}, {Lee}, \& {Luo}}]{Bamba_2011}
{Bamba}, K., {Geng}, C.-Q., {Lee}, C.-C., \& {Luo}, L.-W. 2011, Journal of
  Cosmology and Astroparticle Physics, 1, 021

\bibitem[{{Barreira} {et~al.}(2014{\natexlab{a}}){Barreira}, {Li}, {Baugh}, \&
  {Pascoli}}]{Barreira_2014a}
{Barreira}, A., {Li}, B., {Baugh}, C.~M., \& {Pascoli}, S. 2014{\natexlab{a}},
  \prd, 90, 023528

\bibitem[{{Barreira} {et~al.}(2014{\natexlab{b}}){Barreira}, {Li}, {Baugh}, \&
  {Pascoli}}]{Barreira_2014b}
---. 2014{\natexlab{b}}, Journal of Cosmology and Astroparticle Physics, 8, 059

\bibitem[{{Barreiro} {et~al.}(2000){Barreiro}, {Copeland}, \&
  {Nunes}}]{Barreiro_2000}
{Barreiro}, T., {Copeland}, E.~J., \& {Nunes}, N.~J. 2000, \prd, 61, 127301

\bibitem[{{Basilakos} {et~al.}(2010){Basilakos}, {Plionis}, \&
  {Lima}}]{Basilakos_2010}
{Basilakos}, S., {Plionis}, M., \& {Lima}, J.~A.~S. 2010, \prd, 82, 083517

\bibitem[{{Bean} \& {Dor{\'e}}(2004)}]{Bean_2004}
{Bean}, R., \& {Dor{\'e}}, O. 2004, \prd, 69, 083503

\bibitem[{{Bean} {et~al.}(2008){Bean}, {Flanagan}, {Laszlo}, \&
  {Trodden}}]{Bean_2008}
{Bean}, R., {Flanagan}, {\'E}.~{\'E}., {Laszlo}, I., \& {Trodden}, M. 2008,
  \prd, 78, 123514

\bibitem[{Bekenstein(1973)}]{Bekenstein_1973}
Bekenstein, J.~D. 1973, Phys. Rev. D, 7, 2333

\bibitem[{Bekenstein(1981)}]{Bekenstein_1981}
---. 1981, Phys. Rev. D, 23, 287

\bibitem[{{Bekenstein}(1994)}]{Bekenstein_1994}
{Bekenstein}, J.~D. 1994, \prd, 49, 1912

\bibitem[{{Benaoum}(2002)}]{Benaoum_2002}
{Benaoum}, H.~B. 2002, ArXiv High Energy Physics - Theory e-prints,
  hep-th/0205140

\bibitem[{Bengochea \& Ferraro(2009)}]{Bengochea_2009}
Bengochea, G.~R., \& Ferraro, R. 2009, Phys. Rev. D, 79, 124019

\bibitem[{{Bento} {et~al.}(2002){Bento}, {Bertolami}, \& {Sen}}]{Bento_2002}
{Bento}, M.~C., {Bertolami}, O., \& {Sen}, A.~A. 2002, \prd, 66, 043507

\bibitem[{{Bertschinger} \& {Zukin}(2008)}]{Bertschinger_2008}
{Bertschinger}, E., \& {Zukin}, P. 2008, \prd, 78, 024015

\bibitem[{{Betoule} {et~al.}(2014){Betoule}, {Kessler}, {Guy}, {Mosher},
  {Hardin}, {Biswas}, {Astier}, {El-Hage}, {Konig}, {Kuhlmann}, {Marriner},
  {Pain}, {Regnault}, {Balland}, {Bassett}, {Brown}, {Campbell}, {Carlberg},
  {Cellier-Holzem}, {Cinabro}, {Conley}, {D'Andrea}, {DePoy}, {Doi}, {Ellis},
  {Fabbro}, {Filippenko}, {Foley}, {Frieman}, {Fouchez}, {Galbany}, {Goobar},
  {Gupta}, {Hill}, {Hlozek}, {Hogan}, {Hook}, {Howell}, {Jha}, {Le Guillou},
  {Leloudas}, {Lidman}, {Marshall}, {M{\"o}ller}, {Mour{\~a}o}, {Neveu},
  {Nichol}, {Olmstead}, {Palanque-Delabrouille}, {Perlmutter}, {Prieto},
  {Pritchet}, {Richmond}, {Riess}, {Ruhlmann-Kleider}, {Sako}, {Schahmaneche},
  {Schneider}, {Smith}, {Sollerman}, {Sullivan}, {Walton}, \&
  {Wheeler}}]{Betoule_2014}
{Betoule}, M., {Kessler}, R., {Guy}, J., {et~al.} 2014, \aap, 568, A22

\bibitem[{{Beutler} {et~al.}(2011){Beutler}, {Blake}, {Colless}, {Jones},
  {Staveley-Smith}, {Campbell}, {Parker}, {Saunders}, \&
  {Watson}}]{Beutler_2011}
{Beutler}, F., {Blake}, C., {Colless}, M., {et~al.} 2011, \mnras, 416, 3017

\bibitem[{{Beutler} {et~al.}(2012){Beutler}, {Blake}, {Colless}, {Jones},
  {Staveley-Smith}, {Poole}, {Campbell}, {Parker}, {Saunders}, \&
  {Watson}}]{Beutler_2012}
---. 2012, \mnras, 423, 3430

\bibitem[{{Beutler} {et~al.}(2014){Beutler}, {Saito}, {Seo}, {Brinkmann},
  {Dawson}, {Eisenstein}, {Font-Ribera}, {Ho}, {McBride}, {Montesano},
  {Percival}, {Ross}, {Ross}, {Samushia}, {Schlegel}, {S{\'a}nchez}, {Tinker},
  \& {Weaver}}]{Beutler_2014}
{Beutler}, F., {Saito}, S., {Seo}, H.-J., {et~al.} 2014, \mnras, 443, 1065

\bibitem[{{Bili{\'c}} {et~al.}(2002){Bili{\'c}}, {Tupper}, \&
  {Viollier}}]{Bilic_2002}
{Bili{\'c}}, N., {Tupper}, G.~B., \& {Viollier}, R.~D. 2002, Physics Letters B,
  535, 17

\bibitem[{{Blake} \& {Glazebrook}(2003)}]{Blake_2003}
{Blake}, C., \& {Glazebrook}, K. 2003, \apj, 594, 665

\bibitem[{{Bordag} {et~al.}(2001){Bordag}, {Mohideen}, \&
  {Mostepanenko}}]{Bordag_2001}
{Bordag}, M., {Mohideen}, U., \& {Mostepanenko}, V.~M. 2001, \physrep, 353, 1

\bibitem[{{Bordemann} \& {Hoppe}(1993)}]{Bordemann_1993}
{Bordemann}, M., \& {Hoppe}, J. 1993, Physics Letters B, 317, 315

\bibitem[{{Brax} {et~al.}(2015){Brax}, {Burrage}, {Davis}, \&
  {Gubitosi}}]{Brax_2015}
{Brax}, P., {Burrage}, C., {Davis}, A.-C., \& {Gubitosi}, G. 2015, Journal of
  Cosmology and Astroparticle Physics, 3, 028

\bibitem[{{Cai}(2007)}]{Cai_2007}
{Cai}, R.-G. 2007, Physics Letters B, 657, 228

\bibitem[{{Cai} {et~al.}(2012){Cai}, {Tuo}, {Wu}, \& {Zhao}}]{Cai_2012}
{Cai}, R.-G., {Tuo}, Z.-L., {Wu}, Y.-B., \& {Zhao}, Y.-Y. 2012, \prd, 86,
  023511

\bibitem[{{Cai} {et~al.}(2015){Cai}, {Capozziello}, {De Laurentis}, \&
  {Saridakis}}]{Cai_2015}
{Cai}, Y.-F., {Capozziello}, S., {De Laurentis}, M., \& {Saridakis}, E.~N.
  2015, ArXiv e-prints, arXiv:1511.07586

\bibitem[{{Caldera-Cabral} {et~al.}(2009){Caldera-Cabral}, {Maartens}, \&
  {Schaefer}}]{Caldera-Cabral_2009}
{Caldera-Cabral}, G., {Maartens}, R., \& {Schaefer}, B.~M. 2009, Journal of
  Cosmology and Astroparticle Physics, 7, 027

\bibitem[{{Caldwell}(2002)}]{Caldwell_2002}
{Caldwell}, R.~R. 2002, Physics Letters B, 545, 23

\bibitem[{{Caldwell} {et~al.}(1998){Caldwell}, {Dave}, \&
  {Steinhardt}}]{Caldwell_1998}
{Caldwell}, R.~R., {Dave}, R., \& {Steinhardt}, P.~J. 1998, Physical Review
  Letters, 80, 1582

\bibitem[{{Caldwell} \& {Kamionkowski}(2009)}]{Caldwell_2009}
{Caldwell}, R.~R., \& {Kamionkowski}, M. 2009, Annual Review of Nuclear and
  Particle Science, 59, 397

\bibitem[{{Caldwell} {et~al.}(2003){Caldwell}, {Kamionkowski}, \&
  {Weinberg}}]{Caldwell_2003}
{Caldwell}, R.~R., {Kamionkowski}, M., \& {Weinberg}, N.~N. 2003, Physical
  Review Letters, 91, 071301

\bibitem[{{Capozziello} {et~al.}(2003){Capozziello}, {Cardone}, {Carloni}, \&
  {Troisi}}]{Capozziello_2003}
{Capozziello}, S., {Cardone}, V.~F., {Carloni}, S., \& {Troisi}, A. 2003,
  International Journal of Modern Physics D, 12, 1969

\bibitem[{{Capozziello} \& {Fang}(2002)}]{Capozziello_2002}
{Capozziello}, S., \& {Fang}, L.~Z. 2002, International Journal of Modern
  Physics D, 11, 483

\bibitem[{{Capozziello} {et~al.}(2014){Capozziello}, {Farooq}, {Luongo}, \&
  {Ratra}}]{Capozziello_2014}
{Capozziello}, S., {Farooq}, O., {Luongo}, O., \& {Ratra}, B. 2014, \prd, 90,
  044016

\bibitem[{{Cataneo} {et~al.}(2015){Cataneo}, {Rapetti}, {Schmidt}, {Mantz},
  {Allen}, {Applegate}, {Kelly}, {von der Linden}, \& {Morris}}]{Cataneo_2015}
{Cataneo}, M., {Rapetti}, D., {Schmidt}, F., {et~al.} 2015, \prd, 92, 044009

\bibitem[{{Chen} \& {Ratra}(2011)}]{Chen_2011}
{Chen}, Y., \& {Ratra}, B. 2011, Physics Letters B, 703, 406

\bibitem[{{Chevallier} \& {Polarski}(2001)}]{Chevallier_2001}
{Chevallier}, M., \& {Polarski}, D. 2001, International Journal of Modern
  Physics D, 10, 213

\bibitem[{{Chiba} {et~al.}(2013){Chiba}, {De Felice}, \&
  {Tsujikawa}}]{Chiba_2013}
{Chiba}, T., {De Felice}, A., \& {Tsujikawa}, S. 2013, \prd, 87, 083505

\bibitem[{{Chiba} {et~al.}(1997){Chiba}, {Sugiyama}, \&
  {Nakamura}}]{Chiba_1997}
{Chiba}, T., {Sugiyama}, N., \& {Nakamura}, T. 1997, \mnras, 289, L5

\bibitem[{{Chimento} {et~al.}(2003){Chimento}, {Jakubi}, {Pav{\'o}n}, \&
  {Zimdahl}}]{Chimento_2003}
{Chimento}, L.~P., {Jakubi}, A.~S., {Pav{\'o}n}, D., \& {Zimdahl}, W. 2003,
  \prd, 67, 083513

\bibitem[{{Chuang} {et~al.}(2013){Chuang}, {Prada}, {Cuesta}, {Eisenstein},
  {Kazin}, {Padmanabhan}, {S{\'a}nchez}, {Xu}, {Beutler}, {Manera}, {Schlegel},
  {Schneider}, {Weinberg}, {Brinkmann}, {Brownstein}, \&
  {Thomas}}]{Chuang_2013}
{Chuang}, C.-H., {Prada}, F., {Cuesta}, A.~J., {et~al.} 2013, \mnras, 433, 3559

\bibitem[{{Cognola} {et~al.}(2005){Cognola}, {Elizalde}, {Nojiri}, {Odintsov},
  \& {Zerbini}}]{Cognola_2005}
{Cognola}, G., {Elizalde}, E., {Nojiri}, S., {Odintsov}, S.~D., \& {Zerbini},
  S. 2005, Journal of Cosmology and Astroparticle Physics, 2, 010

\bibitem[{{Cohen} {et~al.}(1999){Cohen}, {Kaplan}, \& {Nelson}}]{Cohen_1999}
{Cohen}, A.~G., {Kaplan}, D.~B., \& {Nelson}, A.~E. 1999, Physical Review
  Letters, 82, 4971

\bibitem[{{Conley} {et~al.}(2011){Conley}, {Guy}, {Sullivan}, {Regnault},
  {Astier}, {Balland}, {Basa}, {Carlberg}, {Fouchez}, {Hardin}, {Hook},
  {Howell}, {Pain}, {Palanque-Delabrouille}, {Perrett}, {Pritchet}, {Rich},
  {Ruhlmann-Kleider}, {Balam}, {Baumont}, {Ellis}, {Fabbro}, {Fakhouri},
  {Fourmanoit}, {Gonz{\'a}lez-Gait{\'a}n}, {Graham}, {Hudson}, {Hsiao},
  {Kronborg}, {Lidman}, {Mourao}, {Neill}, {Perlmutter}, {Ripoche}, {Suzuki},
  \& {Walker}}]{Conley_2011}
{Conley}, A., {Guy}, J., {Sullivan}, M., {et~al.} 2011, \apjs, 192, 1

\bibitem[{{Cooray} \& {Huterer}(1999)}]{Cooray_1999}
{Cooray}, A.~R., \& {Huterer}, D. 1999, \apjl, 513, L95

\bibitem[{{Copeland} {et~al.}(1998){Copeland}, {Liddle}, \&
  {Wands}}]{Copeland_1998}
{Copeland}, E.~J., {Liddle}, A.~R., \& {Wands}, D. 1998, \prd, 57, 4686

\bibitem[{{Copeland} {et~al.}(2006){Copeland}, {Sami}, \&
  {Tsujikawa}}]{Copeland_2006}
{Copeland}, E.~J., {Sami}, M., \& {Tsujikawa}, S. 2006, International Journal
  of Modern Physics D, 15, 1753

\bibitem[{{Costa} {et~al.}(2014){Costa}, {Xu}, {Wang}, {Ferreira}, \&
  {Abdalla}}]{Costa_2014}
{Costa}, A.~A., {Xu}, X.-D., {Wang}, B., {Ferreira}, E.~G.~M., \& {Abdalla}, E.
  2014, \prd, 89, 103531

\bibitem[{{Dalal} {et~al.}(2001){Dalal}, {Abazajian}, {Jenkins}, \&
  {Manohar}}]{Dalal_2001}
{Dalal}, N., {Abazajian}, K., {Jenkins}, E., \& {Manohar}, A.~V. 2001, Physical
  Review Letters, 87, 141302

\bibitem[{{Davis} {et~al.}(2007){Davis}, {M{\"o}rtsell}, {Sollerman}, {Becker},
  {Blondin}, {Challis}, {Clocchiatti}, {Filippenko}, {Foley}, {Garnavich},
  {Jha}, {Krisciunas}, {Kirshner}, {Leibundgut}, {Li}, {Matheson}, {Miknaitis},
  {Pignata}, {Rest}, {Riess}, {Schmidt}, {Smith}, {Spyromilio}, {Stubbs},
  {Suntzeff}, {Tonry}, {Wood-Vasey}, \& {Zenteno}}]{Davis_2007}
{Davis}, T.~M., {M{\"o}rtsell}, E., {Sollerman}, J., {et~al.} 2007, \apj, 666,
  716

\bibitem[{{Dawson} {et~al.}(2016){Dawson}, {Kneib}, {Percival}, {Alam},
  {Albareti}, {Anderson}, {Armengaud}, {Aubourg}, {Bailey}, {Bautista},
  {Berlind}, {Bershady}, {Beutler}, {Bizyaev}, {Blanton}, {Blomqvist},
  {Bolton}, {Bovy}, {Brandt}, {Brinkmann}, {Brownstein}, {Burtin}, {Busca},
  {Cai}, {Chuang}, {Clerc}, {Comparat}, {Cope}, {Croft}, {Cruz-Gonzalez}, {da
  Costa}, {Cousinou}, {Darling}, {de la Macorra}, {de la Torre}, {Delubac}, {du
  Mas des Bourboux}, {Dwelly}, {Ealet}, {Eisenstein}, {Eracleous}, {Escoffier},
  {Fan}, {Finoguenov}, {Font-Ribera}, {Frinchaboy}, {Gaulme}, {Georgakakis},
  {Green}, {Guo}, {Guy}, {Ho}, {Holder}, {Huehnerhoff}, {Hutchinson}, {Jing},
  {Jullo}, {Kamble}, {Kinemuchi}, {Kirkby}, {Kitaura}, {Klaene}, {Laher},
  {Lang}, {Laurent}, {Le Goff}, {Li}, {Liang}, {Lima}, {Lin}, {Lin}, {Lin},
  {Long}, {Lundgren}, {MacDonald}, {Geimba Maia}, {Malanushenko},
  {Malanushenko}, {Mariappan}, {McBride}, {McGreer}, {M{\'e}nard}, {Merloni},
  {Meza}, {Montero-Dorta}, {Muna}, {Myers}, {Nandra}, {Naugle}, {Newman},
  {Noterdaeme}, {Nugent}, {Ogando}, {Olmstead}, {Oravetz}, {Oravetz},
  {Padmanabhan}, {Palanque-Delabrouille}, {Pan}, {Parejko}, {P{\^a}ris},
  {Peacock}, {Petitjean}, {Pieri}, {Pisani}, {Prada}, {Prakash}, {Raichoor},
  {Reid}, {Rich}, {Ridl}, {Rodriguez-Torres}, {Carnero Rosell}, {Ross},
  {Rossi}, {Ruan}, {Salvato}, {Sayres}, {Schneider}, {Schlegel}, {Seljak},
  {Seo}, {Sesar}, {Shandera}, {Shu}, {Slosar}, {Sobreira}, {Streblyanska},
  {Suzuki}, {Taylor}, {Tao}, {Tinker}, {Tojeiro}, {Vargas-Maga{\~n}a}, {Wang},
  {Weaver}, {Weinberg}, {White}, {Wood-Vasey}, {Yeche}, {Zhai}, {Zhao}, {Zhao},
  {Zheng}, {Ben Zhu}, \& {Zou}}]{Dawson_2016}
{Dawson}, K.~S., {Kneib}, J.-P., {Percival}, W.~J., {et~al.} 2016, \aj, 151, 44

\bibitem[{{de Felice} {et~al.}(2011){de Felice}, {Kase}, \&
  {Tsujikawa}}]{de_Felice_2011}
{de Felice}, A., {Kase}, R., \& {Tsujikawa}, S. 2011, \prd, 83, 043515

\bibitem[{{de Felice} \& {Tsujikawa}(2010)}]{de_Felice_2010}
{de Felice}, A., \& {Tsujikawa}, S. 2010, Physical Review Letters, 105, 111301

\bibitem[{{de Felice} \& {Tsujikawa}(2011)}]{de_Felice_2011b}
---. 2011, \prd, 84, 124029

\bibitem[{{Deffayet} {et~al.}(2009{\natexlab{a}}){Deffayet}, {Deser}, \&
  {Esposito-Far{\`e}se}}]{Deffayet_2009}
{Deffayet}, C., {Deser}, S., \& {Esposito-Far{\`e}se}, G. 2009{\natexlab{a}},
  \prd, 80, 064015

\bibitem[{{Deffayet} {et~al.}(2009{\natexlab{b}}){Deffayet},
  {Esposito-Far{\`e}se}, \& {Vikman}}]{Deffayet_2009b}
{Deffayet}, C., {Esposito-Far{\`e}se}, G., \& {Vikman}, A. 2009{\natexlab{b}},
  \prd, 79, 084003

\bibitem[{{Deffayet} {et~al.}(2010){Deffayet}, {Pujol{\`a}s}, {Sawicki}, \&
  {Vikman}}]{Deffayet_2010c}
{Deffayet}, C., {Pujol{\`a}s}, O., {Sawicki}, I., \& {Vikman}, A. 2010, Journal
  of Cosmology and Astroparticle Physics, 10, 026

\bibitem[{{Delubac} {et~al.}(2015){Delubac}, {Bautista}, {Busca}, {Rich},
  {Kirkby}, {Bailey}, {Font-Ribera}, {Slosar}, {Lee}, {Pieri}, {Hamilton},
  {Aubourg}, {Blomqvist}, {Bovy}, {Brinkmann}, {Carithers}, {Dawson},
  {Eisenstein}, {Gontcho}, {Kneib}, {Le Goff}, {Margala}, {Miralda-Escud{\'e}},
  {Myers}, {Nichol}, {Noterdaeme}, {O'Connell}, {Olmstead},
  {Palanque-Delabrouille}, {P{\^a}ris}, {Petitjean}, {Ross}, {Rossi},
  {Schlegel}, {Schneider}, {Weinberg}, {Y{\`e}che}, \& {York}}]{Delubac_2015}
{Delubac}, T., {Bautista}, J.~E., {Busca}, N.~G., {et~al.} 2015, \aap, 574, A59

\bibitem[{{DESI}(2016)}]{DESI_2016}
{DESI}, C. 2016, arXiv:1611.00036

\bibitem[{{Di Valentino} {et~al.}(2017){Di Valentino}, {Melchiorri}, {Linder},
  \& {Silk}}]{Valentino_2017}
{Di Valentino}, E., {Melchiorri}, A., {Linder}, E.~V., \& {Silk}, J. 2017,
  \prd, 96, 023523

\bibitem[{{Di Valentino} {et~al.}(2016){Di Valentino}, {Melchiorri}, \&
  {Silk}}]{Valentino_2016}
{Di Valentino}, E., {Melchiorri}, A., \& {Silk}, J. 2016, Physics Letters B,
  761, 242

\bibitem[{{Doran} \& {Robbers}(2006)}]{Doran_2006}
{Doran}, M., \& {Robbers}, G. 2006, Journal of Cosmology and Astroparticle
  Physics, 6, 026

\bibitem[{{Dutta} \& {Sorbo}(2007)}]{Dutta_2007}
{Dutta}, K., \& {Sorbo}, L. 2007, \prd, 75, 063514

\bibitem[{{Dvali} {et~al.}(2000){Dvali}, {Gabadadze}, \&
  {Porrati}}]{Davli_2000}
{Dvali}, G., {Gabadadze}, G., \& {Porrati}, M. 2000, Physics Letters B, 485,
  208

\bibitem[{{Efstathiou}(2008)}]{Efstathiou_2008}
{Efstathiou}, G. 2008, \mnras, 388, 1314

\bibitem[{Einstein(1928)}]{Einstein_1928}
Einstein, A. 1928, Riemann-Geometrie mit Aufrechterhaltung des Begriffes des
  Fernparallelismus (Wiley Online Library)

\bibitem[{{Eisenstein} {et~al.}(2005){Eisenstein}, {Zehavi}, {Hogg},
  {Scoccimarro}, {Blanton}, {Nichol}, {Scranton}, {Seo}, {Tegmark}, {Zheng},
  {Anderson}, {Annis}, {Bahcall}, {Brinkmann}, {Burles}, {Castander},
  {Connolly}, {Csabai}, {Doi}, {Fukugita}, {Frieman}, {Glazebrook}, {Gunn},
  {Hendry}, {Hennessy}, {Ivezi{\'c}}, {Kent}, {Knapp}, {Lin}, {Loh}, {Lupton},
  {Margon}, {McKay}, {Meiksin}, {Munn}, {Pope}, {Richmond}, {Schlegel},
  {Schneider}, {Shimasaku}, {Stoughton}, {Strauss}, {SubbaRao}, {Szalay},
  {Szapudi}, {Tucker}, {Yanny}, \& {York}}]{Eisenstein_2005}
{Eisenstein}, D.~J., {Zehavi}, I., {Hogg}, D.~W., {et~al.} 2005, \apj, 633, 560

\bibitem[{{Faraoni} \& {Dolgov}(2002)}]{Faraoni_2002}
{Faraoni}, V., \& {Dolgov}, A. 2002, International Journal of Modern Physics D,
  11, 471

\bibitem[{{Farooq} \& {Ratra}(2013)}]{Farooq_2013}
{Farooq}, O., \& {Ratra}, B. 2013, Physics Letters B, 723, 1

\bibitem[{{Farrar} \& {Peebles}(2004)}]{Farrar_2004}
{Farrar}, G.~R., \& {Peebles}, P.~J.~E. 2004, \apj, 604, 1

\bibitem[{{Fay}(2016)}]{Fay_2016}
{Fay}, S. 2016, \mnras, 460, 1863

\bibitem[{{Feeney} {et~al.}(2017){Feeney}, {Mortlock}, \&
  {Dalmasso}}]{Feeney_2017}
{Feeney}, S.~M., {Mortlock}, D.~J., \& {Dalmasso}, N. 2017, ArXiv e-prints,
  arXiv:1707.00007

\bibitem[{Ferraro \& Fiorini(2007)}]{Ferraro_2007}
Ferraro, R., \& Fiorini, F. 2007, Phys. Rev. D, 75, 084031

\bibitem[{Ferraro \& Fiorini(2008)}]{Ferraro_2008}
---. 2008, Phys. Rev. D, 78, 124019

\bibitem[{{Font-Ribera} {et~al.}(2014){Font-Ribera}, {Kirkby}, {Busca},
  {Miralda-Escud{\'e}}, {Ross}, {Slosar}, {Rich}, {Aubourg}, {Bailey},
  {Bhardwaj}, {Bautista}, {Beutler}, {Bizyaev}, {Blomqvist}, {Brewington},
  {Brinkmann}, {Brownstein}, {Carithers}, {Dawson}, {Delubac}, {Ebelke},
  {Eisenstein}, {Ge}, {Kinemuchi}, {Lee}, {Malanushenko}, {Malanushenko},
  {Marchante}, {Margala}, {Muna}, {Myers}, {Noterdaeme}, {Oravetz},
  {Palanque-Delabrouille}, {P{\^a}ris}, {Petitjean}, {Pieri}, {Rossi},
  {Schneider}, {Simmons}, {Viel}, {Yeche}, \& {York}}]{Font-Ribera_2014}
{Font-Ribera}, A., {Kirkby}, D., {Busca}, N., {et~al.} 2014, Journal of
  Cosmology and Astroparticle Physics, 5, 027

\bibitem[{{Foreman-Mackey} {et~al.}(2013){Foreman-Mackey}, {Hogg}, {Lang}, \&
  {Goodman}}]{Foreman-Mackey_2013}
{Foreman-Mackey}, D., {Hogg}, D.~W., {Lang}, D., \& {Goodman}, J. 2013, \pasp,
  125, 306

\bibitem[{{Freese} \& {Lewis}(2002)}]{Freese_2002}
{Freese}, K., \& {Lewis}, M. 2002, Physics Letters B, 540, 1

\bibitem[{{Frieman} {et~al.}(1995){Frieman}, {Hill}, {Stebbins}, \&
  {Waga}}]{Frieman_1995}
{Frieman}, J.~A., {Hill}, C.~T., {Stebbins}, A., \& {Waga}, I. 1995, Physical
  Review Letters, 75, 2077

\bibitem[{{Frith}(2004)}]{Frith_2004}
{Frith}, W.~J. 2004, \mnras, 348, 916

\bibitem[{{Gao} {et~al.}(2009){Gao}, {Wu}, {Chen}, \& {Shen}}]{Gao_2009}
{Gao}, C., {Wu}, F., {Chen}, X., \& {Shen}, Y.-G. 2009, \prd, 79, 043511

\bibitem[{{Gehrels}(1986)}]{Gehrels_1986}
{Gehrels}, N. 1986, \apj, 303, 336

\bibitem[{{Giannantonio} \& {Melchiorri}(2006)}]{Gaiannantonio_2006}
{Giannantonio}, T., \& {Melchiorri}, A. 2006, Classical and Quantum Gravity,
  23, 4125

\bibitem[{{God{\l}owski} \& {Szyd{\l}owski}(2006)}]{Gaodlowski_2006}
{God{\l}owski}, W., \& {Szyd{\l}owski}, M. 2006, Physics Letters B, 642, 13

\bibitem[{Goodman \& Weare(2010)}]{GW_2010}
Goodman, J., \& Weare, J. 2010, Communications in applied mathematics and
  computational science, 5, 65

\bibitem[{{Gott} \& {Slepian}(2011)}]{Gott_2011}
{Gott}, J.~R., \& {Slepian}, Z. 2011, \mnras, 416, 907

\bibitem[{Hawking(1976)}]{Hawking_1976}
Hawking, S.~W. 1976, Phys. Rev. D, 13, 191

\bibitem[{Hayashi \& Shirafuji(1979)}]{Hayashi_1979}
Hayashi, K., \& Shirafuji, T. 1979, Phys. Rev. D, 19, 3524

\bibitem[{Hinshaw {et~al.}(2013)Hinshaw, Larson, Komatsu, Spergel, Bennett,
  Dunkley, Nolta, Halpern, Hill, Odegard, Page, Smith, Weiland, Gold, Jarosik,
  Kogut, Limon, Meyer, Tucker, Wollack, \& Wright}]{Hinshaw_2009}
Hinshaw, G., Larson, D., Komatsu, E., {et~al.} 2013, The Astrophysical Journal
  Supplement Series, 208, 19

\bibitem[{{Holsclaw} {et~al.}(2010){Holsclaw}, {Alam}, {Sans{\'o}}, {Lee},
  {Heitmann}, {Habib}, \& {Higdon}}]{Holsclaw_2010}
{Holsclaw}, T., {Alam}, U., {Sans{\'o}}, B., {et~al.} 2010, Physical Review
  Letters, 105, 241302

\bibitem[{{Hsu}(2004)}]{Hsu_2004}
{Hsu}, S.~D.~H. 2004, Physics Letters B, 594, 13

\bibitem[{{Hu} \& {Sawicki}(2007)}]{Hu_2007}
{Hu}, W., \& {Sawicki}, I. 2007, \prd, 76, 064004

\bibitem[{{Huang} \& {Li}(2004)}]{Huang_2004}
{Huang}, Q.-G., \& {Li}, M. 2004, Journal of Cosmology and Astroparticle
  Physics, 8, 013

\bibitem[{{Huang} {et~al.}(2011){Huang}, {Bond}, \& {Kofman}}]{Huang_2011}
{Huang}, Z., {Bond}, J.~R., \& {Kofman}, L. 2011, \apj, 726, 64

\bibitem[{{Huterer} \& {Starkman}(2003)}]{Huterer_2003}
{Huterer}, D., \& {Starkman}, G. 2003, Physical Review Letters, 90, 031301

\bibitem[{{Huterer} {et~al.}(2015){Huterer}, {Kirkby}, {Bean}, {Connolly},
  {Dawson}, {Dodelson}, {Evrard}, {Jain}, {Jarvis}, {Linder}, {Mandelbaum},
  {May}, {Raccanelli}, {Reid}, {Rozo}, {Schmidt}, {Sehgal}, {Slosar}, {van
  Engelen}, {Wu}, \& {Zhao}}]{Huterer_2015}
{Huterer}, D., {Kirkby}, D., {Bean}, R., {et~al.} 2015, Astroparticle Physics,
  63, 23

\bibitem[{{Jassal} {et~al.}(2005){Jassal}, {Bagla}, \&
  {Padmanabhan}}]{Jassal_2005}
{Jassal}, H.~K., {Bagla}, J.~S., \& {Padmanabhan}, T. 2005, \mnras, 356, L11

\bibitem[{{Joyce} {et~al.}(2015){Joyce}, {Jain}, {Khoury}, \&
  {Trodden}}]{Joyce_2015}
{Joyce}, A., {Jain}, B., {Khoury}, J., \& {Trodden}, M. 2015, \physrep, 568, 1

\bibitem[{{Kaiser}(1987)}]{Kaiser_1987}
{Kaiser}, N. 1987, \mnras, 227, 1

\bibitem[{{Kamenshchik} {et~al.}(2001){Kamenshchik}, {Moschella}, \&
  {Pasquier}}]{Kamenshchik_2001}
{Kamenshchik}, A., {Moschella}, U., \& {Pasquier}, V. 2001, Physics Letters B,
  511, 265

\bibitem[{Karolyhazy(1966)}]{Karolyhazy_1966}
Karolyhazy, F. 1966, Il Nuovo Cimento A (1971-1996), 42, 390

\bibitem[{Kawarabayashi \& Ohta(1980)}]{KAWARABAYASHI_1980}
Kawarabayashi, K., \& Ohta, N. 1980, Nuclear Physics B, 175, 477

\bibitem[{{Kawarabayashi} \& {Ohta}(1981)}]{Kawarabayashi_1981}
{Kawarabayashi}, K., \& {Ohta}, N. 1981, Progress of Theoretical Physics, 66,
  1789

\bibitem[{{Kimura} {et~al.}(2012){Kimura}, {Kobayashi}, \&
  {Yamamoto}}]{Kimura_2012}
{Kimura}, R., {Kobayashi}, T., \& {Yamamoto}, K. 2012, \prd, 85, 123503

\bibitem[{{Kimura} \& {Yamamoto}(2011)}]{Kimura_2011}
{Kimura}, R., \& {Yamamoto}, K. 2011, Journal of Cosmology and Astroparticle
  Physics, 4, 025

\bibitem[{{Levi} {et~al.}(2013){Levi}, {Bebek}, {Beers}, {Blum}, {Cahn},
  {Eisenstein}, {Flaugher}, {Honscheid}, {Kron}, {Lahav}, {McDonald}, {Roe},
  {Schlegel}, \& {representing the DESI collaboration}}]{Levi_2013}
{Levi}, M., {Bebek}, C., {Beers}, T., {et~al.} 2013, ArXiv e-prints,
  arXiv:1308.0847

\bibitem[{Lewis \& Bridle(2002)}]{Lewis_2002ah}
Lewis, A., \& Bridle, S. 2002, Phys. Rev., D66, 103511

\bibitem[{Lewis {et~al.}(2000)Lewis, Challinor, \& Lasenby}]{Lewis:1999bs}
Lewis, A., Challinor, A., \& Lasenby, A. 2000, Astrophys. J., 538, 473

\bibitem[{{Li} \& {Barrow}(2007)}]{Li_2007}
{Li}, B., \& {Barrow}, J.~D. 2007, \prd, 75, 084010

\bibitem[{{Li}(2004)}]{Li_2004}
{Li}, M. 2004, Physics Letters B, 603, 1

\bibitem[{Li {et~al.}(2013)Li, Li, Ma, Zhang, \& Zhang}]{Li_2013}
Li, M., Li, X.-D., Ma, Y.-Z., Zhang, X., \& Zhang, Z. 2013, Journal of
  Cosmology and Astroparticle Physics, 2013, 021

\bibitem[{{Li} {et~al.}(2009){Li}, {Li}, {Wang}, \& {Zhang}}]{Li_2009}
{Li}, M., {Li}, X.-D., {Wang}, S., \& {Zhang}, X. 2009, Journal of Cosmology
  and Astroparticle Physics, 6, 036

\bibitem[{{Li} {et~al.}(2013){Li}, {Wang}, {Li}, \& {Zhang}}]{Li_2013_HDE}
{Li}, Y.-H., {Wang}, S., {Li}, X.-D., \& {Zhang}, X. 2013, Journal of Cosmology
  and Astroparticle Physics, 2, 033

\bibitem[{{Liddle}(2004)}]{Liddle_2004}
{Liddle}, A.~R. 2004, \mnras, 351, L49

\bibitem[{{Linder}(2003)}]{Linder_2003}
{Linder}, E.~V. 2003, Physical Review Letters, 90, 091301

\bibitem[{{Linder}(2008)}]{Linder_2008}
---. 2008, Reports on Progress in Physics, 71, 056901

\bibitem[{Linder(2009)}]{Linder_2009}
Linder, E.~V. 2009, Phys. Rev. D, 80, 123528

\bibitem[{Linder(2010)}]{Linder_2010}
---. 2010, Phys. Rev. D, 81, 127301

\bibitem[{{Lombriser} {et~al.}(2009){Lombriser}, {Hu}, {Fang}, \&
  {Seljak}}]{Lombriser_2009}
{Lombriser}, L., {Hu}, W., {Fang}, W., \& {Seljak}, U. 2009, \prd, 80, 063536

\bibitem[{{Lombriser} {et~al.}(2012){Lombriser}, {Slosar}, {Seljak}, \&
  {Hu}}]{Lombriser_2012}
{Lombriser}, L., {Slosar}, A., {Seljak}, U., \& {Hu}, W. 2012, \prd, 85, 124038

\bibitem[{Lu {et~al.}(2015)Lu, Geng, Xu, Wu, \& Liu}]{Lu_2015}
Lu, J., Geng, D., Xu, L., Wu, Y., \& Liu, M. 2015, Journal of High Energy
  Physics, 2015, 1

\bibitem[{Maluf(1994)}]{Maluf_1994}
Maluf, J.~W. 1994, Journal of Mathematical Physics, 35

\bibitem[{{Maziashvili}(2007{\natexlab{a}})}]{Maziashvili_2007b}
{Maziashvili}, M. 2007{\natexlab{a}}, Physics Letters B, 652, 165

\bibitem[{{Maziashvili}(2007{\natexlab{b}})}]{Maziashvili_2007}
---. 2007{\natexlab{b}}, International Journal of Modern Physics D, 16, 1531

\bibitem[{{Murgia} {et~al.}(2016){Murgia}, {Gariazzo}, \&
  {Fornengo}}]{Murgia_2016}
{Murgia}, R., {Gariazzo}, S., \& {Fornengo}, N. 2016, Journal of Cosmology and
  Astroparticle Physics, 4, 014

\bibitem[{{Nesseris} {et~al.}(2013){Nesseris}, {Basilakos}, {Saridakis}, \&
  {Perivolaropoulos}}]{Nesseris_2013}
{Nesseris}, S., {Basilakos}, S., {Saridakis}, E.~N., \& {Perivolaropoulos}, L.
  2013, \prd, 88, 103010

\bibitem[{{Nesseris} {et~al.}(2010){Nesseris}, {de Felice}, \&
  {Tsujikawa}}]{Nesseris_2010}
{Nesseris}, S., {de Felice}, A., \& {Tsujikawa}, S. 2010, \prd, 82, 124054

\bibitem[{{Neupane}(2009)}]{Neupane_2009}
{Neupane}, I.~P. 2009, Physics Letters B, 673, 111

\bibitem[{{Neveu} {et~al.}(2016){Neveu}, {Ruhlmann-Kleider}, {Astier},
  {Besan{\c c}on}, {Guy}, {M{\"o}ller}, \& {Babichev}}]{Neveu_2016}
{Neveu}, J., {Ruhlmann-Kleider}, V., {Astier}, P., {et~al.} 2016, ArXiv
  e-prints, arXiv:1605.02627

\bibitem[{{Nicolis} {et~al.}(2009){Nicolis}, {Rattazzi}, \&
  {Trincherini}}]{Nicolis_2009}
{Nicolis}, A., {Rattazzi}, R., \& {Trincherini}, E. 2009, \prd, 79, 064036

\bibitem[{{Nojiri} \& {Odintsov}(2003)}]{Nojiri_2003}
{Nojiri}, S., \& {Odintsov}, S.~D. 2003, \prd, 68, 123512

\bibitem[{{Ohta}(1981)}]{Ohta_1981}
{Ohta}, N. 1981, Progress of Theoretical Physics, 66, 1408

\bibitem[{{Ohta}(2011)}]{Ohta_2011}
---. 2011, Physics Letters B, 695, 41

\bibitem[{{Paliathanasis} {et~al.}(2014){Paliathanasis}, {Tsamparlis}, \&
  {Basilakos}}]{Paliathanasis_2014}
{Paliathanasis}, A., {Tsamparlis}, M., \& {Basilakos}, S. 2014, \prd, 90,
  103524

\bibitem[{{Park} {et~al.}(2010){Park}, {Hwang}, {Park}, \& {Noh}}]{Park_2010}
{Park}, C.-G., {Hwang}, J.-C., {Park}, J., \& {Noh}, H. 2010, \prd, 81, 063532

\bibitem[{{Paul} \& {Thakur}(2013)}]{Paul_2013}
{Paul}, B.~C., \& {Thakur}, P. 2013, Journal of Cosmology and Astroparticle
  Physics, 11, 052

\bibitem[{{Pavlov} {et~al.}(2014){Pavlov}, {Farooq}, \& {Ratra}}]{Pavlov_2014}
{Pavlov}, A., {Farooq}, O., \& {Ratra}, B. 2014, \prd, 90, 023006

\bibitem[{{Peebles} \& {Ratra}(2003)}]{Peebles_2003}
{Peebles}, P.~J., \& {Ratra}, B. 2003, Reviews of Modern Physics, 75, 559

\bibitem[{{Peebles} \& {Ratra}(1988)}]{Peebles_1988}
{Peebles}, P.~J.~E., \& {Ratra}, B. 1988, \apjl, 325, L17

\bibitem[{{Peebles} \& {Yu}(1970)}]{Peebles_1970}
{Peebles}, P.~J.~E., \& {Yu}, J.~T. 1970, \apj, 162, 815

\bibitem[{{Perlmutter} {et~al.}(1999){Perlmutter}, {Aldering}, {Goldhaber},
  {Knop}, {Nugent}, {Castro}, {Deustua}, {Fabbro}, {Goobar}, {Groom}, {Hook},
  {Kim}, {Kim}, {Lee}, {Nunes}, {Pain}, {Pennypacker}, {Quimby}, {Lidman},
  {Ellis}, {Irwin}, {McMahon}, {Ruiz-Lapuente}, {Walton}, {Schaefer}, {Boyle},
  {Filippenko}, {Matheson}, {Fruchter}, {Panagia}, {Newberg}, {Couch}, \&
  {Project}}]{Perlmutter_1999}
{Perlmutter}, S., {Aldering}, G., {Goldhaber}, G., {et~al.} 1999, \apj, 517,
  565

\bibitem[{{Planck Collaboration} {et~al.}(2014){Planck Collaboration}, {Ade},
  {Aghanim}, {Armitage-Caplan}, {Arnaud}, {Ashdown}, {Atrio-Barandela},
  {Aumont}, {Baccigalupi}, {Banday}, \& et~al.}]{Planck_2013}
{Planck Collaboration}, {Ade}, P.~A.~R., {Aghanim}, N., {et~al.} 2014, \aap,
  571, A16

\bibitem[{{Planck Collaboration} {et~al.}(2015{\natexlab{a}}){Planck
  Collaboration}, {Ade}, {Aghanim}, {Arnaud}, {Ashdown}, {Aumont},
  {Baccigalupi}, {Banday}, {Barreiro}, {Bartlett}, \& et~al.}]{Planck_2015}
---. 2015{\natexlab{a}}, ArXiv e-prints, arXiv:1502.01589

\bibitem[{{Planck Collaboration} {et~al.}(2015{\natexlab{b}}){Planck
  Collaboration}, {Ade}, {Aghanim}, {Arnaud}, {Ashdown}, {Aumont},
  {Baccigalupi}, {Banday}, {Barreiro}, {Bartolo}, \& et~al.}]{Planck_2015b}
---. 2015{\natexlab{b}}, ArXiv e-prints, arXiv:1502.01590

\bibitem[{{Pogosian} \& {Silvestri}(2008)}]{Pogosian_2008}
{Pogosian}, L., \& {Silvestri}, A. 2008, \prd, 77, 023503

\bibitem[{Randall \& Sundrum(1999)}]{Randall_1999b}
Randall, L., \& Sundrum, R. 1999, Phys. Rev. Lett., 83, 4690

\bibitem[{{Randall} \& {Sundrum}(1999)}]{Randall_1999}
{Randall}, L., \& {Sundrum}, R. 1999, Physical Review Letters, 83, 3370

\bibitem[{Rapetti {et~al.}(2013)Rapetti, Blake, Allen, Mantz, Parkinson, \&
  Beutler}]{Rapetti_2013}
Rapetti, D., Blake, C., Allen, S.~W., {et~al.} 2013, Monthly Notices of the
  Royal Astronomical Society, 432, 973

\bibitem[{Ratra \& Peebles(1988)}]{Ratra_1988}
Ratra, B., \& Peebles, P. J.~E. 1988, Phys. Rev. D, 37, 3406

\bibitem[{{Reid} {et~al.}(2012){Reid}, {Samushia}, {White}, {Percival},
  {Manera}, {Padmanabhan}, {Ross}, {S{\'a}nchez}, {Bailey}, {Bizyaev},
  {Bolton}, {Brewington}, {Brinkmann}, {Brownstein}, {Cuesta}, {Eisenstein},
  {Gunn}, {Honscheid}, {Malanushenko}, {Malanushenko}, {Maraston}, {McBride},
  {Muna}, {Nichol}, {Oravetz}, {Pan}, {de Putter}, {Roe}, {Ross}, {Schlegel},
  {Schneider}, {Seo}, {Shelden}, {Sheldon}, {Simmons}, {Skibba}, {Snedden},
  {Swanson}, {Thomas}, {Tinker}, {Tojeiro}, {Verde}, {Wake}, {Weaver},
  {Weinberg}, {Zehavi}, \& {Zhao}}]{Reid_2012}
{Reid}, B.~A., {Samushia}, L., {White}, M., {et~al.} 2012, \mnras, 426, 2719

\bibitem[{{Riess} {et~al.}(1998){Riess}, {Filippenko}, {Challis},
  {Clocchiatti}, {Diercks}, {Garnavich}, {Gilliland}, {Hogan}, {Jha},
  {Kirshner}, {Leibundgut}, {Phillips}, {Reiss}, {Schmidt}, {Schommer},
  {Smith}, {Spyromilio}, {Stubbs}, {Suntzeff}, \& {Tonry}}]{Riess_1998}
{Riess}, A.~G., {Filippenko}, A.~V., {Challis}, P., {et~al.} 1998, \aj, 116,
  1009

\bibitem[{{Riess} {et~al.}(2016){Riess}, {Macri}, {Hoffmann}, {Scolnic},
  {Casertano}, {Filippenko}, {Tucker}, {Reid}, {Jones}, {Silverman},
  {Chornock}, {Challis}, {Yuan}, {Brown}, \& {Foley}}]{Riess_2016}
{Riess}, A.~G., {Macri}, L.~M., {Hoffmann}, S.~L., {et~al.} 2016, ArXiv
  e-prints, arXiv:1604.01424

\bibitem[{Rosenzweig {et~al.}(1980)Rosenzweig, Schechter, \&
  Trahern}]{Rosenzweig_1980}
Rosenzweig, C., Schechter, J., \& Trahern, C.~G. 1980, Phys. Rev. D, 21, 3388

\bibitem[{{Ross} {et~al.}(2015){Ross}, {Samushia}, {Howlett}, {Percival},
  {Burden}, \& {Manera}}]{Ross_2015}
{Ross}, A.~J., {Samushia}, L., {Howlett}, C., {et~al.} 2015, \mnras, 449, 835

\bibitem[{Sakharov(1966)}]{Sakharov_1966}
Sakharov, A.~D. 1966, Sov. Phys. JETP, 22, 241

\bibitem[{{Sako} {et~al.}(2014){Sako}, {Bassett}, {Becker}, {Brown},
  {Campbell}, {Cane}, {Cinabro}, {D'Andrea}, {Dawson}, {DeJongh}, {Depoy},
  {Dilday}, {Doi}, {Filippenko}, {Fischer}, {Foley}, {Frieman}, {Galbany},
  {Garnavich}, {Goobar}, {Gupta}, {Hill}, {Hayden}, {Hlozek}, {Holtzman},
  {Hopp}, {Jha}, {Kessler}, {Kollatschny}, {Leloudas}, {Marriner}, {Marshall},
  {Miquel}, {Morokuma}, {Mosher}, {Nichol}, {Nordin}, {Olmstead}, {Ostman},
  {Prieto}, {Richmond}, {Romani}, {Sollerman}, {Stritzinger}, {Schneider},
  {Smith}, {Wheeler}, {Yasuda}, \& {Zheng}}]{Sako_2014}
{Sako}, M., {Bassett}, B., {Becker}, A.~C., {et~al.} 2014, ArXiv e-prints,
  arXiv:1401.3317

\bibitem[{{Salvatelli} {et~al.}(2013){Salvatelli}, {Marchini}, {Lopez-Honorez},
  \& {Mena}}]{Salvatelli_2013}
{Salvatelli}, V., {Marchini}, A., {Lopez-Honorez}, L., \& {Mena}, O. 2013,
  \prd, 88, 023531

\bibitem[{{Salvatelli} {et~al.}(2014){Salvatelli}, {Said}, {Bruni},
  {Melchiorri}, \& {Wands}}]{Salvatelli_2014}
{Salvatelli}, V., {Said}, N., {Bruni}, M., {Melchiorri}, A., \& {Wands}, D.
  2014, Physical Review Letters, 113, 181301

\bibitem[{{Samushia} {et~al.}(2012){Samushia}, {Percival}, \&
  {Raccanelli}}]{Samushia_2012}
{Samushia}, L., {Percival}, W.~J., \& {Raccanelli}, A. 2012, \mnras, 420, 2102

\bibitem[{{Samushia} \& {Ratra}(2008)}]{Samushia_2008}
{Samushia}, L., \& {Ratra}, B. 2008, \apjl, 680, L1

\bibitem[{{Santos} {et~al.}(2016){Santos}, {Chandrachani Devi}, \&
  {Alcaniz}}]{Santos_2016}
{Santos}, B., {Chandrachani Devi}, N., \& {Alcaniz}, J.~S. 2016, ArXiv
  e-prints, arXiv:1603.06563

\bibitem[{{Sawicki} \& {Hu}(2007)}]{Sawicki_2007}
{Sawicki}, I., \& {Hu}, W. 2007, \prd, 75, 127502

\bibitem[{Schwarz {et~al.}(1978)}]{Schwarz_1978}
Schwarz, G., {et~al.} 1978, The annals of statistics, 6, 461

\bibitem[{{Scoccimarro}(2004)}]{Scoccimarro_2004}
{Scoccimarro}, R. 2004, \prd, 70, 083007

\bibitem[{{Seikel} {et~al.}(2012){Seikel}, {Clarkson}, \&
  {Smith}}]{Seikel_2012}
{Seikel}, M., {Clarkson}, C., \& {Smith}, M. 2012, Journal of Cosmology and
  Astroparticle Physics, 6, 036

\bibitem[{{Sen} \& {Sen}(2003)}]{Sen_2003}
{Sen}, S., \& {Sen}, A.~A. 2003, \apj, 588, 1

\bibitem[{{Senovilla} {et~al.}(1998){Senovilla}, {Sopuerta}, \&
  {Szekeres}}]{Senovilla_1998}
{Senovilla}, J.~M.~M., {Sopuerta}, C.~F., \& {Szekeres}, P. 1998, General
  Relativity and Gravitation, 30, 389

\bibitem[{{Seo} \& {Eisenstein}(2003)}]{Seo_2003}
{Seo}, H.-J., \& {Eisenstein}, D.~J. 2003, \apj, 598, 720

\bibitem[{{Shafieloo} {et~al.}(2012){Shafieloo}, {Kim}, \&
  {Linder}}]{Shafieloo_2012}
{Shafieloo}, A., {Kim}, A.~G., \& {Linder}, E.~V. 2012, \prd, 85, 123530

\bibitem[{{Sharov}(2016)}]{Sharov_2016}
{Sharov}, G.~S. 2016, Journal of Cosmology and Astroparticle Physics, 6, 023

\bibitem[{{Shi} \& {Baugh}(2016)}]{Shi_2016}
{Shi}, D., \& {Baugh}, C.~M. 2016, \mnras, 459, 3540

\bibitem[{{Shi} {et~al.}(2012){Shi}, {Huang}, \& {Lu}}]{Shi_2012}
{Shi}, K., {Huang}, Y.~F., \& {Lu}, T. 2012, \mnras, 426, 2452

\bibitem[{{Silvestri} \& {Trodden}(2009)}]{Silverstri_2009}
{Silvestri}, A., \& {Trodden}, M. 2009, Reports on Progress in Physics, 72,
  096901

\bibitem[{{Slepian} {et~al.}(2014){Slepian}, {Gott}, \& {Zinn}}]{Slepian_2014}
{Slepian}, Z., {Gott}, J.~R., \& {Zinn}, J. 2014, \mnras, 438, 1948

\bibitem[{{Sola} {et~al.}(2017{\natexlab{a}}){Sola}, {de Cruz Perez}, \&
  {Gomez-Valent}}]{Sola_2017b}
{Sola}, J., {de Cruz Perez}, J., \& {Gomez-Valent}, A. 2017{\natexlab{a}},
  ArXiv e-prints, arXiv:1703.08218

\bibitem[{{Sola} {et~al.}(2017{\natexlab{b}}){Sola}, {Gomez-Valent}, \& {de
  Cruz Perez}}]{Sola_2017a}
{Sola}, J., {Gomez-Valent}, A., \& {de Cruz Perez}, J. 2017{\natexlab{b}},
  \apj, 836, 43

\bibitem[{{Sola} {et~al.}(2017{\natexlab{c}}){Sola}, {Gomez-Valent}, \& {de
  Cruz Perez}}]{Sola_2017c}
---. 2017{\natexlab{c}}, ArXiv e-prints, arXiv:1705.06723

\bibitem[{{Song} {et~al.}(2007){Song}, {Hu}, \& {Sawicki}}]{Song_2007}
{Song}, Y.-S., {Hu}, W., \& {Sawicki}, I. 2007, \prd, 75, 044004

\bibitem[{Sotiriou \& Faraoni(2010)}]{Sotiriou_2010}
Sotiriou, T.~P., \& Faraoni, V. 2010, Rev. Mod. Phys., 82, 451

\bibitem[{{Starobinsky}(2007)}]{Starobinsky_2007}
{Starobinsky}, A.~A. 2007, Soviet Journal of Experimental and Theoretical
  Physics Letters, 86, 157

\bibitem[{Sunyaev \& Zeldovich(1970)}]{Sunyaev_1970}
Sunyaev, R.~A., \& Zeldovich, Y.~B. 1970, Astrophysics and Space Science, 7, 3

\bibitem[{{Susskind}(1995)}]{Susskind_1995}
{Susskind}, L. 1995, Journal of Mathematical Physics, 36, 6377

\bibitem[{{'t Hooft}(1993)}]{tHooft_1993}
{'t Hooft}, G. 1993, ArXiv General Relativity and Quantum Cosmology e-prints,
  gr-qc/9310026

\bibitem[{{'t Hooft}(2001)}]{tHooft_2001}
{'t Hooft}, G. 2001, in Basics and Highlights in Fundamental Physics, ed.
  A.~{Zichichi}, 72--100

\bibitem[{{Tocchini-Valentini} \& {Amendola}(2002)}]{Tocchini-Valentini_2002}
{Tocchini-Valentini}, D., \& {Amendola}, L. 2002, \prd, 65, 063508

\bibitem[{{Tsujikawa}(2007)}]{Tsujikawa_2007}
{Tsujikawa}, S. 2007, \prd, 76, 023514

\bibitem[{{Turner} \& {White}(1997)}]{Turner_1997}
{Turner}, M.~S., \& {White}, M. 1997, \prd, 56, R4439

\bibitem[{{Unzicker} \& {Case}(2005)}]{Unzicker_2005}
{Unzicker}, A., \& {Case}, T. 2005, ArXiv Physics e-prints, physics/0503046

\bibitem[{{Urban} \& {Zhitnitsky}(2009{\natexlab{a}})}]{Urban_2009}
{Urban}, F.~R., \& {Zhitnitsky}, A.~R. 2009{\natexlab{a}}, \prd, 80, 063001

\bibitem[{{Urban} \& {Zhitnitsky}(2009{\natexlab{b}})}]{Urban_2009JCAP}
---. 2009{\natexlab{b}}, Journal of Cosmology and Astroparticle Physics, 9, 018

\bibitem[{{Urban} \& {Zhitnitsky}(2010{\natexlab{a}})}]{Urban_2010}
---. 2010{\natexlab{a}}, Physics Letters B, 688, 9

\bibitem[{{Urban} \& {Zhitnitsky}(2010{\natexlab{b}})}]{Urban_2010NuPhB}
---. 2010{\natexlab{b}}, Nuclear Physics B, 835, 135

\bibitem[{Veneziano(1979)}]{VENEZIANO_1979}
Veneziano, G. 1979, Nuclear Physics B, 159, 213

\bibitem[{{Wang} {et~al.}(2003){Wang}, {Freese}, {Gondolo}, \&
  {Lewis}}]{Wang_2003}
{Wang}, Y., {Freese}, K., {Gondolo}, P., \& {Lewis}, M. 2003, \apj, 594, 25

\bibitem[{{Wang} {et~al.}(2013){Wang}, {Wands}, {Xu}, {De-Santiago}, \&
  {Hojjati}}]{Wang_2013}
{Wang}, Y., {Wands}, D., {Xu}, L., {De-Santiago}, J., \& {Hojjati}, A. 2013,
  \prd, 87, 083503

\bibitem[{Wasserstein \& Lazar(2016)}]{Ronald_2016}
Wasserstein, R.~L., \& Lazar, N.~A. 2016, The American Statistician, 70, 129

\bibitem[{{Watson} \& {Scherrer}(2003)}]{Watson_2003}
{Watson}, C.~R., \& {Scherrer}, R.~J. 2003, \prd, 68, 123524

\bibitem[{{Wei} \& {Cai}(2008)}]{Wei_2008}
{Wei}, H., \& {Cai}, R.-G. 2008, Physics Letters B, 660, 113

\bibitem[{{Weinberg} {et~al.}(2013){Weinberg}, {Mortonson}, {Eisenstein},
  {Hirata}, {Riess}, \& {Rozo}}]{Weinberg_2013}
{Weinberg}, D.~H., {Mortonson}, M.~J., {Eisenstein}, D.~J., {et~al.} 2013,
  \physrep, 530, 87

\bibitem[{Weinberg(1989)}]{Weinberg_1989}
Weinberg, S. 1989, Rev. Mod. Phys., 61, 1

\bibitem[{{Weller} \& {Albrecht}(2002)}]{Weller_2002}
{Weller}, J., \& {Albrecht}, A. 2002, \prd, 65, 103512

\bibitem[{Wetterich(1988)}]{Wetterich_1988}
Wetterich, C. 1988, Nuclear Physics B, 302, 668

\bibitem[{Witten(1979)}]{WITTEN_1979}
Witten, E. 1979, Nuclear Physics B, 156, 269

\bibitem[{{Wu} \& {Huterer}(2017)}]{Wu_2017}
{Wu}, H.-Y., \& {Huterer}, D. 2017, \mnras, 471, 4946

\bibitem[{{Wu} \& {Yu}(2011)}]{Wu_2011}
{Wu}, P., \& {Yu}, H. 2011, European Physical Journal C, 71, 1552

\bibitem[{{Xu}(2012)}]{Xu_2012}
{Xu}, L. 2012, European Physical Journal C, 72, 2134

\bibitem[{{Xu}(2013)}]{Xu_2013}
---. 2013, \prd, 87, 043525

\bibitem[{{Xu}(2014)}]{Xu_2014}
---. 2014, Journal of Cosmology and Astroparticle Physics, 2, 048

\bibitem[{{Xu} \& {Zhang}(2016)}]{Xu_2016}
{Xu}, Y.-Y., \& {Zhang}, X. 2016, ArXiv e-prints, arXiv:1607.06262

\bibitem[{{Zhang} {et~al.}(2017){Zhang}, {Childress}, {Davis}, {Karpenka},
  {Lidman}, {Schmidt}, \& {Smith}}]{Zhang_2017}
{Zhang}, B.~R., {Childress}, M.~J., {Davis}, T.~M., {et~al.} 2017, \mnras, 471,
  2254

\bibitem[{{Zhang} {et~al.}(2014){Zhang}, {Zhao}, {Cui}, \&
  {Zhang}}]{Zhang_2014}
{Zhang}, J.-F., {Zhao}, M.-M., {Cui}, J.-L., \& {Zhang}, X. 2014, European
  Physical Journal C, 74, 3178

\bibitem[{{Zhang} {et~al.}(2015){Zhang}, {Zhao}, {Li}, \& {Zhang}}]{Zhang_2015}
{Zhang}, J.-F., {Zhao}, M.-M., {Li}, Y.-H., \& {Zhang}, X. 2015, Journal of
  Cosmology and Astroparticle Physics, 4, 038

\bibitem[{{Zhang} \& {Wu}(2007)}]{Zhang_2007}
{Zhang}, X., \& {Wu}, F.-Q. 2007, \prd, 76, 023502

\bibitem[{{Zhang} {et~al.}(2007){Zhang}, {Li}, {Wu}, {Wei}, \&
  {Cai}}]{Zhang_2007ADE}
{Zhang}, Y., {Li}, H., {Wu}, X., {Wei}, H., \& {Cai}, R.-G. 2007, ArXiv
  e-prints, arXiv:0708.1214

\bibitem[{{Zhao} {et~al.}(2011){Zhao}, {Li}, \& {Koyama}}]{Zhao_2011}
{Zhao}, G.-B., {Li}, B., \& {Koyama}, K. 2011, \prd, 83, 044007

\bibitem[{{Zhao} {et~al.}(2016){Zhao}, {Wang}, {Ross}, {Shandera}, {Percival},
  {Dawson}, {Kneib}, {Myers}, {Brownstein}, {Comparat}, {Delubac}, {Gao},
  {Hojjati}, {Koyama}, {McBride}, {Meza}, {Newman}, {Palanque-Delabrouille},
  {Pogosian}, {Prada}, {Rossi}, {Schneider}, {Seo}, {Tao}, {Wang}, {Y{\`e}che},
  {Zhang}, {Zhang}, {Zhou}, {Zhu}, \& {Zou}}]{Zhao_2016}
{Zhao}, G.-B., {Wang}, Y., {Ross}, A.~J., {et~al.} 2016, \mnras, 457, 2377

\bibitem[{{Zhao} {et~al.}(2017){Zhao}, {Raveri}, {Pogosian}, {Wang},
  {Crittenden}, {Handley}, {Percival}, {Beutler}, {Brinkmann}, {Chuang},
  {Cuesta}, {Eisenstein}, {Kitaura}, {Koyama}, {L'Huillier}, {Nichol}, {Pieri},
  {Rodriguez-Torres}, {Ross}, {Rossi}, {S{\'a}nchez}, {Shafieloo}, {Tinker},
  {Tojeiro}, {Vazquez}, \& {Zhang}}]{Zhao_2017}
{Zhao}, G.-B., {Raveri}, M., {Pogosian}, L., {et~al.} 2017, Nature Astronomy,
  1, 627

\bibitem[{{Zhu} \& {Fujimoto}(2002)}]{Zhu_2002}
{Zhu}, Z.-H., \& {Fujimoto}, M.-K. 2002, \apj, 581, 1

\bibitem[{{Zhu} \& {Fujimoto}(2003)}]{Zhu_2003}
---. 2003, \apj, 585, 52

\bibitem[{{Zhu} {et~al.}(2004){Zhu}, {Fujimoto}, \& {He}}]{Zhu_2004}
{Zhu}, Z.-H., {Fujimoto}, M.-K., \& {He}, X.-T. 2004, \apj, 603, 365

\bibitem[{Zimdahl {et~al.}(2001)Zimdahl, Pavón, \& Chimento}]{Zimdahl_2001}
Zimdahl, W., Pavón, D., \& Chimento, L.~P. 2001, Physics Letters B, 521, 133

\end{thebibliography}

\end{document}